# FINAL REPORT
# WORKLOAD ANALYSIS OF BLUE WATERS
# (ACI 1650758)

Matthew D. Jones, Joseph P. White, Martins Innus, Robert L. DeLeon, Nikolay Simakov, Jeffrey T. Palmer, Steven M. Gallo, and Thomas R. Furlani (furlani@buffalo.edu), Center for Computational Research, University at Buffalo, SUNY

Michael Showerman, Robert Brunner, Andry Kot, Gregory Bauer, Brett Bode, Jeremy Enos, and William Kramer (wtkramer@illinois.edu), National Center for Supercomputing Applications (NCSA), University of Illinois at Urbana Champaign

January13, 2017

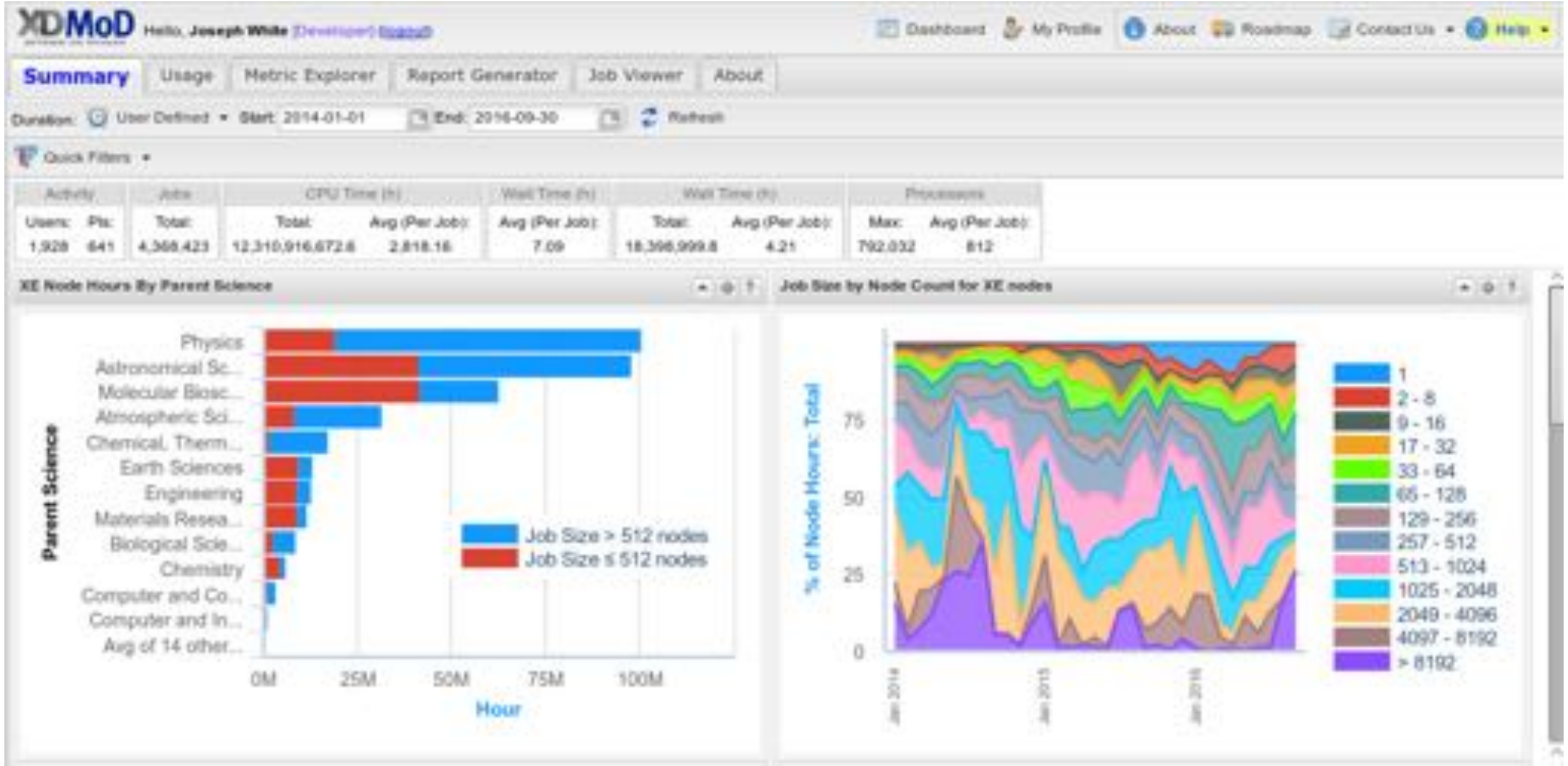

# EXECUTIVE SUMMARY

Blue Waters is a Petascale-level supercomputer whose mission is to enable the national scientific and research community to solve "grand challenge" problems that are orders of magnitude more complex than can be carried out on other high performance computing systems. Given the important and unique role that Blue Waters plays in the U.S. research portfolio, it is important to have a detailed understanding of its workload in order to guide performance optimization both at the software and system configuration level as well as inform architectural balance tradeoffs. Furthermore, understanding the computing requirements of the Blue Water's workload (memory access, IO, communication, etc.), which is comprised of some of the most computationally demanding scientific problems, will help drive changes in future computing architectures, especially at the leading edge. With this objective in mind, the project team carried out a detailed workload analysis of Blue Waters.

The workload analysis itself was a challenging computational problem – requiring more than 35,000 node hours (over 1.1 million core hours) on Blue Waters to analyze roughly 95 TB of input data from over 4.5M jobs that ran on Blue Waters during the period of our analysis (April 1, 2013 – September 30, 2016) that spans the beginning to Full Service Operations for Blue Waters to the recent past. In the process, approximately 250 TB of data across 100M files was generated. This data was subsequently entered into MongoDB and a MySQL data warehouse to allow rapid searching, analysis and display in Open XDMoD. A workflow pipeline was established so that data from all future Blue Waters jobs will be automatically ingested into the Open XDMoD data warehouse, making future analyses much easier.

The workload analysis sought to understand a wide variety of use patterns and performance requirements for the jobs running on Blue Waters, including differences that may exist between the computational needs of the various fields of science such as average job size, use of accelerator technology, I/O, communication, and memory use. Historical trends for almost the entire operational life of Blue Waters to date were also studied to determine if these requirements are changing over time since this may impact future architecture designs. File system performance, the use of numerical libraries, and the types of parallelism employed by users (MPI, threads, etc.) were also studied. Highlights of the analysis follow.

**Science and Engineering fields and Application Codes**

- The overall historical trend for all NSF directorates is toward increasing use of Blue Waters, albeit at much different overall scales of utilization.
- The MPS and Biological Sciences directorates are the leading consumers of node hours, typically accounting for more than 2/3 of all node hours used.
- The number of fields of science represented in the Blue Waters portfolio has increased in each year of its operation – more than doubling since its first year of operation, providing further evidence of the growing diversity of its research base.
- The applications run on Blue Waters represent an increasingly diverse mix of disciplines, ranging from broad use of community codes to more specific scientific sub-disciplines.
- The top 10 applications consume about 2/3 of all node hours, with the top 5 (NAMD, CHROMA, MILC, AMBER, and CACTUS) consuming about 50%.

- Common algorithms, as characterized by Colella's original seven dwarfs, are roughly equally represented within the applications run on Blue Waters aside from unstructured grids and Monte Carlo methods, which exhibit a much smaller fraction.

- Optimized numerical library usage, essential for high performance, has been used throughout in areas in which such libraries are applicable.

**Concurrency and Parallelism**

- Blue Waters supports a diverse mix of job sizes from single node jobs to jobs that use in excess of 20,000 nodes in a single application. The patterns of usage differ between the XE and XK (GPU) nodes.

- Single node jobs, some of which may be attributable to high throughput computing, represent a small fraction (less than 2%) of the total node hours consumed on Blue Waters.

- For XE node jobs, all of the major science areas (> 1 million node hours) run a mix of job sizes and all have very large jobs (> 4096 nodes). The relative proportions of job size vary between different parent science areas. The job size distribution weighted by node hours consumed peaks at 1025 – 2048 for XE jobs. The largest 3% of the jobs (by node hours) account for 90% of the total node-hours consumed.

- The majority of XE node hours on the machine are spent running parallel jobs that use some form of message passing for inter-process communication. At least 25% of the workload uses some form of threading, however the larger jobs (> 4096 nodes) mostly use message passing with no threading. There is no obvious trend in the variation of thread usage over time, however, thread usage information is only available for a short time period.

- For the XK (GPU) nodes, the parent sciences Molecular Biosciences, Chemistry and Physics are the largest users with NAMD and AMBER the two most prevalent applications. The job size distribution weighted by node hours consumed peaks at 65 – 128 nodes for the XK jobs. Similarly to the XE nodes, the largest 7% of the jobs (by node-hour) account for 90% of the node-hours consumed on the XK nodes.

- The aggregate GPU utilization (efficiency) varies significantly by application, with MELD achieving over 90% utilization and GROMACS, NAMD, and MILC averaging less than 30% GPU utilization. However, for each of the applications, the GPU utilization can vary significantly from job to job.

**Memory Usage**

- Most jobs that run on the XE nodes use less than 50% of the memory available on the node. However, the distribution of memory use has a substantial tail to higher memory usage.

- Most jobs that run on the XK nodes use less than 25% of the available memory in the node with a short tail to higher usage.

- GPU memory usage is very small with few jobs using more than 1GB per GPU.

- The XE and XK nodes show no historical differences in memory use from year to year.

- For almost all applications and parent fields of science, memory usage has not changed over time.

**Storage and I/O**

- On average Blue Waters' three filesystems have a balanced reads/writes ratio with large fluctuations. The volume of traffic on the largest filesystem peaks at 10PB per month.

- Users' jobs exhibit a wide range of I/O patterns.

- Overall there is a tendency to use a very large number of small files, which is a challenge for many parallel file systems.

- Read and write rates stay significantly below possible peak filesystem performance.

- User jobs spend a very small fraction of time in filesystem I/O operations (0.04% of runtime for 90% of jobs).

- Many jobs utilize specialized libraries for their I/O operations (about 20% use MPI-IO, HDF5, NetCDF).

# TABLE OF CONTENTS

# 1.0 INTRODUCTION AND BACKGROUND

Blue Waters is a Petascale-level supercomputer whose mission is to greatly accelerate insight to the most challenging computational and data analysis problems by enabling the national scientific and research community to solve "frontier" (aka "grand challenge") problems that are orders of magnitude more complex than can be done on other systems. The Blue Waters Project also has strong educational and outreach programs, advanced support for science teams and for improving system and applications software. At any point in time, Blue Waters is serving a diverse user community of 800 to 1,000 active users and about 120-130 distinct projects with different allocation categories currently from 47 states. The workload is comprised of a wide range of applications to serve the diverse needs of the U.S. scientific community.

Given the important and unique role that Blue Waters plays in the U.S. research portfolio, it is important to have a detailed understanding of its workload and how it is evolving. An understanding of workload properties sheds light on resource utilization and can be used to guide performance optimization both at the software and system configuration level as well as guidelines for resource allocation processes. Furthermore, understanding the computing requirements of the workload (memory access, IO, communication, etc.) will help drive changes in future computing architectures that will result in better HPC systems in future generation systems. Workload analysis also informs the architectural balance tradeoffs of systems. Accordingly, this project team carried out a workload analysis of Blue Waters.

## *1.1 Workload Analysis Goals*

This analysis, which was modeled in part after the 2014 NERSC Workload Analysis [1], specifically targets the following high-level questions:

1. What is the proportional mix of sub-disciplines and how are the proportions growing/shrinking over the life-time of Blue Waters, including the use of different types of nodes (XE and XK)?
2. What are the top representative algorithms on Blue Waters that consume a majority of the node hours including the use of different types of nodes (XE and XK)?
    a) What is the distribution of job sizes by application and Field of Science (FoS)?
    b) Some sampling and analysis of the communication/compute, IO/compute, and memory/compute ratios for different applications as feasible.
3. How much of Blue Waters is consumed by high throughput (HT) applications, and is this changing over time?
4. Are jobs sizes, in terms of numbers of nodes in a job, changing over time? Are there differences of job size by disciplines/FoS?
5. Analysis of application I/O patterns, if any, and how is this evolving over time? Does this differ between disciplines, job type, etc.?
6. Identification of current and potential I/O performance bottlenecks?
7. Is job/application memory usage increasing/decreasing over time including the use of different types of nodes (XE and XK)? Are there specific discipline differences? Are there different memory use profiles based on problems being address by applications.

8. Are there any data analytical frameworks being used on Blue Waters? What is the rate of growth, if any, for these frameworks?

This workload study provides extensive analysis toward answering many of the questions posed above. However, based on the data available for the analysis, the project team was unable to fully determine the extent to which data analytical frameworks are being used on Blue Waters and therefore this goal (8) is not addressed in this report. This is not to say that there are not data analytic frameworks running on Blue Waters. There are a growing set of data analytics projects currently running on Blue Waters, such as the Dark Energy Survey analysis and the ArcticDEM framework for image analysis. Rather than formulate a (almost certainly too simplistic) definition of data analytical frameworks, we instead focus on the data usage patterns themselves.

Blue Waters has already published extensive analysis of system and component failure and resiliency rates, so that and other performance and use metrics are not considered here [2-6]. Likewise, the Blue Waters project has a large body of assessment information for its user/PI satisfaction and its Education and Workforce Development programs so that is also not covered here.

## 1.2 Blue Waters System Summary

The Blue Waters system [7] is a Cray XE6/XK7 system that has two types of nodes connected via a single Cray Gemini High Speed Network in a large-scale 3D Torus topology. The two different types of nodes, XE6 and XK7 are described in the table below. Blue Waters has a high performance on-line storage system with over 25 PB of usable storage (36 PB raw) and over 1 TB/s sustained performance [8]. The system is summarized in Table 1.0-1 and Figure 1.0-2 below.

*Table 1.0-1 Blue Waters hardware overview.*

**System Totals**

| | |
|---|---|
| Total Cabinets | 288 |
| Total Peak Performance | 13.34 PF |
| Total System Memory | 1.634 PB |
| | |
| XE Compute Cabinets | 237 |
| XE Peak Performance | 7.1 PF |
| XE Compute Nodes | 22,636 |
| XE Bulldozer modules | 362,240 |
| XE System Memory | 1.382 PB |
| | |
| XK Compute Cabinets | 44 |
| XK Peak Performance (CPU+GPU) | 6.24 PF |
| XK Compute Nodes | 4,228 |
| XK Bulldozer modules | 33,792 |
| XK Kepler K20X Accelerators (GPU) | 4,228 |
| XK System Memory (CPU) | 135 TB |
| XK Accelerator Memory (GPU) | 25 TB |

**Interconnect**

| | |
|---|---|
| Architecture | 3D Torus |
| Topology | 24x24x24 |
| Compute nodes per Gemini | 2 |
| Peak Node Injection Bandwidth per Gemini | 9.6 GB/s |

## Online Storage

| | | |
|---|---|---|
| Total Usable Storage | | 26.4 PB |
| Total Raw Storage | | 34.0 PB |
| Aggregate Measured I/O Bandwidth | | > 1.1 TB/s |
| File System | Size (PB) | # of OSTs |
| home | 2.2 | 36 (was 144) |
| projects | 2.2 | 36 (was 144) |
| scratch | 22 | 360 (was 1440) |

## Near-line Storage

| | |
|---|---|
| Archive Software | HPSS |
| Online disk cache for HPPS | 1.2 PB |
| Aggregate Bandwidth to tape | 58 GB/s |
| Raw capacity assuming all slots filled | 250+ PB |

## XE Compute Node

| | |
|---|---|
| AMD 6276 Interlagos Processors | 2 |
| Bulldozer module | 16 |
| Integer Scheduling Units “integer core” | 32 |
| Memory per Bulldozer module | 4 GB |
| Total Node Memory | 64 GB |
| Peak Performance | 313.6 GF |
| Memory Bandwidth | 102.4 GB/s |

## XK Compute Node

| | |
|---|---|
| AMD 6276 Interlagos Processors | 1 |
| Bulldozer module | 8 |
| Integer Scheduling Units “integer core” | 16 |
| Memory per Bulldozer module | 4 GB |
| Node System Memory | 32 GB |
| GPU Memory | 6 GB |
| Peak CPU Performance | 156.8 GF |
| CPU Memory Bandwidth | 51.2 GB/s |
| CUDA cores | 2688 |
| Peak GPU Performance (DP) | 1.31 TF |
| GPU Memory Bandwidth (ECC off) | 250 GB/s |

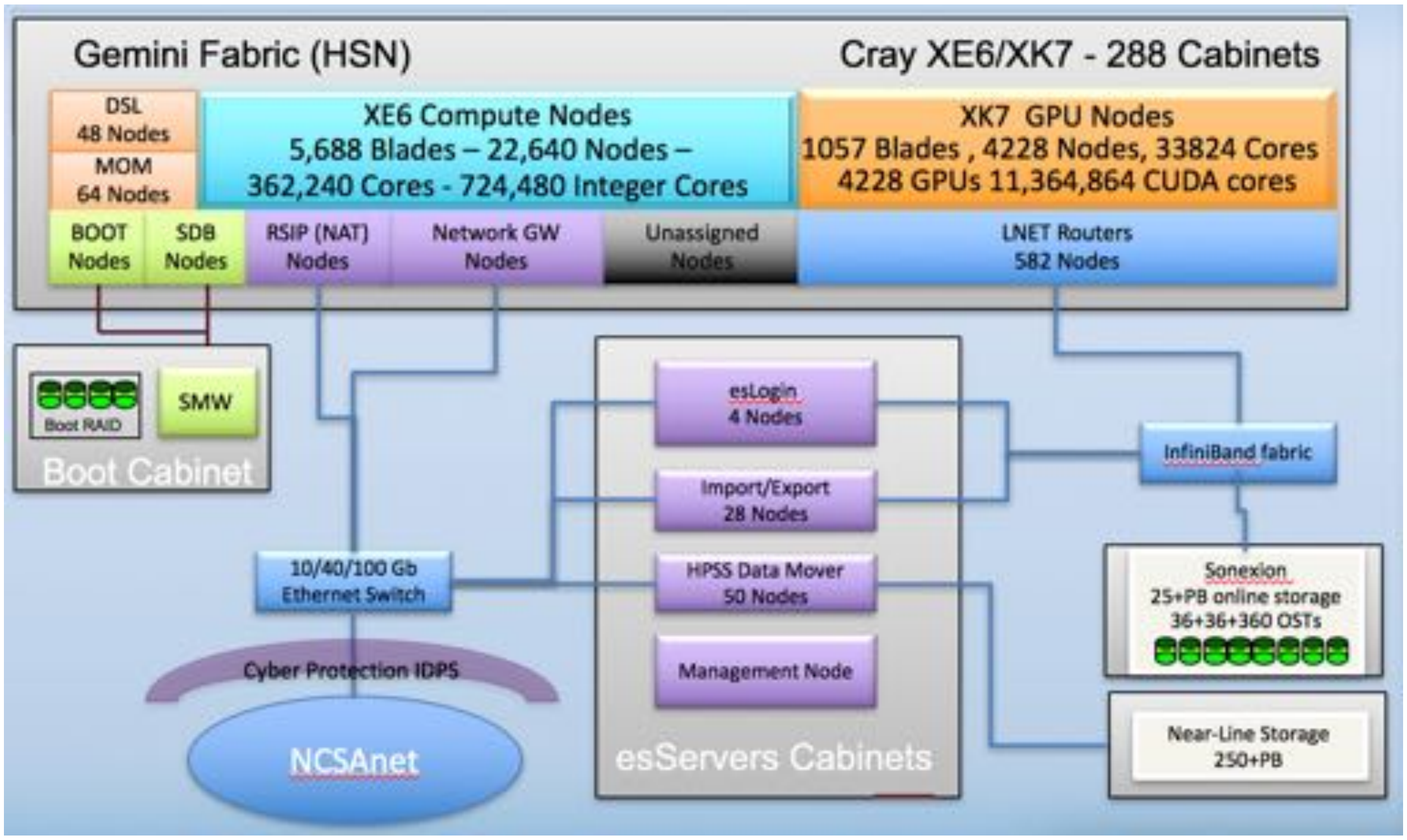

*Figure 1.0-2 A schematic diagram summarizing the Blue Waters System hardware architecture*

## 1.3 Open XDMoD and Data Collection Methods

The primary tool utilized in this analysis was Open XDMoD [9]. Open XDMoD is an open source tool for the comprehensive management of HPC systems. Metrics provided by Open XDMoD include comprehensive statistics on: number and type of computational jobs run, resources (computation, memory, disk, network, etc.) consumed, job wait times, and quality of service. The web interface is intuitive, allowing one to chart various metrics and interactively drill down to access additional related information. An important capability of Open XDMoD, particularly for the analysis carried out here, is centered around its ability to provide detailed job level performance data for all user jobs running on a given HPC resource.

Data for this study was collected from the Blue Waters system in a variety of formats and provided by a variety of tools. As a result, the ability to analyze various subsystems of the Blue Waters machine depends on the availability of the collected data and coverage across the time periods during which the machine has been in service. Below we describe the various data sources, the date availability of the individual datasets and which analyses cannot be performed when those data are missing. Also, the configuration of the various collectors has changed over time and we call out some instances where this has affected the analysis.

Job level performance data in Open XDMoD was constructed by analyzing a wide variety of data sources from the Blue Waters system. These included LDMS (device data), MSR (performance counters), APRUN (application launch data), DARSHAN (I/O), XALT/ALTD (library usage) and Torque/PBS (accounting logs); all of which are described in greater detail below. The process of analyzing and combining this data was itself a challenging computational problem - requiring running large batch jobs on Blue Waters. This included 35,000 node hours to analyze the roughly 95 TB of input data that were aggregated from the above data sources. In the process, we generated approximately 250 TB of data across 100M files. This data was then summarized and stored in a non-relational database with one or two records per job depending on the quantity of source data. This resulted in roughly 7.2M records to support the 4.6M jobs that ran on Blue Waters during the time period of interest. From the non-relational database, the data was normalized into a MySQL data warehouse to allow rapid searching and display in the Open XDMoD interface.

As a part of this project, a workflow pipeline was developed in which data from all future Blue Waters jobs will be automatically ingested into the Open XDMoD data warehouse, including the job level performance data that provided the basis for much of the analysis carried out in this report. This will greatly facilitate future workload analyses as well as allow Blue Waters support personnel to easily obtain detailed performance data for any desired job.

### 1.3.1 System Accounting Logs

Job information is provided by the torque resource manager, which operates in coordination with the Cray ALPS resource manager and the MOAB job scheduler. Log entries are created in files when a job on Blue Waters passes through various states of execution. This includes job submission, changes of scheduler status (queued, eligible to run, start/running, etc.), and completion. Job completion is accompanied by different Linux exit code status conditions that indicate whether the job completed normally (exit code = 0), or other termination codes. The logs also provide information of command changes to the jobs, such as a job may be deleted from the queue system before it has begun execution, or held. In this analysis we only include jobs that have begun execution on the system. We ignore jobs that may have been queued but deleted from the system prior to start. From the accounting logs we determine job runtime, account to charge, node allocation and job type (XE vs. XK).

Data Coverage Comments:

- 2013-03 to 2013-05 & 2013-07
  - Node usage information is unavailable for some periods in the accounting logs. During this time, we cannot confirm some jobs as to whether they were running on XE or XK nodes. Analysis that requires this information has been excluded from these time periods.

### 1.3.2 LDMS

The Lightweight Distributed Metric Service (LDMS, https://ovis.ca.sandia.gov) is software that provides data collection and transport of metrics from the compute nodes. The LDMS data collection provides information on several subsystems of the Blue Waters machine. Metrics on load average, memory usage, limited filesystem data transfers, network utilization are collected with one-minute granularity at the compute node level. For the XK nodes, GPU utilization and GPU memory utilization is also collected. These metrics make up one portion of the information used to construct the job level performance realm of Open XDMoD. LDMS collects a variety of data including the machine-specific registers discussed in the next section.

Data Coverage Comments:

- Available from 2014-03-28
  - Prior to this date, no analysis of memory utilization can be carried out.

- GPU coverage began and continued after 2015-01-18.
    - Missing for several months in the middle of 2015
    - During these times, analysis on the usage of the GPUs cannot be performed
- Lustre client data is valid only after 2016-01-15.
    - This is due to a bug in the collector that undercounted byte transfers

### 1.3.3 Machine Specific Register Data

The AMD CPU processors on the XE and XK compute nodes have programmable hardware performance counters. These counters are configured to count various events within the CPU such as the number clock ticks, the number of instructions retired and the number of floating point operations performed. The hardware counters are accessed via machine-specific registers (MSRs). Throughout this document we use the abbreviation 'MSR' to refer to the information read from the hardware performance counters on the compute nodes. The CPU clock tick metrics are used in the job level performance realm of Open XDMoD to determine processor activity and estimate utilization. Also, combined with the APRUN data, we can determine the allocation and use of cores in the parallelism and concurrency section of the analysis.

Data Coverage Comments:

- Available from 2015-09-04
- Could become deprogrammed prior to 2016-01-05.
    - When unavailable, we cannot reliably determine processor utilization on the nodes.

### 1.3.4 Darshan

Darshan (http://www.mcs.anl.gov/research/projects/darshan/) is an HPC I/O characterization tool that is designed to capture an accurate picture of application MPI I/O behavior, including properties such as patterns of access within files, with minimum overhead. Darshan provides very detailed reports for file-system access for each executable compiled with Darshan support. To make Blue Waters jobs analysis feasible, the Darshan data was summarized for each job and includes total bytes written and read, the number of opened files, the time spent in file system I/O operations as well as the number of opened files in HDF5 and NetCDF formats.

Data Coverage Comments:

- Only a small portion of Darshan data is available starting in March, 2014, the most likely reason being that compiling with the Darshan library was optional. After January 2015, more Darshan data became available since it was the default option, with an overall coverage of 26% of jobs as measured by node hours.

### 1.3.5 XALT/ALTD

ALTD and XALT tools [10] provide tracking capabilities for utilization of statically and dynamically linked libraries. XALT is the continuation of the ALTD project. ALTD and XALT allow one to analyze which libraries and modules were used by each job.

Data Coverage Comments:

- ALTD data is available from 2013-08-01 to 2016-05-18. During 2014 and 2015, the complete data coverage was 51% by node hours, and an additional 31% has partial ALTD data.

- XALT data is available from 2015-11-24. During this period, only 4% (by node hours) of Blue Waters jobs have XALT data.

### 1.3.6 APRUN

APRUN logs provide the primary means to determine what application(s) was executed during the job. In most cases, this information includes the executable path, the number of nodes requested and the layout of the tasks on the cores of the nodes. This information is used to map the job to an application algorithm, and, in conjunction with the MSR data, to determine concurrency/parallelism.

Data Coverage Comments:

- There are approximately 6 million node hours (over 287,000 jobs) where the executable information is unavailable (approximately 1% of all node hours for the period covered).
- In some cases, a job launcher/wrapper (e.g. scheduler.x) is used and the executable is hidden.

### 1.3.7 Job Level Performance Data

The job level performance data ingested by Open XDMoD was validated in several ways.

- LDMS
  - Included in Appendix I/II
- Memory
  - Sample jobs were run, and memory reporting was validated against data reported by the application
- Load Average
  - Data was validated to be in the range as expected to be reported by the kernel.
  - Compared to CPU utilization on a subset of jobs where MSR data was available
- Lustre
  - Compared with DARSHAN I/O data
- MSR
  - Control registers were validated to be programmed as documented
  - Data was discarded when mis-programmed counters were determined
  - Data was validated to be in the range as expected to be reported by the kernel.

### 1.3.8 Time Periods in the Study

Unless otherwise indicated, the time period covered in all subsequent analyses refers to the production period of Blue Waters to the time of study, 2013-04-01 to 2016-09-30.

Weighted averages by node hours are often used to avoid skewing results coming from many short (often small, testing, or exploratory) jobs.

# 2.0 WORKLOAD ANALYSIS BY FIELD OF SCIENCE, APPLICATION, ALGORITHM

*BW Analysis Goal 1: What is the proportional mix of sub-disciplines and how are the proportions growing/shrinking over the life-time of Blue Waters, including the use of different types of nodes (XE and XK)?*

*BW Analysis Goal 2: What are the top representative algorithms on Blue Waters that consume a majority of the node hours including the use of different types of nodes (XE and XK)?*

- *What is the distribution of job sizes by application and Field of Science (FoS)?*
- *Some sampling and analysis of the communication/compute, IO/compute, and memory/compute ratios for different applications as feasible.*

## 2.1 Historical Trends: NSF Directorates and Parent Science

In this section we present a historical usage analysis of the NSF directorates and the parent sciences within each directorate served by Blue Waters. We begin with the NSF directorates, which are shown in both linear scale (Figure 2.1-1) and log scale (Figure 2.1-2) in total node hours delivered. *It is important to note that at the time of this report, usage data for 2016 consisted only of the first nine months, and therefore one would not expect the total node hours to be as large as in previous years. Likewise, 2013, the first-year BW became operational, is comprised of nine months of usage data from April 1 to the end of the year.* From Figure 2.1-1 it is evident that the Biological and Engineering directorates show an increase in use over time. Given the magnitude of node hours consumed by the heavier-using directorates such as MPS, it is difficult on a linear scale to see trends for the directorates that consume a smaller number of node hours. However, Figure 2.1-2, which uses a log scale, shows the trend for all directorates. Although the level of use varies widely, and there are yearly fluctuations, the overall trend for the directorates is toward increasing utilization.

In analyzing the historical trends for the parent sciences, we find it convenient, due to the large number of parent sciences, to consider the usage trends within each directorate broken out by parent science. For the same reasons as given above, we employ a log scale in node hours for this series of plots. The general trend in all these plots is one of increasing diversity, albeit at much different use scales.

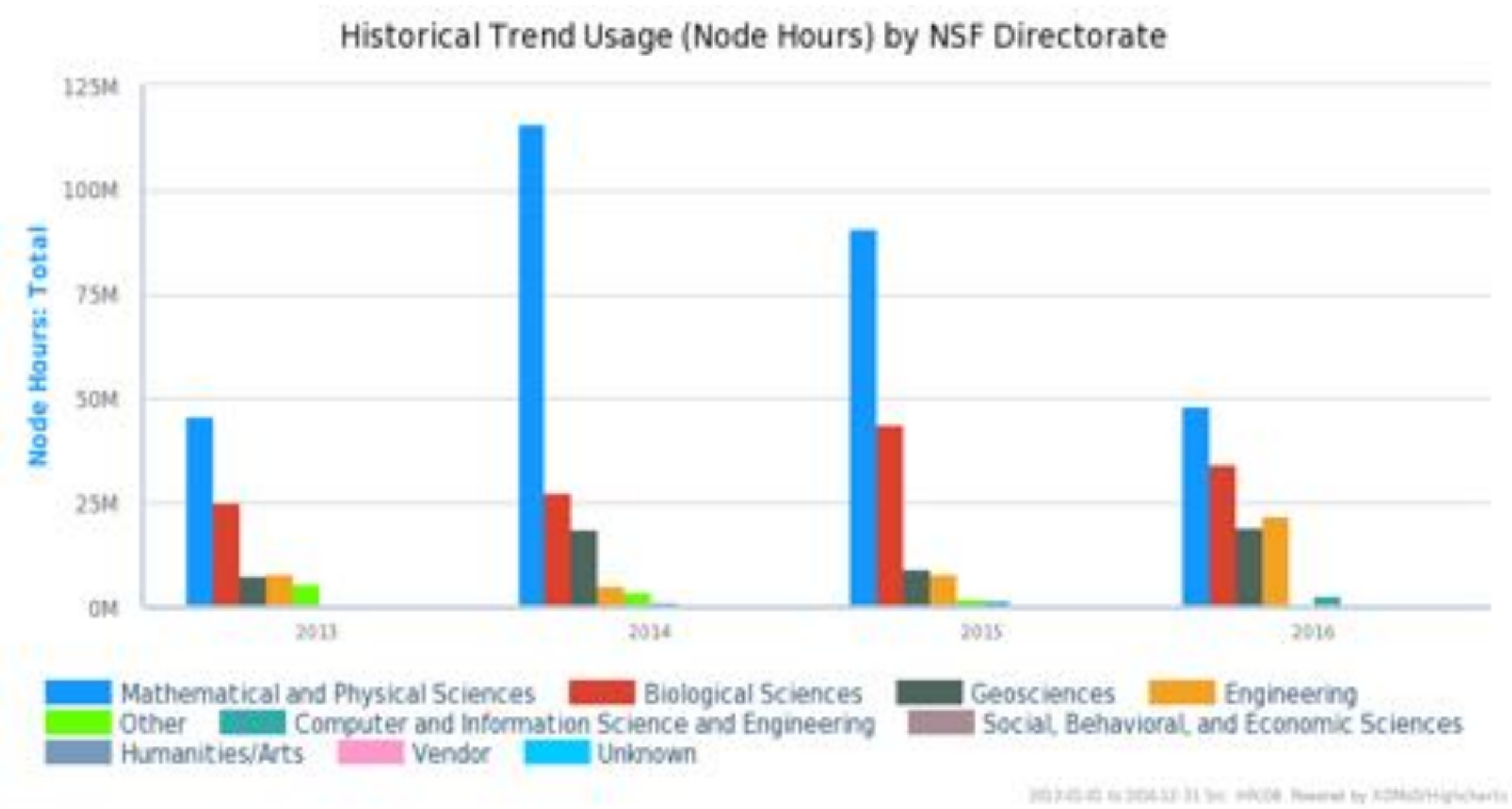

*Figure 2.1-1 Linear scale in node hours consumed by directorate (2013, 2014, 2015, and 2016). Note: For 2016, the data represents usage only in the first 3 quarters. Likewise, 2013, the first year BW became operational, is comprised of usage data only from April 1 to the end of the year.*

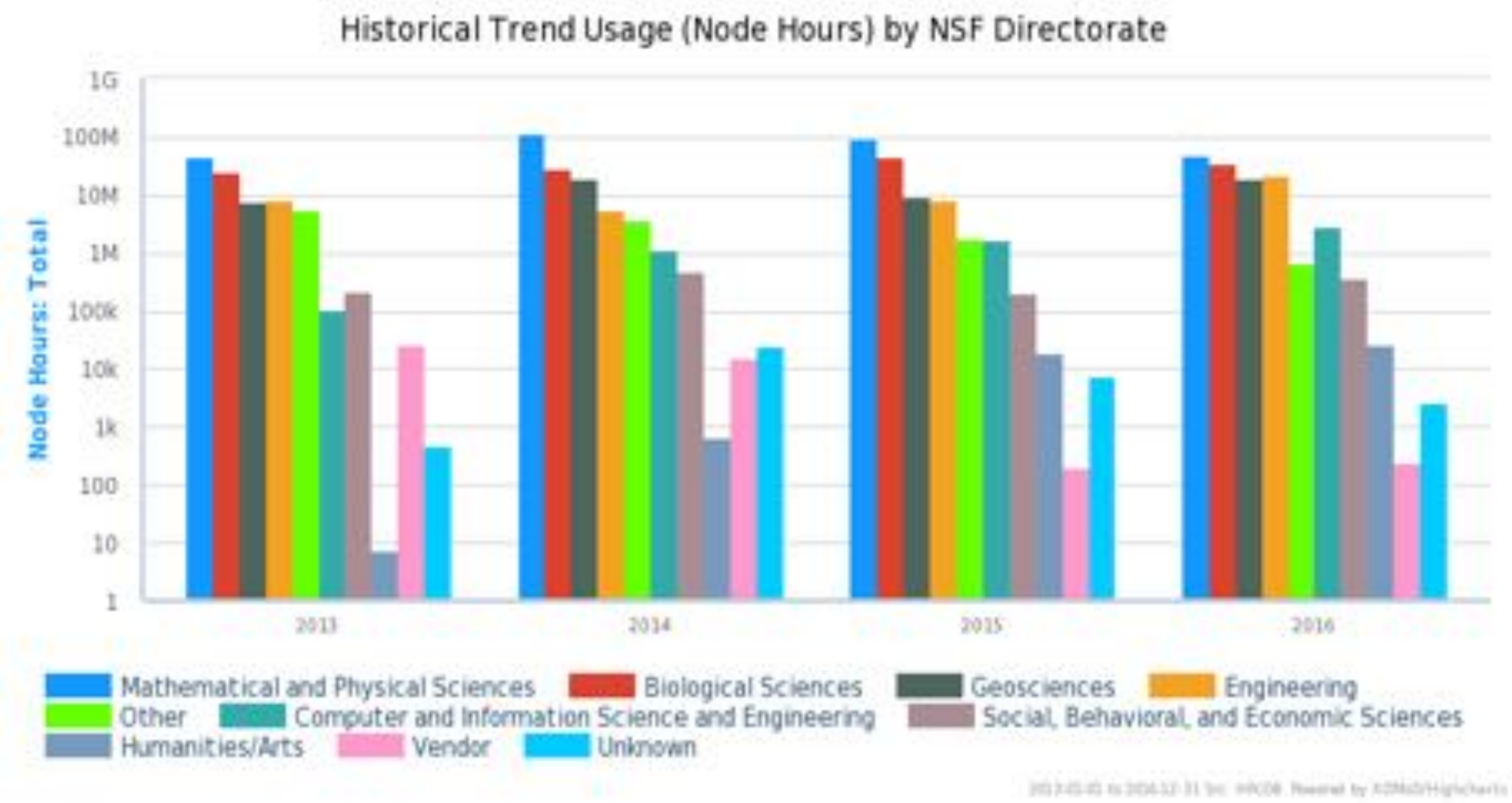

*Figure 2.1-2 Log scale in node hours consumed by directorate (2013, 2014, 2015, and 2016)*

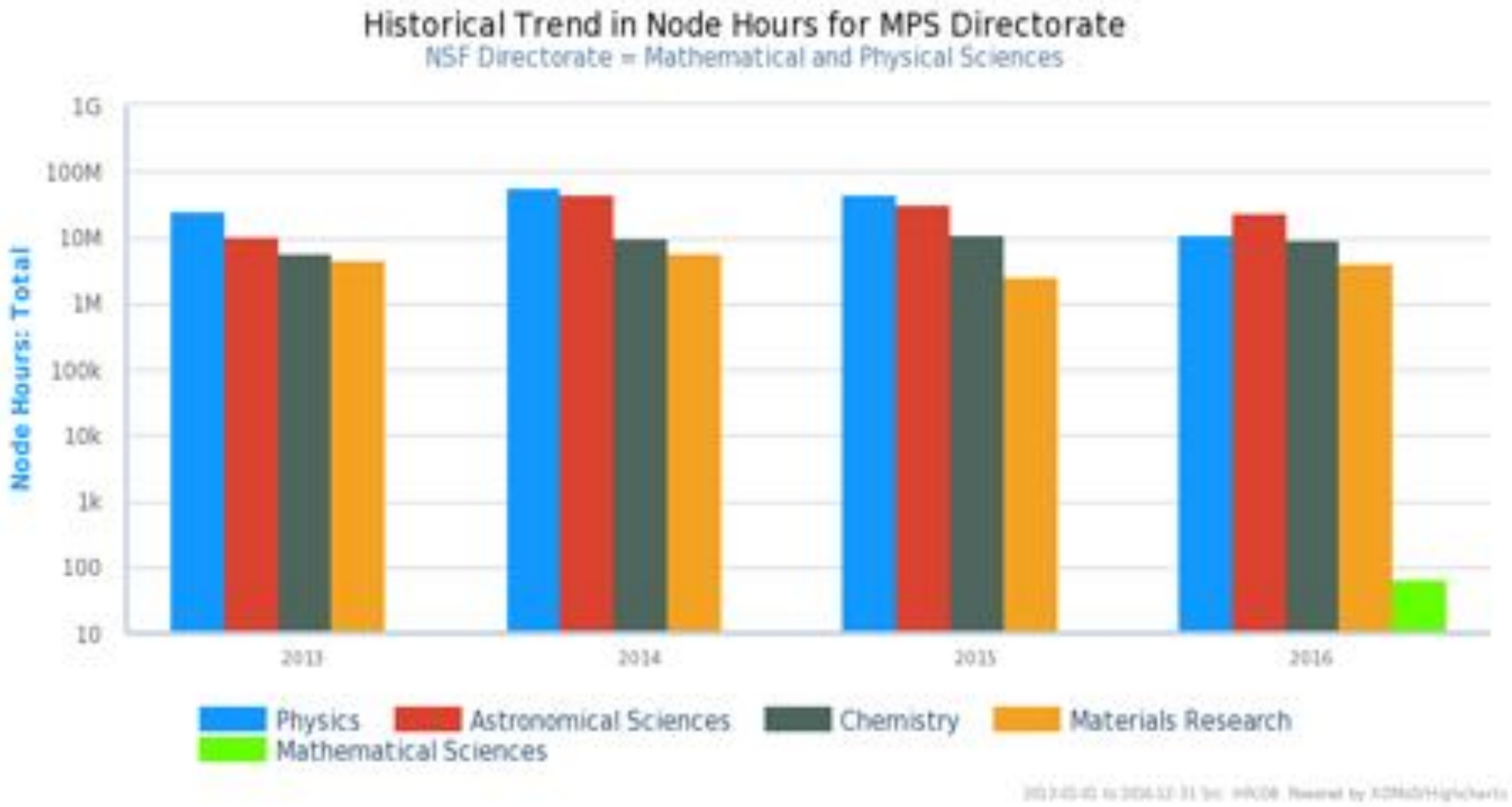

*Figure 2.1-3 Log scale in node hours consumed by parent sciences within the MPS directorate (2013, 2014, 2015, and 2016).*

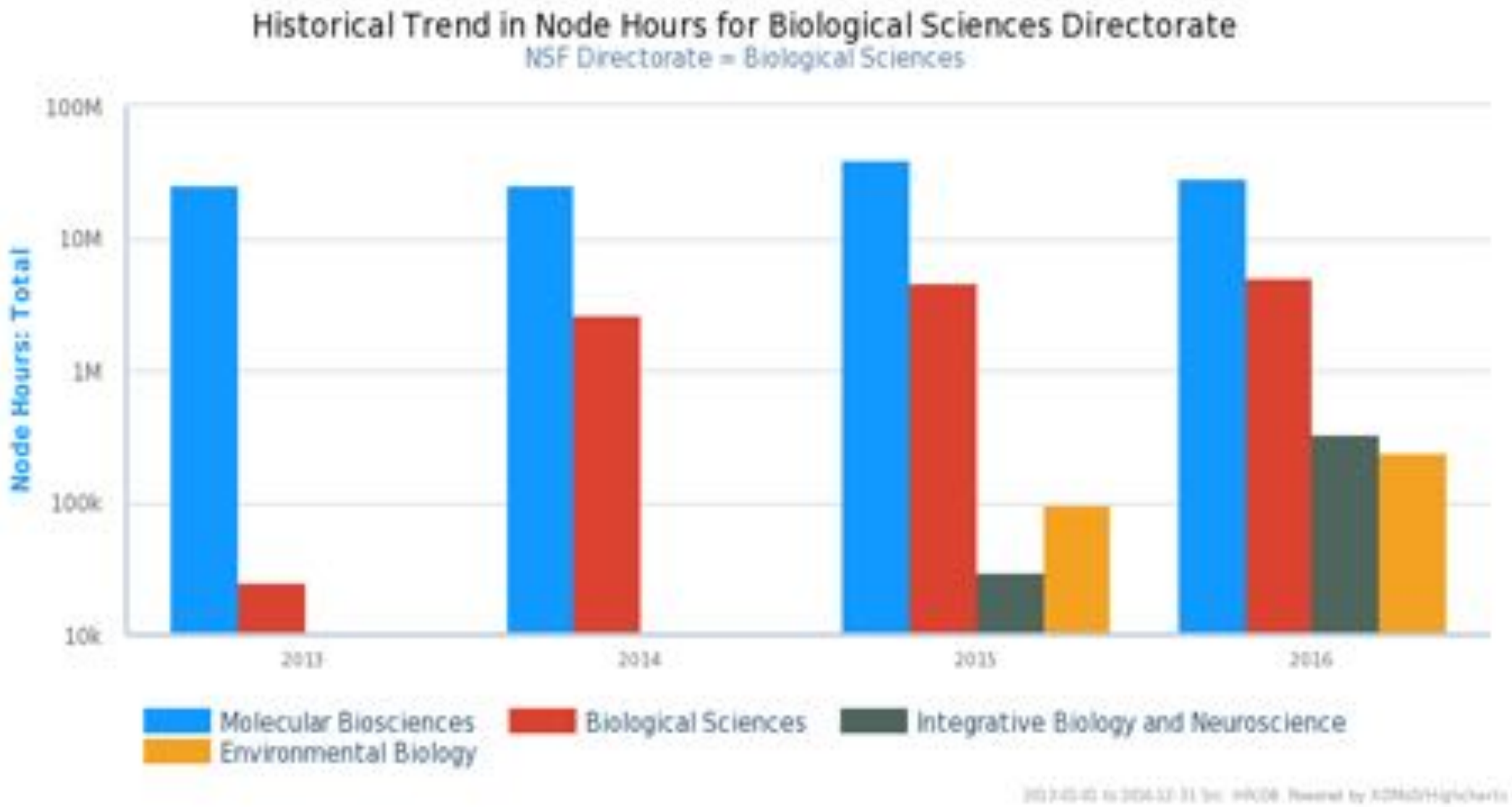

*Figure 2.1-4 Log scale in node hours consumed by parent sciences within the Biological directorate (2013, 2014, 2015, and 2016).*

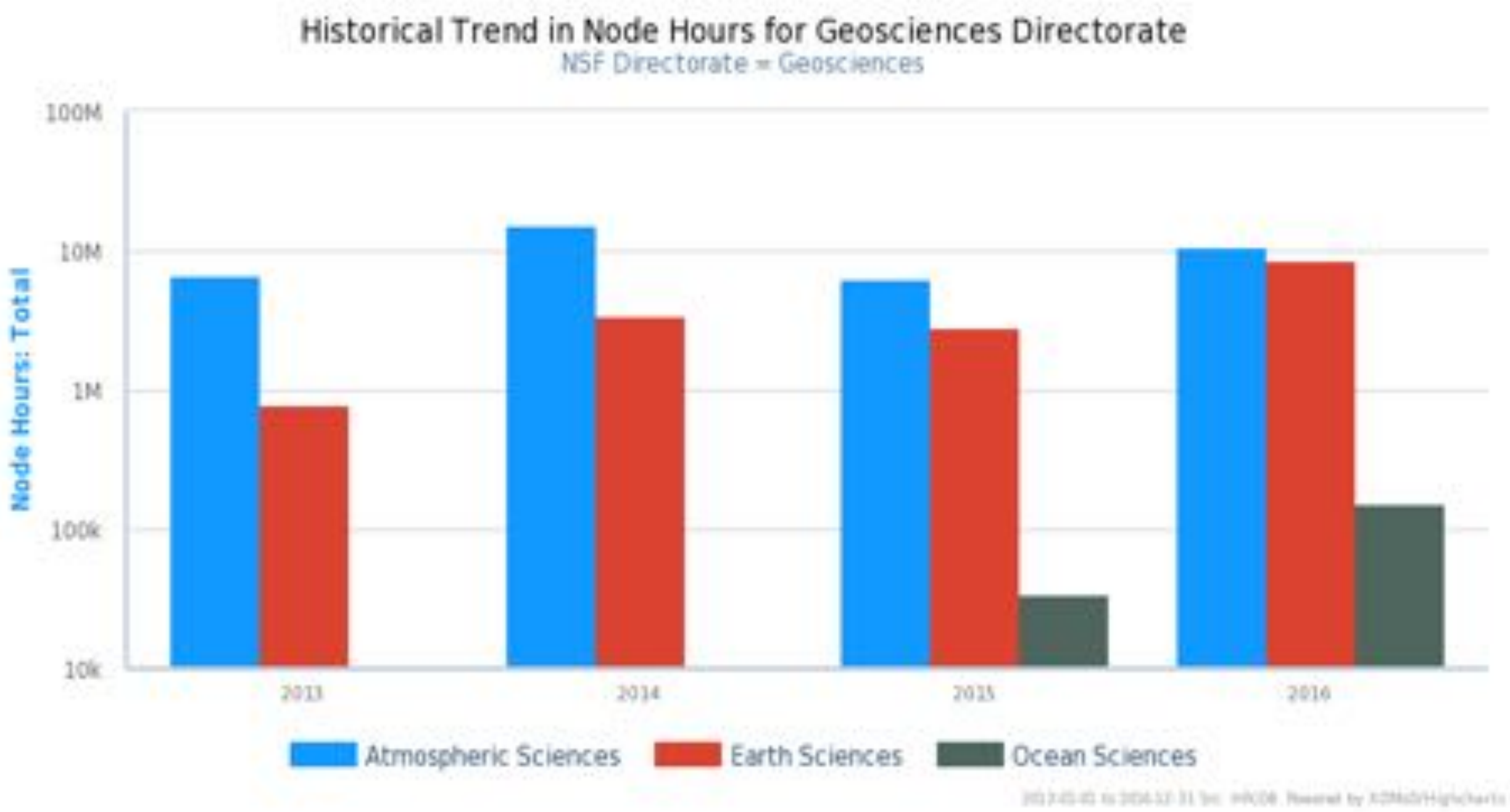

*Figure 2.1-5 Log scale in node hours consumed by parent sciences within the Geoscience directorate (2013, 2014, 2015, and 2016).*

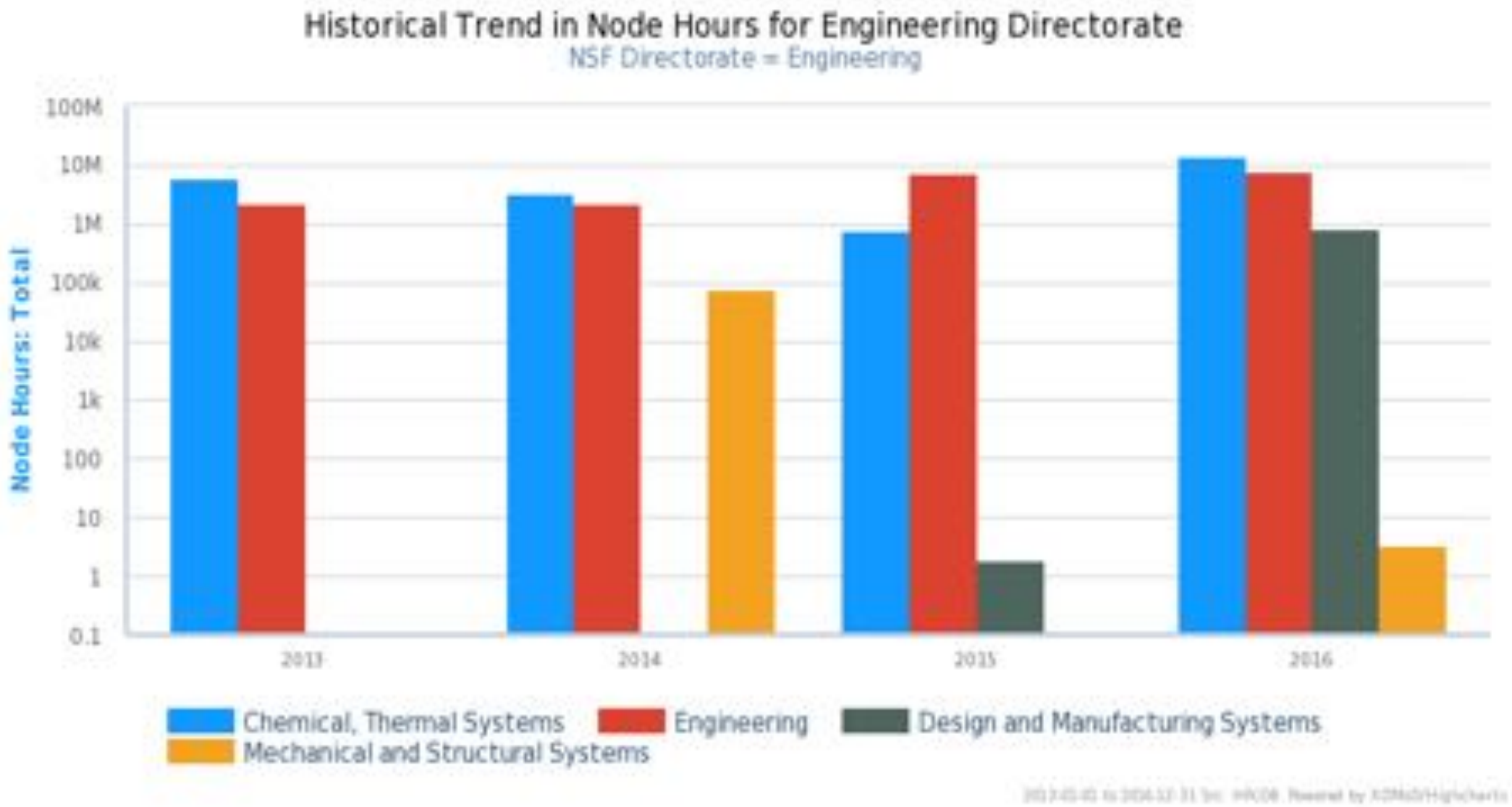

*Figure 2.1-6 Log scale in node hours consumed by parent sciences within the Engineering directorate (2013, 2014, 2015, and 2016).*

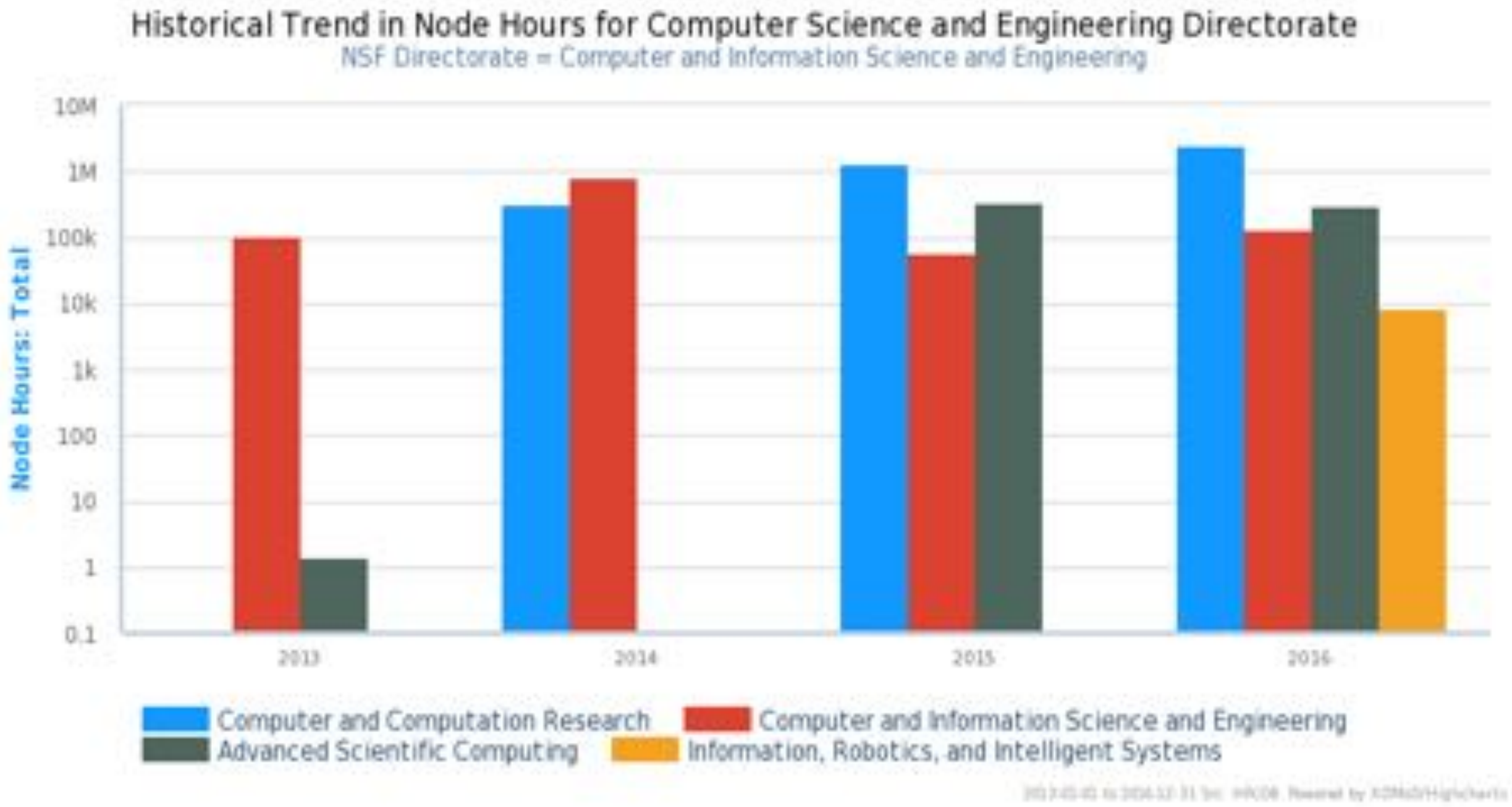

*Figure 2.1-7 Log scale in node hours consumed by parent sciences within the Computer and Information Science and Engineering directorate (2013, 2014, 2015, and 2016).*

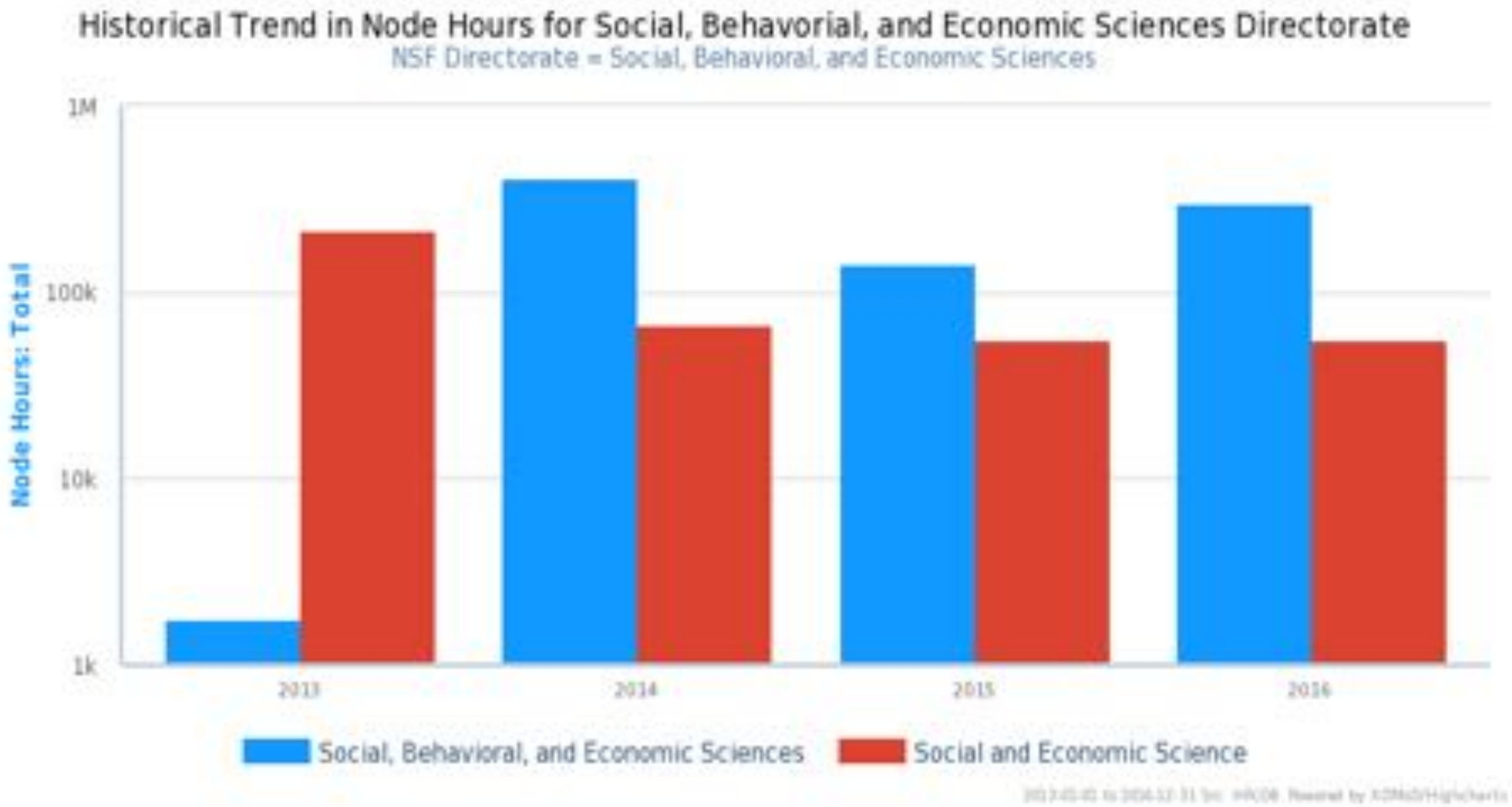

*Figure 2.1-8 Log scale in node hours consumed by parent sciences within the Social Behavioral, and Economic Sciences directorate (2013, 2014, 2015, and 2016).*

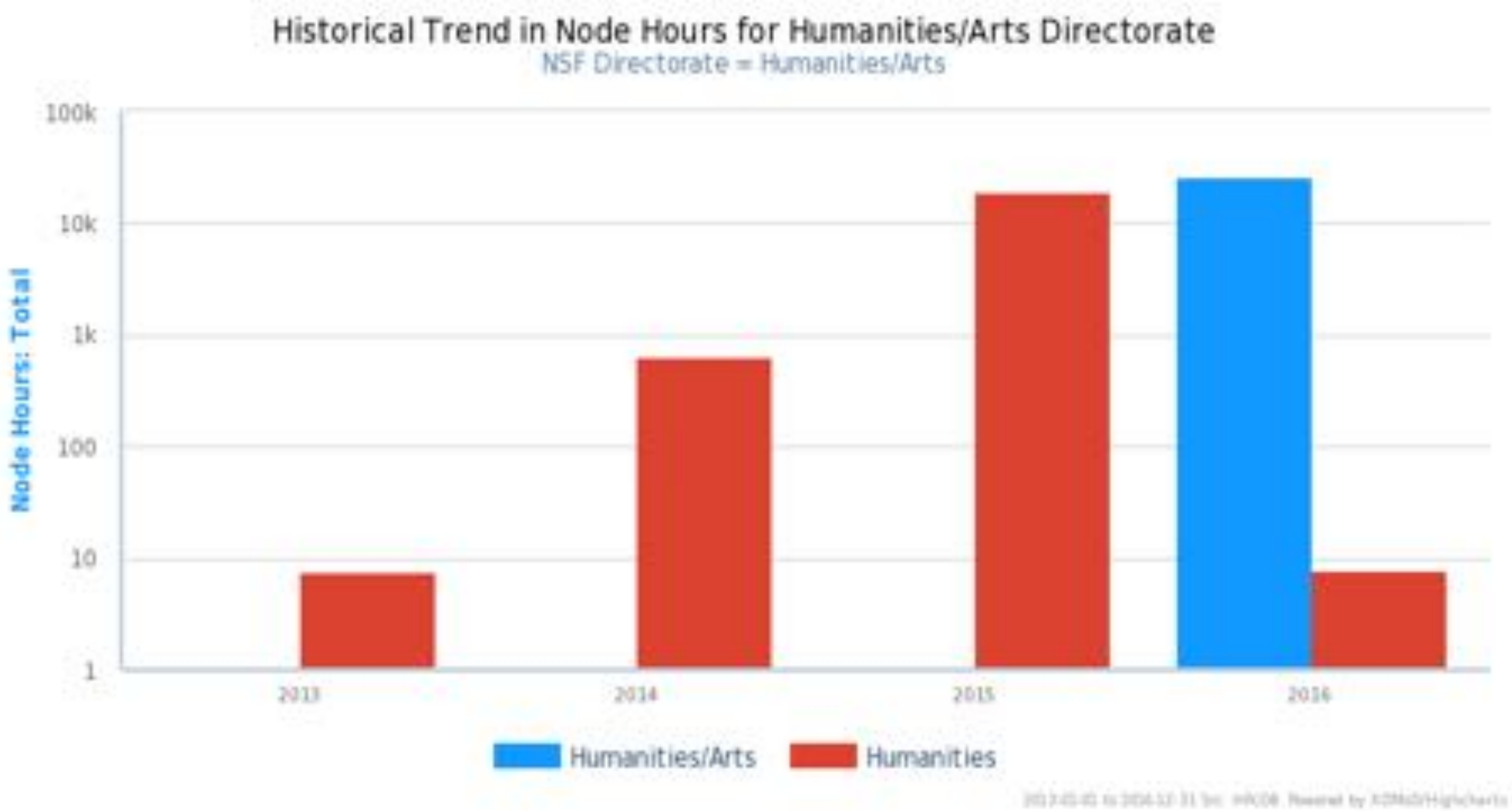

*Figure 2.1-9 Log scale in node hours consumed by parent sciences within the Humanities/Arts directorate (2013, 2014, 2015, and 2016).*

Another way to characterize the historical trends is to simply look at the increase in the number of fields of science (FOS) that have received allocations over time on Blue Waters. Since over 50 of the more than 120 fields of science identified by NSF are represented on Blue Waters at some point in the three plus years of full service, it is not practical to effectively graph the evolution of the actual fields of science over time. Instead we show the historical trend in terms of the number of fields of science during each of Blue Waters' 4 years. This is shown in Figure 2.1-10, which shows doubling of the number of FOS over Blue Water's lifetime. Appendix V presents the full list of these fields of science.

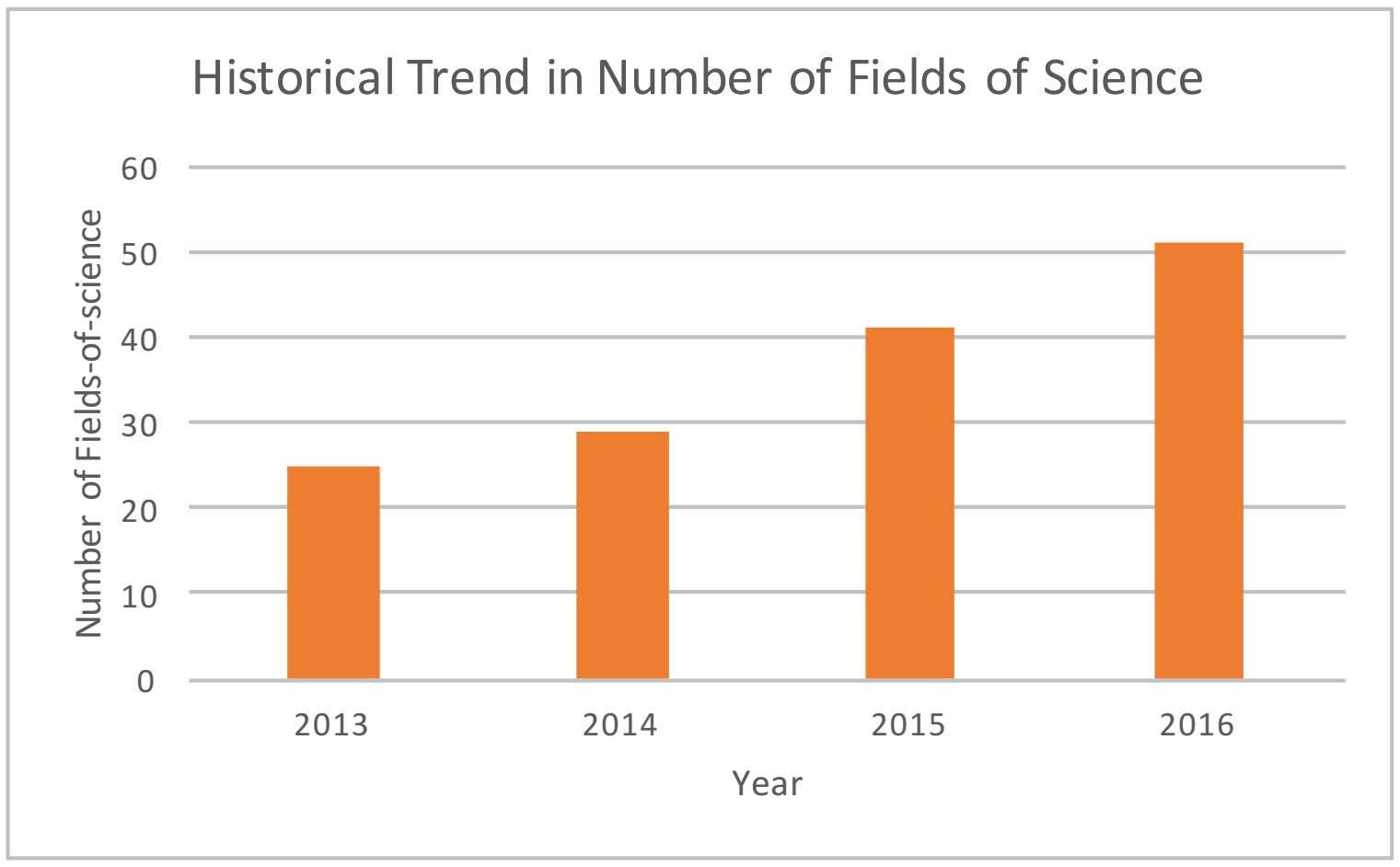

*Figure 2.1-10 The number of fields of science represented in Blue Waters' portfolio increases with every year of full service operation.*

In addition to a historical increase in the number of fields of science served by Blue Waters over time, the overall usage is spread out fairly well across the disciplines. This is shown in Figure 2.1-11, which is a plot of the actual usage of Blue Waters by discipline from June 1, 2016 – August 31, 2016. Note that no single FOS consumed more than sixteen percent of the available cycles.

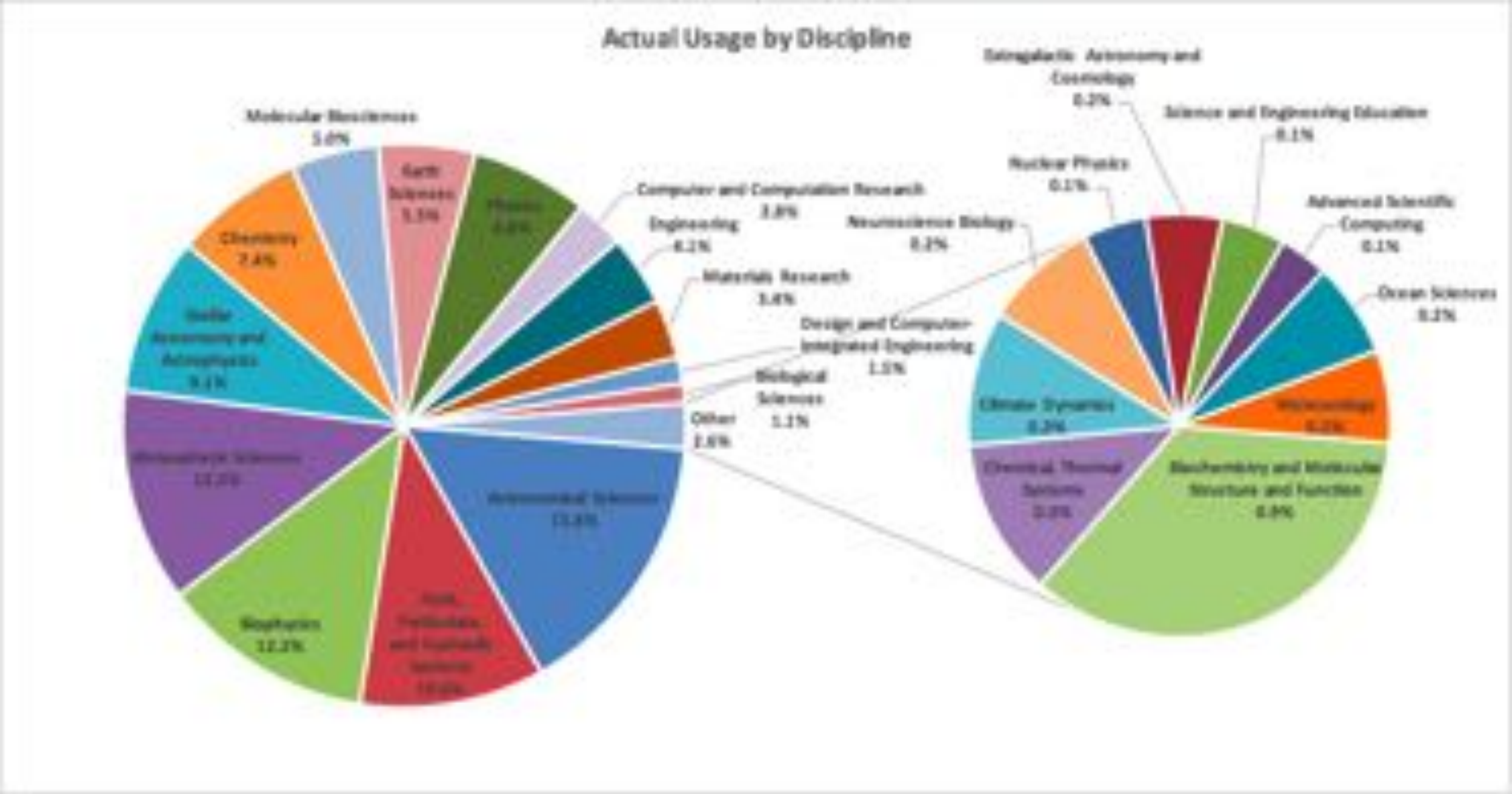

*Figure 2.1-11 Blue Waters usage by discipline from June 1, 2016 to August 31, 2016.*

## 2.2 Applications

Here we show application usage over the lifetime of Blue Waters. Unless otherwise noted, the charts are for the full period from start of production 2013-04-01 to 2016-09-30. Owing to the large number of applications and the broad range of usage, we have found it convenient to break the usage into multiple plots. Figure 2.2-1 shows the node hours consumed for the top 24 applications, and Figure 2.2-2 shows the node hours consumed for the next 24 applications. Please note that not all runs on Blue Waters had an identifiable application. As described in Appendix II, some runs had incomplete or missing data from the aprun task launcher (about 4M node hours which corresponds to 0.9% of total node hours), and not all application names were recognized by Open XDMoD's application name recognition algorithm (approximately 35M node hours corresponding to 7.8% of total node hours). Many of these unrecognized names are user-specific and quite varied. Some categories, such as python (for the python interpreter), and INDUSTRY (for proprietary commercial applications) are more general. Overall however, more than 90% (by usage in node-hours) of application names were categorized. Figure 2.2-3 is a pie chart showing the percentage of node hours consumed by the top applications, with NAMD consuming about 18% of all cycles.

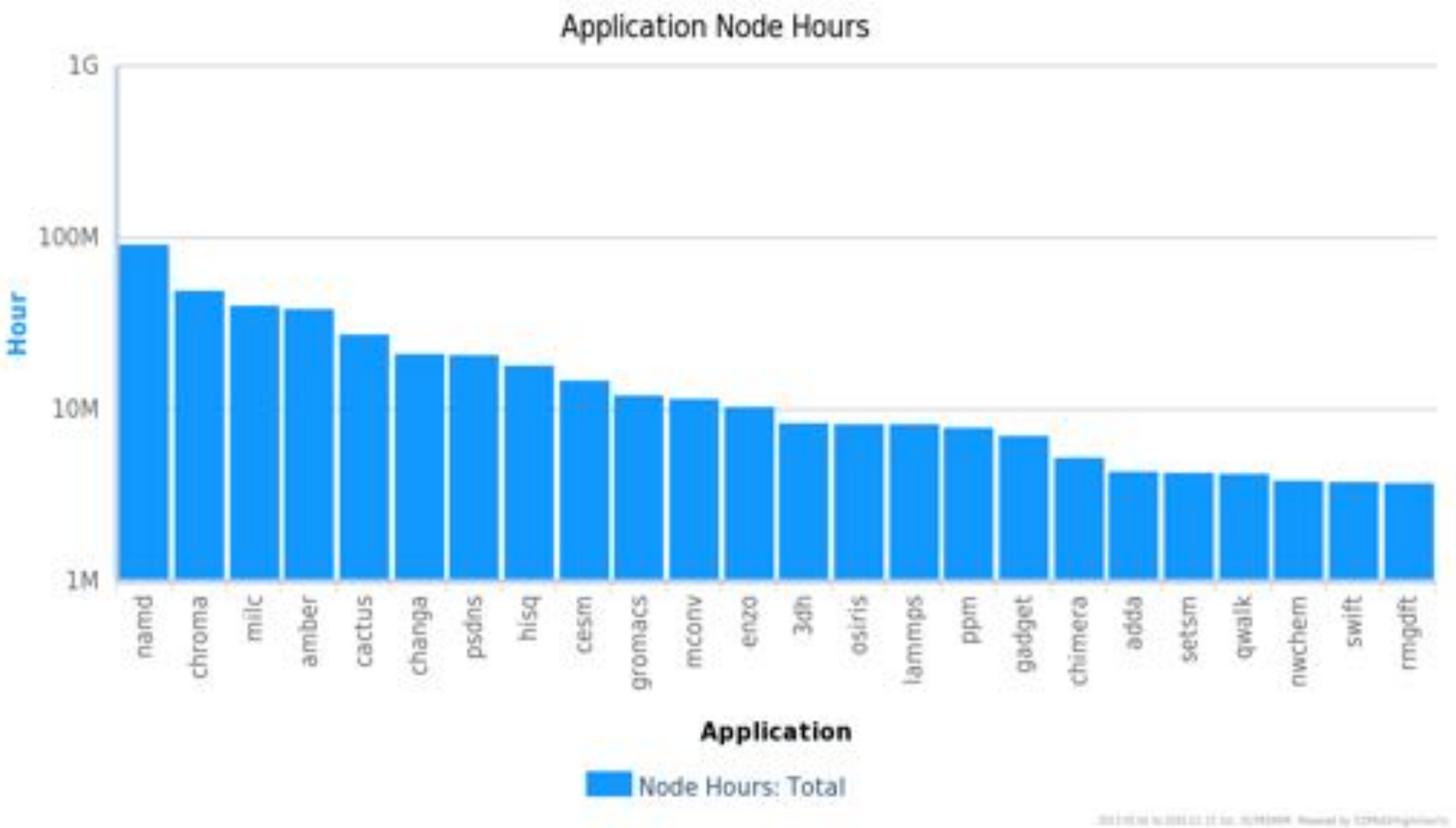

*Figure 2.2-1 Application usage for the top 24 applications by node hours consumed on Blue Waters*

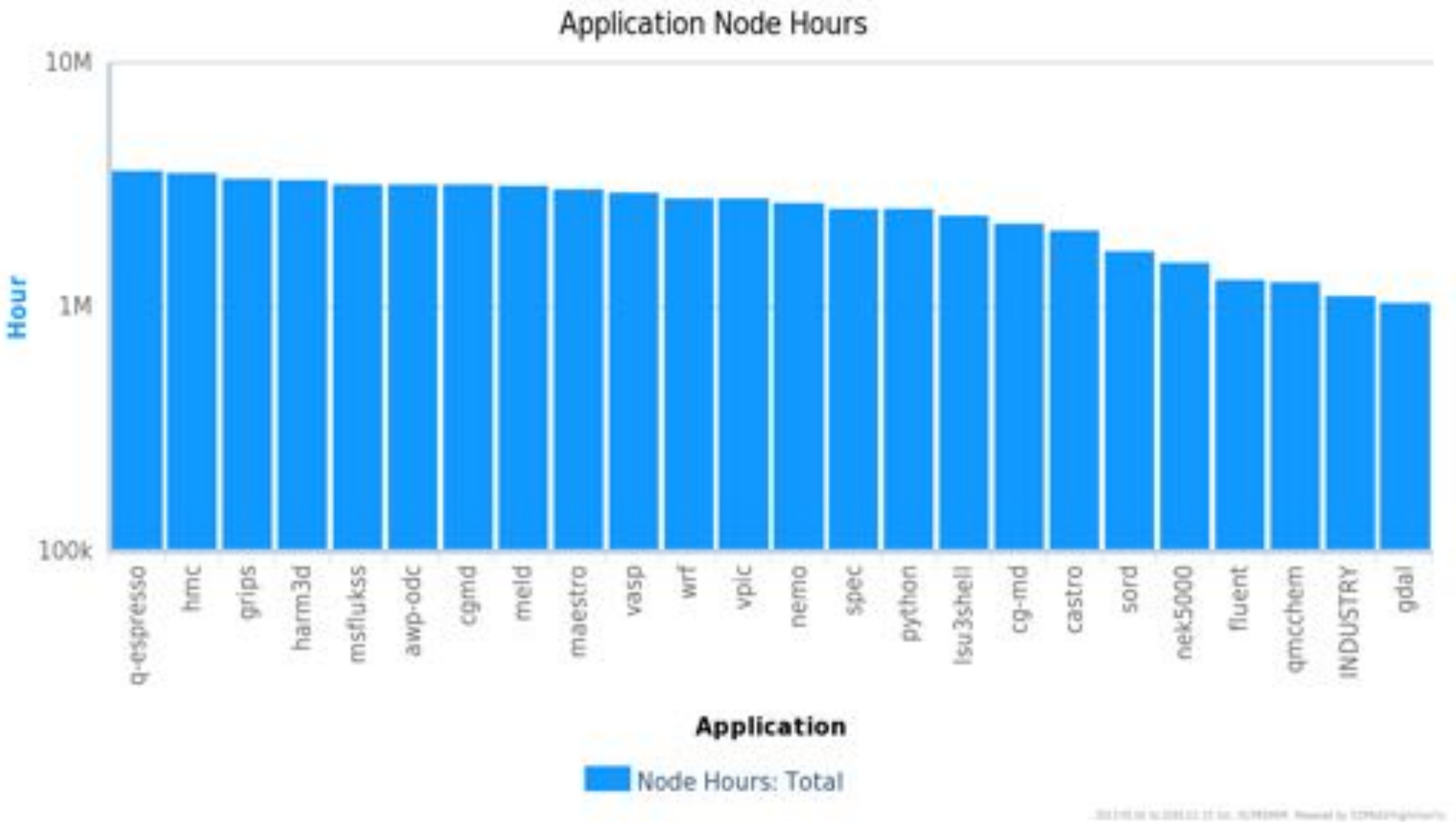

*Figure 2.2-2 Application usage for the next 24 most heavily used applications on Blue Waters*

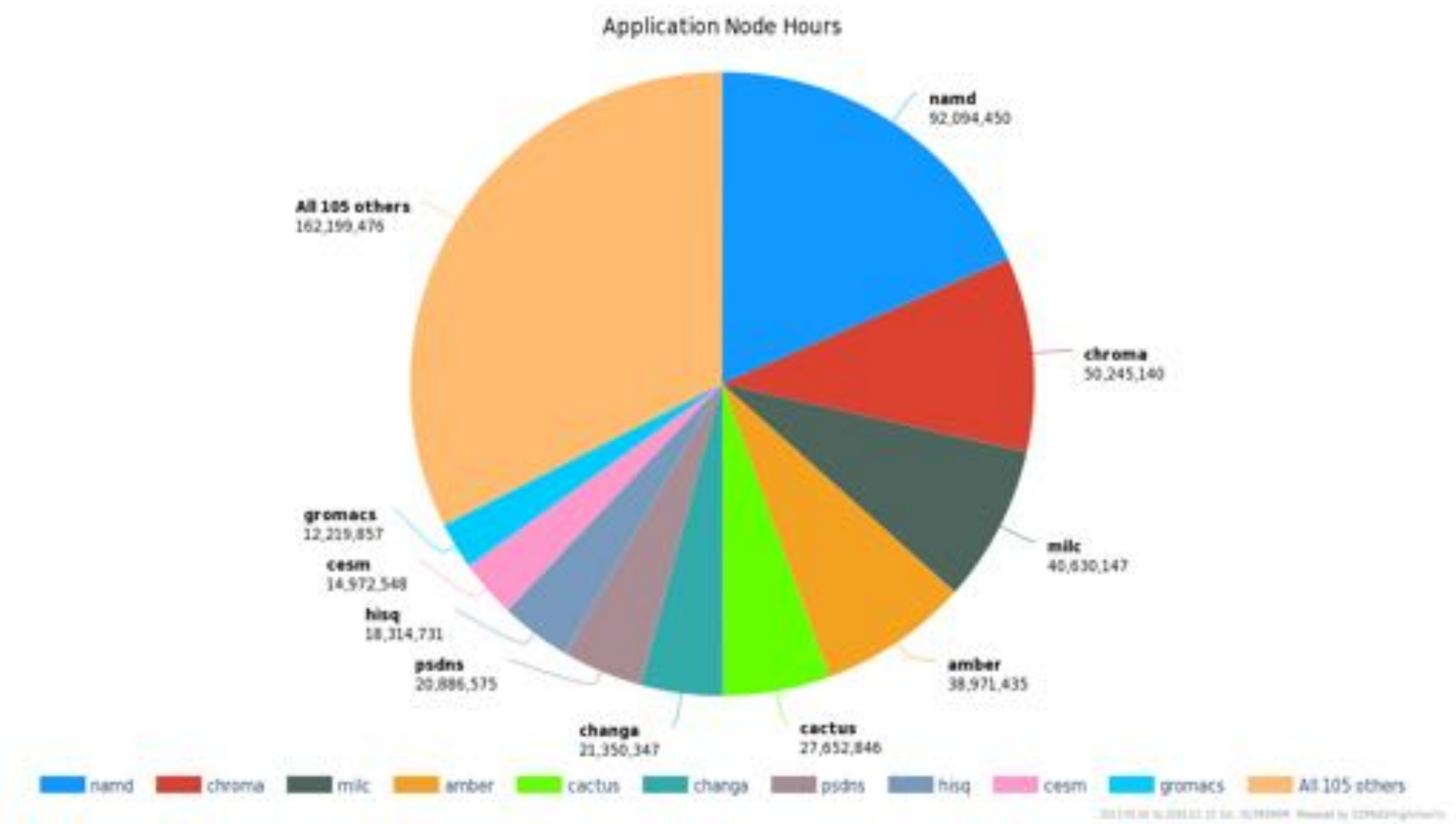

*Figure 2.2-3 Top 10 applications running on Blue Waters. (NAMD is about 18% of recognized usage by node hours and is used by multiple teams.*

The usage of most applications is dominated by a particular science area (likely due to the larger allocation policies on Blue Waters), but some popular community applications have a broader base of users. If we focus on the top four applications by used node hours, and organize them by NSF parent science[1], we note that NAMD and AMBER, two popular molecular mechanics applications, are used by multiple projects (largely still within biological science areas). As shown in Figure 2.2-4, CHROMA and MILC, two quantum chromodynamics codes, are largely used by single projects.

[1] The "Field of Science", "Parent Science" and "NSF Directorate" dimensions in Open XDMoD all come from the allocation information from the project that the job ran against.

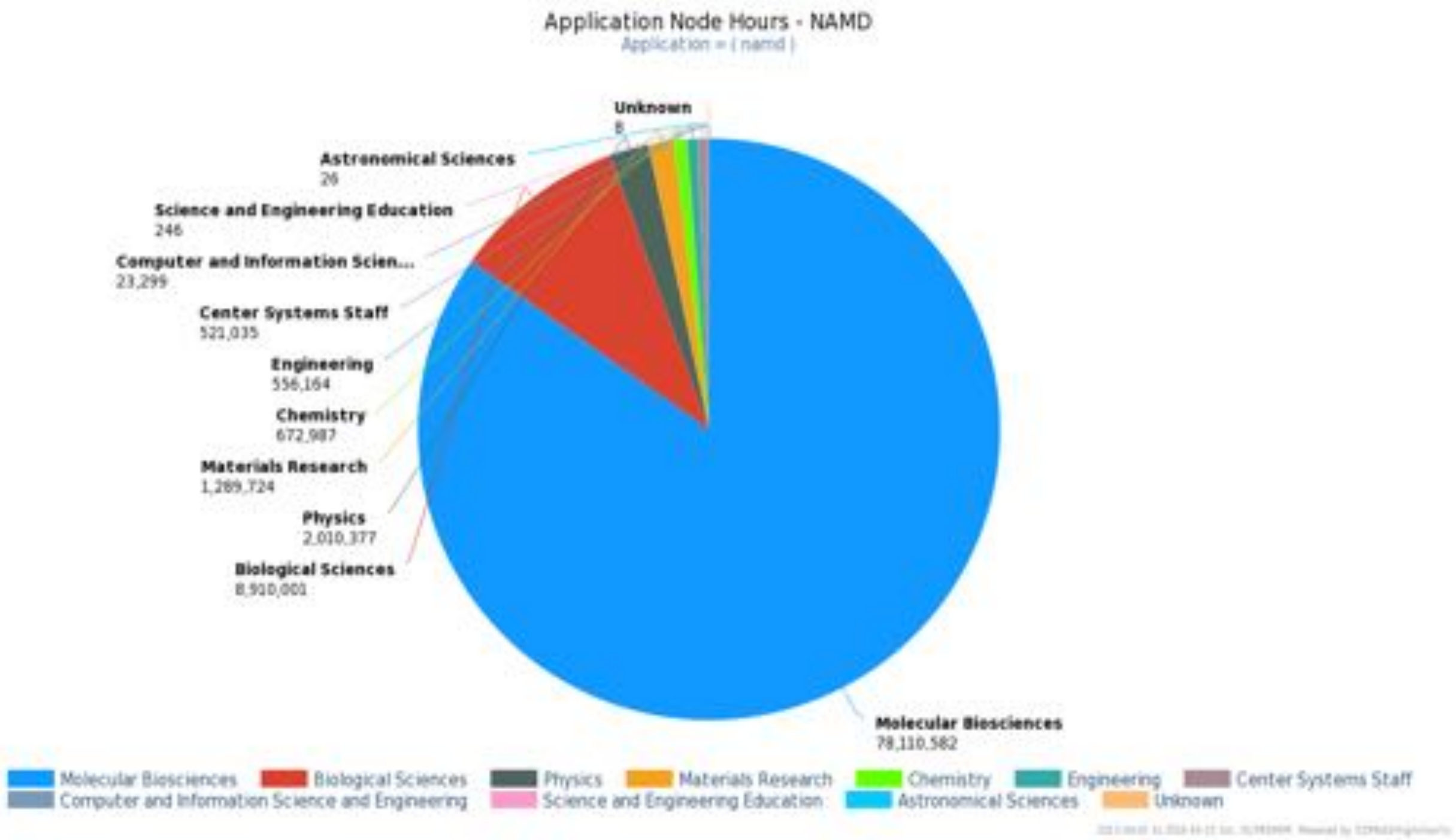

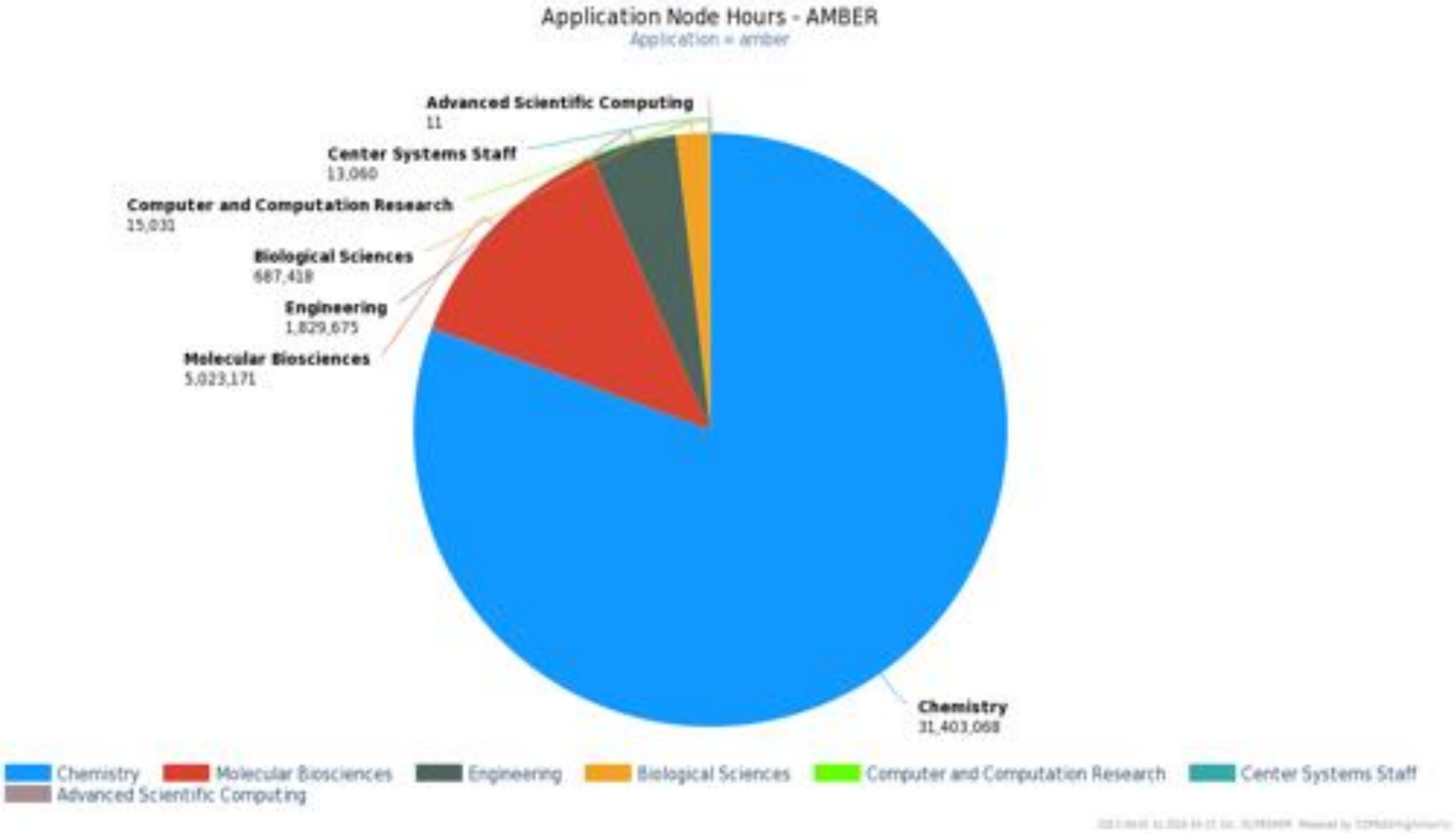

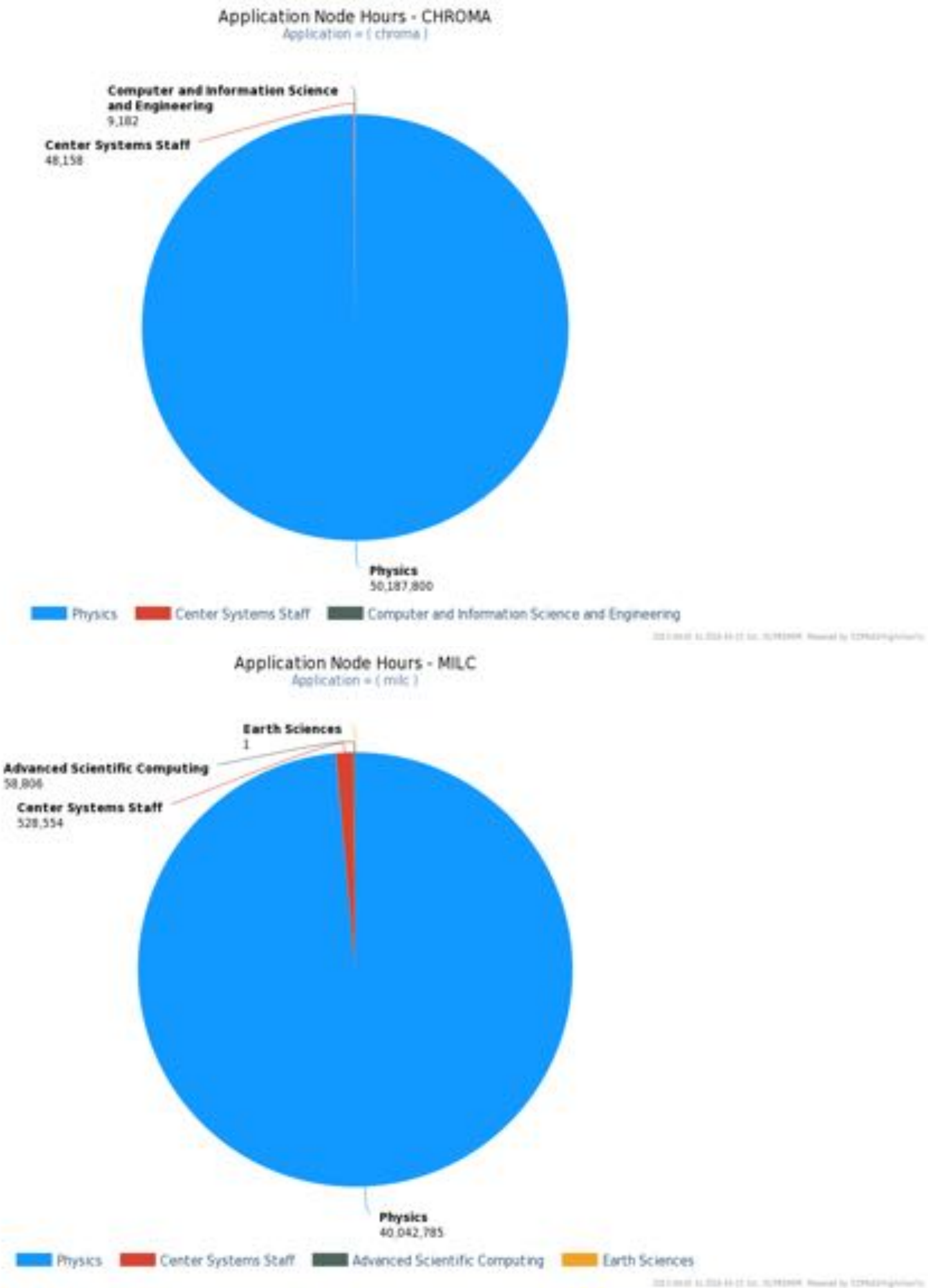

*Figure 2.2-4 Breakdown of the top 4 applications (NAMD, AMBER, CHROMA, and MILC) by Parent Science. NAMD and AMBER have a greater utilization across parent science than do CHROMA and MILC.*

## 2.3 Algorithms

Given the mix of applications described in the previous section, we can build a mapping of the most common algorithms used in these applications. The close cooperation between the Blue Waters science team and investigators led to the categorization of the major applications in terms of their internal algorithms, described in Table 11.0-1 of Appendix IV. Table 11.0-1 breaks down applications in terms of their representative use of the Colella "dwarfs," seven historically important algorithm areas, namely grid-based (structured and unstructured), matrix-based (dense and sparse), N-body, Monte Carlo, and Fourier transform algorithms [11]. Applications that rely heavily on input/output (not part of the original seven dwarfs) are also noted. Equal weight was applied to each algorithm used in a particular application, as it is not possible to clearly distinguish the internal representation of each algorithmic contribution for the applications used in all simulations. Figure 2.3-1 shows the breakdown done in this manner for Blue Waters.

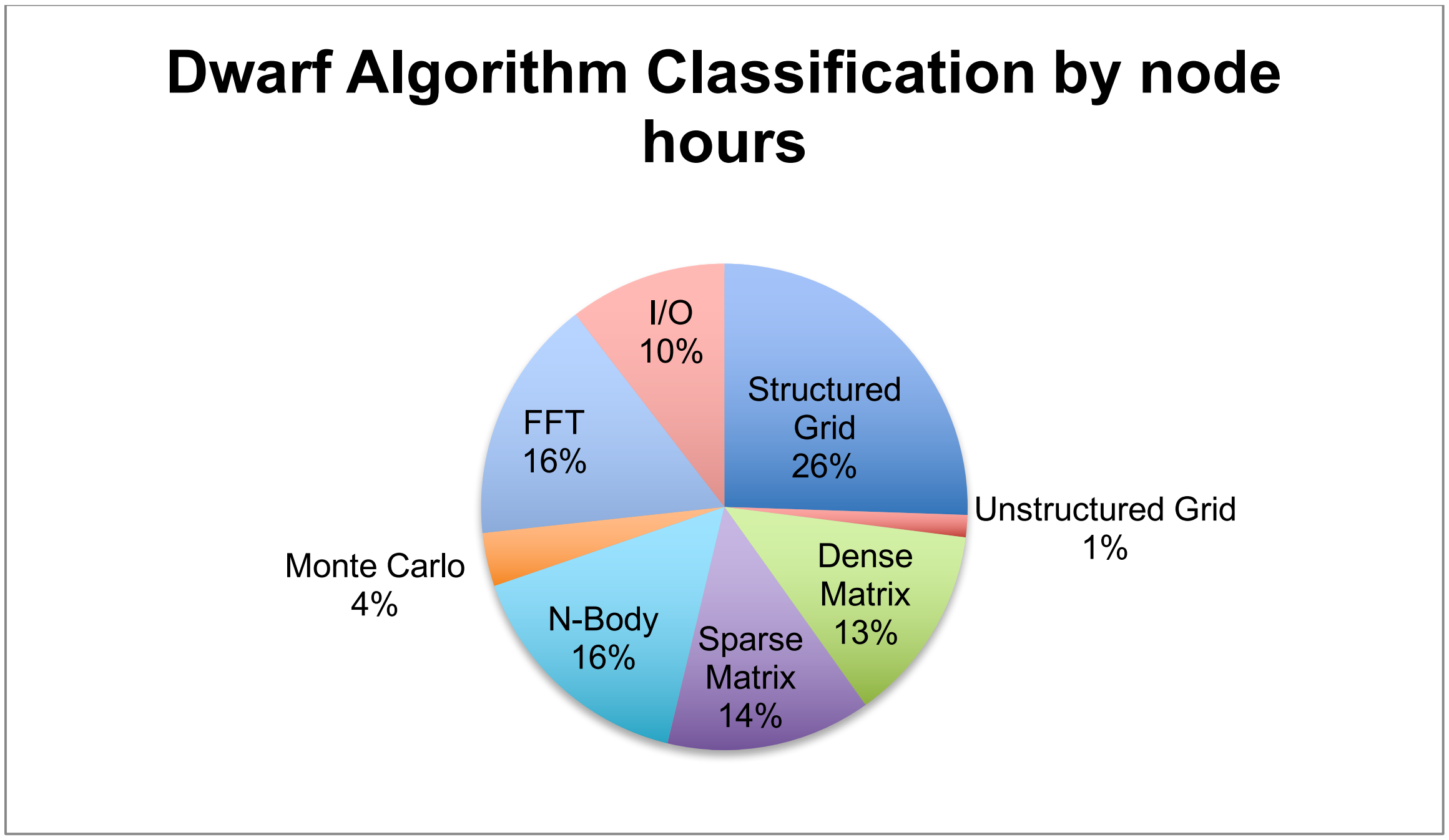

*Figure 2.3-1 Colella's seven dwarf classification of recognized applications run on Blue Waters (by total node hours) in the study period, assuming equal weighting if an application is using more than one algorithm in Table 10.0-1 in Appendix IV.*

## 2.4 Numerical Libraries

### 2.4.1 Data Coverage and Limitations

Utilization of libraries was tracked using ALTD and XALT. Both tools track utilization of statically and dynamically linked libraries but cannot detect the usage of real-time loaded libraries. Although real-time loaded libraries are currently rarely used in scientific and engineering application, their use is growing with wider usage of scripting languages such as python. Library utilization is measured indirectly by tracing the execution of the applications linked against the library. Thus, even though a given library might not be used during execution of the application, it will be marked as used in our analysis. The reported node hours consumed by the libraries correspond to the total node hours consumed by the applications linked to the libraries.

Figure 2.4-1 shows the data coverage by ALTD and XALT versus the total number of jobs run. Due to small XALT and ALTD coverage available in 2016, further analysis is carried out for 2014 and 2015 using ALTD data only (green and blue traces). ALTD tracks only jobs that use aprun to start execution. Furthermore, for database storage reasons, ALTD truncates the linking line to 4 KB, further limiting available library information since most large applications have a long linking line and unfortunately the linked libraries are located at the end of the linking line. Thus, truncating the linking line can result in loss of library data, which can lead to erroneous conclusions regarding library use. For example, only 0.004% of all NAMD jobs have a complete link line showing FFT library utilization, while 85% have an incomplete link line without any indication of FFT library usage. The remaining NAMD jobs have no library usage information. The limited library usage information might lead one to erroneously conclude that the majority of NAMD jobs do not use FFT, which is incorrect. Because there is useful information to be gleaned, even in the truncated lines, they were included in the analysis below. However, extra care is needed during comparison of different library usage due to the incompleteness of the data and its qualitative nature.

During 2014 and 2015, ALTD complete data coverage was 25% by job count and 51% by node hours, partial ALTD was available for 48% of all jobs by job count and 82% by node hours. Figures 2.4-2 and 2.4-3 show the coverage grouped by parent science and application name. Fortunately, only a few of the parent sciences have little or no ALTD coverage and most applications have at least some ALTD coverage.

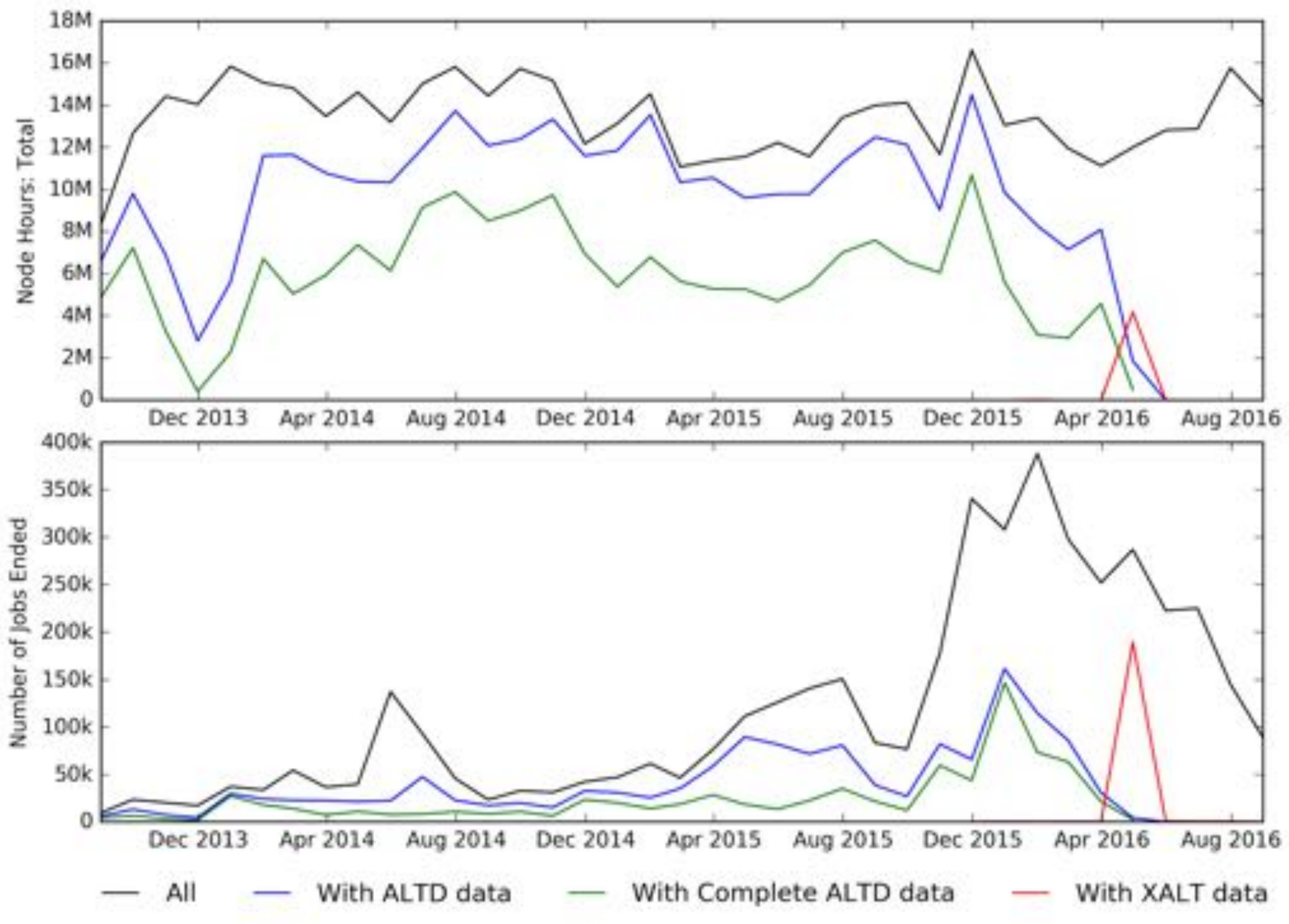

*Figure 2.4-1. Coverage of Blue Waters jobs with ALTD and XALD data. Black line shows all jobs executed on resource, blue line shows jobs with any ALTD, green line shows jobs with complete ALTD data and red line shows jobs with available XALT data.*

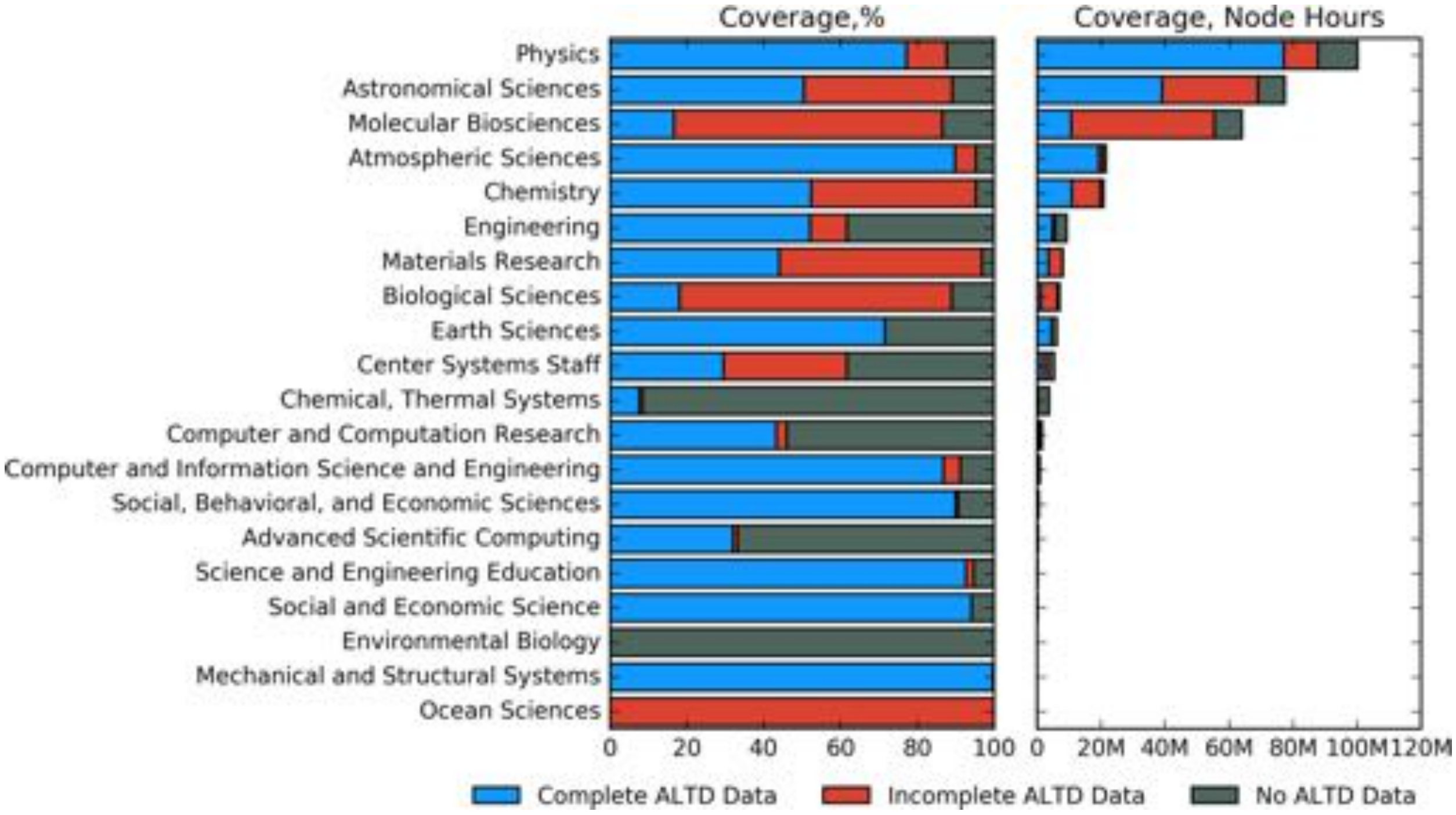

*Figure 2.4-2. Coverage of Blue Waters jobs (node hours) by parent science for 2014 and 2015 years*

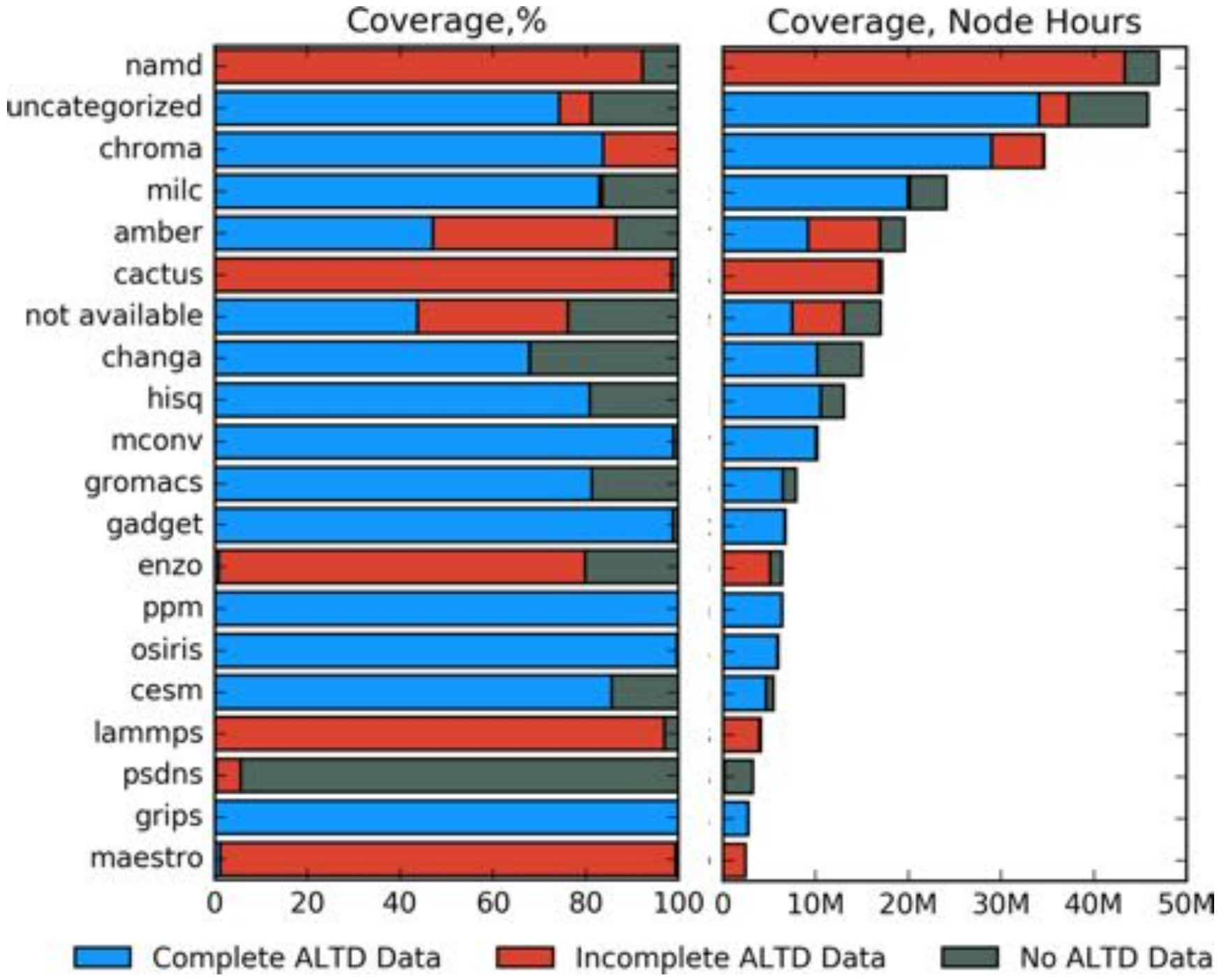

*Figure 2.4-3. Coverage of Blue Waters jobs (node hours) by application for 2014 and 2015 years*

### 2.4.2 Libraries Usage

Here we measure library usage in several different ways – by unique users, node hours, number of jobs, and BLAS, LAPACK, and FFT usage by parent field of science. We also measured the utilization of the libraries that provide BLAS, LAPACK, and FFT functionality since a large number of applications across all fields of science are dependent on these. Usage of the file storage libraries was also examined. We begin with usage by users, node hours and number of jobs.

Figure 2.4-4 shows the top 30 libraries/modules (excluding system, MPI, compilers, debugging and profiling libraries) used on Blue Waters as measured by the number of unique users, jobs, and node hours based on ALTD/XALT data from 2014 to 2015. The data covers about 81% of node hours and of this, about 38% of the jobs have incomplete data.

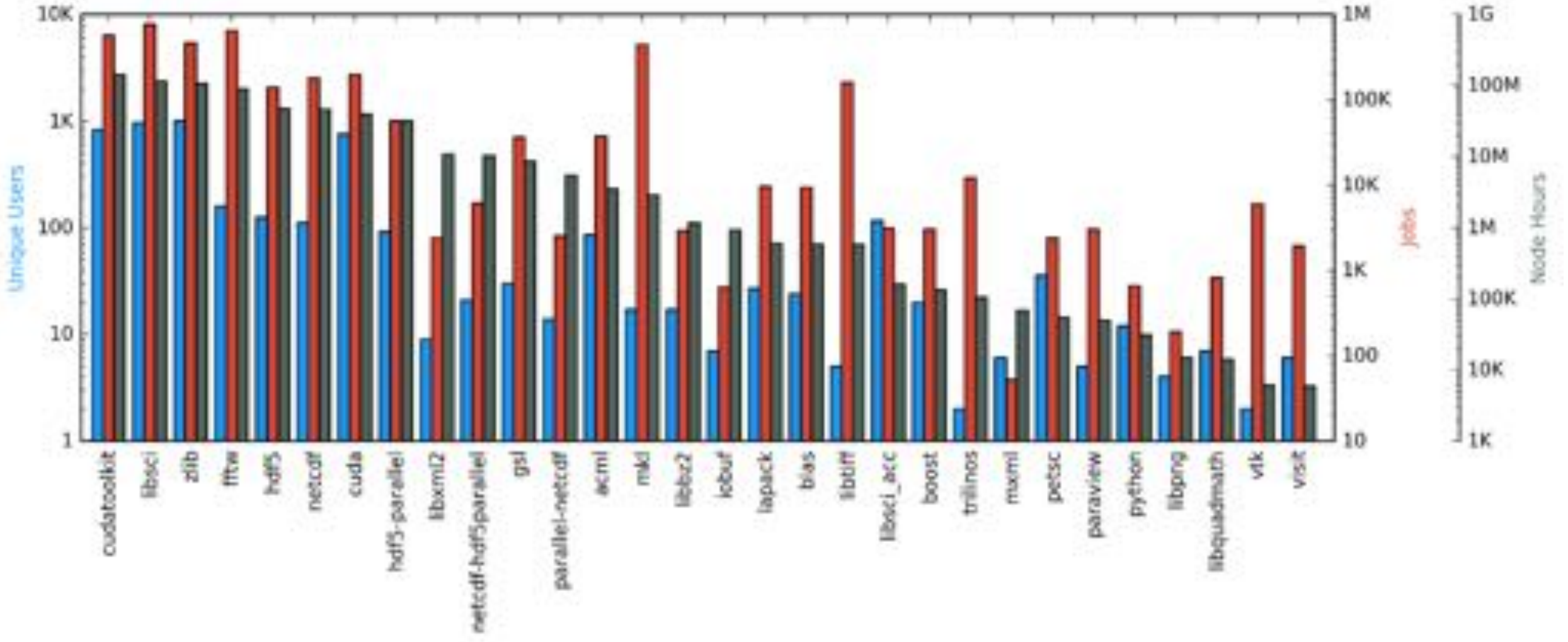

*Figure 2.4-4. Module and library usage on Blue Waters as measured by the number of unique users, number of jobs and node hours utilized by corresponding applications.*

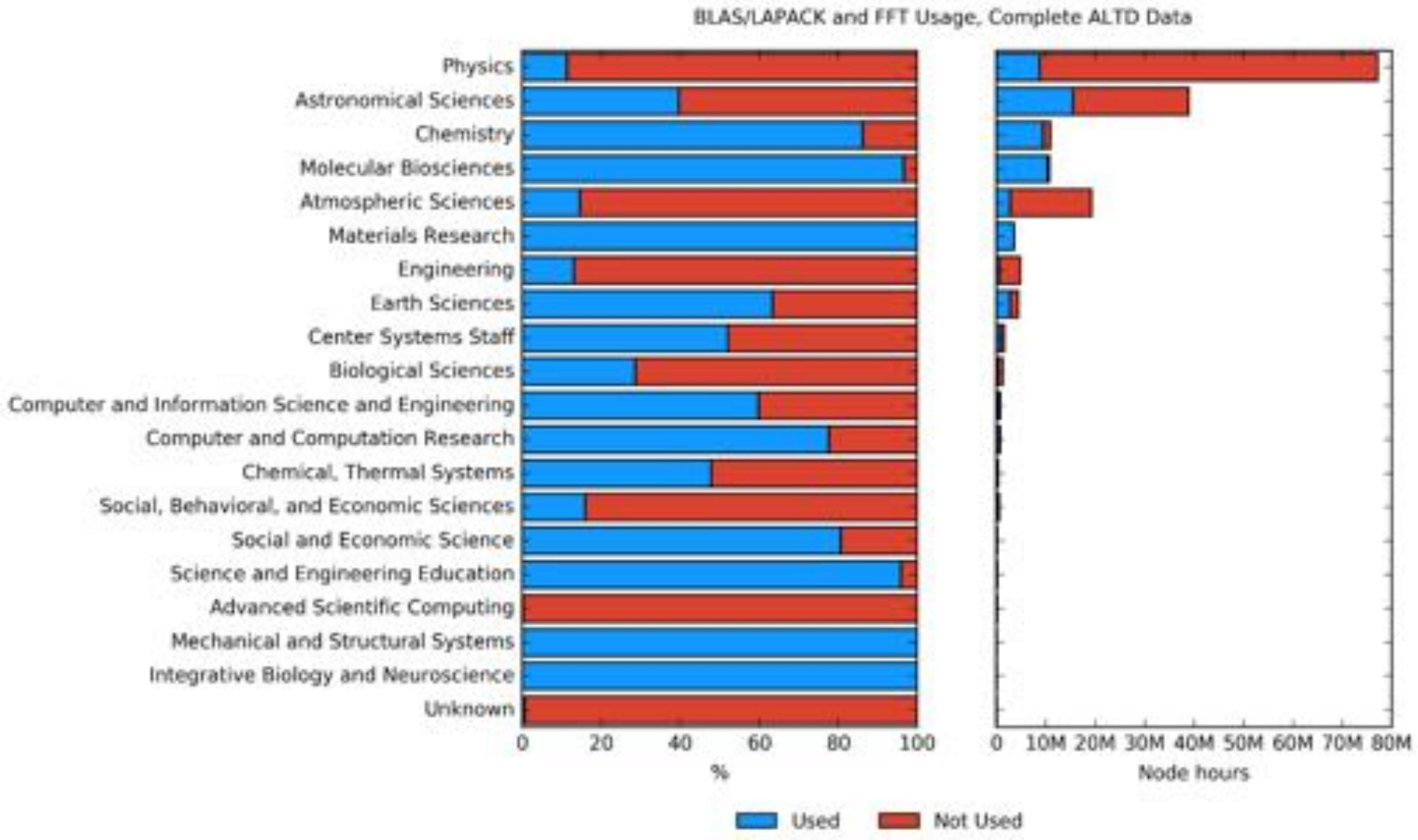

*Figure 2.4-5. Utilization of BLAS, LAPACK or FFT by parent field of science using only complete ALTD data.*

Figure 2.4-5 shows the utilization of BLAS, LAPACK and FFT over the parent field of science and is based on those jobs for which we had complete ALTD data. The inclusion of all available ALTD data leads to roughly the same results in most parent fields of science with few exceptions namely Molecular Biophysics and Biological Science. The majority of the discrepancies are most likely an artifact of truncation of the link line rather than actual change in libraries usage ratio. As discussed previously, if the link line is too long (often the case with many large programs) the link line is truncated in order to fit into the ALTD database. Given that libraries are most often located at the end of the link line, it is very likely that they were removed. Overall, based on the limited data available, a large number of applications across wide range of fields of science (Figure 2.4-5) use libraries that provide BLAS, LAPACK and FFT functionality.

More in depth analysis of the usage of such libraries is shown on Figure 2.4-6. The left most plot in this figure (linear scale) shows that LibSci (optimized for CRAY) is the most heavily used, followed by FFTW and ACML (which is optimized for AMD). Intel's MKL library shows relatively little use, partially due to the fact MKL was installed on the system in 2015. The middle plot is the same data but with a log scale. There are a very limited utilization of non-optimized BLAS and LAPACK libraries (labeled as lapack, blas and cblas). Also included in Figure 2.4-6 is the utilization of the data file storage libraries, with HDF5 and NETCDF most heavily utilized, and the parallel version of HDF5 also showing significant utilization.

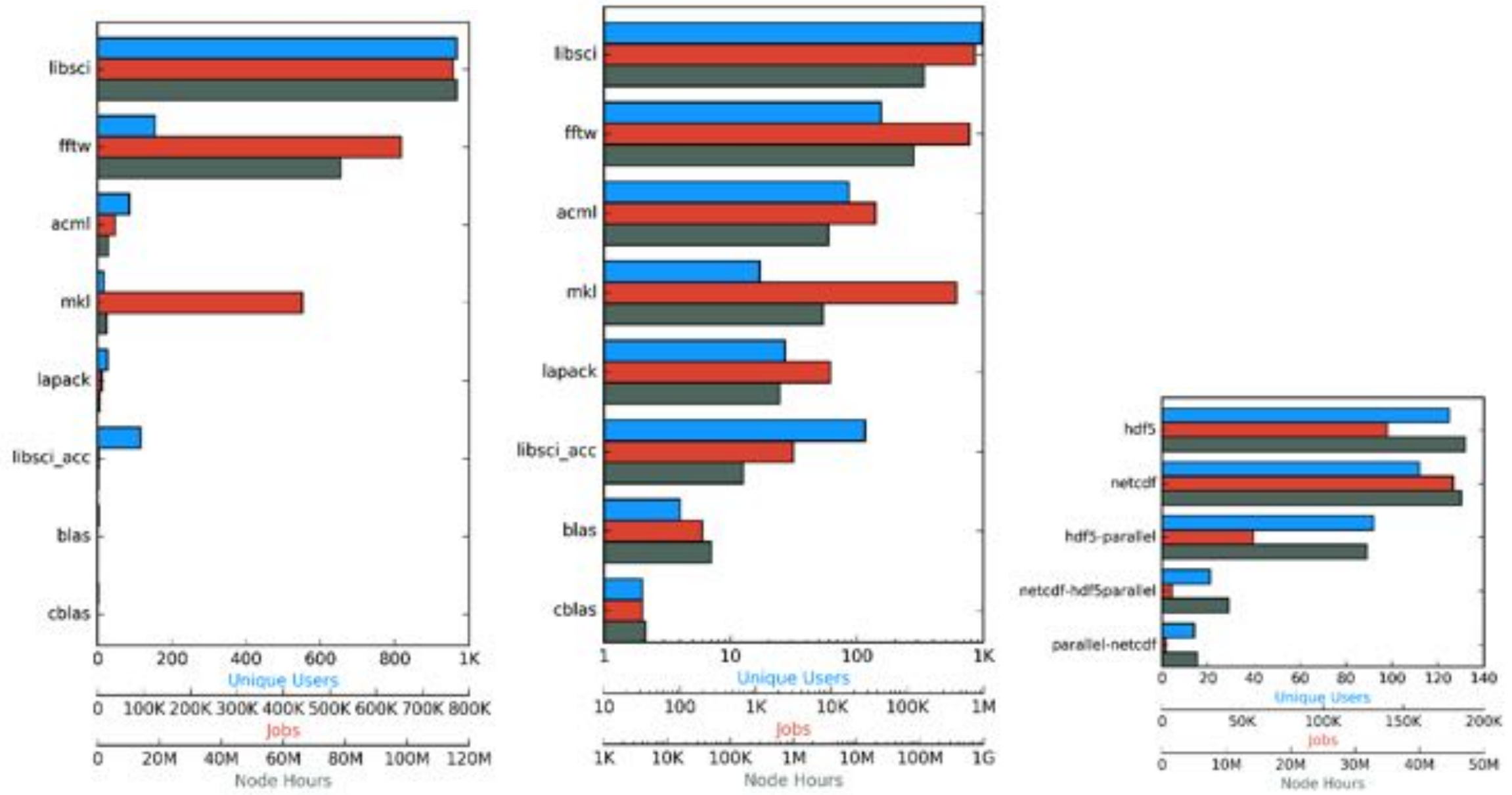

*Figure 2.4-6. The left plot shows Utilization of numeric libraries providing BLAS, LAPACK and FFT functionality. The center chart shows the same data as the left but with log scale. The right chart indicates utilization of data file storage libraries.*

## 2.5 Interconnect Usage

Figure 2.5-1 shows the average interconnect usage by application for the top twenty applications in terms of *average interconnect usage weighted by node hours*. Weighted averages by node hours are often used to avoid skewing results coming from many short (often small, testing, or exploratory) jobs. Note that in Figure 2.5-1 the weighted average is over the jobs in each individual application. The purpose of Figure 2.5-1 is to rank the various applications by interconnect usage and not to determine the overall machine average. Therefore, these do not represent the interconnect usage for the top 20 applications by node hours consumed.

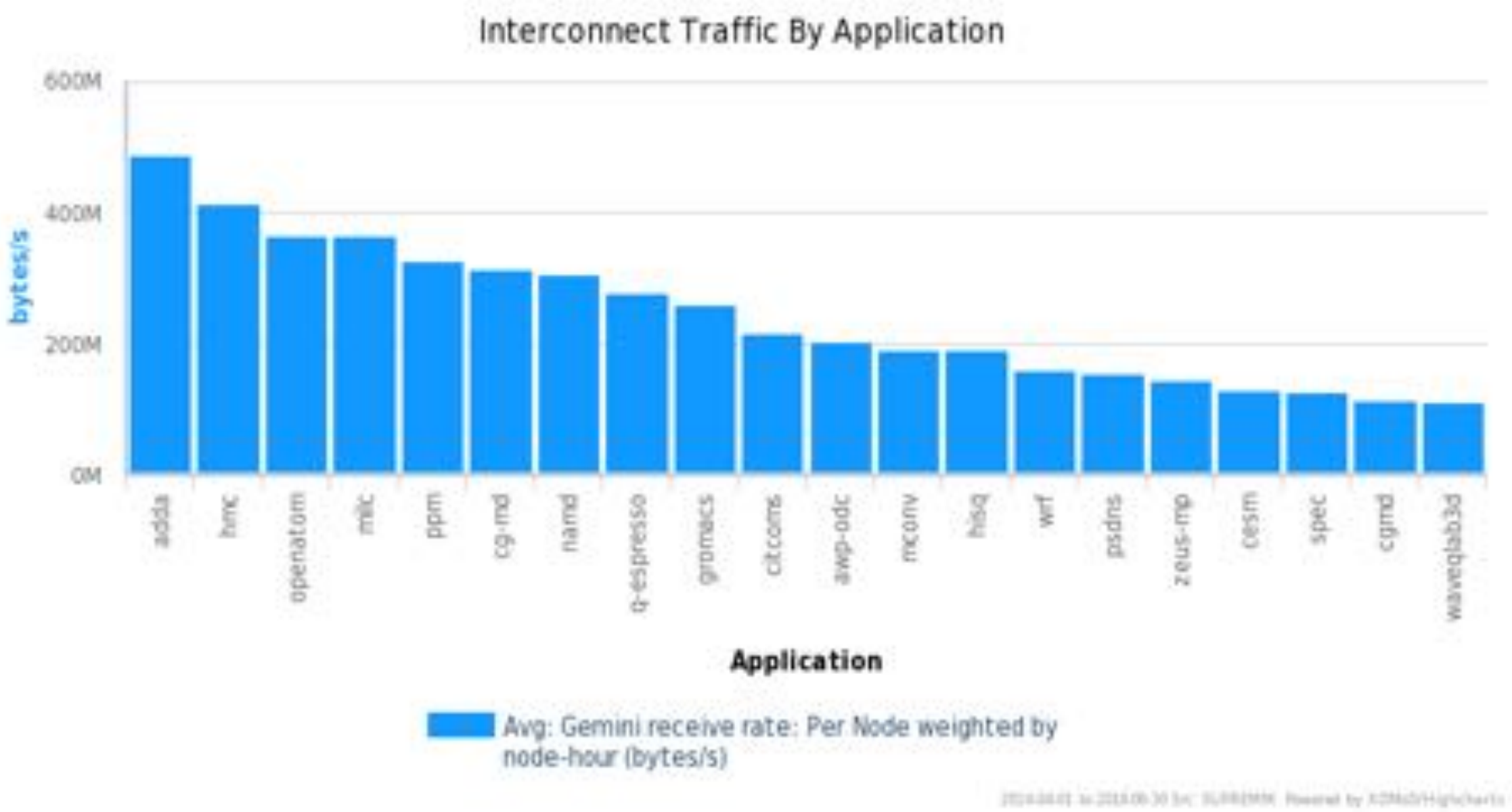

Figure 2.5-1. Interconnect usage (receives) by application, average per node weighted by node hours.

Note that we cannot cleanly distinguish system, filesystem, and message passing traffic, hence the data shown in this section represents the combined traffic. Receives only are shown (rather than sends and receives) due to problems with the collected send data.

Figures 2.5-2 shows the breakdown of the average interconnect use in terms of bandwidth for the top interconnect usage application ADDA, and Figures 2.5-3 and 2.5-4 for MILC and NAMD, respectively. On each of these three figures the y-axis is proportional to the fraction of node hours consumed for the specific application. That is, each figure only shows the relative distribution of GB/s received for the given application. Only the shapes of the distributions and not the relative size can be compared across the three figures. NAMD is the most frequently used individual application (by node hours).

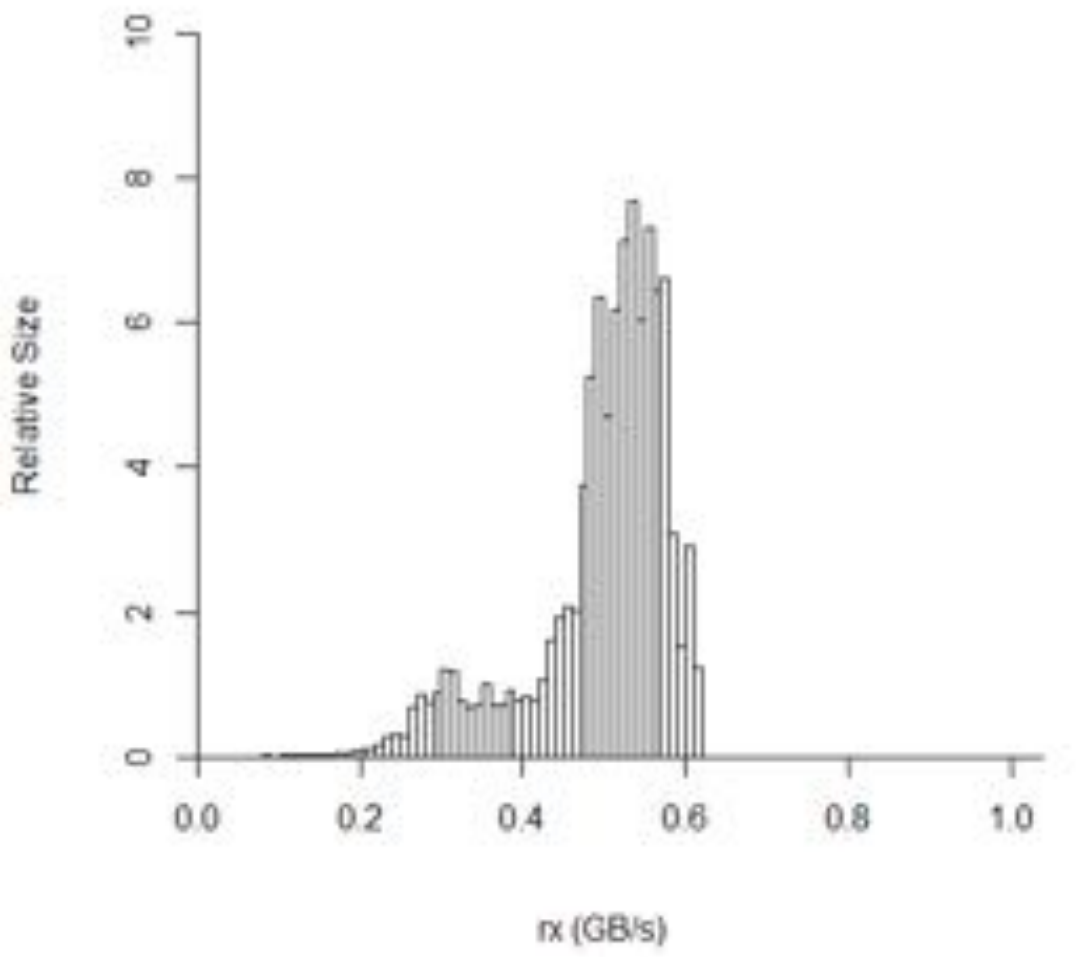

*Figure 2.5-2 ADDA per-node interconnect receive usage broken down by rate.*

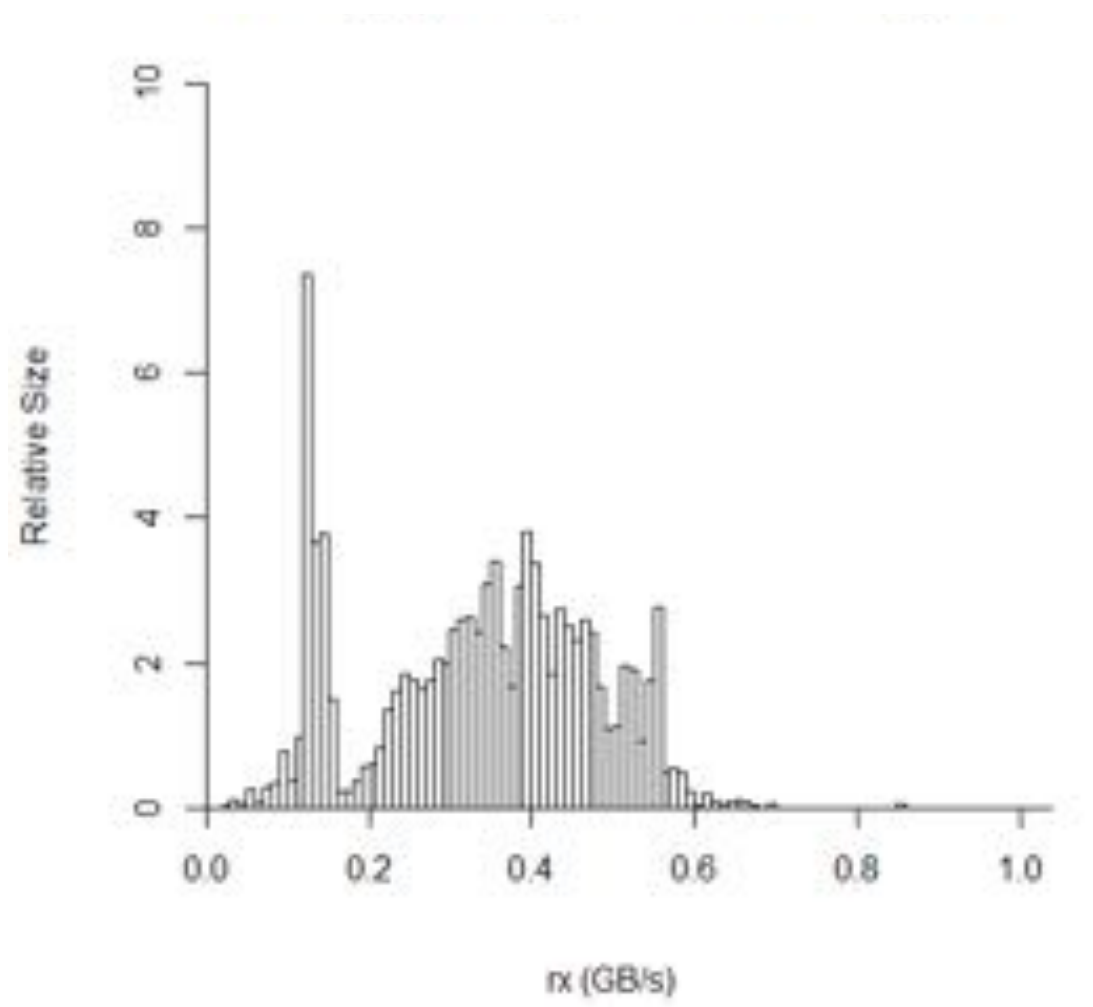

*Figure 2.5-3 MILC per-node interconnect receive usage broken down by rate.*

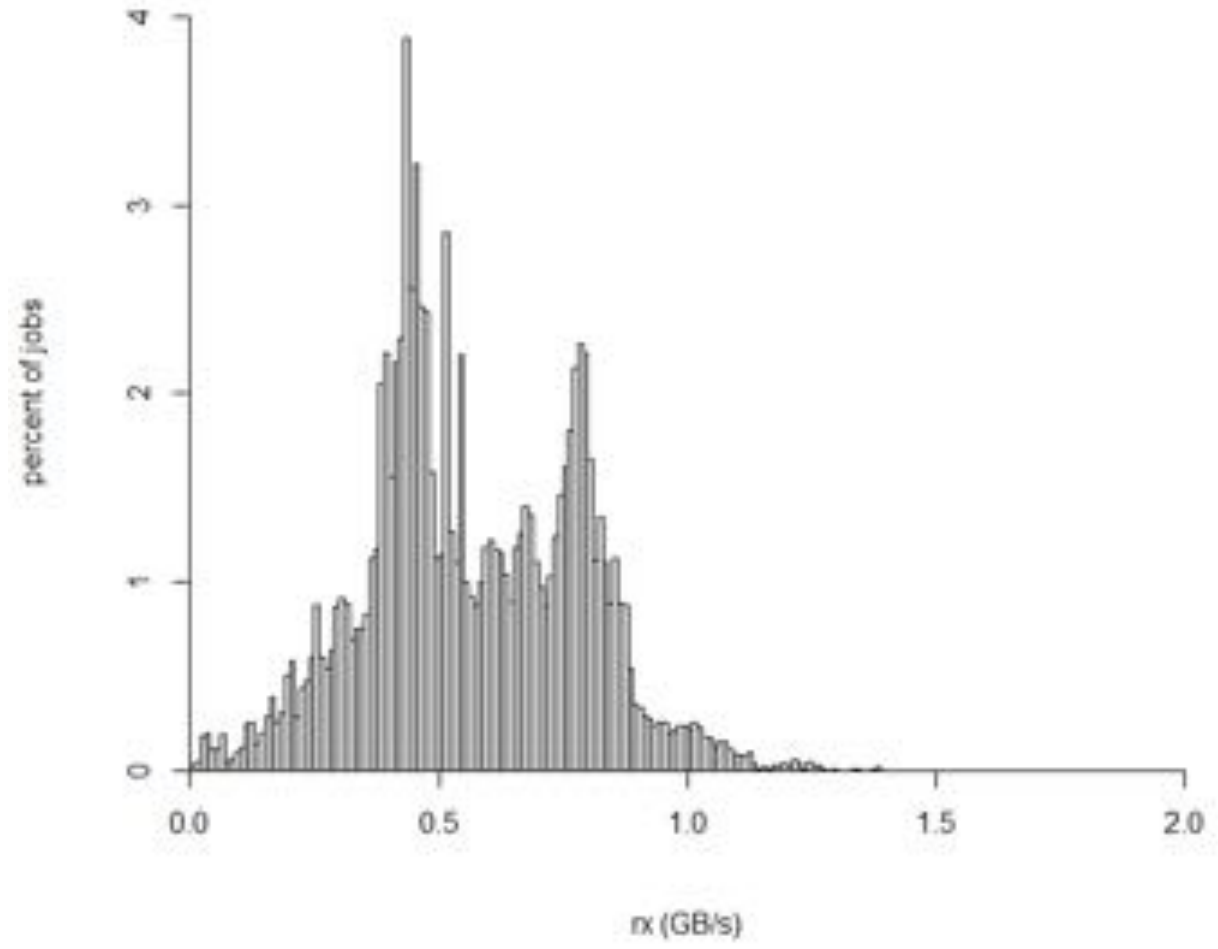

*Figure 2.5-4 NAMD per-node interconnect receive usage broken down by rate.*

The breakdown of interconnect usage exemplified by ADDA and MILC shows a distribution that peaks around 20-30% higher than the overall average, while NAMD is broader with longer tails. Note that these charts include the combination of input/output and inter-process communication, and averages over the collection samples (jobs, 1 minute sampling rate), *hence higher burst rates may well be occurring that are not reflected in the plots*.

## 2.6 Summary: Science and Engineering Fields and Application Areas

Overall the applications run on Blue Waters represent an increasingly diverse mix of disciplines, ranging from broad use of community codes to more specific scientific sub-discipline codes. Optimized library usage, essential for high performance, has been used throughout areas in which such libraries are applicable. Common algorithms (as characterized by Colella's original seven dwarfs) are roughly equally represented aside from unstructured grids (which may reflect the complexity of scaling such applications on systems with a very large number of cores) and Monte Carlo methods (which may be indicative of the relatively smaller number of researchers using such methods, particularly in comparison with those using density functional theory, for example). Interconnect usage by application (with combined input/output and inter-process communication) shows a fairly broad distribution in terms of receive rates.

# 3.0 CONCURRENCY AND PARALLELISM

*BW Analysis Goal 2: What are the top representative algorithms on Blue Waters that consume a majority of the node hours including the use of different types of nodes (XE and XK)?*

- *What is the distribution of job sizes by application and Field of Science (FoS)?*
- *Some sampling and analysis of the communication/compute, IO/compute, and memory/compute ratios for different applications as feasible.*

*BW Analysis Goal 3: How much of Blue Waters is consumed by high throughput (HT) applications, and is this changing over time?*

*BW Analysis Goal 4: Are jobs sizes, in terms of numbers of nodes in a job, changing over time? Are there differences of job size by disciplines/FoS?*

In this section we study trends related to the level of concurrency and types of parallelism achieved by users on Blue Waters. For concurrency we consider the historical trends as well as any differences there are across the various scientific disciplines. For parallelism, we consider the type of parallelism employed (message passing, threading, serial, etc.). We begin by considering how the distribution in job size, weighted by total node hours, has varied over time as shown in Figure 3.0-1. Blue Waters supports a diverse mix of job sizes from single node jobs to jobs that use in excess of 20,000 nodes. During the first year of full service operation, the majority of the usage of Blue Waters was taken up by large jobs greater than 512 nodes[2]. There has been a steady increase in smaller jobs < 64 nodes from less than 2M node hours per quarter in 2013 to approximately 8M node hours Q3 2016, reflecting the projects receiving allocations.

The concurrency of a job is influenced by several factors. One is the ability of the code to efficiently scale to large node counts either in a weak or strong scaling manner. Another factor is the problems being addressed and their size. Other factors include the projects receiving allocations, the amount of allocation the project has, the job wait and turnaround times and other factors. The data shows Blue Waters is capable of, and science teams do regularly run, at very high concurrency. The changes in the amount of time being awarded to projects using modest parallelism is the primary reason the job sizes changes over this time.

[2] 512 nodes is 16,384 integer core-equivalents, which by most other standards is extremely large.

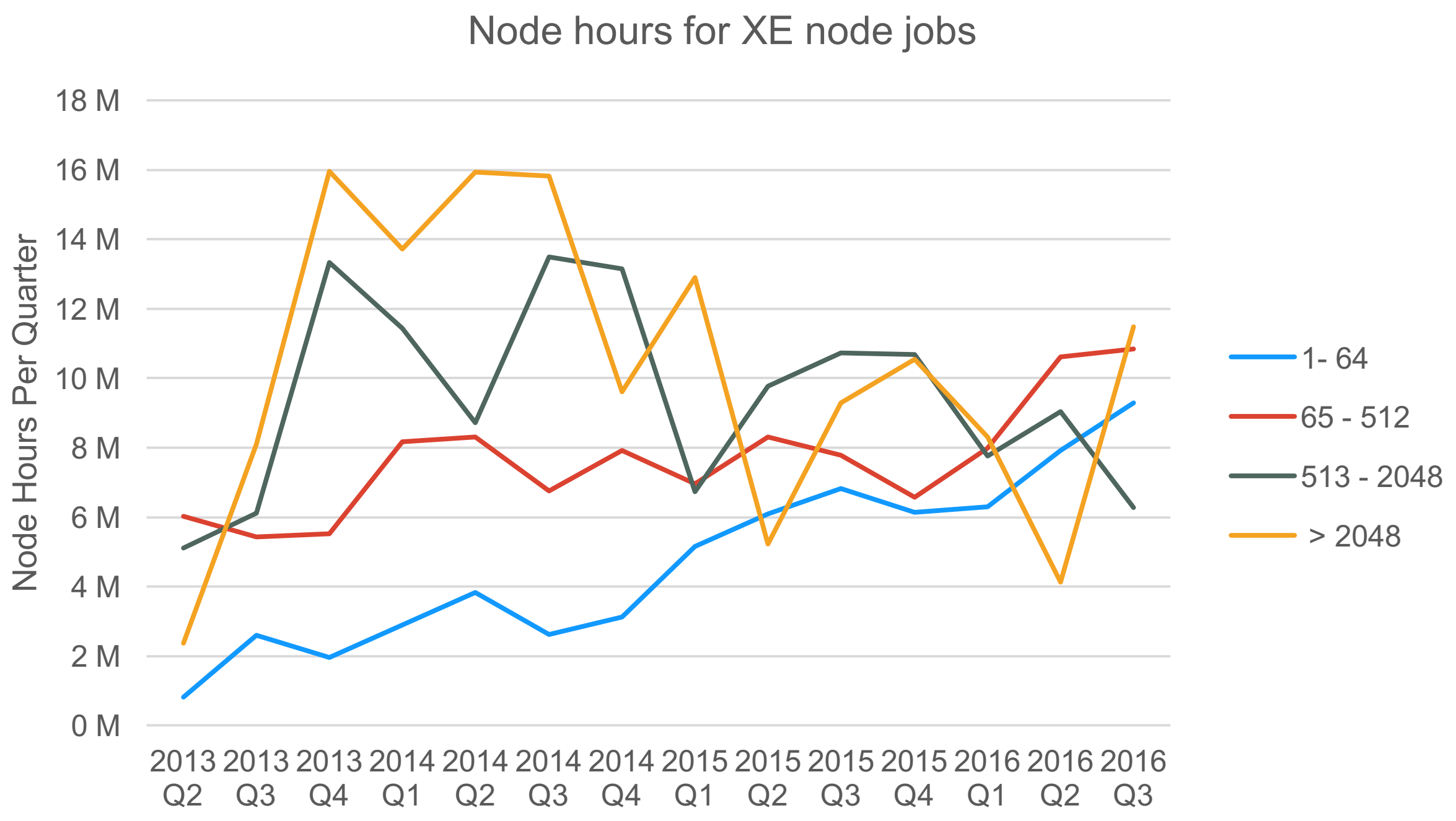

*Figure 3.0-1 Variation in job size over time on Blue Waters.*

Figures 3.0-2 - 3.0-5 show the variation in job size within the various NSF defined parent science areas for XE and XK nodes. For XE node jobs, all of the major science areas (> 1 million node hours) run a mix of job sizes and all have very large jobs (> 4,096 nodes). Some of the science areas, such as Chemical, Thermal Systems, have a much higher percentage of very large jobs (> 4,096 nodes), while others such as Chemistry and Earth Sciences typically run jobs requiring fewer than 256 nodes.

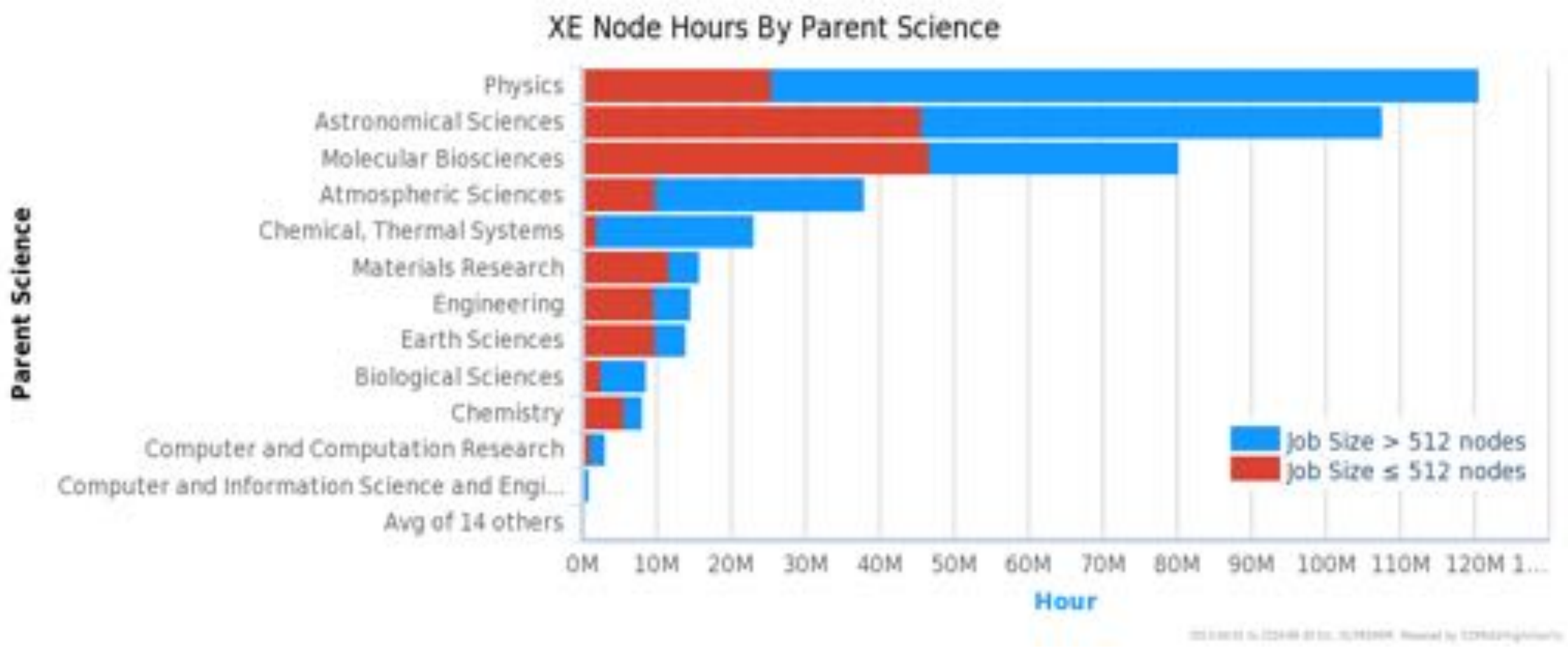

*Figure 3.0-2 Distribution of the size of XE node jobs by parent science over two bin sizes: jobs > 512 nodes and jobs ≤ 512 nodes. The top 12 science areas with the highest usage are shown.*

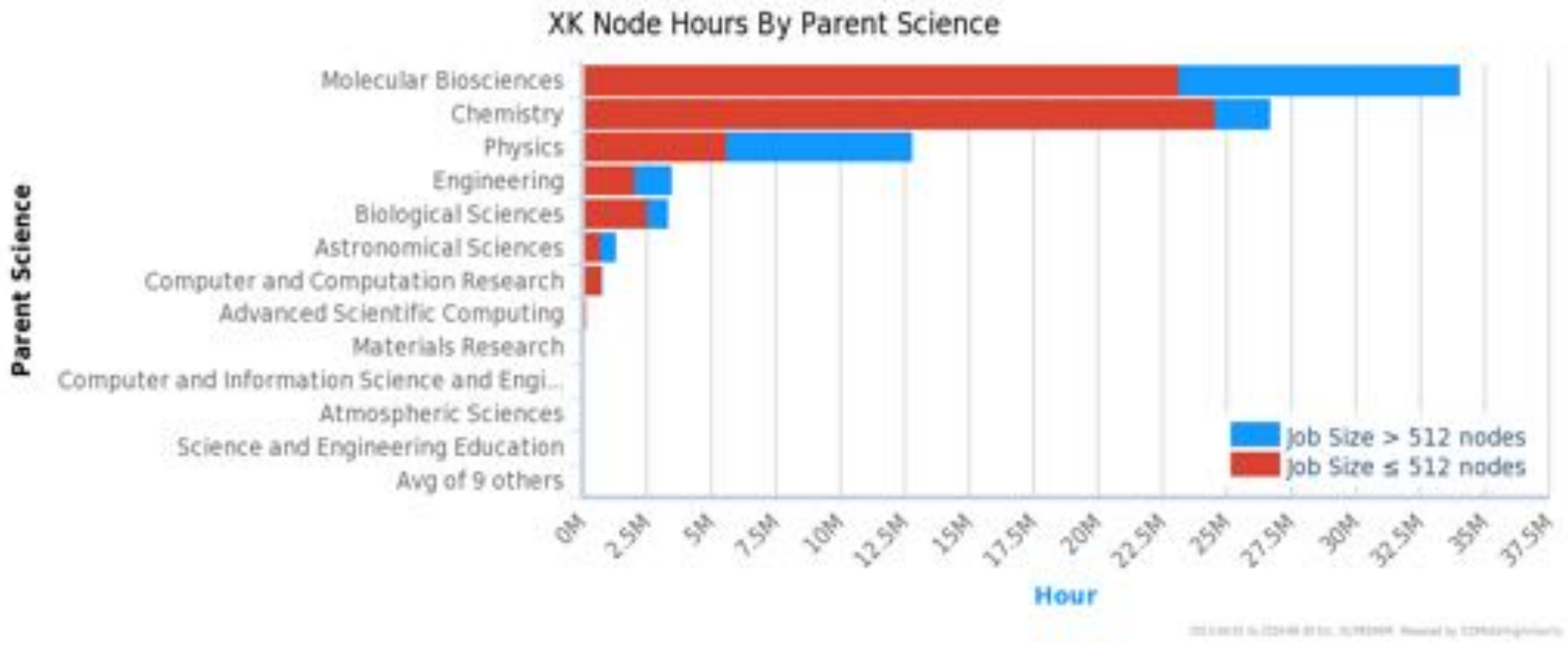

*Figure 3.0-3 Distribution of the size of XK node jobs by parent science over two bin sizes: Jobs > 512 nodes and jobs ≤ 512 nodes. The top 12 science areas with the highest usage are shown. Note the number of XK nodes in Blue Waters is less than 1/6$^{th}$ the total node count so jobs. Hence a 512 node job is using more than 12% of the entire XK region.*

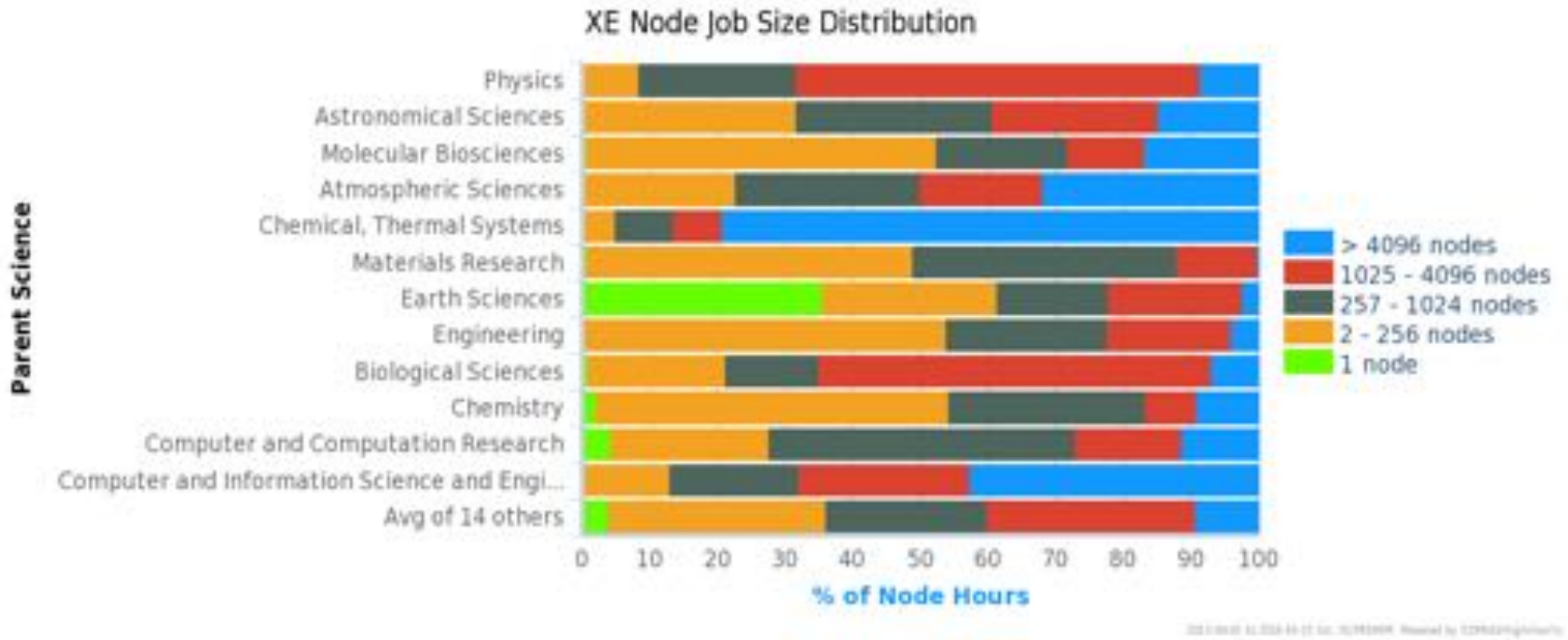

*Figure 3.0-4 Distribution of the size of XE node jobs by parent science in terms of percent of node hours over several bin sizes. There are 22,640 XE nodes available.*

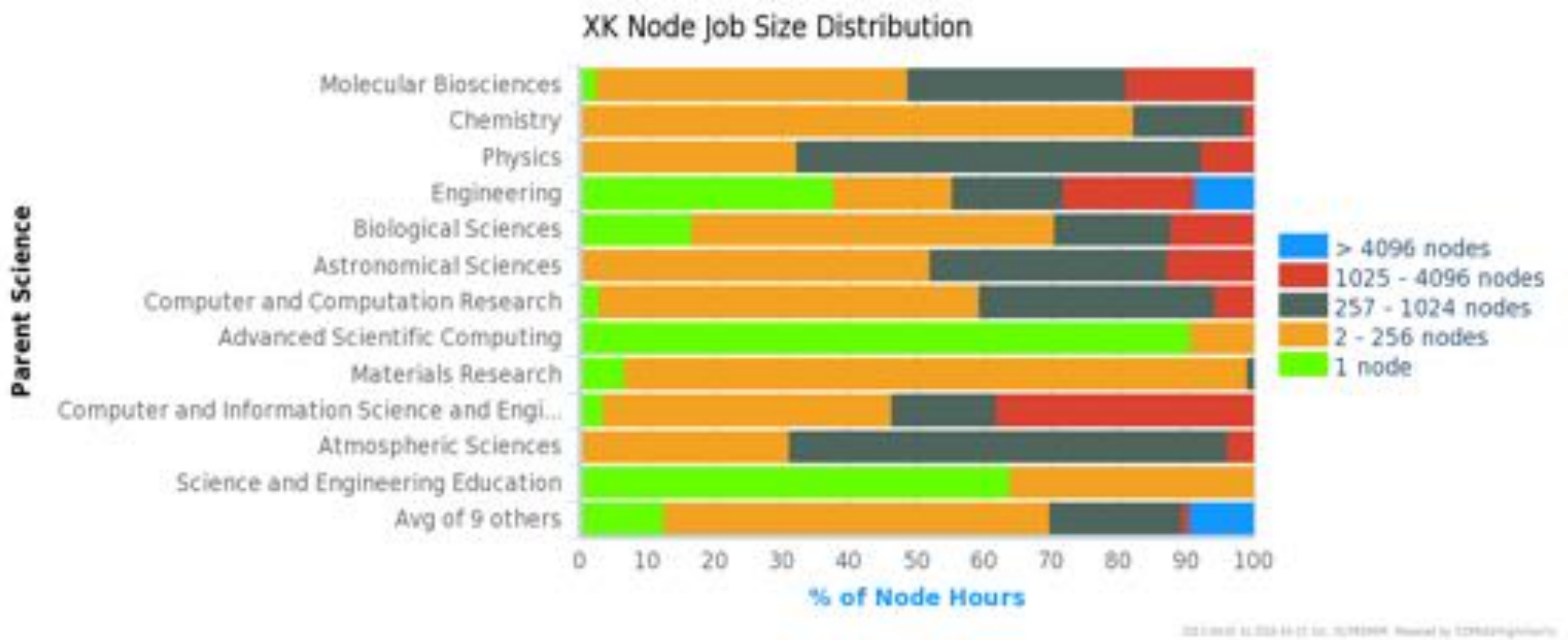

*Figure 3.0-5 Distribution of the size of XK node jobs by parent science in terms of percent of node hours over several bin sizes. There are 4,228 XK nodes available.*

## 3.1 Compute node core usage

The XE nodes have two AMD Interlagos model processors. These processors each have 16 integer cores and 8 floating point units for a total of 32 integer cores and 16 floating point units per compute node. We note that the design of the Interlagos processor is such that floating point intensive software may run more efficiently (i.e. at a higher FLOPS) if only scheduled on half the integer cores. The benefit of this flexibility is that application developers can decide the mode that is most effective for their application. The XK nodes are populated with a single AMD Interlagos processor and one NVIDIA GK110 "Kepler" accelerator K20X card. In this document the term 'CPU core' or 'core' refers to an integer core (32 per XE node, 16 per XK node).

The CPU core usage on the compute nodes was estimated by using the values of the hardware counter that records the number of core clock ticks. These values are read from the model specific registers (MSRs) on the processors. We infer that an integer core was in use if the clock tick rate was above a certain threshold[3].

Figures 3.1-1 and 3.1-2 show the utilization of the cores on nodes for XE and XK nodes respectively. In the case of XE nodes, the majority of node hours were spent utilizing all 32 available cores with a secondary peak at 16 cores which is likely due to users explicitly choosing to run with a 1:1 ratio of floating point units to processes (or threads).

On the XK nodes, the bulk of the floating point processing power is in the GPU device. The GPU usage is discussed in section 3.2 below. The majority of node hours use only a single CPU core, which is likely due to the software application designed to do the bulk of the processing on the GPU. The secondary peaks at eight and nine cores are almost entirely due to the CUDA version of the NAMD software. This software is able to use both the GPU and CPU cores.

[3] The clock tick rate threshold is an average of 100MHz over the job. This value was chosen by manual inspection of the clock tick data for jobs with known CPU core usage.

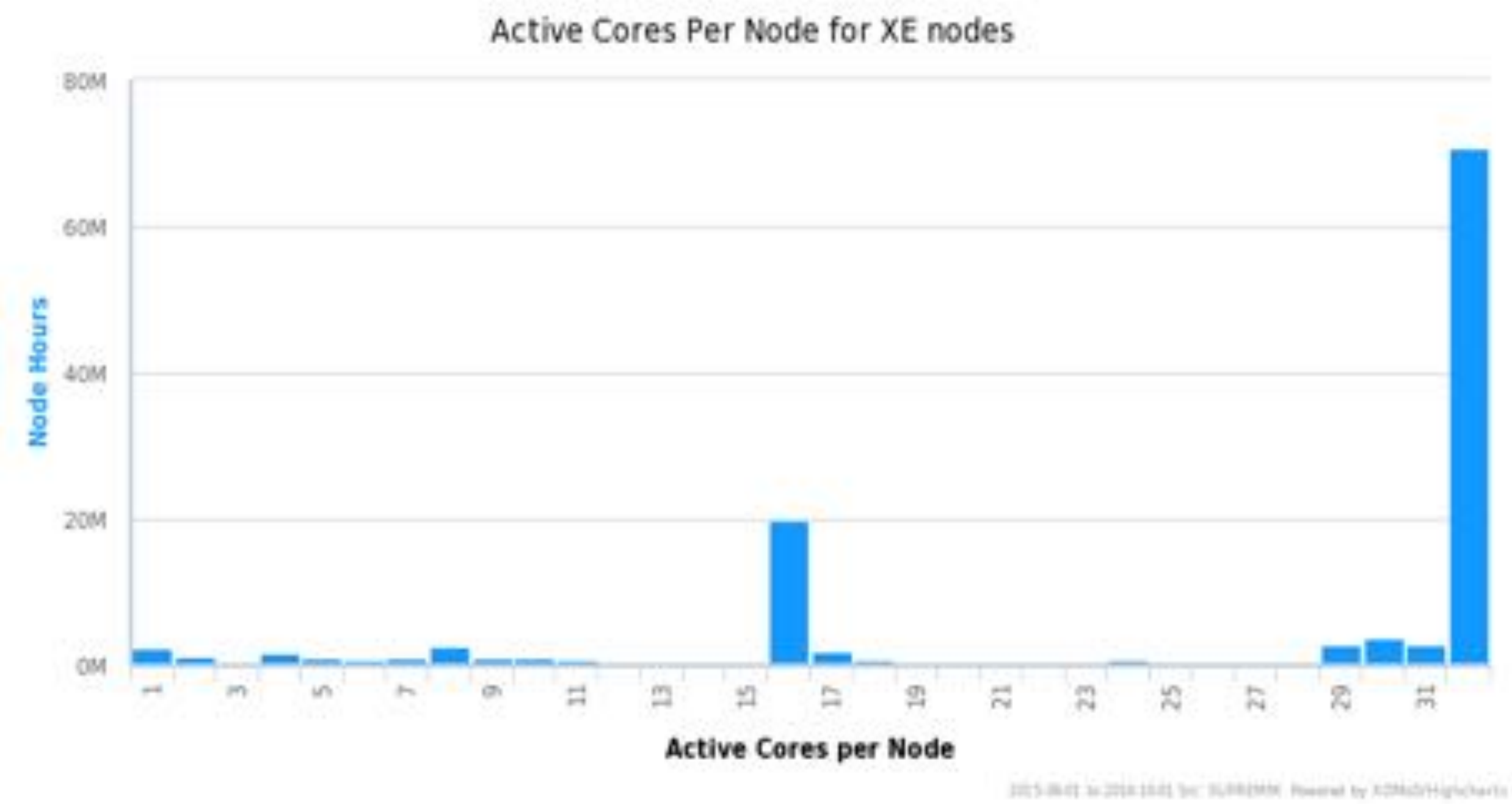

*Figure 3.1-1 Distribution of cores utilized per XE node by total node hours from 2015-09-01 to 2016-09-30. There were 25 million node hours during this period with no MSR data (not shown in this figure).*

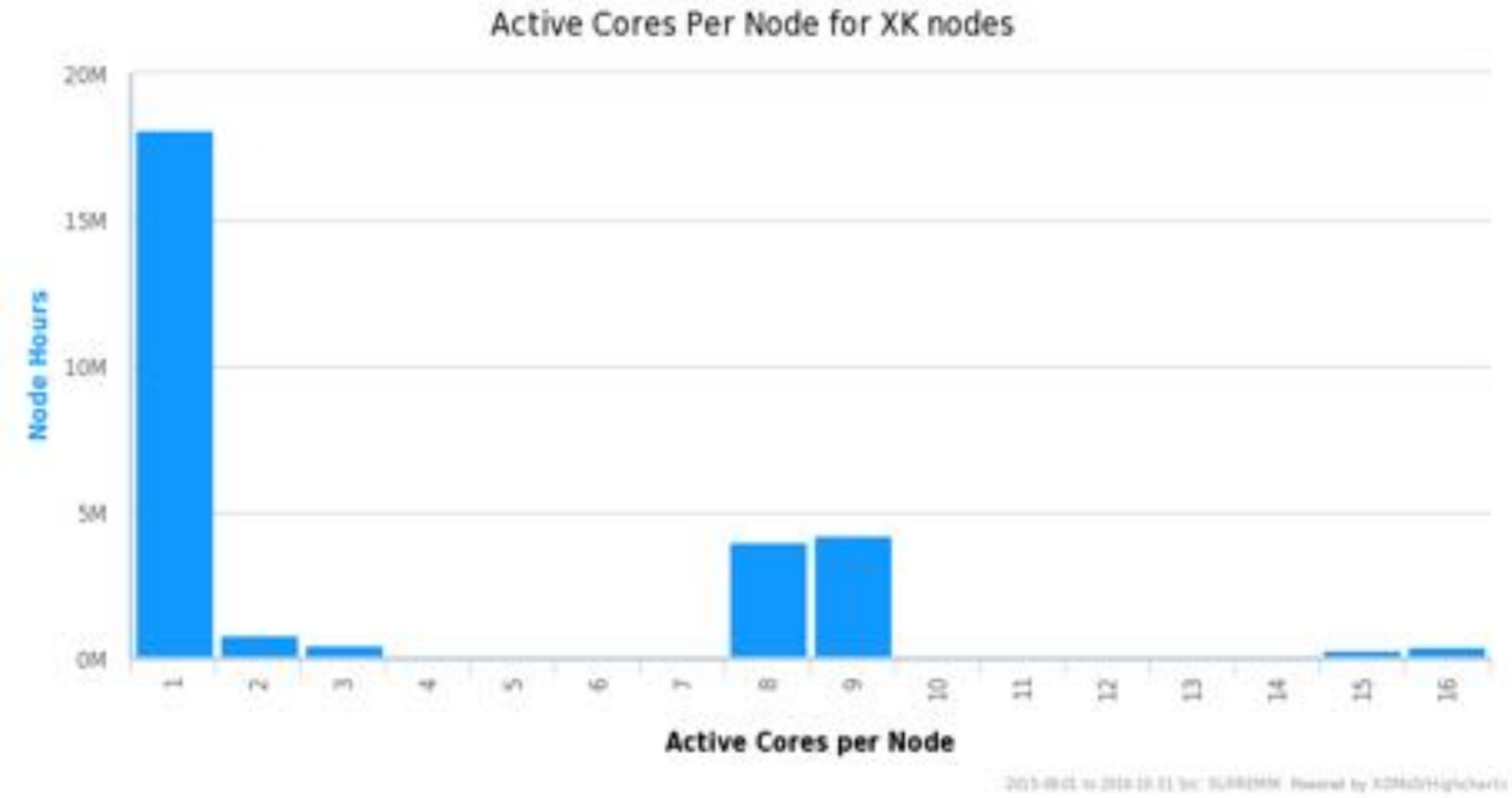

*Figure 3.1-2 Distribution of cores utilized per XK node by total node hours from 2015-09-01 to 2016-09-30. There were 1.3 million node hours during this period with no MSR data (not shown in this figure).*

We categorized the jobs by concurrency type. The four categorized types are defined below:

- message passing: A job is categorized as message passing if it used aprun to launch multiple processing elements and the number of active CPU cores was equal to the number of processing elements.
- message passing + threads: A job is categorized as message passing + threads if it used aprun to launch multiple processing elements and the number of active CPU cores was greater than the number of processing elements.
- serial: A job is categorized as serial if aprun was used to launch a single processing element. (Multiple apruns may run in the same ‘serial’ job).
- threads: A job is categorized as thread if aprun was used to launch a single processing element and the aprun options specify that multiple cores be reserved for the processing element. (Multiple apruns may run in the same ‘threads’ job).

The two unknown types are:

- NA: The aprun data is unavailable for the job.
- message passing + unknown: The aprun data indicates that multiple processing elements were launched with multiple cores per processing element. However, the MSR data is unavailable so it is not possible to determine whether the processing elements spawned threads or were just sparsely allocated.

We note that this simple classification algorithm has the potential to misidentify certain jobs. For example, it is possible to use aprun to launch multiple processing elements that do not have any interprocess communication. We assume that this situation is rare based on our knowledge of the mix of applications that we know run on Blue Waters.

Looking at the types of parallelism employed by Blue Waters users, we see in Figure 3.1-3 and Figure 3.1-4 that the majority of jobs, based on node hours, run multiple processing elements per job. The ratio of jobs in each category appears to be approximately stable over time. The jobs categorized as ‘threads’ in the period from 2015-10-01 to 2016-07-01 are mostly due to a single project. Figure 3.1-4 shows that overall, the majority of the node hours, ~64%, are message passing with ~26% of node hours spent running jobs that use message passing and threads. Please note that message passing in this context is inferred from aprun tasks, and includes mpi, CAF, charm++, etc. Also, threads can be implemented in a number of ways with two of the most common are OpenMP and pthreads.

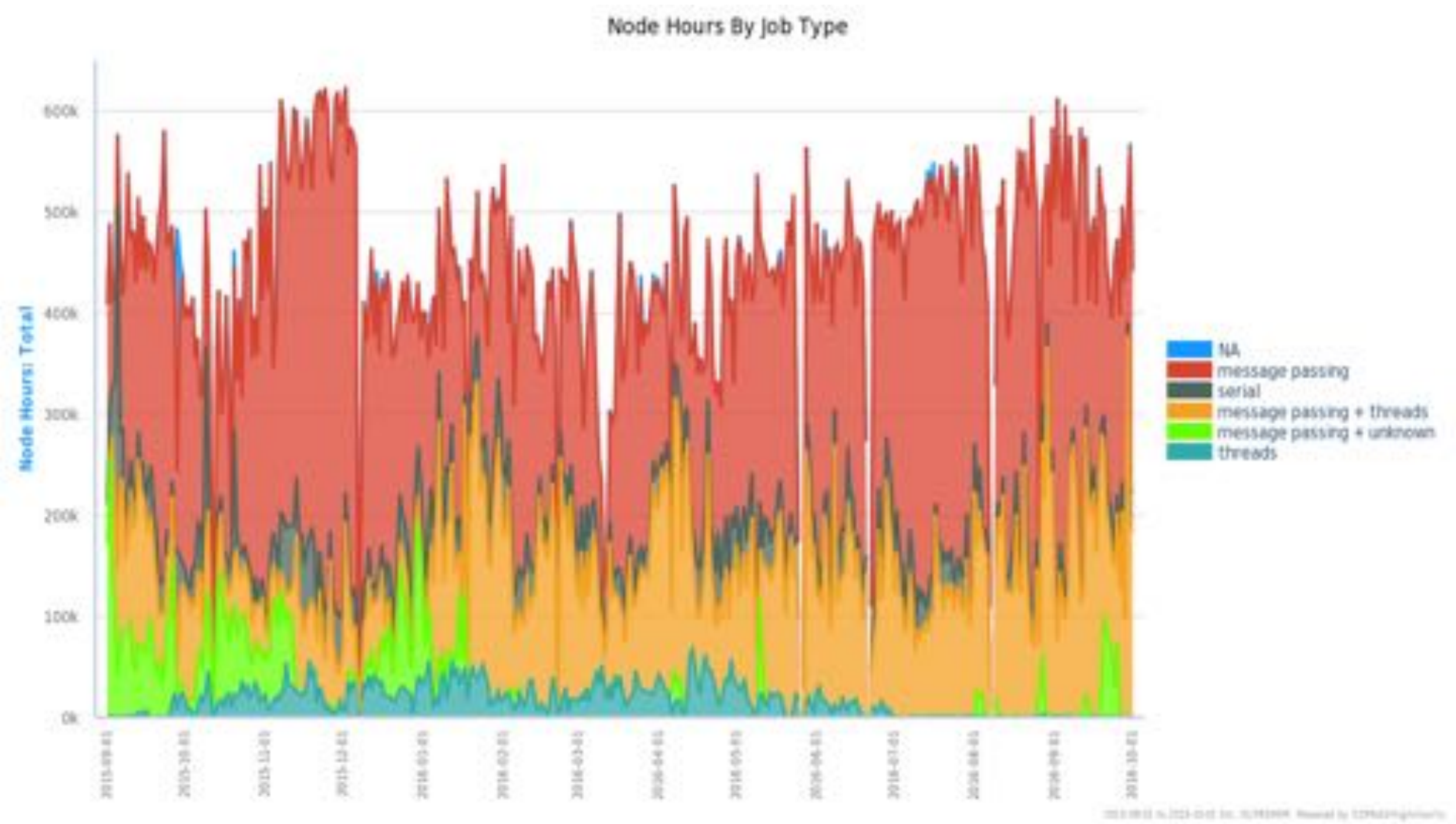

*Figure 3.1.3 Historical analysis of the type of parallelism employed on Blue Waters from 2015-09-01 to 2016-09-30, using a stacked chart. The NA category indicates that aprun information for the job is not available.*

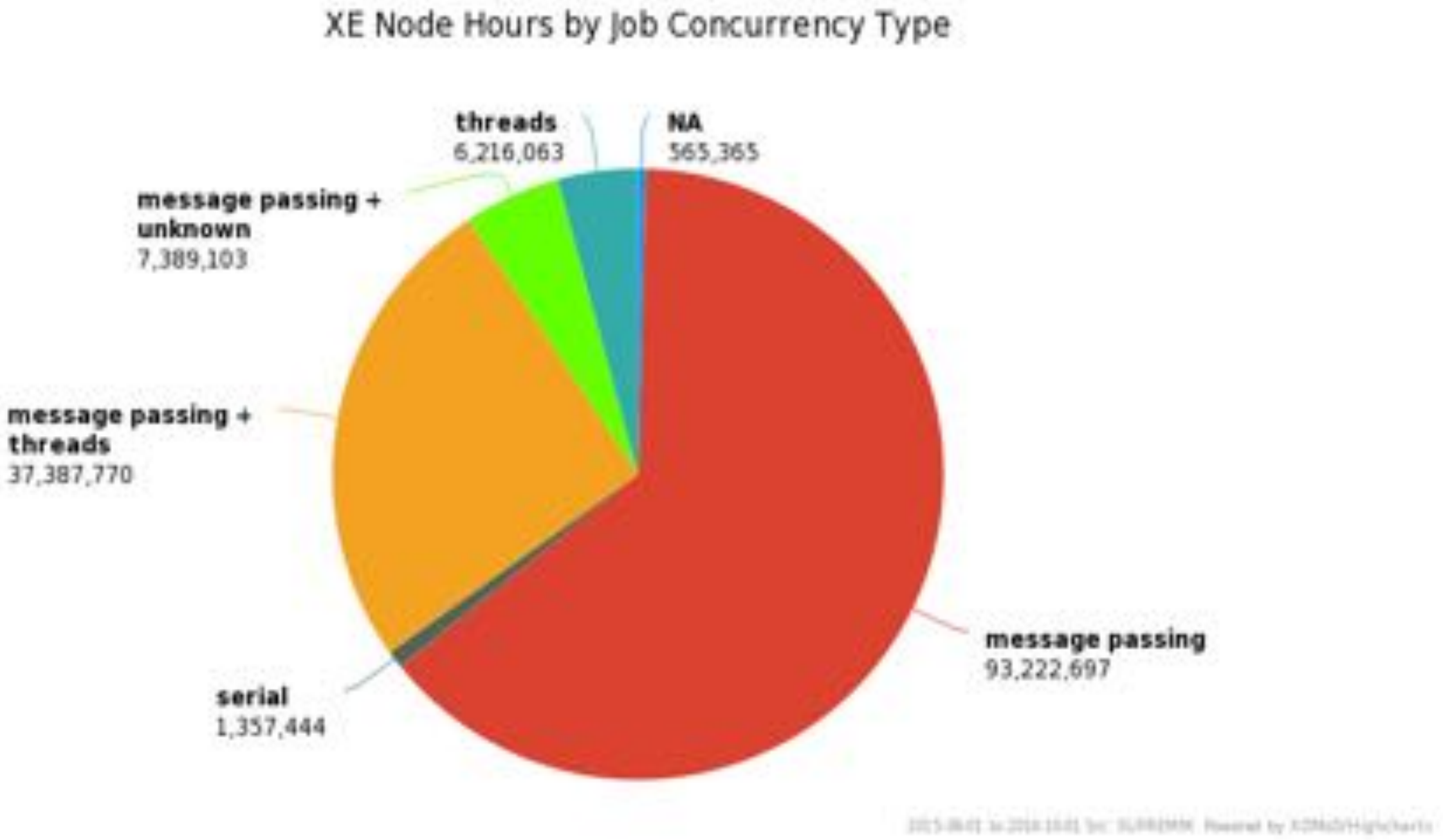

*Figure 3.1-4 Percentage of job by type of parallelism employed on Blue Waters from 2015-09-01 to 2016-09-30. The NA category indicates that aprun information for the job is not available.*

Given that the majority of the XE node jobs are parallel jobs that use message passing or message passing with threads, the breakdown of parallelism by parent science as shown in Figures 3.1.5 not surprisingly shows a preponderance of that type of parallelism across almost all fields of science. The jobs run under the Earth Sciences category show a much higher percentage of non-parallel threaded application code usage than the other sciences. This is due to a single project that runs a multithreaded single node code.

The same information for XK node jobs is shown in Figure 3.1.6. There is overall a much smaller relative usage of parallel codes on this hardware.

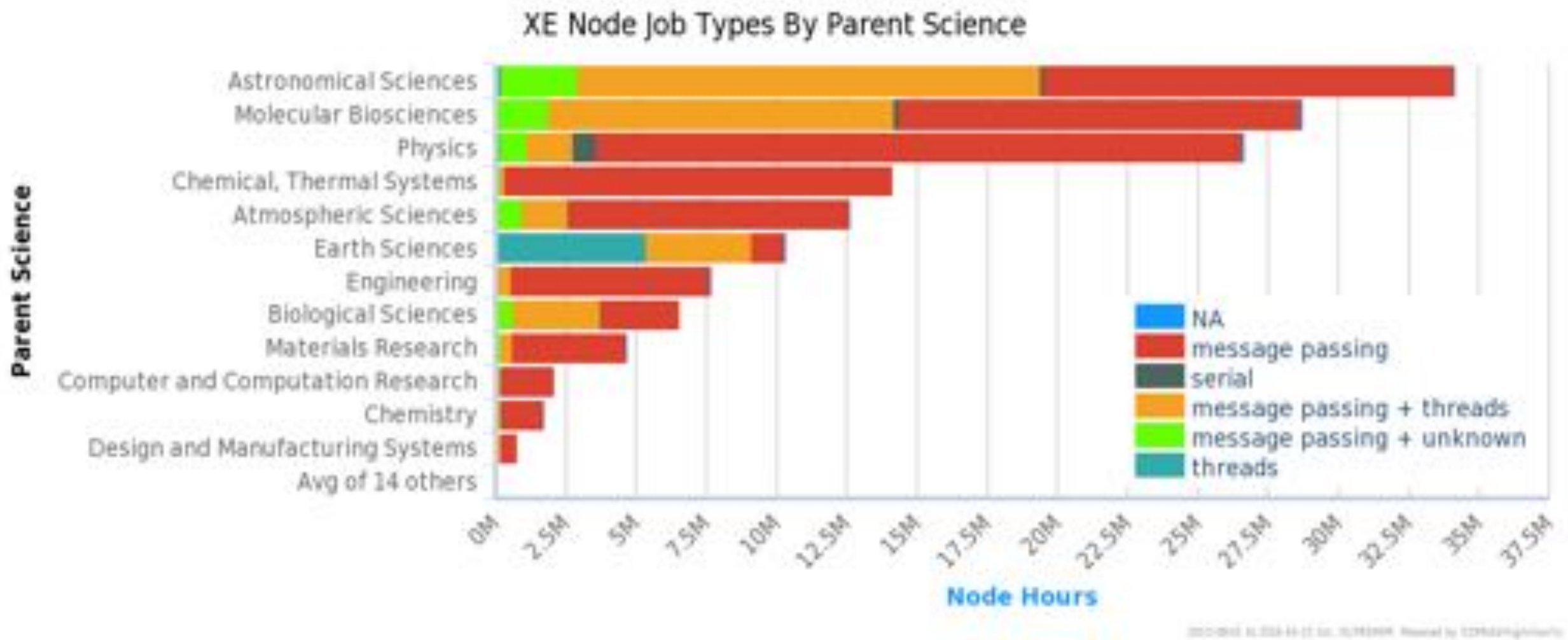

*Figure 3.1-5 Breakdown of the job parallelism types by parent science for jobs that ran on XE nodes between 2015-09-01 and 2016-09-30.*

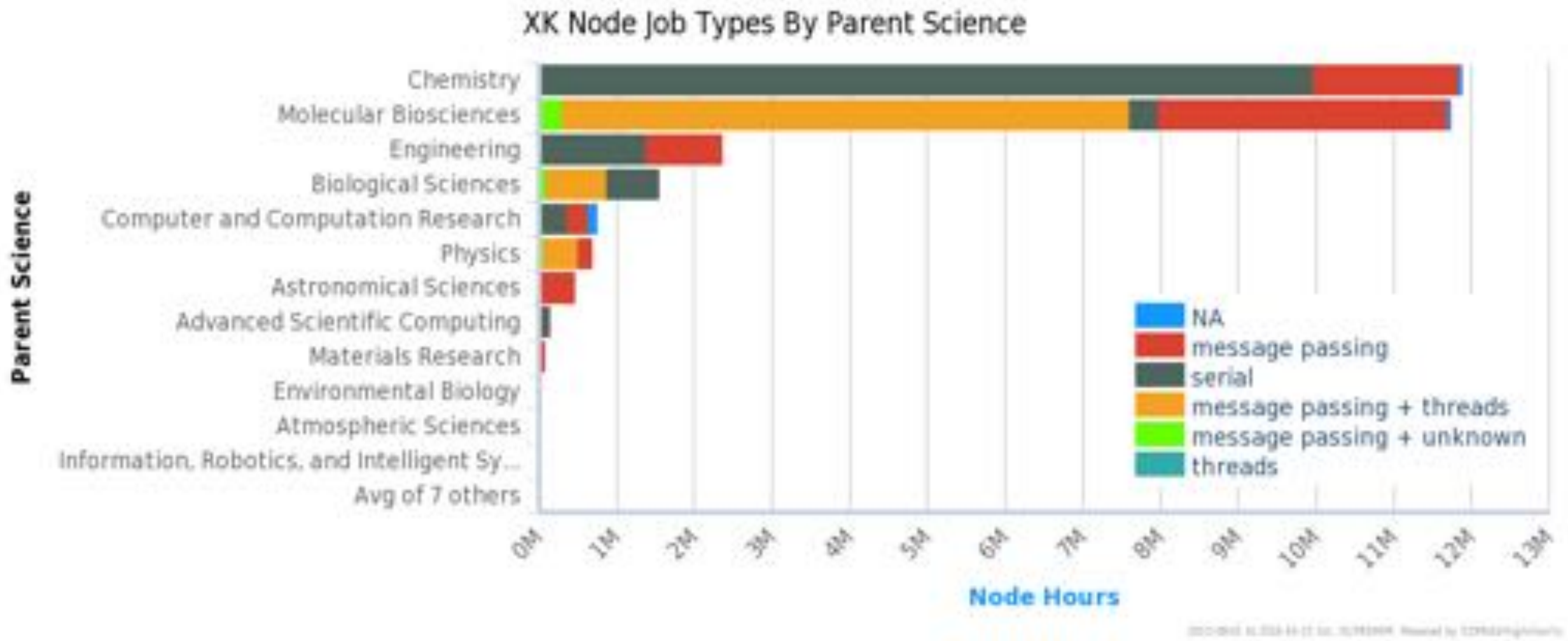

*Figure 3.1-6 Breakdown of the job parallelism types by parent science for jobs that ran on XK nodes between 2015-09-01 and 2016-09-30.*

Figure 3.1-7 breaks down the parallelism by job size, showing that very large jobs primarily use message passing. The majority of single node jobs use a single process with multiple threads to utilize the CPU cores.

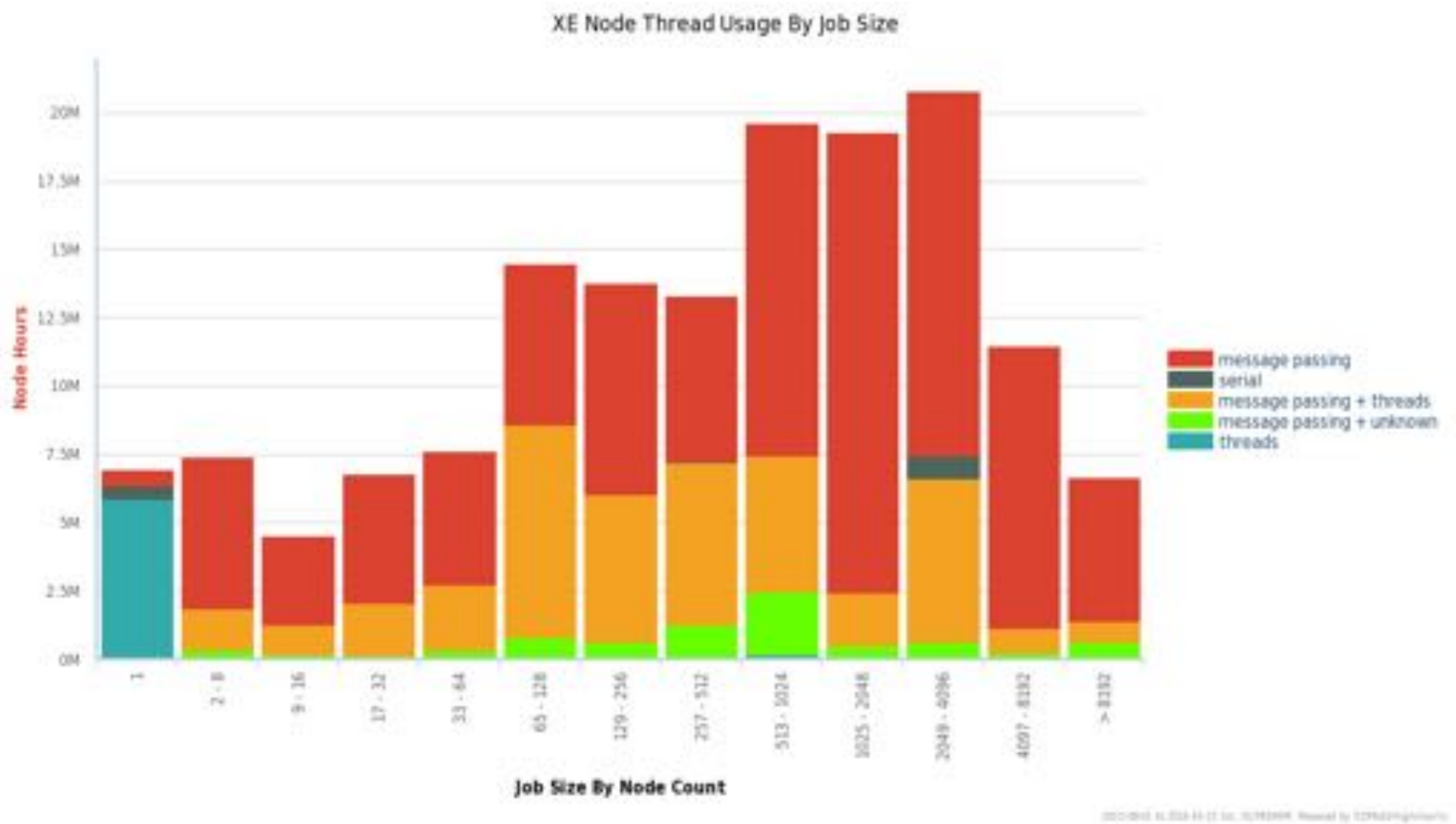

*Figure 3.1-7 Thread usage of XE node jobs by job size for jobs that ran between 2015-09-01 and 2016-09-30.*

Figure 3.1-8 shows the historical utilization of XE cores over time. The majority of codes utilize all 32 integer cores per node. There does not appear to be any significant variation in usage over time. However, the relatively short time period that we have core usage data means that we cannot draw any conclusions about the trend over the full life of the machine.

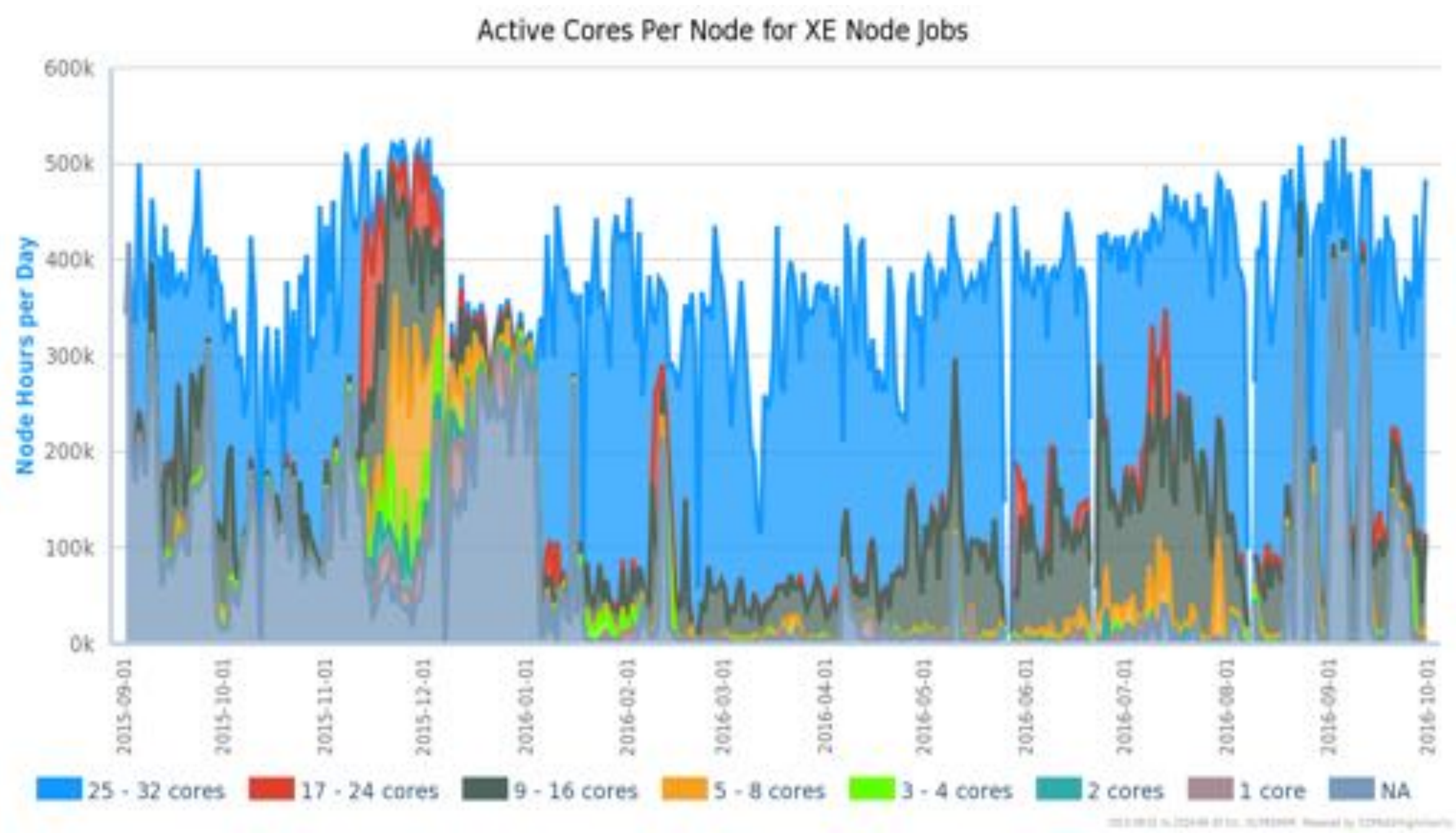

*Figure 3.1-8 Stacked time series plot of node hours for XE node jobs grouped by the number of active cores.*

Figure 3.1-9 shows the distribution of thread counts for XE node jobs. These data were calculated from the ratio of the number of active CPU cores to the number of processing elements. The most popular thread count for multi-threaded jobs is 8 threads per processing element. This is likely due to users running one processing element per non-uniform memory access (NUMA) node.

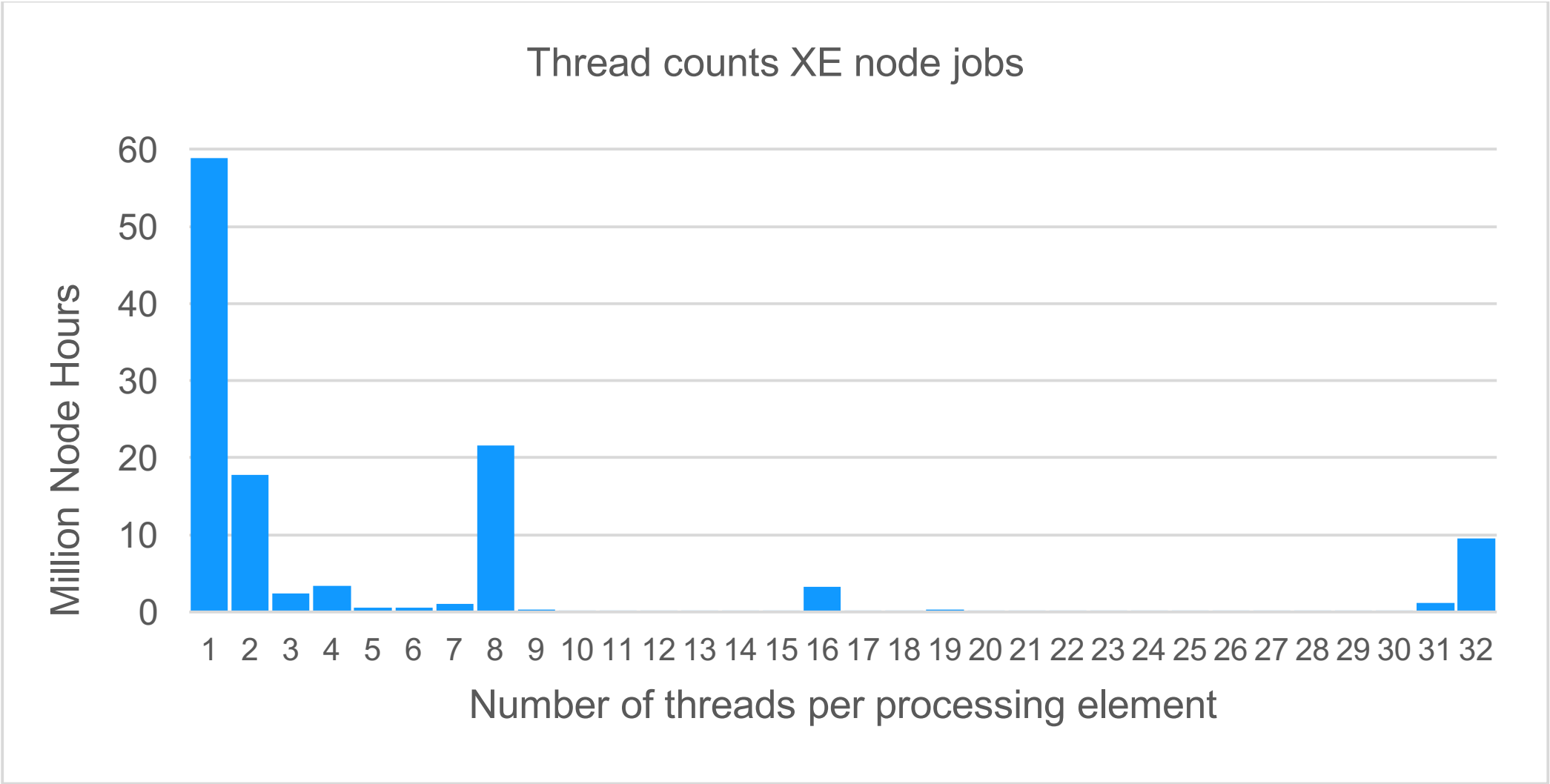

*Figure 3.1-9 The number node hours binned by the number of integer cores active per processing element for jobs running on XE nodes between 2015-09-01 and 2016-09-30. This plot only includes the data for jobs where the MSR data was available.*

Blue Waters uses the aprun utility to launch applications on compute nodes. Each job may run one or more aprun commands and multiple apruns may run concurrently. We analysed the aprun logs to determine how jobs were run and categorized them as follows:

- Single aprun: One apun command was used in the job.
- Sequential apruns: Multiple aprun commands were executed one after the other.
- Concurrent apruns: At some point during the job there was more than one aprun command running on the compute nodes.

Tables 3.1-1 and 3.1-2 show the breakdown of job usage by job launch type for XE and XK node jobs respectively. The majority of node hours are spent with jobs that use a single aprun command to launch the software.

*Table 3.1-1 Breakdown of job usage by job launch type for XE nodes.*

| Job type | Job count | Node hours |
|---|---:|---:|
| Single aprun | 4369222 | 483 M |
| Sequential apruns | 1731556 | 53.1 M |
| Concurrent apruns | 21248 | 65.7 M |

*Table 3.1-2 Breakdown of job usage by job launch type for XK nodes.*

| Job type | Job count | Node hours |
|---|---:|---:|
| Single aprun | 746777 | 65.2 M |
| Sequential apruns | 97041 | 5.25 M |
| Concurrent apruns | 37694 | 43.3 M |

## 3.2 GPU Utilization

We collect GPU utilization information as reported by the NVIDIA driver and collected by the LDMS system. Overall GPU utilization weighted by node hour on XK nodes oscillates around 55%, for those jobs that use the GPU. Aggregate GPU utilization by application and parent science shows higher utilization for several categories.

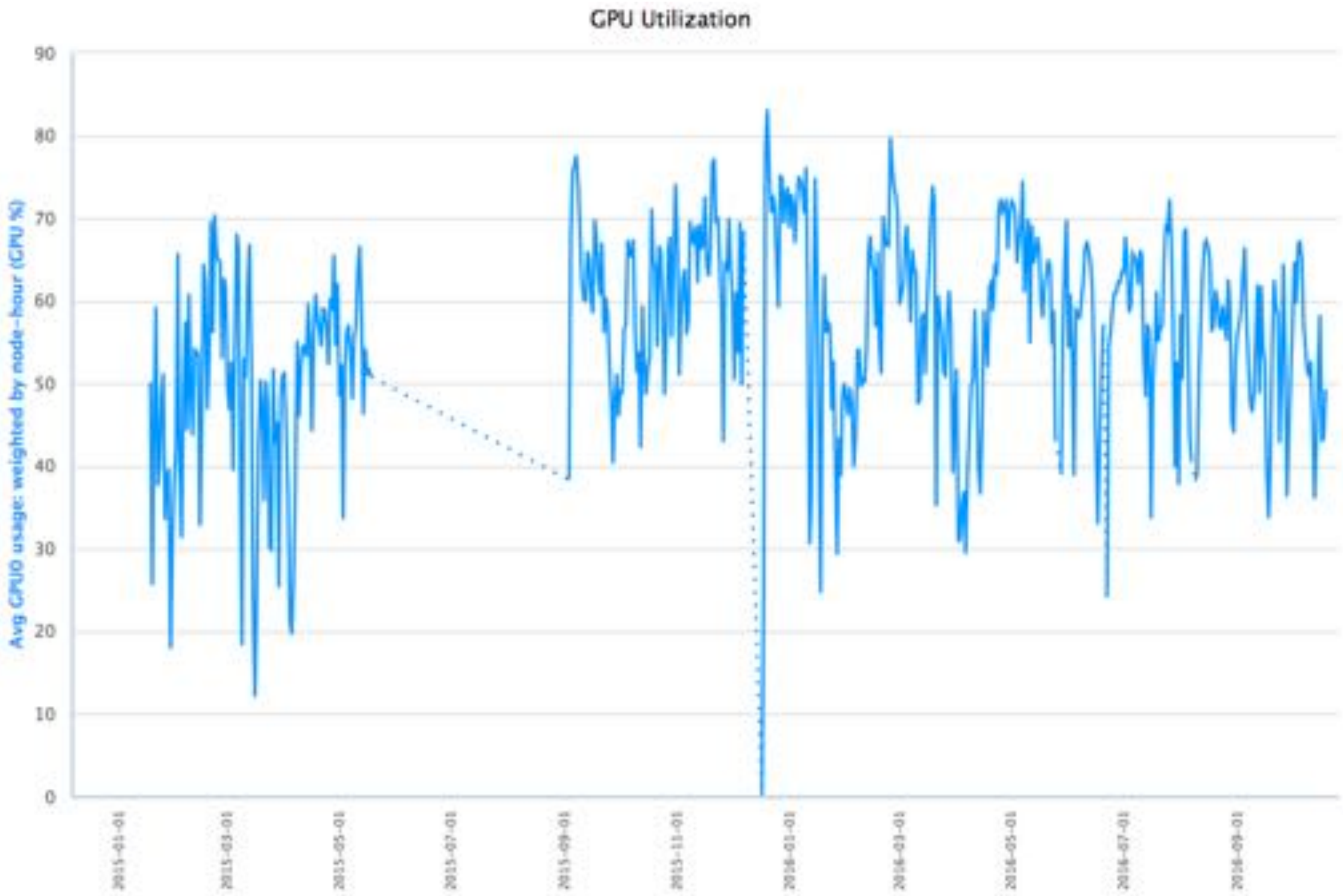

*Figure 3.2-1 Historical analysis of GPU activity over time on jobs using XK nodes.*

Figure 3.2-1 shows GPU utilization on the XK nodes on Blue Waters over time. Data is available starting from 2015-01-18. On the plot, there are several other areas denoted by a dashed line where data is not available.

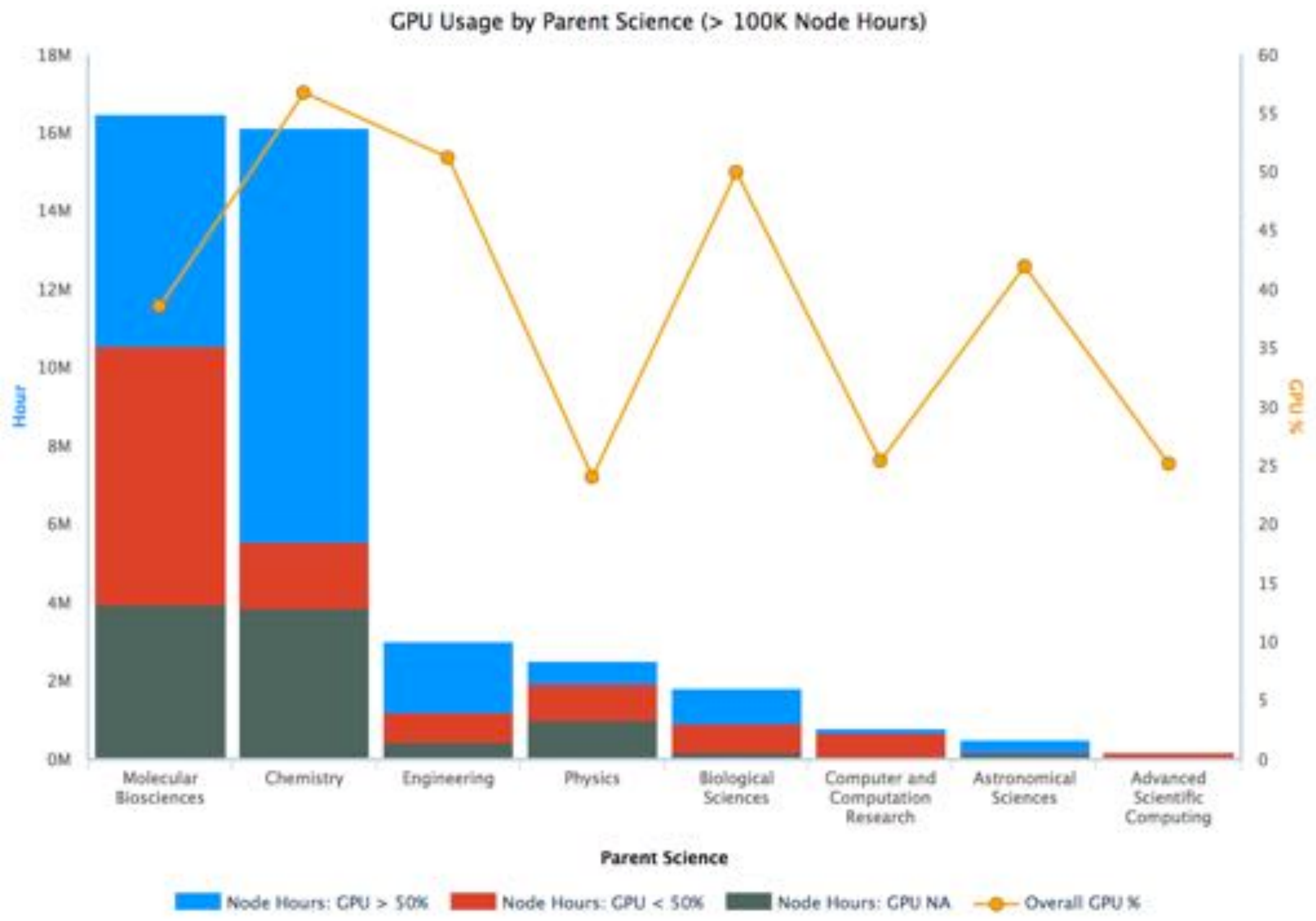

*Figure 3.2-2 Aggregate GPU utilization by parent science for jobs using XK nodes.*

Figure 3.2-2 shows GPU utilization as broken down by parent science on XK nodes. The plot contains all parent science categories that consumed > 100K node hours on the XK nodes during the life of the machine. Each bar has 3 categories to denote the extent of the GPU usage by that grouping. GPU NA portion of the bar indicates that raw GPU counter information was not available, due to the data coverage described in Figure 3.2-1. The red areas indicate the fraction of node hours where the GPU was <50% utilized as reported by NVIDIA's driver. The blue areas indicate where the GPU was >50% utilized. The orange line represents the overall node hour weighted average GPU utilization for the parent science.

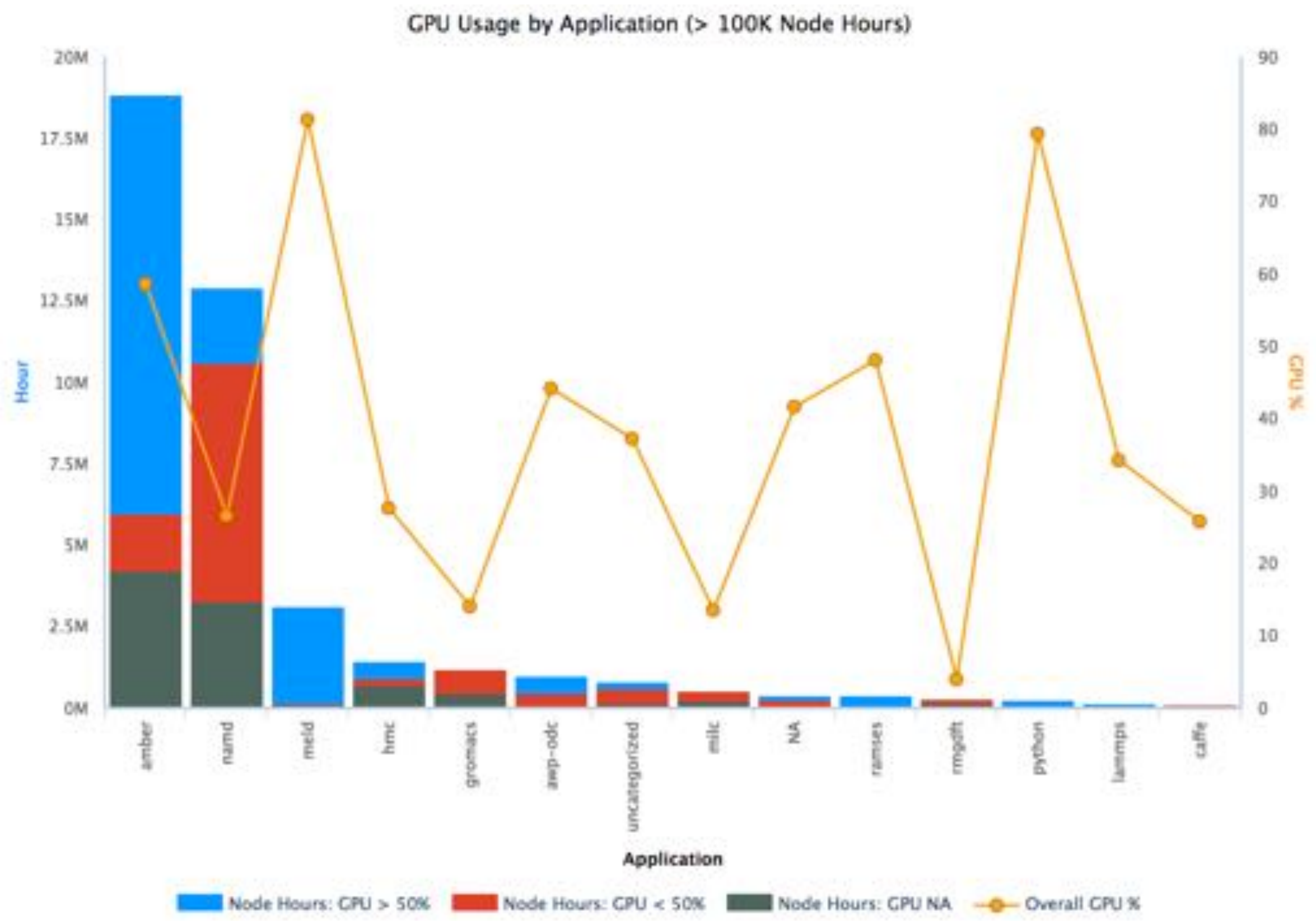

*Figure 3.2-3 Aggregate GPU utilization by application on XK nodes*. NA for the application name means that the executable name was not available (likely due to missing aprun data).

Figure 3.2-3 shows GPU utilization as broken down by application on XK nodes. The plot contains all applications that consumed > 100K node hours on the XK nodes during the life of the machine. Each bar has 3 categories to denote the extent of the GPU usage by that grouping. GPU NA indicates that raw GPU counter information was not available, due to the data coverage described in Figure 3.2-1. The red areas indicate the fraction of node hours where the GPU was <50% utilized. The blue areas indicate where the GPU was >50% utilized. The orange line represents the overall node hour weighted average GPU utilization for the application.

## *3.3 High Throughput Applications*

The data collected on Blue Waters does not allow us to directly identify high throughput (HT) applications, however we can categorize jobs based on the job concurrency and the application. We therefore use the following criteria to identify XE node jobs that may be HT application usage:

- Jobs that used a single compute node;
- Jobs that had concurrent apruns, each of which used a single compute node;
- Jobs that used the aggregate job launcher software scheduler.x (https://github.com/ncsa/Scheduler).

Overall there is very little relative usage of these single node jobs on Blue Waters. Jobs that used a single compute node jobs used approximately 7.9 million node hours, concurrent single node jobs used a few hundred thousand node hours and jobs that used an aggregate launcher used approximately 1.6 million node hours.

Figures 3.3-1 and 3.3-2 show the breakdown of these jobs by Parent Science area and by application respectively. The setsm and gdal applications are used by projects in the Earth Sciences and are primarily single node jobs.

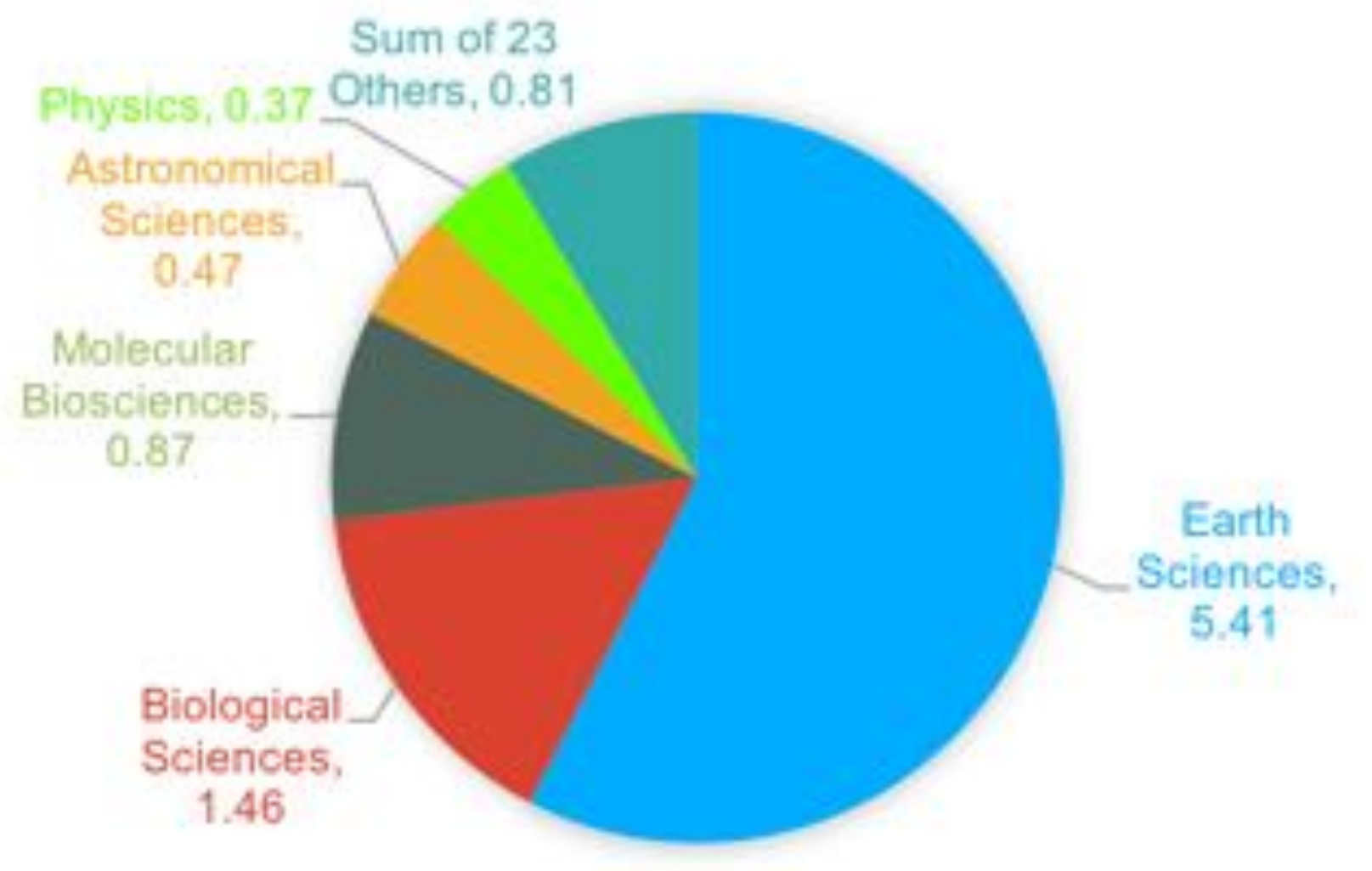

*Figure 3.3-1 Breakdown of job usage by parent science in millions of node hours for jobs where all aprun commands used a single node or an aggregate job launcher was used.*

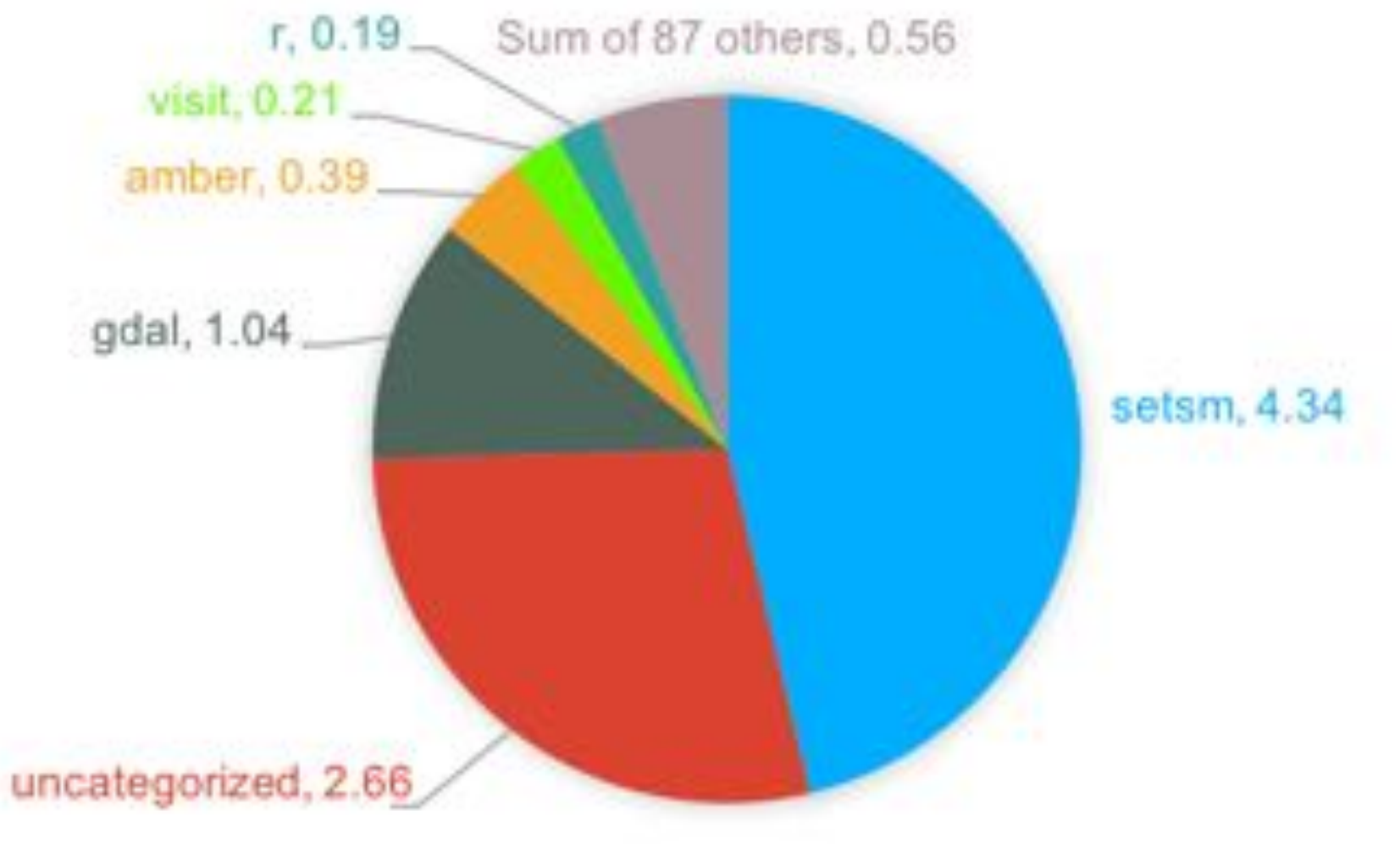

*Figure 3.3-2 Breakdown of job usage by application in millions of node hours for jobs where all aprun commands used a single node or an aggregate job launcher was used. The jobs that used an aggregate job launcher are shown in the "uncategorized" segment.*

## 3.4 Summary: Concurrency and Parallelism

Blue Waters supports a diverse mix of job sizes from single node jobs to jobs that use in excess of 20,000 nodes for a single run. The patterns of usage differ between the XE and XK nodes. For XE node jobs, all of the major science areas (> 1 million node hours) run a mix of job sizes and all have very large jobs (> 4,096 nodes). The relative proportions of job size vary between different parent science areas.

The majority of XE node hours on the system are spent running parallel jobs that use some form of message passing for inter-process communication. At least 25% of the workload uses some form of threading, however the larger jobs (> 4,096 nodes) mostly use message passing with no threading.

There is no obvious trend in the variation of thread usage over time, however, since the thread usage information is only available for a single year we cannot draw any conclusions about the trend over the full life of the system.

Overall, the core usage in the XE nodes is high with the majority of node hours spent using all 32 integer cores per node.

The XK nodes show a different usage pattern from the XE nodes. Although the XK nodes are used by a range of applications and fields of science, the usage is dominated by two main applications, NAMD and AMBER. The AMBER jobs use a single CPU core per XK node and do the majority of the computation on the GPU device. The NAMD jobs typically use eight or nine CPU cores per XK node in addition to the GPU device.

Single node jobs, some of which may be attributable to high throughput computing, represent a small fraction (less than 2%) of the total node hours consumed on Blue Waters.

# 4.0 XK AND XE USAGE

*BW Analysis Goal 1: What is the proportional mix of sub-disciplines and how are the proportions growing/shrinking over the life-time of Blue Waters, including the use of different types of nodes (XE and XK)?*

*BW Analysis Goal 2: What are the top representative algorithms on Blue Waters that consume a majority of the node hours including the use of different types of nodes (XE and XK)?*

- *What is the distribution of job sizes by application and Field of Science (FoS)?*
- *Some sampling and analysis of the communication/compute, IO/compute, and memory/compute ratios for different applications as feasible.*

## 4.1 XE and XK Utilization

The daily fractional utilization of the XE and XK nodes was computed, see Figures 4.1-1 and 4.1-2. In these figures a simple 30 day centered moving average is used to guide the user in the interpretation of the daily utilization. Note that on January 13, 2015 Blue Waters changed to a topology aware scheduler [12]. This resulted in a major improvement in application performance, as has been documented, at the cost of the small difference in average throughput that is observable in comparing 2013-2014 utilization to that of 2015-2016 in Figures 4.1-1 and 4.1-2. There were a couple of periods when the XK node utilization dropped off, but otherwise there seems to be a fairly uniform pattern of usage of the XE and XK nodes.

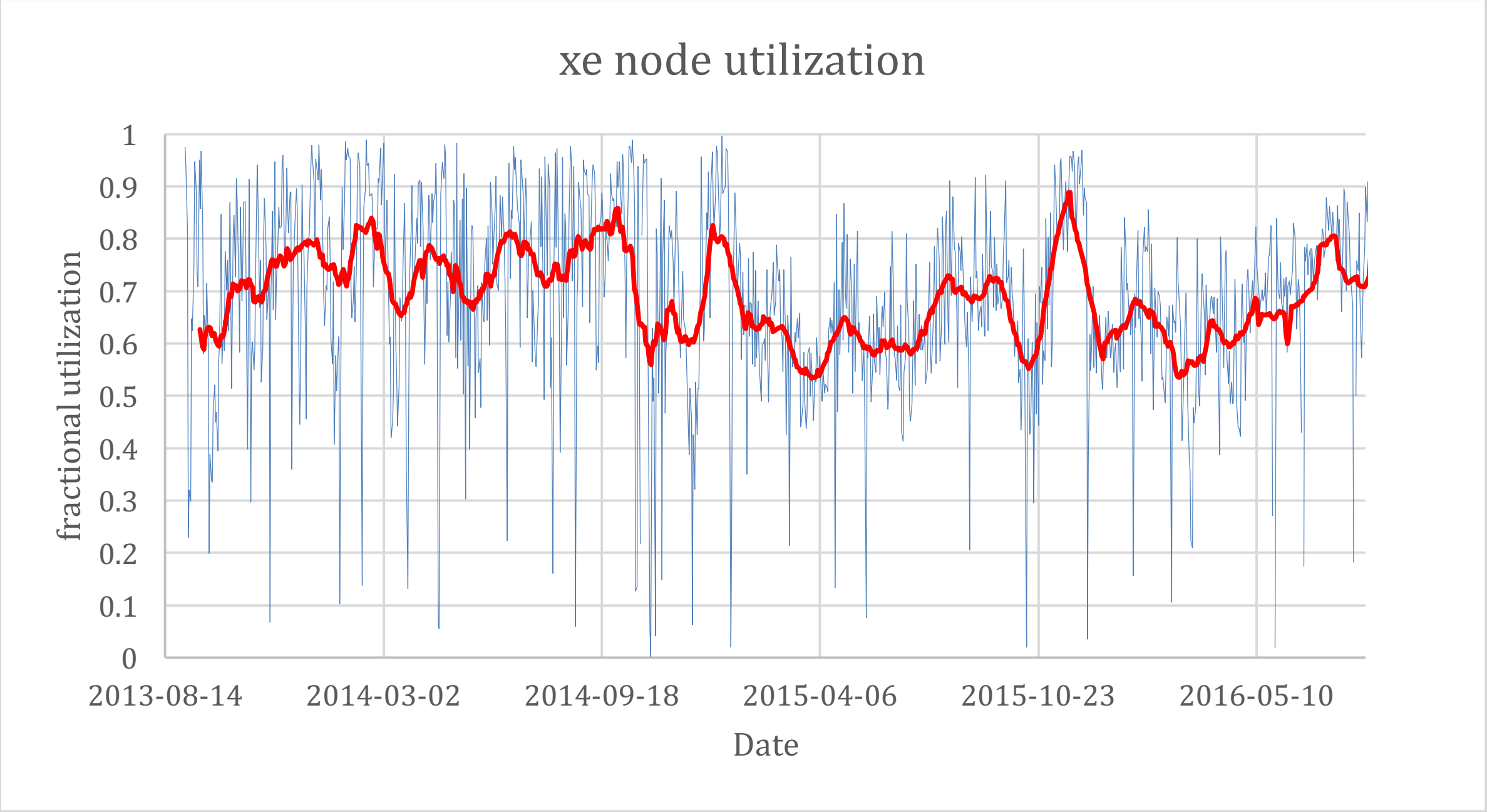

*Figure 4.1-1 Fractional gross utilization[4] of XE nodes (blue) per day. The red line is the 30 day average. The dips in utilization from Sept 2014 to December 2014 are several long system outages.*

[4] Gross utilization is the amount of time nodes are assigned to running jobs divided by the total number of nodes of that type times 24 hours.

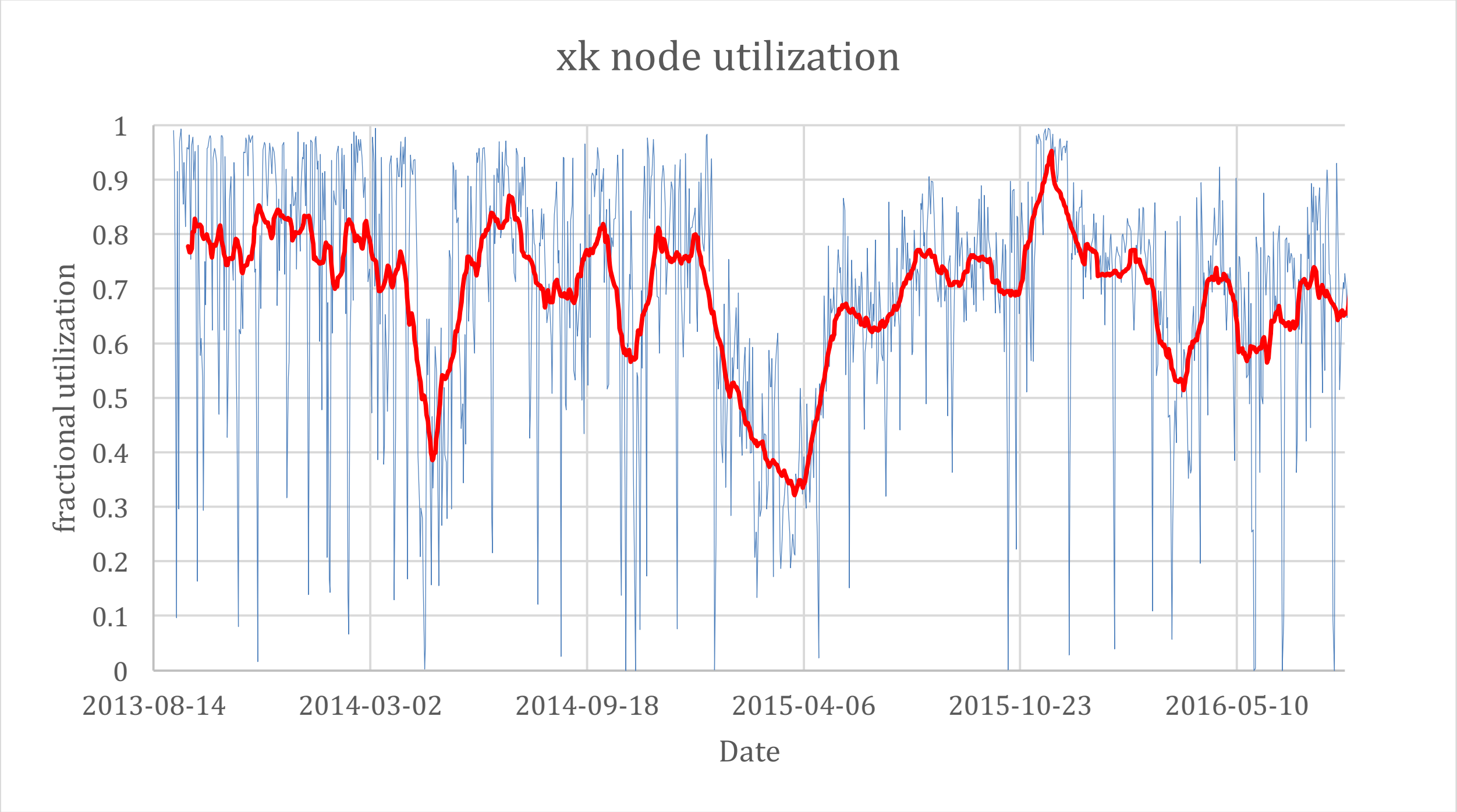

*Figure 4.1-2 Fractional gross utilization of XK nodes (blue) per day. The red line is the 30 day average. The dip in utilization for April 2014 was due to of several very large allocations expiring. The dip in utilization in February and April 2015 is due tuning of the topology aware scheduling algorithms and a major reconfiguration of the XK region of the system.*

## *4.2 XE and XK Utilization by Discipline*

We have compared the XE and XK node usage analyzed by job size (number of nodes), application, field-of-science or both size and field of science or application. We concentrated on the node hours consumed in this analysis. Figure 4.2-1 shows an Open XDMoD plot of the distribution of XE and XK node hours by job size. Not surprisingly, the XE node distribution peaks at larger sizes than the XK nodes. The XK nodes job size maximizes at 65-128 and has tails both to higher and lower sizes. The XE node distribution peaks at 1025-2048 nodes with tails to zero and very large job sizes. Figure 4.2-2 shows a similar plot except that we are examining the distribution of the number of jobs by job size as opposed the distribution of node hours; note the log scale. The figure is dominated by the small jobs which consume very few node hours.

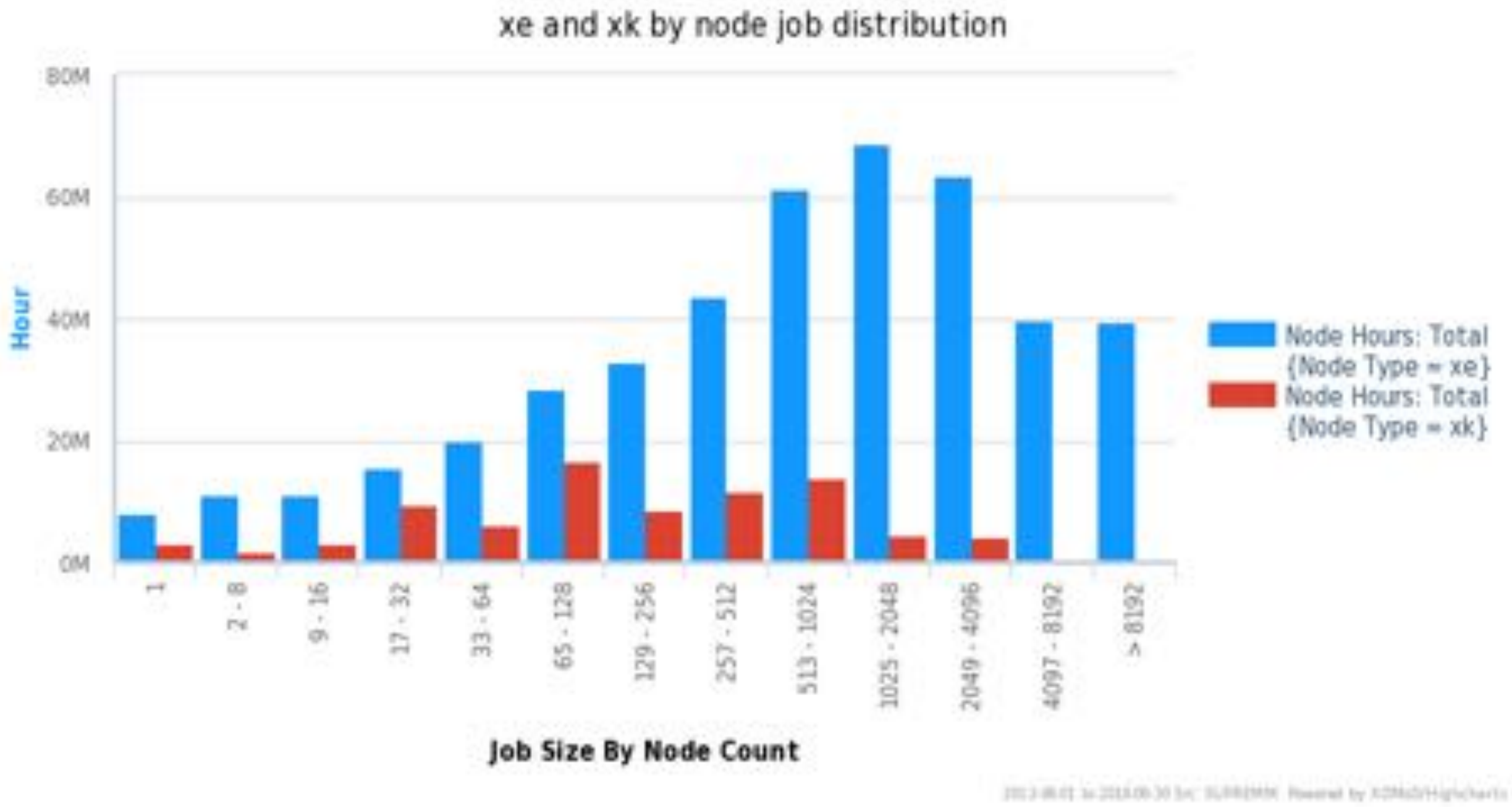

*Figure 4.2-1 Distribution of node hours by job size (node-count) for XE and XK nodes.*

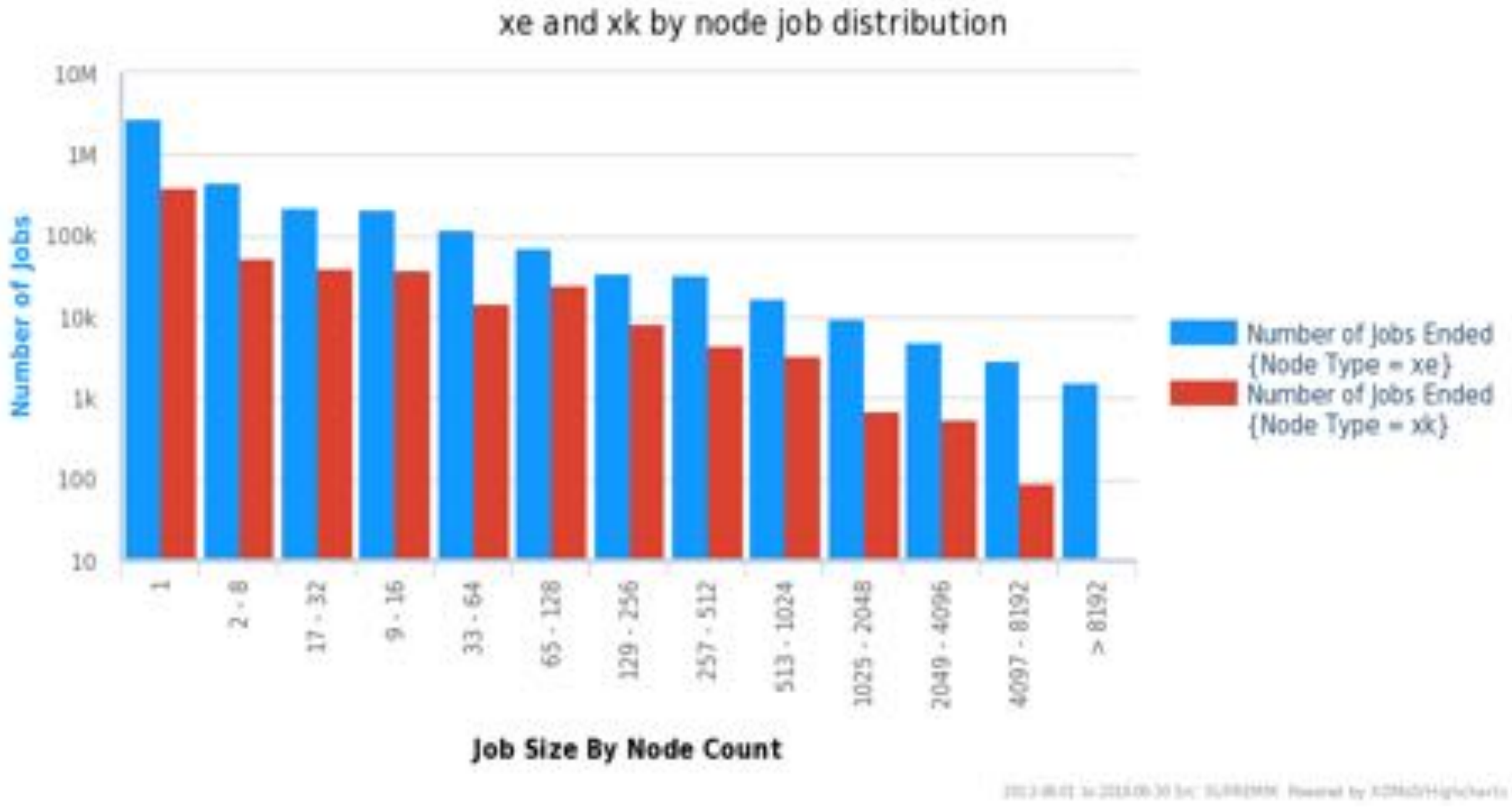

*Figure 4.2-2 Distribution of number of jobs by job size (node-count) for XE and XK nodes. Note the log scale. Without the log scale only the first few bins (below 33-64) would be visible.*

The distribution of the node hours over the Parent Sciences is shown in Figure 4.2-3, with the relative breakdown by node type, XE or XK shown in Figure 4.2-4. There is a large difference in the XK node usage among the fields of science with Chemistry, Biophysics and Molecular Sciences being the leading users of the accelerator nodes.

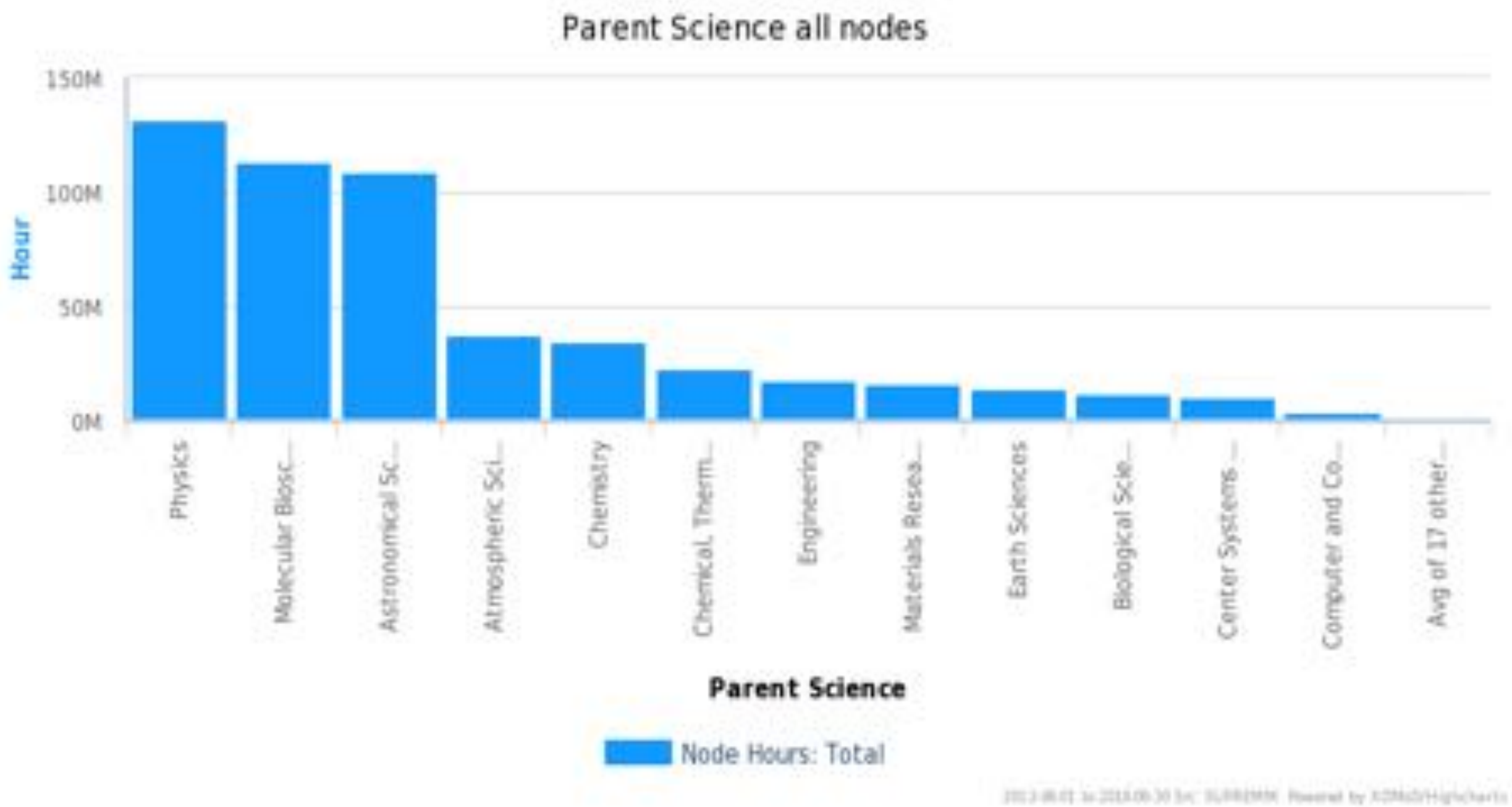

*Figure 4.2-3 Parent Science distribution by node hours.*

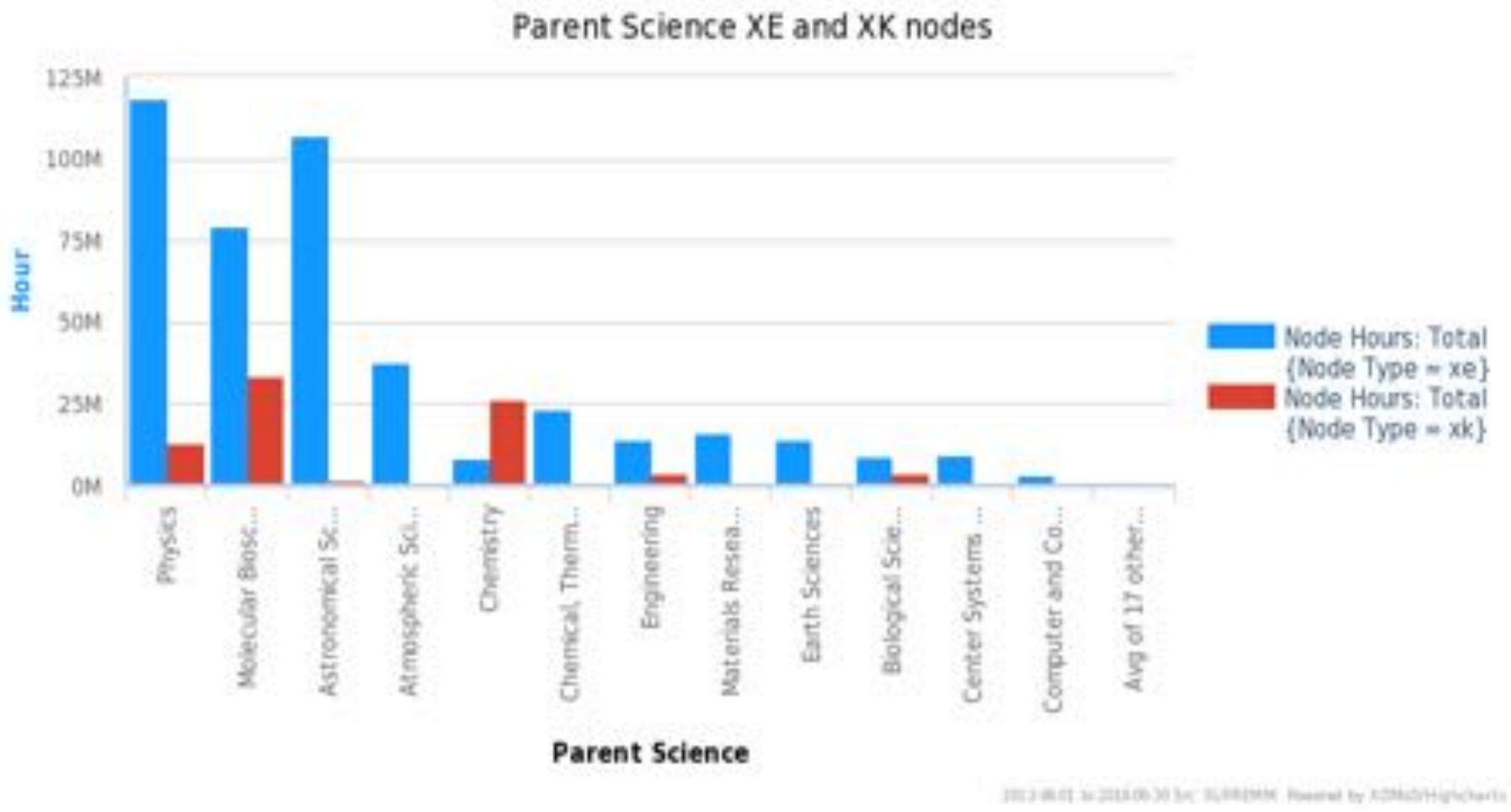

*Figure 4.2-4 Parent Science distribution for XE (blue) and XK (red) nodes.*

Figure 4.2-5 analyzes XE and XK usage by application. Here we see that the parent science usage can be explained by the heavy use of the accelerator nodes by NAMD and AMBER, both of which perform molecular dynamics simulations of biological systems. Figures 4.2-6 a,b show the parent

science analyzed by job size (node count) for the top five applications for XE nodes (Figures 4.2-6 a) and XK nodes (Figures 4.2-6 b). Physics has the largest peak job size. Molecular Biosciences has the smallest peak job size but a widespread distribution. Astronomical Sciences also has a large peak size and the widest flattest distribution. The other top parent sciences have wide flat job size distributions. The XK nodes (Figures 4.2-6 b) are predominantly used for Molecular Biosciences, Chemistry and Physics with much smaller usage by the other sciences. The peak job size is higher for the XE nodes than for the XK nodes. Figures 4.2-7 a,b shows a similar size analysis except for application usage rather than parent science; the five top XE applications (NAMD, CHROMA, MILC, Cactus, CHANGA) are shown. The XK usage is mainly confined to AMBER and NAMD with a small contribution from other applications. Comparing Figures 4.2-6 a and b with Figures 4.2-7 a and b we can see that the AMBER and NAMD usage accounts for the great majority of the Chemistry and Molecular Biosciences parent science usage respectively on the XK nodes. On the XE nodes, NAMD is responsible for most of the Molecular Biosciences usage. CHROMA and MILC are responsible for much of the elementary particle physics usage, with about equal amounts of usage on the XE nodes. CHROMA has more XK usage. The other applications cut across fields of science.

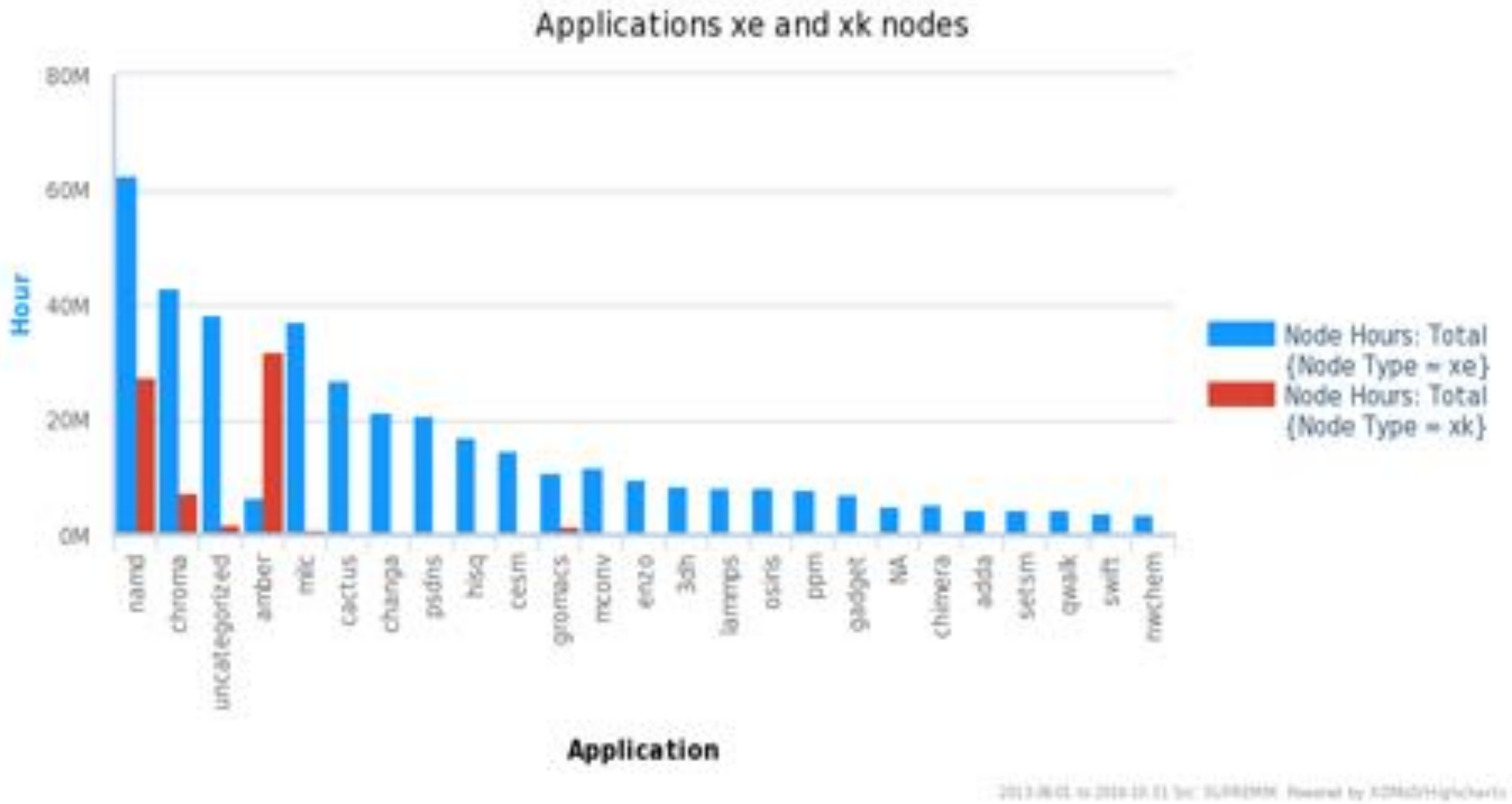

*Figure 4.2-5 Application distribution for both XE (blue) and XK nodes (red), sorted by total node hours used (summed over both XE and XK).*

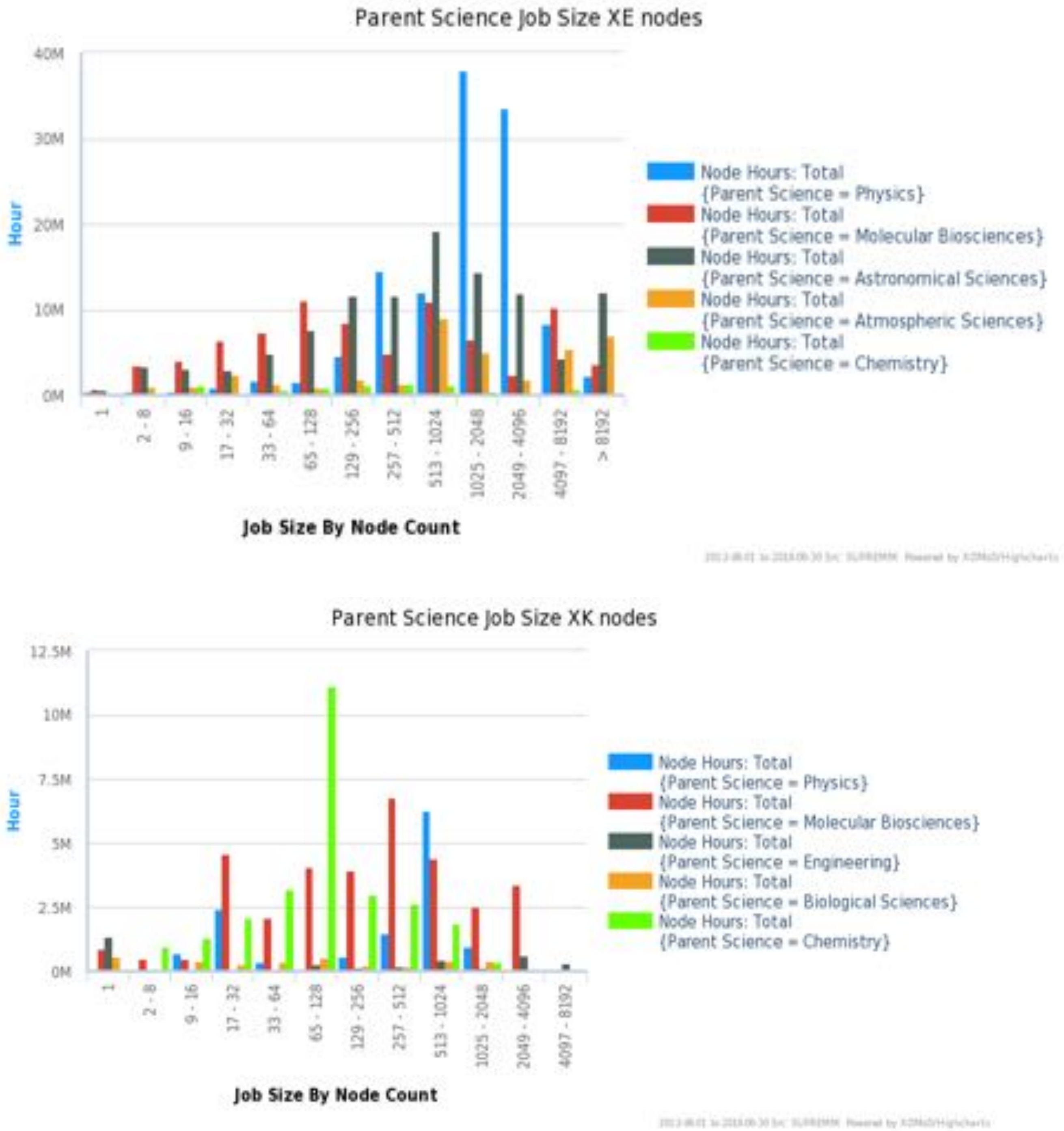

*Figure 4.2-6 a, b Distribution of node hours by job size (node-count) for XE and XK nodes analyzed for the top 5 Parent Sciences for each node type.*

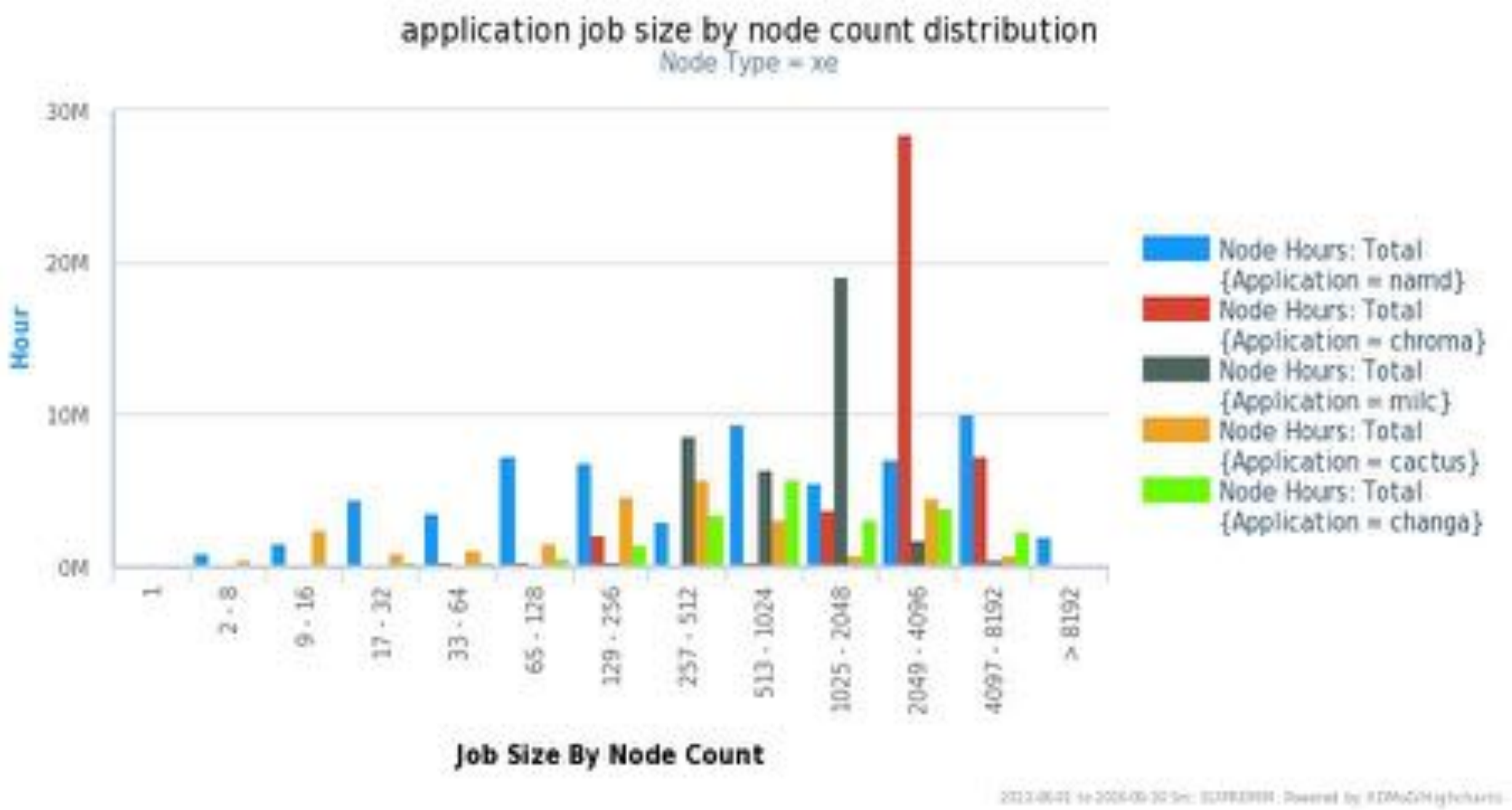

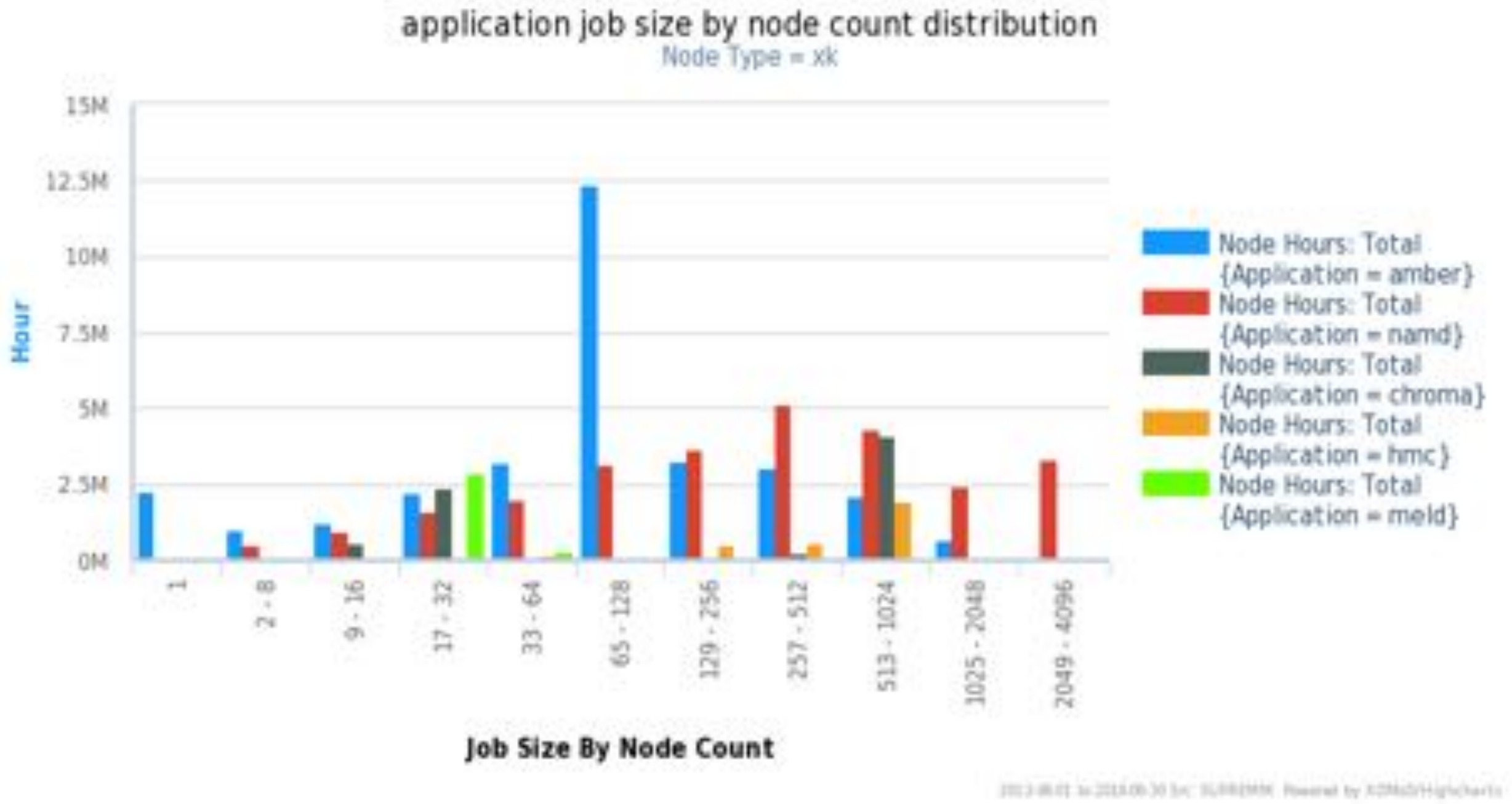

*Figure 4.2-7 a,b Distribution of node hours by job size (node-count) for XE and XK nodes analyzed for the top 5 applications for each node type.*

## 4.3 Summary: XK and XE Usage

The XE and XK nodes show a modest amount of variation in the gross node utilization due to occasional scheduled and unscheduled shut-downs, project allocation cycles, and the influence of the topologically-aware scheduler but both types of nodes are otherwise heavily utilized. There is little difference in the gross utilization pattern between the XE and XK nodes. Large jobs dominate the job size distribution by node hours of the XE nodes peaking in the 1,025-2,048 nodes (32,768-65,536 core equivalents) per job range. The much smaller set of XK nodes has a proportionally smaller peak job size. XK nodes are used mainly for the Molecular Biosciences and Chemistry, for NAMD and AMBER in particular. The top applications for the XE nodes are NAMD, CHROMA, MILC, CACTUS, CHANGA, and PSDNS. This is directly aligned with the amount of allocation the teams using these codes receive.

# 5.0 MEMORY

*BW Analysis Goal 4: Are jobs sizes, in terms of numbers of nodes in a job, changing over time? Are there differences of job size by disciplines/FoS?*

*BW Analysis Goal 7: Is job/application memory usage increasing/decreasing over time including the use of different types of nodes (XE and XK)? Are there specific discipline differences? Are there different memory use profiles based on problems being address by applications.*

## 5.1 Memory Usage and Historical Trends

We did a full analysis of memory usage including calculating distributions for the maximum memory (which is the highest value used by any node in the job at any time) and the mean memory usage (which is the average value over of all of the nodes in the job and all of the job time). We calculated the difference between the maximum and the mean in two ways: the simple or absolute difference and a relative difference defined as (max-mean)/mean which is, by construction, constrained to range from 0 to 1. We analyzed memory usage in two different ways. The simplest was looking at the memory usage distribution per job. However, this simple analysis gives equal weight to small and large jobs and therefore does not reflect the actual resource usage. To correct this we did an analysis in which the jobs were weighted by node hour. Upon comparison it was found that the two analyses were similar and the memory usage conclusions are independent of which scheme was used. In this report we have presented the node hour weighted results since we feel that they are more representative of the way memory is used in Blue Waters. We did the analysis for all nodes, for the XE nodes only and for the XK nodes only. Finally, we compared the GPU mean memory usage with the CPU memory usage for the XK nodes. We based our conclusions on all of these 26 different analyses, however, we will only show the selected plots that we think are most illustrative.

Figures 5.1-1, 5.1-2 and 5.1-3 display histograms weighted by node hours showing the memory usage of the XE nodes. Figure 5.1-1 shows the maximum memory usage per node per job weighted by node hour. The x-axis is the memory used in GB binned in 1 GB increments. The y-axis is the fraction of node hours in each 1GB bin. The x-axis and y-axis for the memory distribution plots will always have this standard definition unless otherwise indicated. Note that most of the jobs use less than one-half of the memory but the distribution has a significant tail to high usage. Many jobs use less than 20GB with the distribution tapering down beyond this. The simple histogram of jobs (not shown) displays a similar trend except with an even larger sub 20 GB peak. Figure 5.1-2 shows the mean memory usage weighted by node hours. It is slightly less than the maximum, as expected, with a larger fraction of the usage under 20 GB and a much smaller tail out to higher memory usage. Figure 5.1-3 shows the distribution of difference between the maximum memory used and the average memory (max-mean) used for each job. Note that most jobs show very little difference with a long tail of jobs that show a substantial difference between the max and the mean stretching out to 30GB. The simple histogram of the difference weighted by number of jobs (not shown) displays a nearly identical trend.

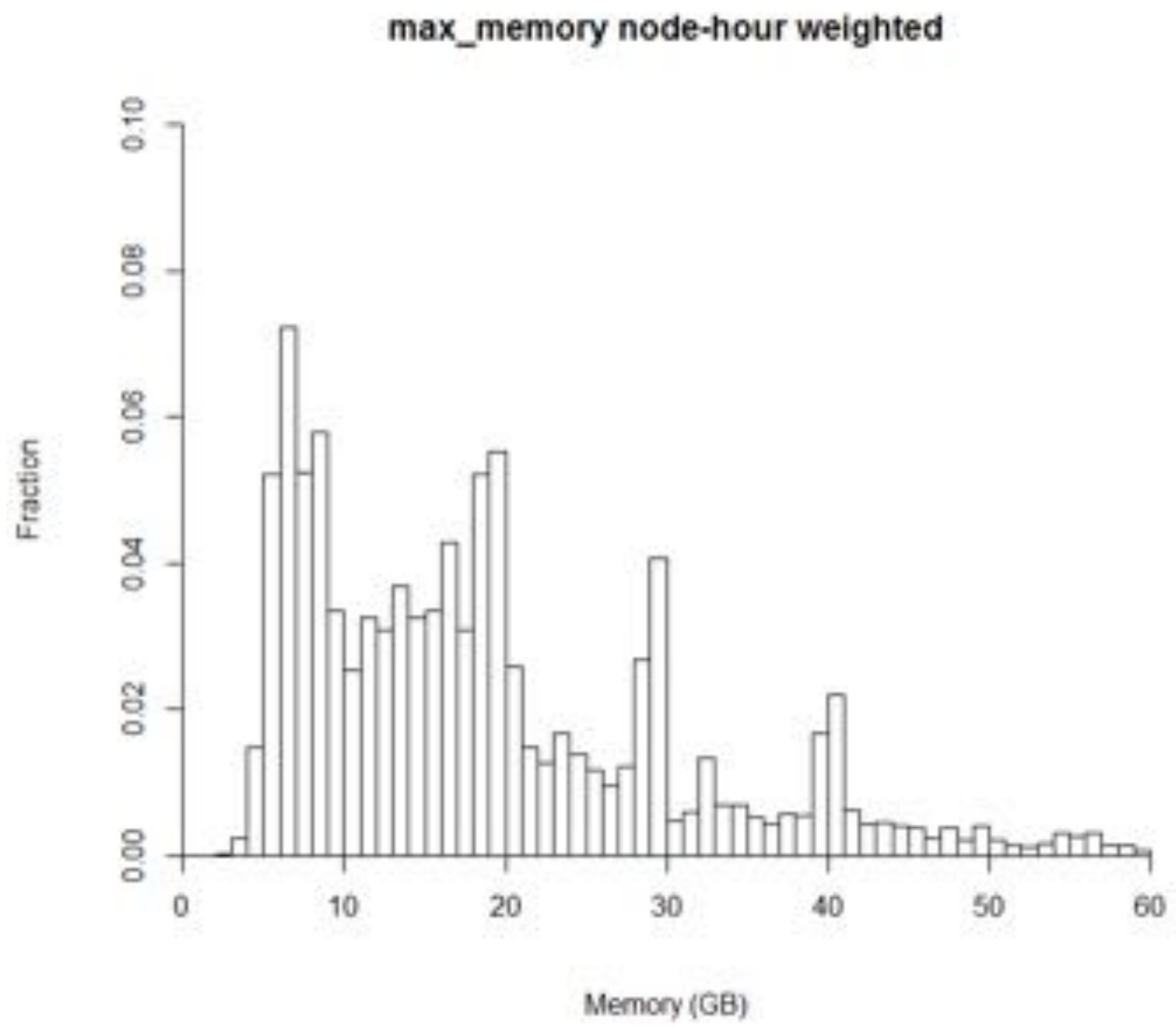

*Figure 5.1-1 Distribution of the maximum memory used per node weighted by node hour for the XE nodes. The y-axis is the fraction of node hours.*

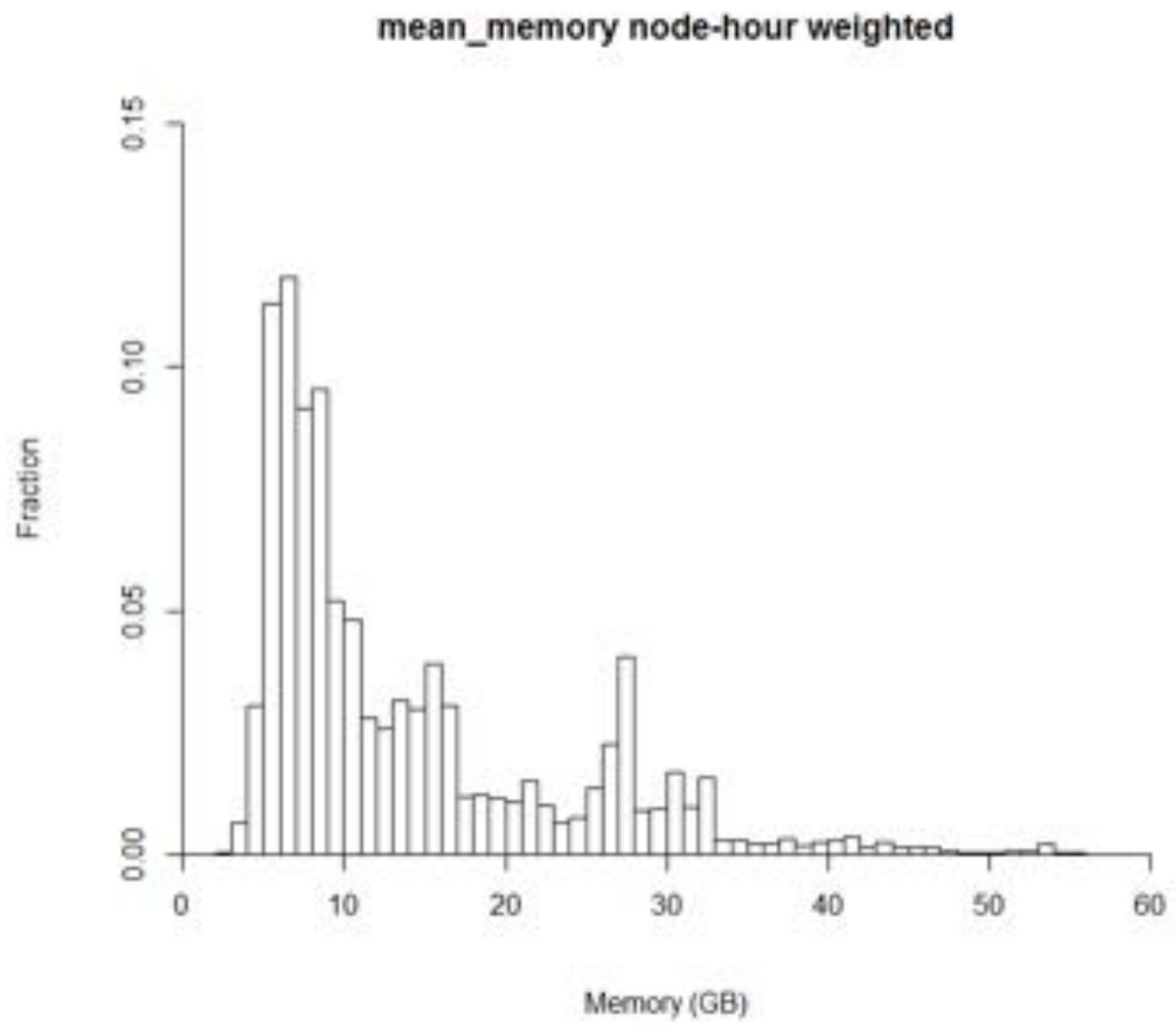

*Figure 5.1-2 Distribution of the mean memory used per node weighted by node hour for the XE nodes. The y-axis is the fraction of node hours.*

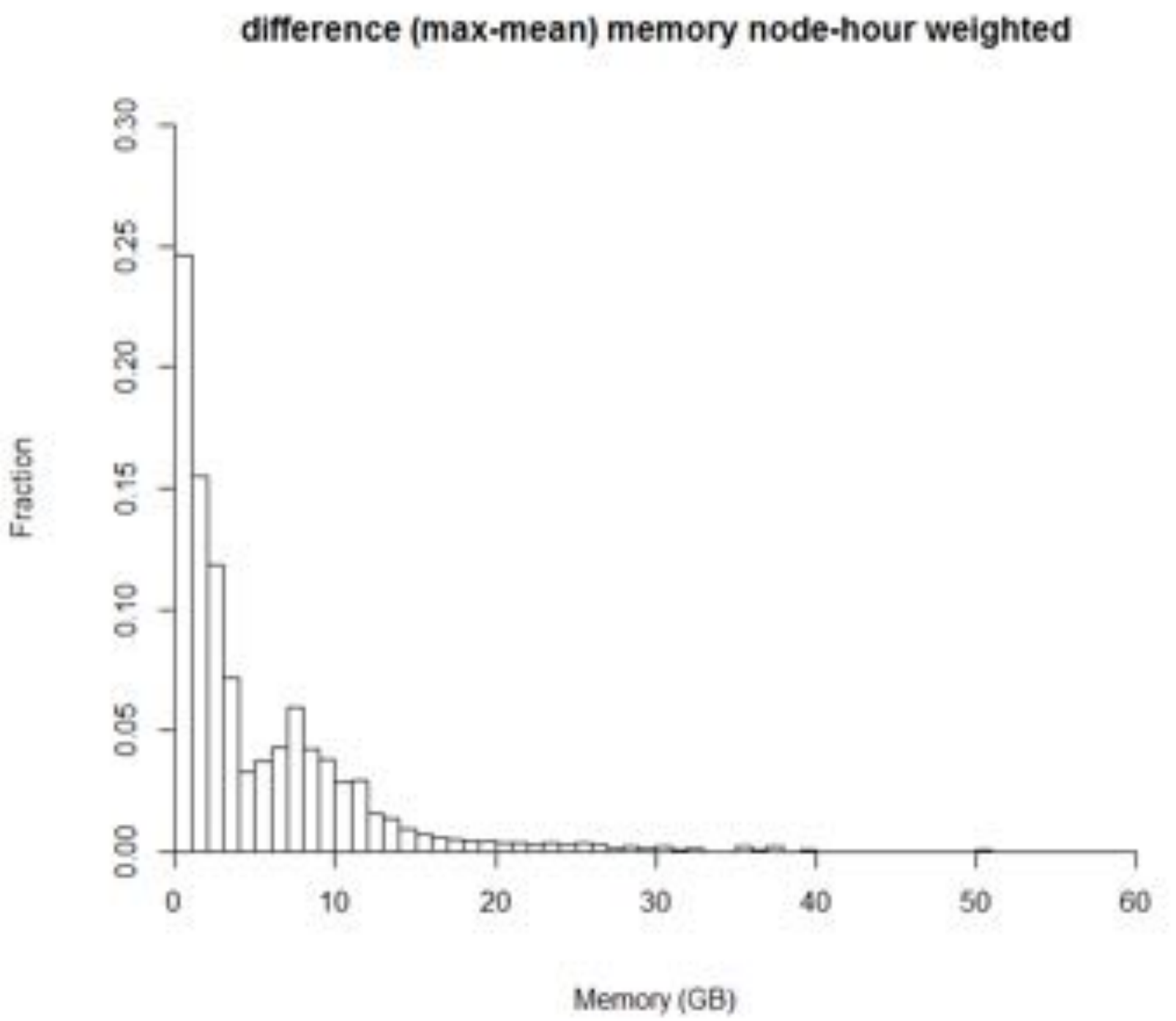

*Figure 5.1-3 Distribution of the difference (max-mean) in the memory used per node weighted by node hour for the XE nodes. The y-axis is the fraction of node hours.*

Figures 5.1-4, 5.1-5 and 5.1-6 display histograms weighted by node hours showing the memory usage of the XK nodes. The memory trends exhibited by the XE node memory usage are even more pronounced in the XK node memory usage. Figure 5.1-4 shows that the maximum memory usage of the XK nodes is less than 15 GB for the majority of jobs with only a few in the 15-32GB range. The mean memory usage for XK nodes, Figure 5.1-5, is by definition less and is closer to zero with very few jobs in the high memory tail. Figure 5.1-6, the difference (max-mean) shows that most jobs are within 1 GB of the maximum memory usage with only a minor tail extending out to ~10 GB.

Figure 5.1-7 shows the mean GPU memory usage for the XK nodes weighted by node hour. Most jobs use 1 GB or less of the ~6 GB available GPU memory with a long tail of jobs that use up to the maximum. Figure 5.1-8 shows a non-normalized version of this plot with more bins. This finer grain histogram and simple job weighted histograms (not shown) confirm generally low memory usage by most jobs with a small number of larger GPU memory usage jobs out in the tail of the distribution.

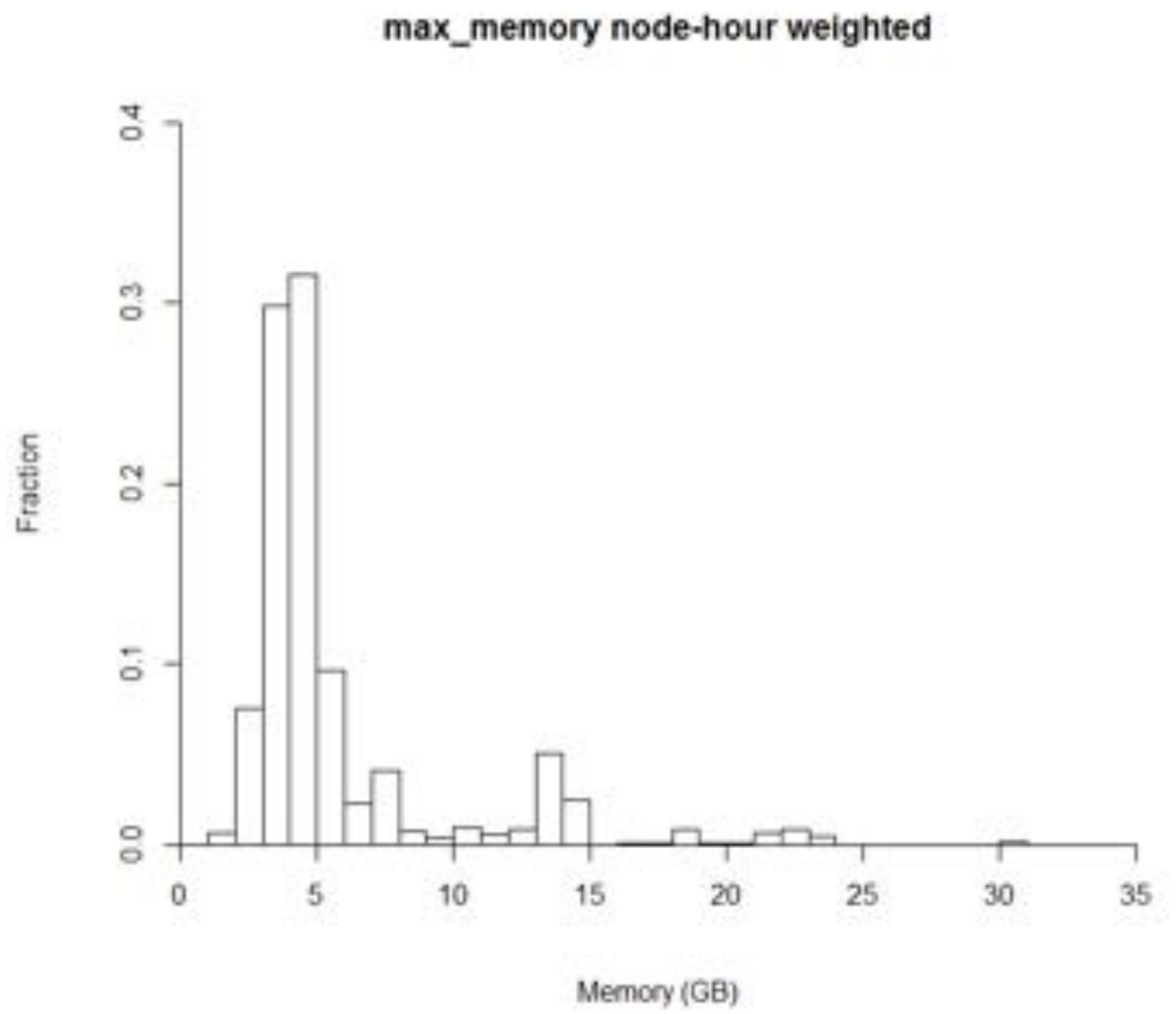

*Figure 5.1-4 Distribution of the maximum memory used per node weighted by node hour for the XK nodes. The y-axis is the fraction of node hours.*

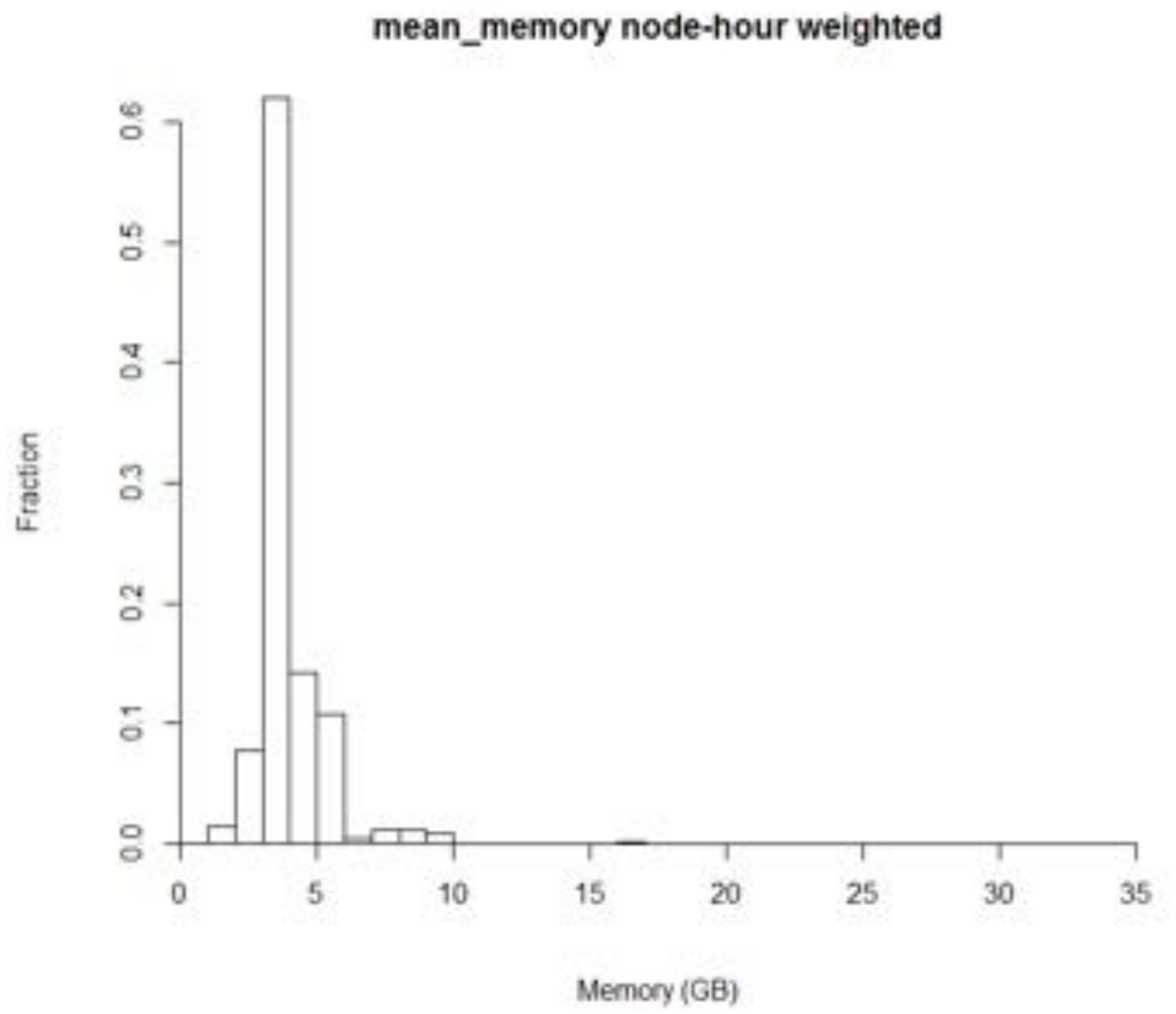

*Figure 5.1-5 Distribution of the mean memory used per node weighted by node hour for the XK nodes. The y-axis is the fraction of node hours.*

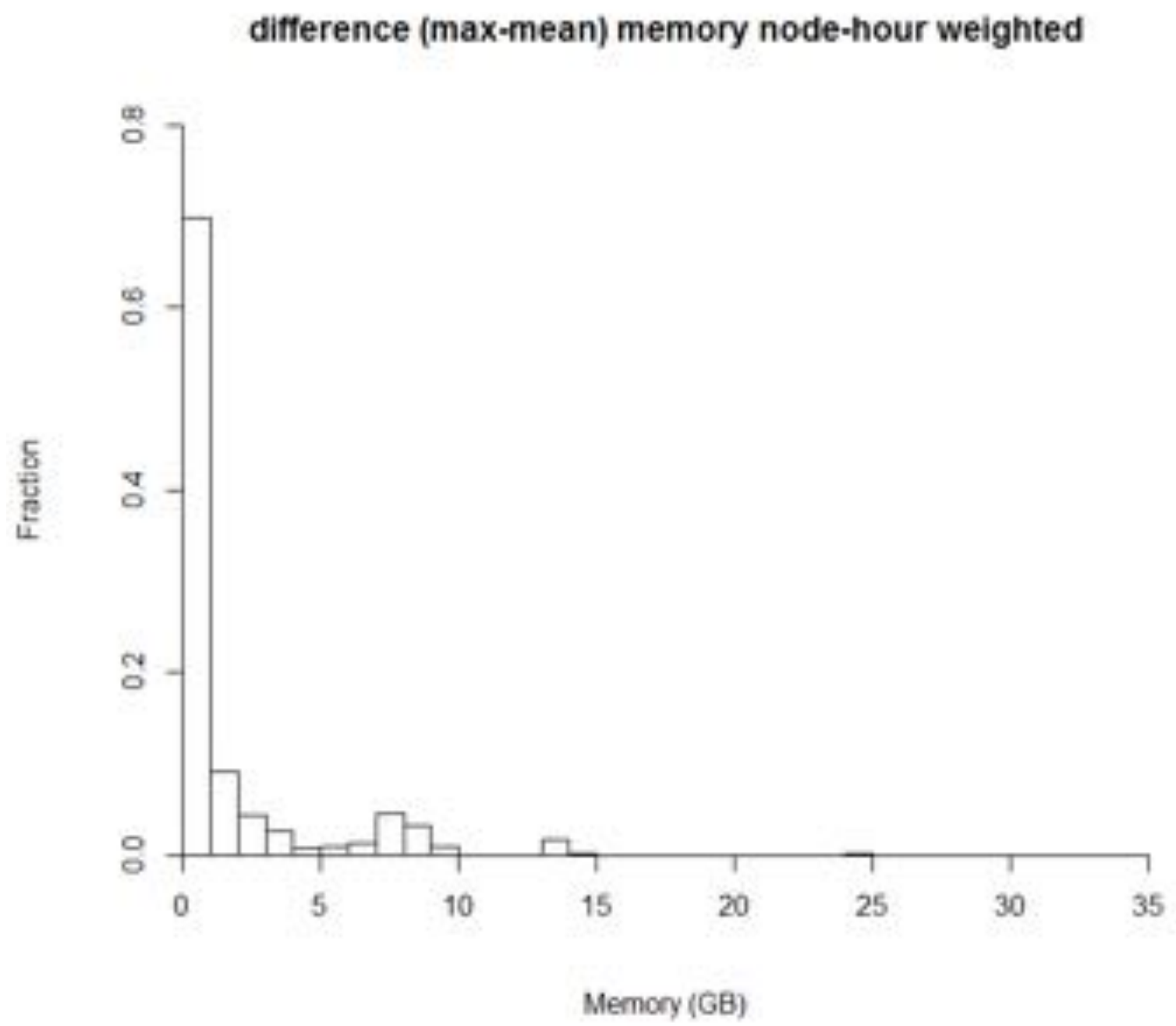

*Figure 5.1-6 Distribution of the difference (max-mean) in the memory used per node weighted by node hour for the XK nodes. The y-axis is the fraction of node hours.*

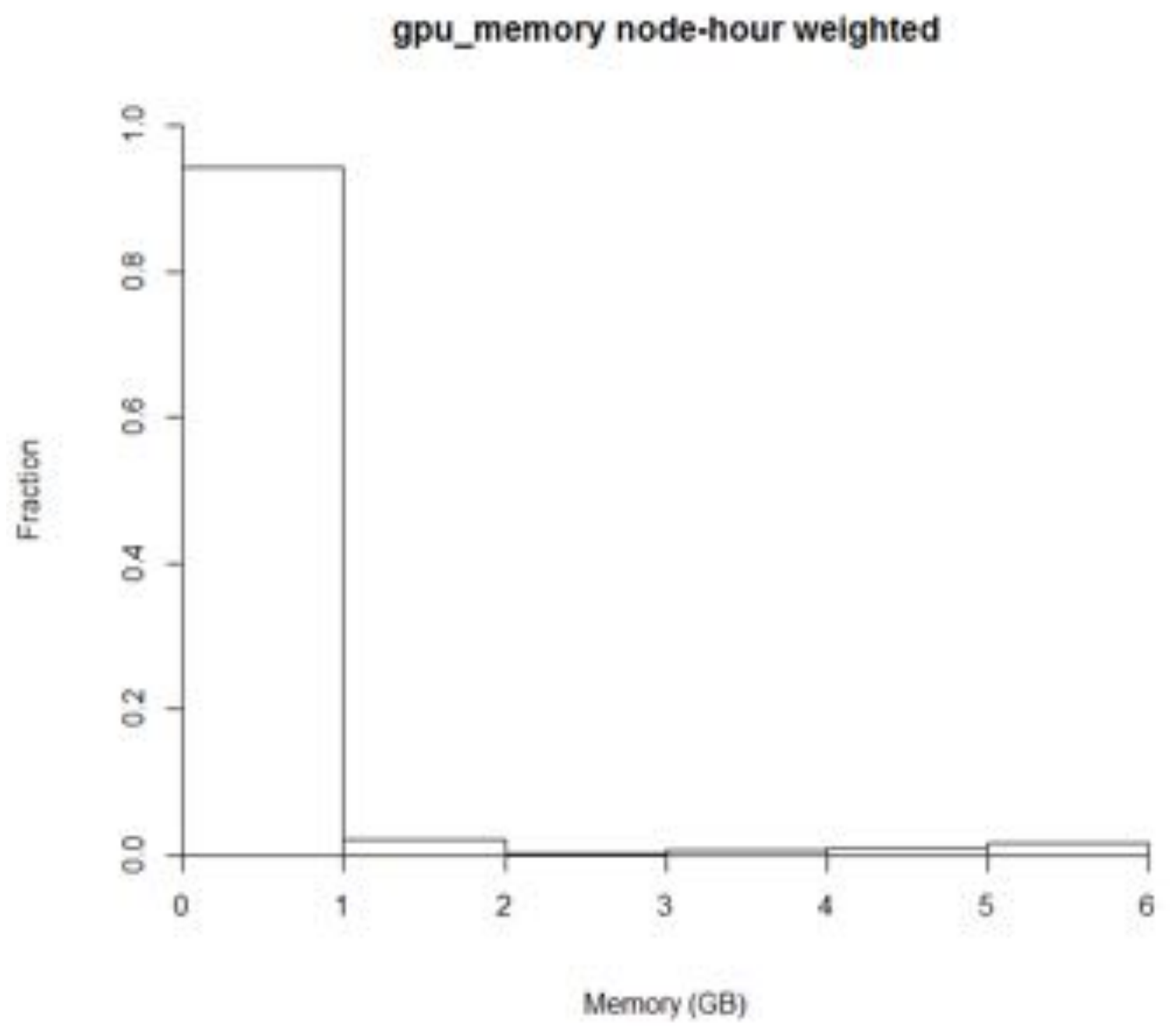

*Figure 5.1-7 Distribution of the mean GPU memory used per node weighted by node hour for the XK nodes. The y-axis is the fraction of node hours.*

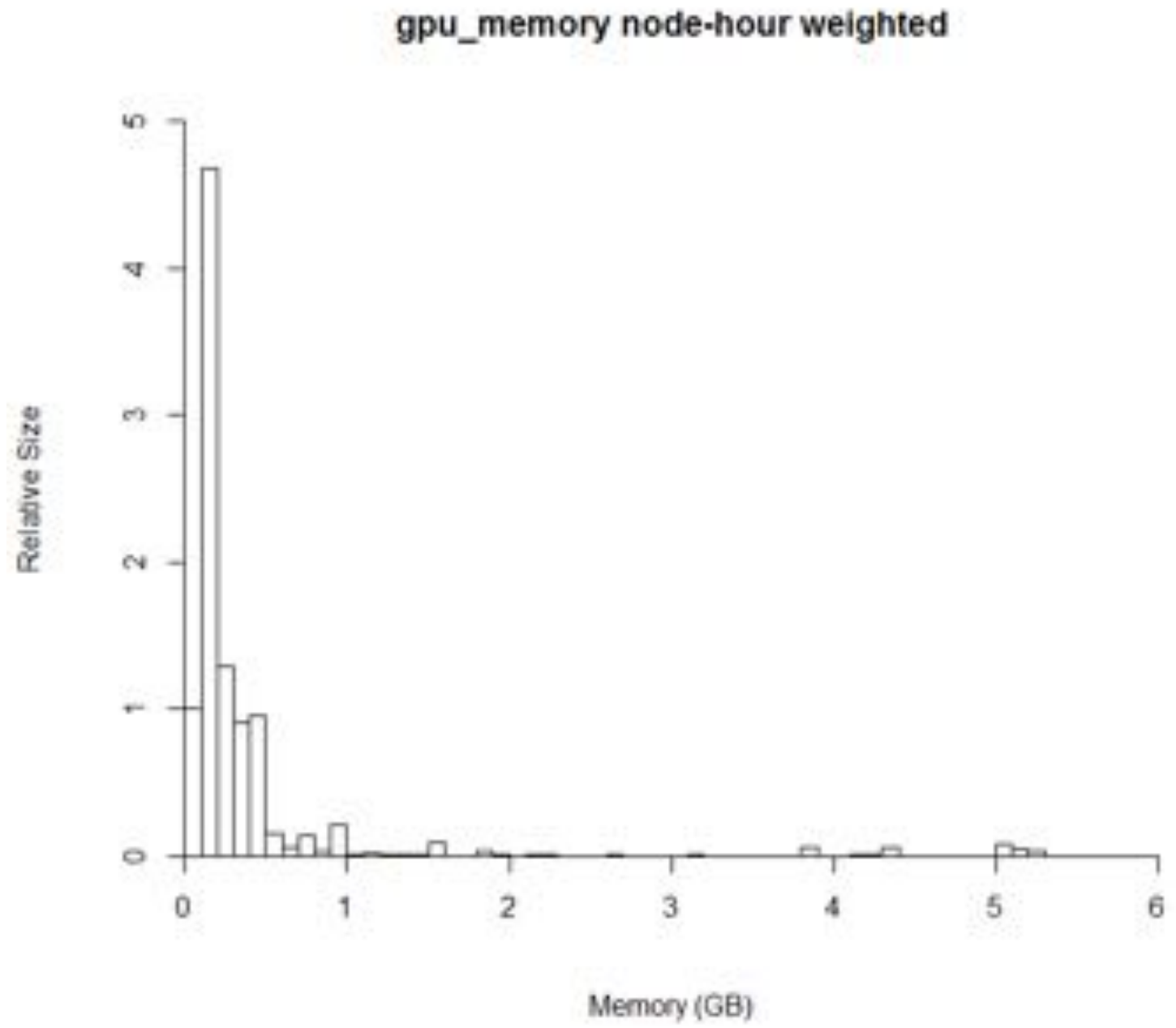

*Figure 5.1-8 Distribution of the mean GPU memory used per node weighted by node hour for the XK nodes. More bins were used to show the detail in the first 1G. Now the y-axis is relative size rather than fraction of node hours.*

The overall memory usage of large jobs is illustrated in Figure 5.1-9. This binned scatter plot shows XE node jobs where the memory usage information is available excluding jobs that were run by the vendor or by Blue Waters staff members. The job size is plotted on the x-axis and total peak memory usage of the job on the y-axis. The color of each bin indicates the total wall hours for jobs in the bin. Note the log scale on the point color code. For example, a point on the top right of the plot indicates one or more jobs ran with approx. 22,600 nodes and used around 1.4 PetaBytes of memory. Assuming the application uses the same amount of memory in each node, the memory used in each node is 61.95 GB that is 97% of the entire memory in the node. The solid diagonal line indicates the points corresponding to 64GB per node. We note that a subset of XE compute nodes (~100) were upgraded to 128GB memory in August 2016. The jobs that ran on these nodes are indicated by the points entirely above the diagonal line near the origin of the plot.

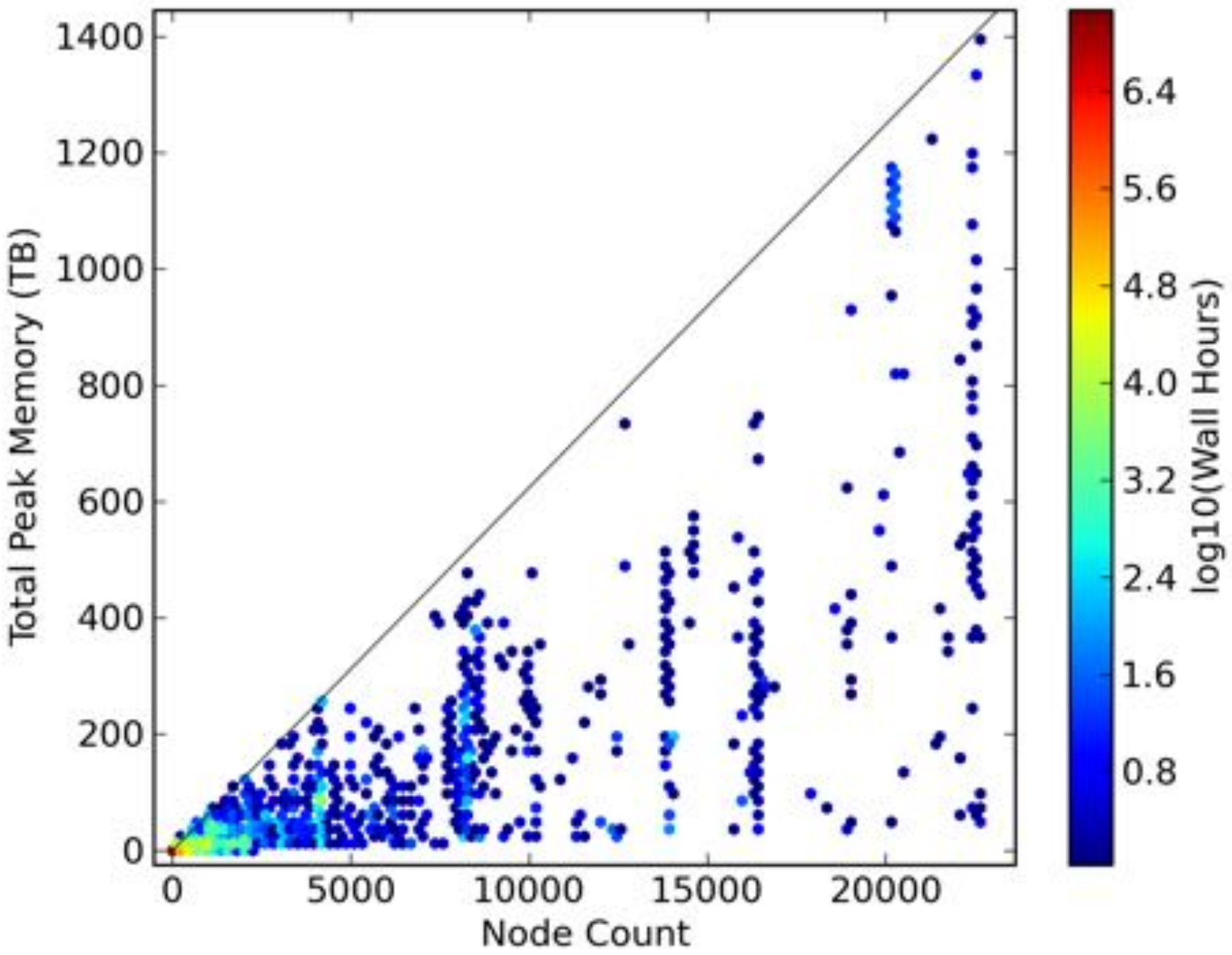

*Figure 5.1-9 2D binned scatter plot showing the total peak memory usage for XE node jobs. The color coding shows the total wall hours for the jobs in each bin using a log scale. The diagonal line indicates the points corresponding to 64 GB per node.*

The historical trend in memory usage on the XE nodes is shown in Figure 5.1-10. There is little systematic change from year to year. A similar plot of XK nodes is shown in Figure 5.1-11; note that there is also no systematic change in the memory usage on the XK nodes.

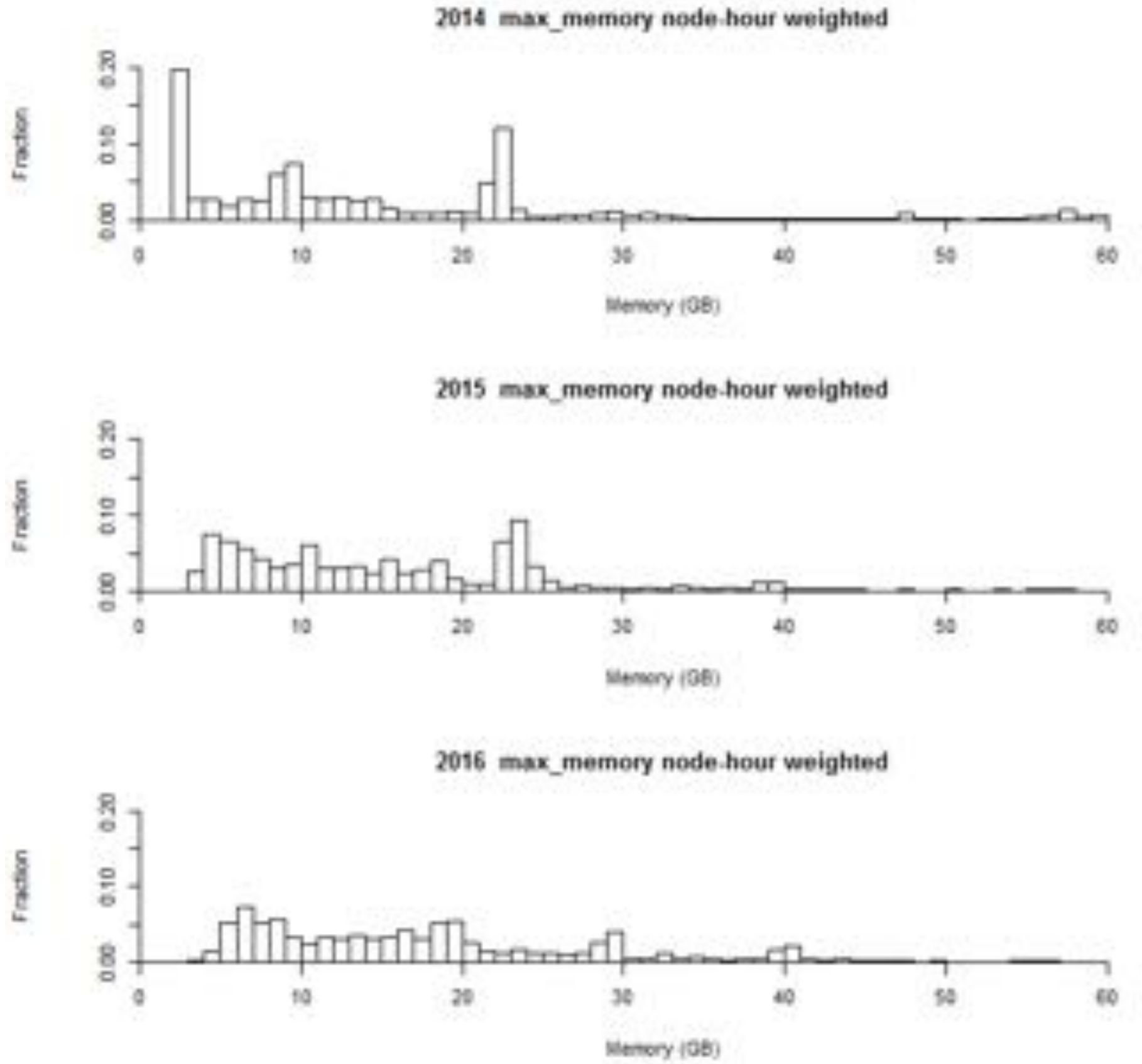

*Figure 5.1-10 Distribution of the mean memory used per node weighted by node hour for the XE nodes. The y-axis is the fraction of node hours. The historical trend is shown in the three plots for 2014, 2015, 2016 (top to bottom).*

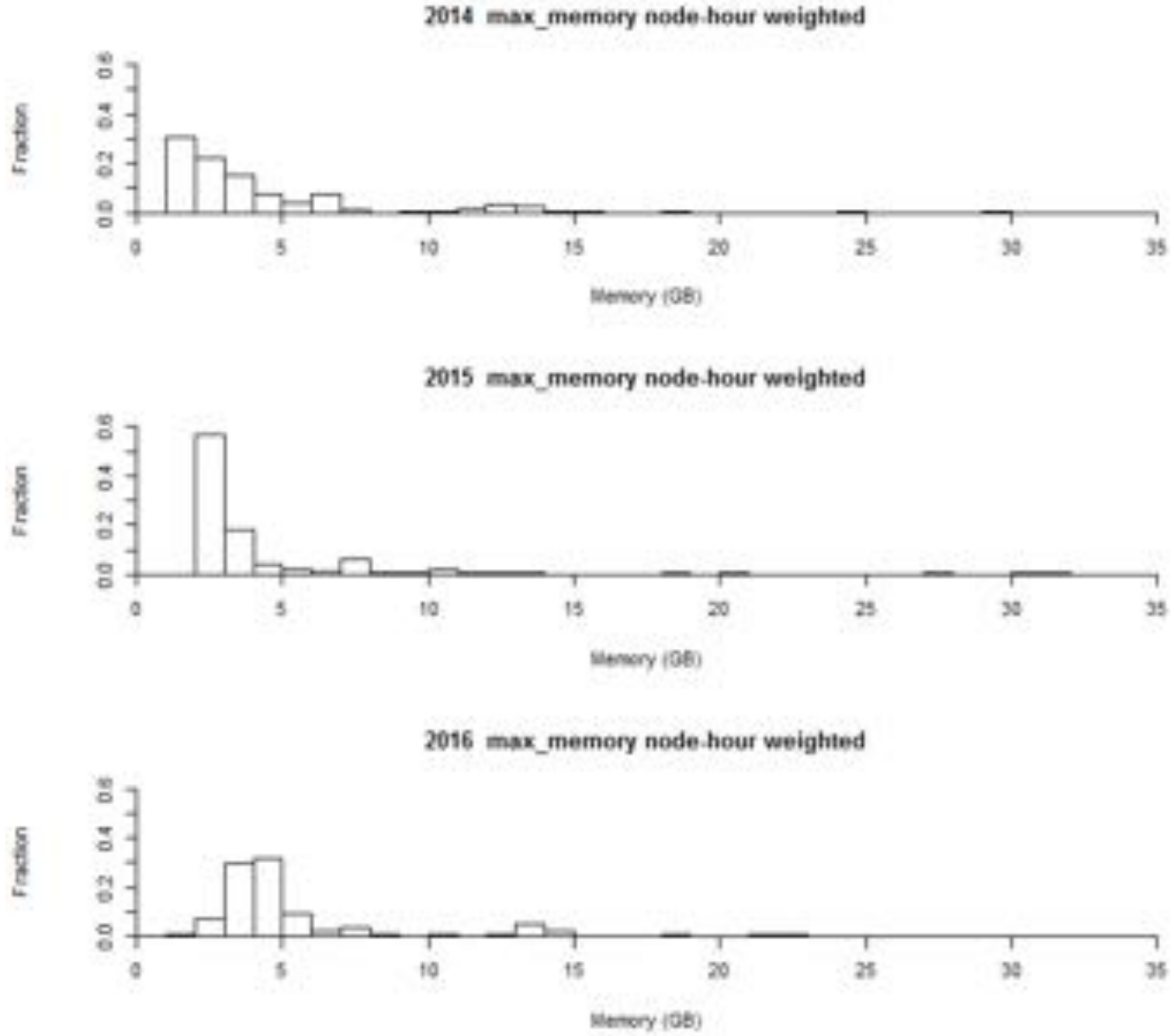

*Figure 5.1-11 Distribution of the mean memory used per node weighted by node hour for the XK nodes. The y-axis is the fraction of node hours. The time trend is shown in the three plots for 2014, 2015, 2016 (top to bottom).*

## 5.2 Memory Usage by Discipline

We examined the memory usage analyzed by either parent science or by application for both the XE and XK nodes. For the most part, there is very little systematic trend in memory usage either by parent science or by application. Tables 10.0-1 and 10.0-2 (Appendix III) show a complete list of the maximum memory used per XE node for each application and each parent science area, respectively. Tables 10.0-3 and 10.0-4 (Appendix III) show a complete list of the maximum memory used per XK node for each application and each parent science area, respectively. In these 4 tables the maximum memory used for each job is weighted by node hours consumed to obtain mean maximum memory usage by application or parent science for the applicable node type.

We did individual analyses of the time trend of the memory distribution for the XE nodes for the top 20 applications that used the most node hours on Blue Waters. Figures 5.2-11, 5.2-12 and 5.2-13 show the trend for three applications NAMD, MILC and CESM. NAMD is the most used application on Blue Waters by node hour. It is a relatively low memory usage application. Examining Figure 5.2-15, one sees no systematic trend towards higher memory usage with time. MILC is a relatively high memory usage application. Examining Figure 5.2-16, one sees no systematic trend towards higher memory usage with time. CESM is an exception in that Figure 5.2-17 does show a trend towards higher memory usage in 2016 than in the earlier years.

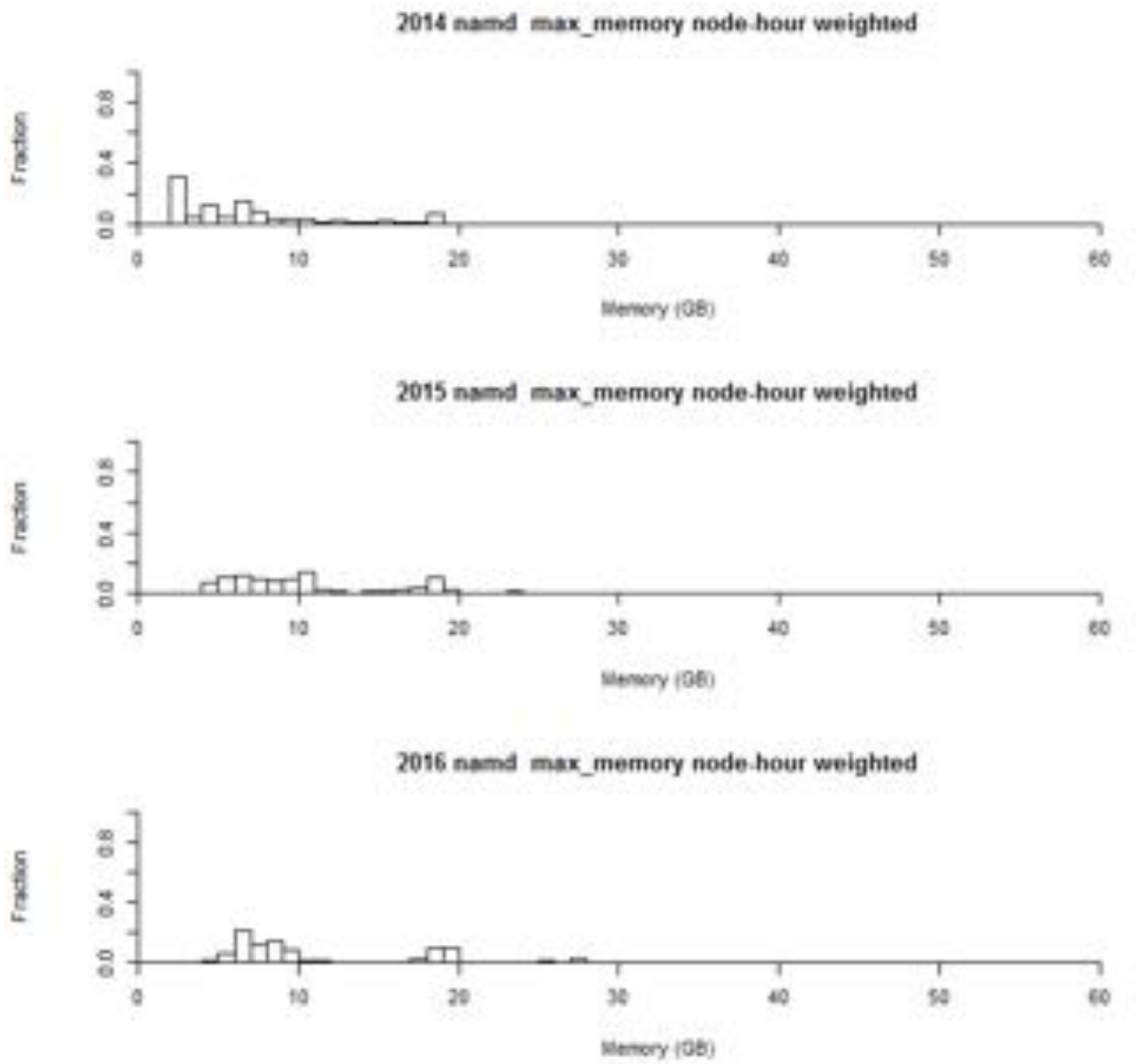

*Figure 5.2-11 Distribution of the mean memory used per node weighted by node hour for the NAMD application running on the XE nodes. The y-axis is the fraction of node hours. The time trend is shown in the three plots for 2014, 2015, 2016 (top to bottom). NAMD is a low memory usage application. There is no systematic trend towards higher memory usage with time.*

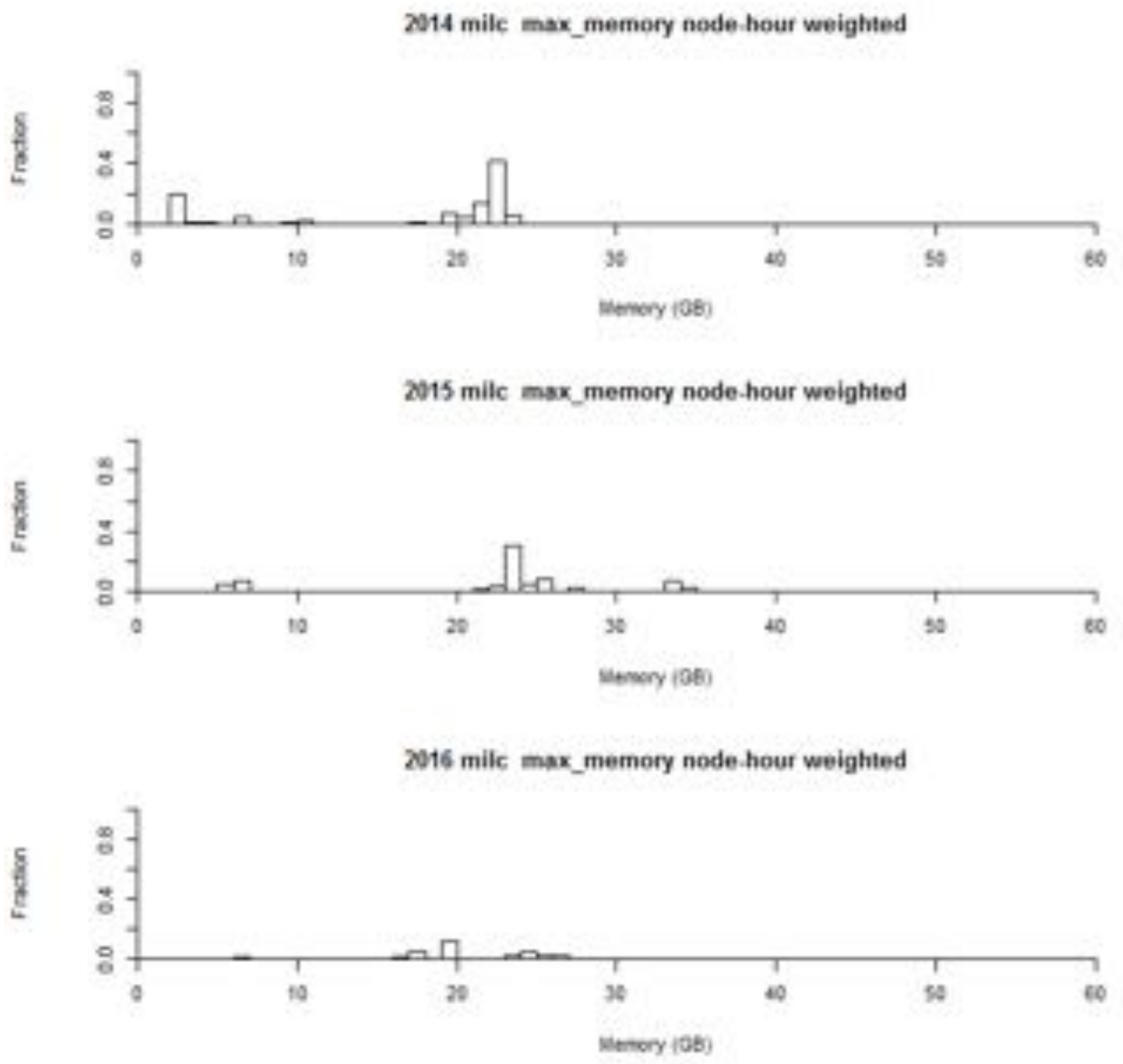

*Figure 5.2-12 Distribution of the mean memory used per node weighted by node hour for the milc application running on the XE nodes. The y-axis is the fraction of node hours. The time trend is shown in the three plots for 2014, 2015, 2016 (top to bottom). MILC is a relatively high memory usage application. There is no systematic trend towards higher memory usage with time.*

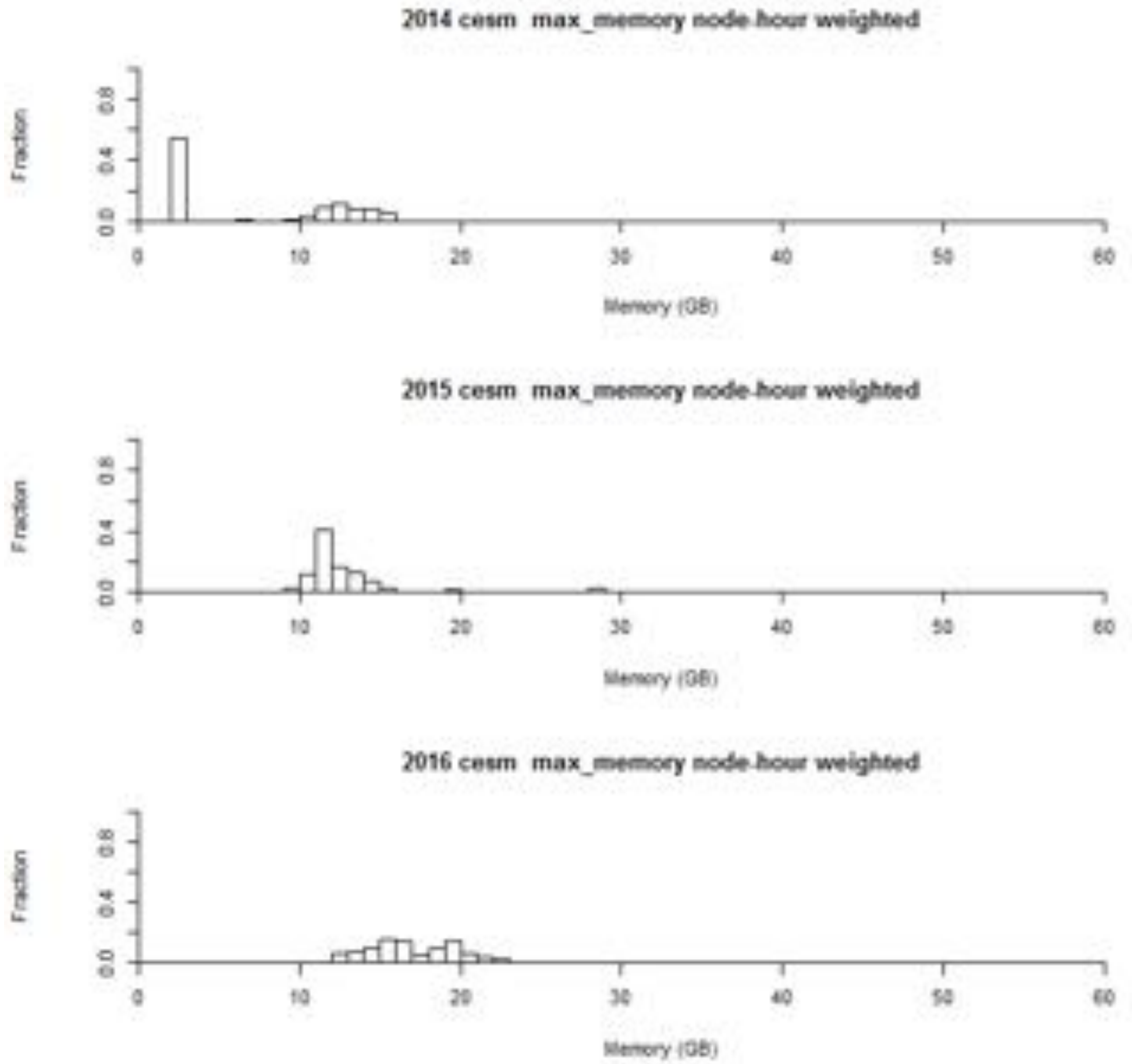

*Figure 5.2-13 Distribution of the mean memory used per node weighted by node hour for the CESM application running on the XE nodes. The y-axis is the fraction of node hours. The time trend is shown in the three plots for 2014, 2015, 2016 (top to bottom). There is a slight trend towards higher memory usage in 2016.*

## 5.3 Memory Correlations

Memory correlations were computed for a variety of metrics. Correlations were run for the total memory used on each job for the XE nodes with: interconnect usage (receive and send); CPU (cpu_user) usage; wall_time; and Flops. Both Pearson correlations and Spearman correlations (non-parametric) were computed. No significant correlations between memory used and other characteristics were determined.

A further analysis was done where the XE node jobs were divided into narrower categories by job type including: the message passing jobs (~636K jobs), the message passing jobs with threading (~75k jobs), serial jobs (~373K jobs), or threads (~1.5M jobs). There was a moderate Spearman correlation coefficient of ρ = 0.41 and 0.53, 0.42, -0.39 respectively for the correlation of total memory used with interconnect usage; see Table 5.3-1. That is, for these classes of jobs, there is a moderate increase in the memory used with the number of bytes sent or received on the network except for the threads, which show a decrease in memory used with network usage.

Table 5.3-1 Correlation of memory usage with network usage by job category

| Jobtype | Number of jobs | Spearman coefficient |
|---|---|---|
| message passing | ~636K | +0.41 |
| message passing w/ threading | ~75K | +0.53 |
| serial jobs | ~373K | +0.42 |
| threads | ~1.5M | -0.39 |

## *5.4 Summary: Memory Usage*

On the XE nodes, most jobs have a maximum memory usage less than 50% of the memory but there is a substantial tail in the distribution with higher memory usage. For the mean memory usage, the peak is lower with a thinner tail as expected. Overall, there is not a significant difference between the maximum and the mean memory usage. On the XK nodes, most jobs use less than 25% of the CPU memory with a short tail to higher usage. Once again, the maximum and mean memory usage is not greatly different indicating most applications do not change their memory profile once they start running. GPU memory usage is typically less than 1 GB out of 6 GB.

For both the XE and XK nodes, there is no trend to higher memory usage from year to year. There also seems to be no trend in the memory usage by application or parent science with year. One exception is that CESM memory usage seems to have increased somewhat in 2016 over that of 2014-2015.

# 6.0 LARGE JOB USAGE

*BW Analysis Goal 2: What are the top representative algorithms on Blue Waters that consume a majority of the node hours including the use of different types of nodes (XE and XK)?*

- *What is the distribution of job sizes by application and Field of Science (FoS)?*
- *Some sampling and analysis of the communication/compute, IO/compute, and memory/compute ratios for different applications as feasible.*

Figure 6.0-1 shows the usage of the XE nodes by all jobs over the course of the entire study period. Ninety percent of the node hours (red line in figure) are consumed by the largest 3% of jobs (by node hour). The mean size of these 3% largest jobs is 712 nodes and 3,428 node hours. Figure 6.0-2 shows the equivalent plot for the XK nodes. The rise is slightly less steep but still only the 7% largest jobs (by node hour) account for 90% of the XK node hours (red line). The mean size of these 7% largest jobs is 178 nodes and 1,808 node hours per job.

Other resources (metrics) such as memory, I/O and network can also be plotted. Figure 6.0-3 below shows a cumulative plot of total memory usage on the XE nodes rather than node hours. Note the shape is completely different and is much more linear than the node hour plot indicating a broad range of memory usage, as seen in the memory usage distributions. If the memory jobs are weighted by node hour, see Figure 6.0-4, the cumulative memory plot basically mimics the node hour plot (Figure 6.0-1) and the memory aspect is only secondary.

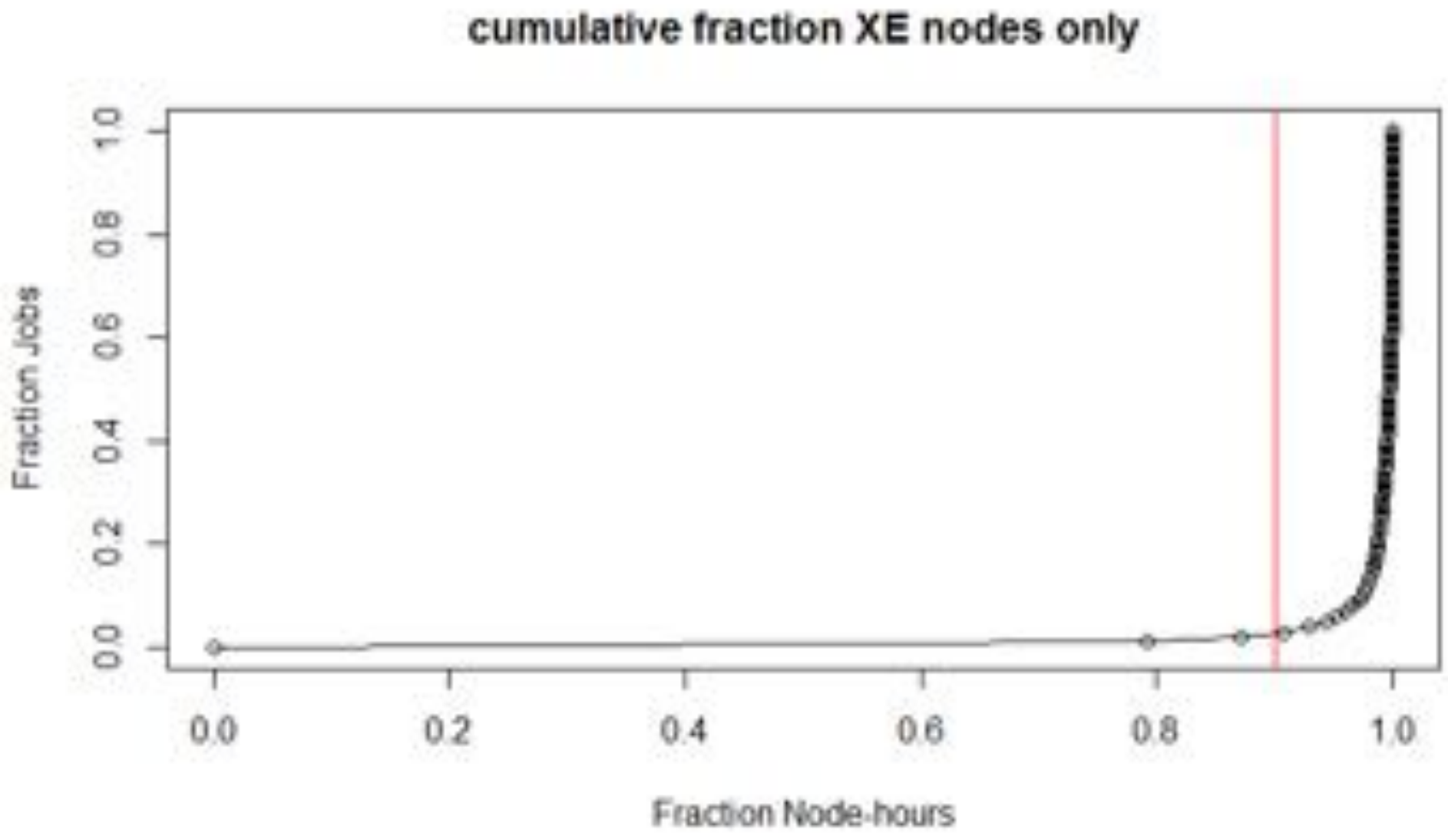

*Figure 6.0-1 Cumulative plot of the fraction of node hours as a function of the fraction of jobs on the XE nodes. Note that the largest 3% of the jobs (by node hour) account for 90% of the node hours consumed. The red line indicates 90% in node hours.*

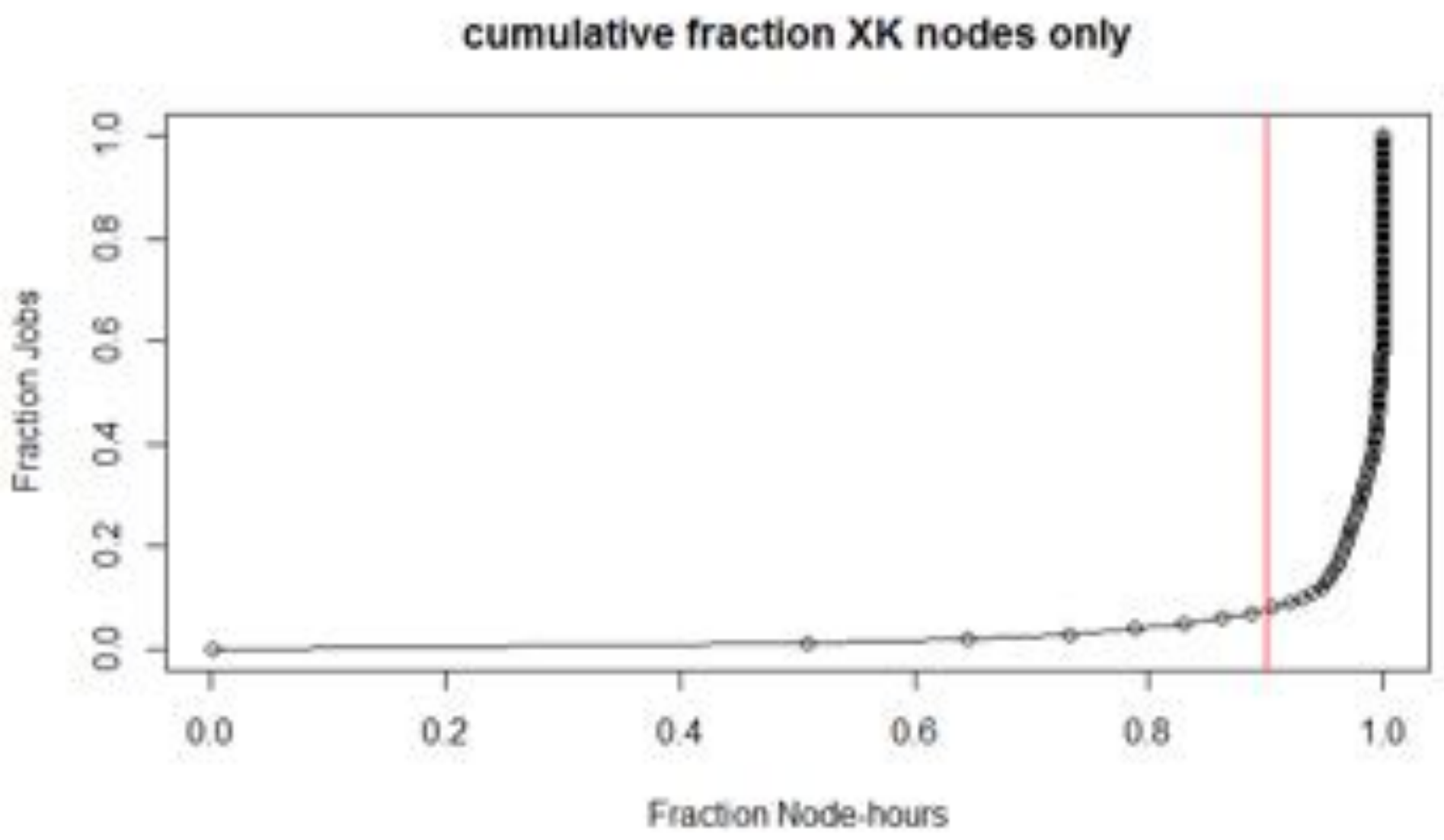

*Figure 6.0-2 Cumulative plot of the fraction of node hours as a function of the fraction of jobs on the XK. Note that the largest 7% of the jobs (by node hour) account for 90% of the node hours (indicated by red line).*

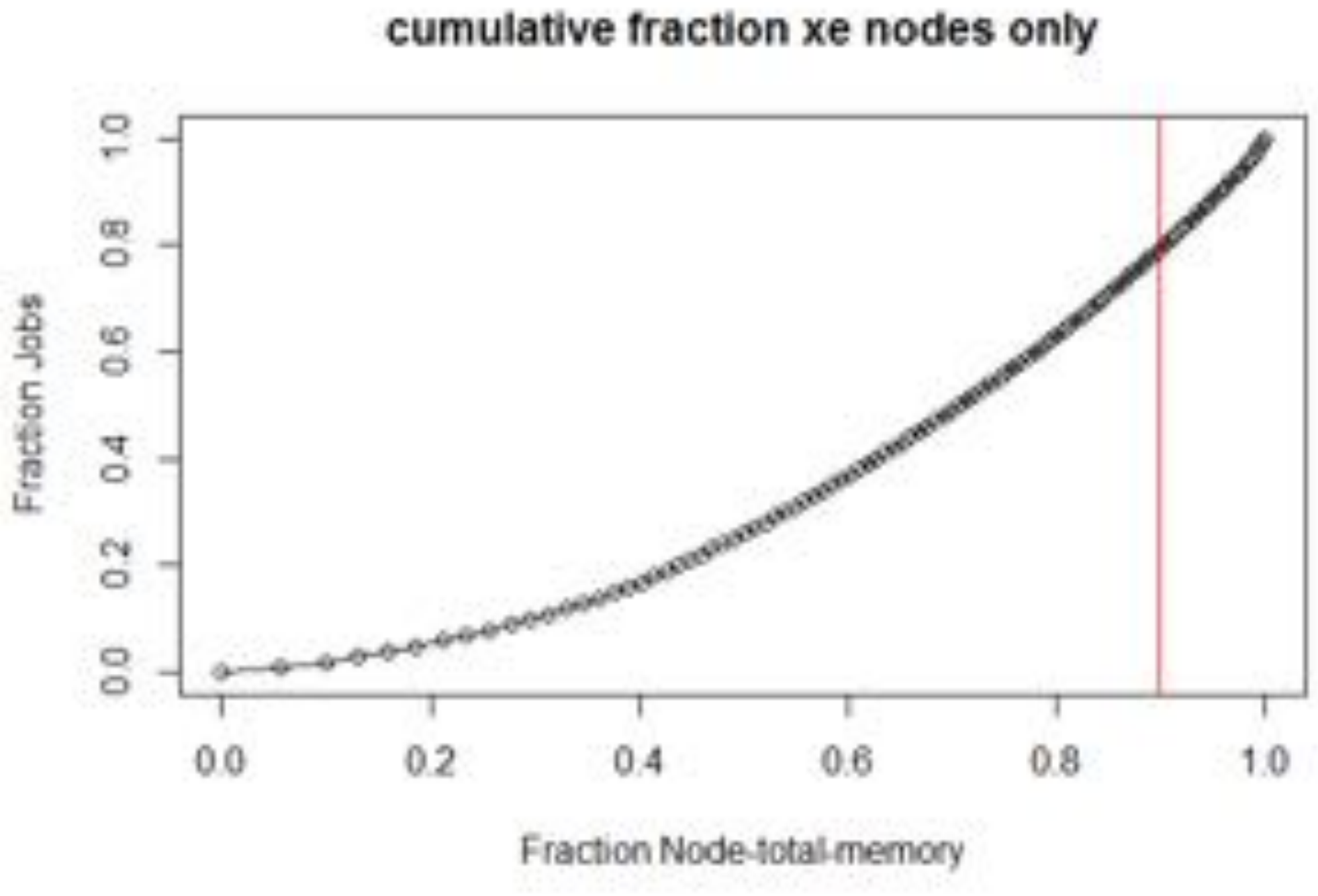

*Figure 6.0-3 Cumulative plot of the fraction of node-total-memory as a function of the fraction of jobs. Note that the plot is much more linear than the node hour plot. The red line indicates 90% in total memory.*

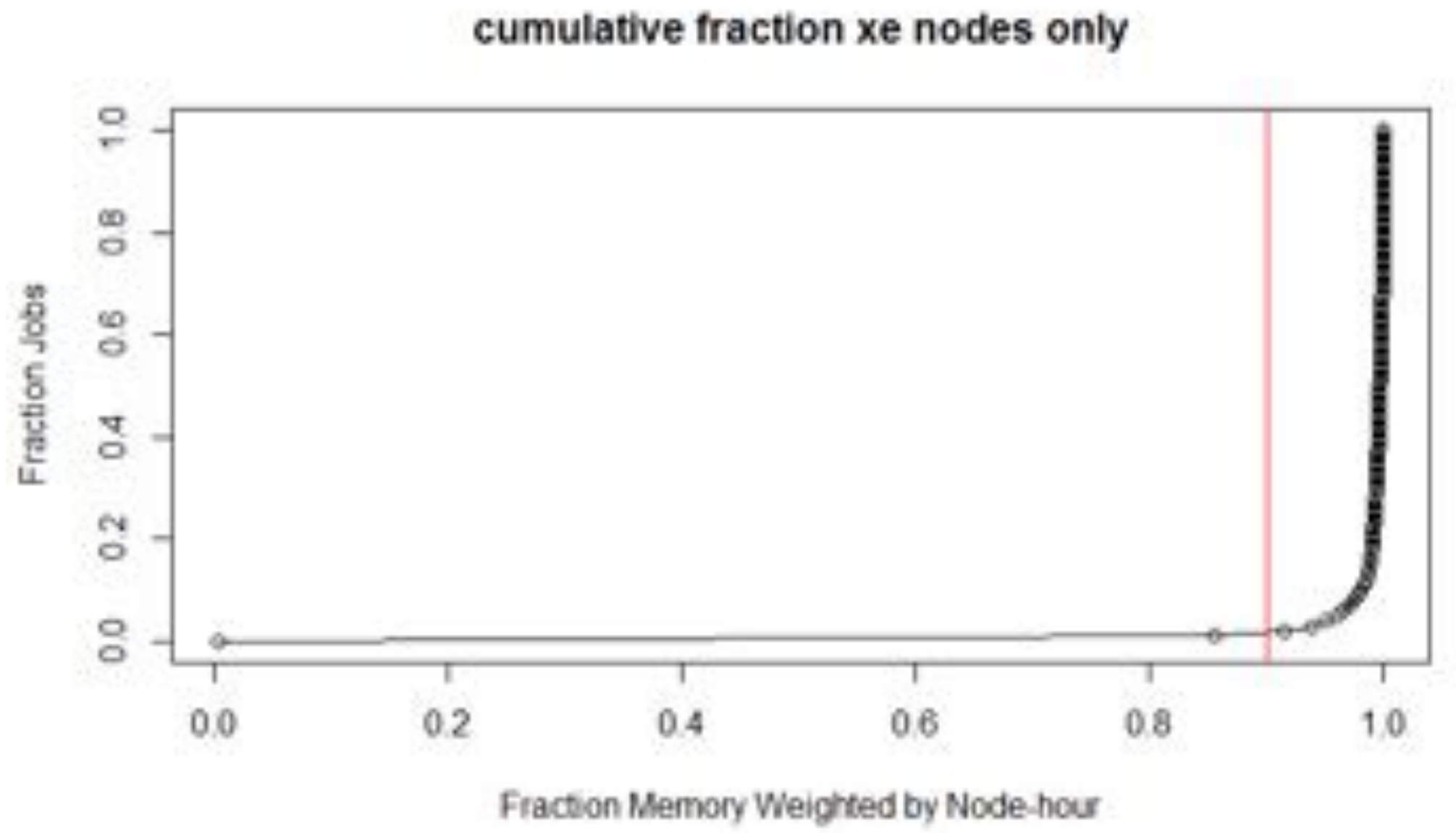

*Figure 6.0-4 Cumulative plot of the fraction of node-total-memory as a function of the fraction of jobs. Note that the plot basically reproduces the node-time cumulative plot; the memory aspect is washed out. The red line indicates 90% in node hour weighted memory.*

## 6.1 Summary: Large Job Usage

On the XE nodes, the largest 3% of the jobs (by node hour) account for 90% of the node hours used. On the XK nodes, the largest 7% of the jobs (by node hour) account for 90% of the node hours used. The memory usage is significantly more linear relative to job size when analyzed simply per job, but there is a substantial fraction of the jobs that use more than 50% of the available memory.

# 7.0 STORAGE AND I/O

*BW Analysis Goal 5: Analysis of application I/O patterns, if any, and how is this evolving over time? Does this differ between disciplines, job type, etc.?*

*BW Analysis Goal 6: Identification of current and potential I/O performance bottlenecks are?*

## 7.1 File System Overview

The monitoring of the I/O subsystems by the Lustre Management System on the server side, the OVIS/LDMSD on the client side and with an application level I/O logging library (Darshan) provides a unique and in-depth view into the use of the file systems on Blue Waters. Figures 7.1-1 through 7.1-3 show the overall filesystem activity for Blue Waters. As a reminder the /scratch file systems is over 21 PB in usable capacity, and the /home and /project are each 2.1PB of usable capacity.

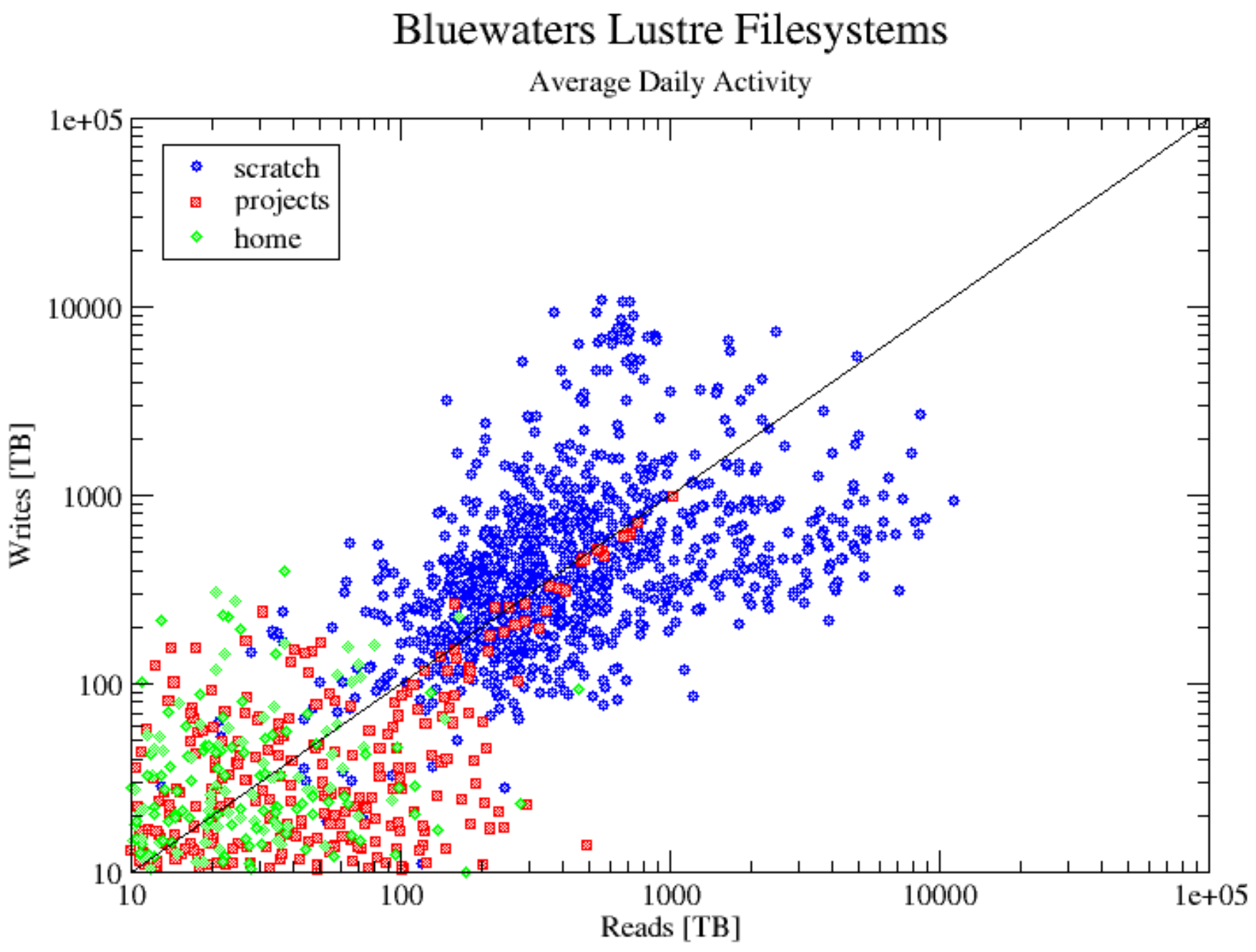

*Figure 7.1-1 Home, projects, and scratch file system activity in 30 day averages, from 2014-11 to 2016-08, as measured by the Lustre Management System. Solid line is an equal amount of reads/writes provided as a guide to the eye.*

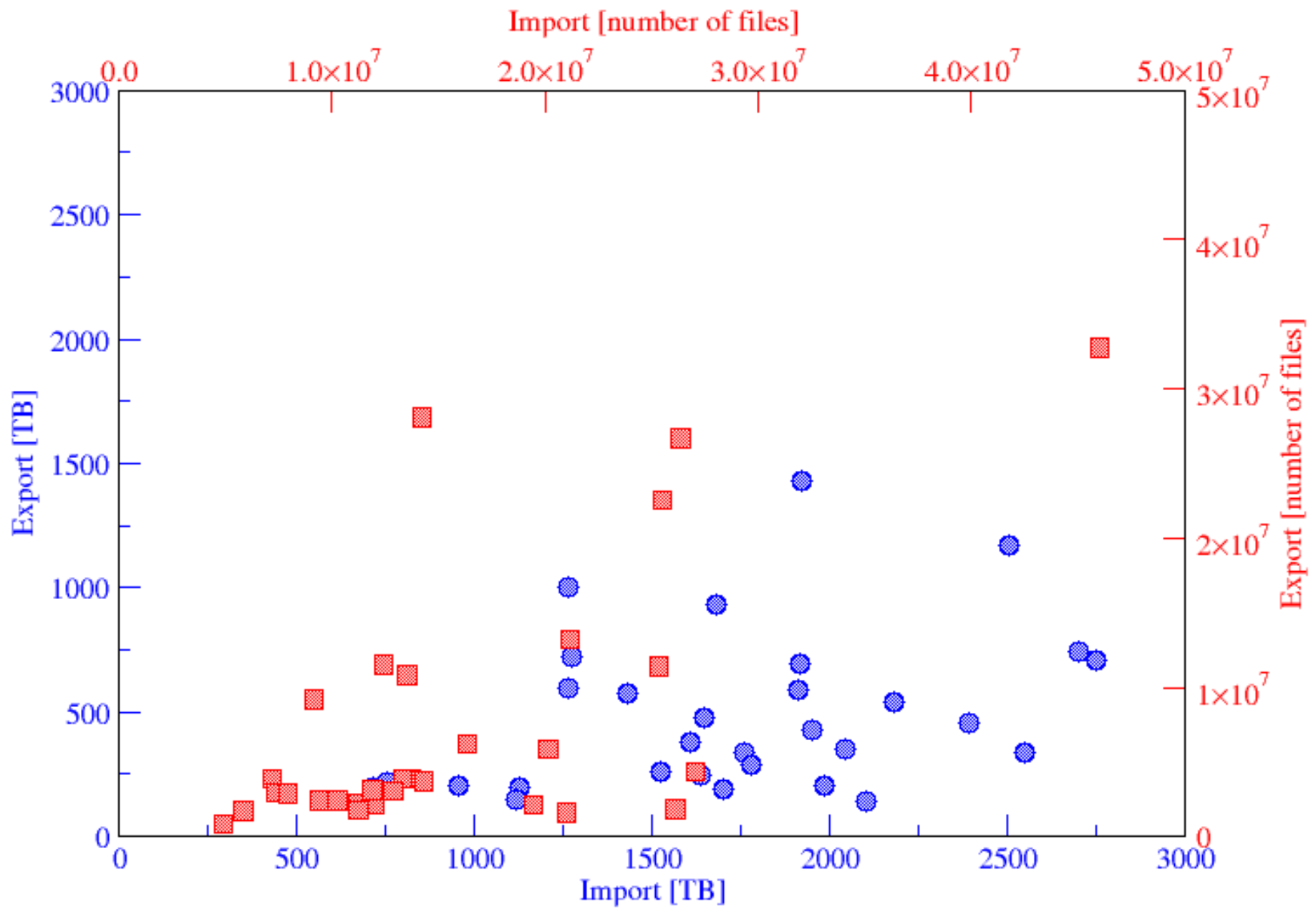

*Figure 7.1-2 External file activity balance to (import) and from (export) Blue Waters, in monthly averages, from March, 2014 to August 2016 as measured by the Lustre Management System. Blue circles represent the volume in TB, red squares the file counts.*

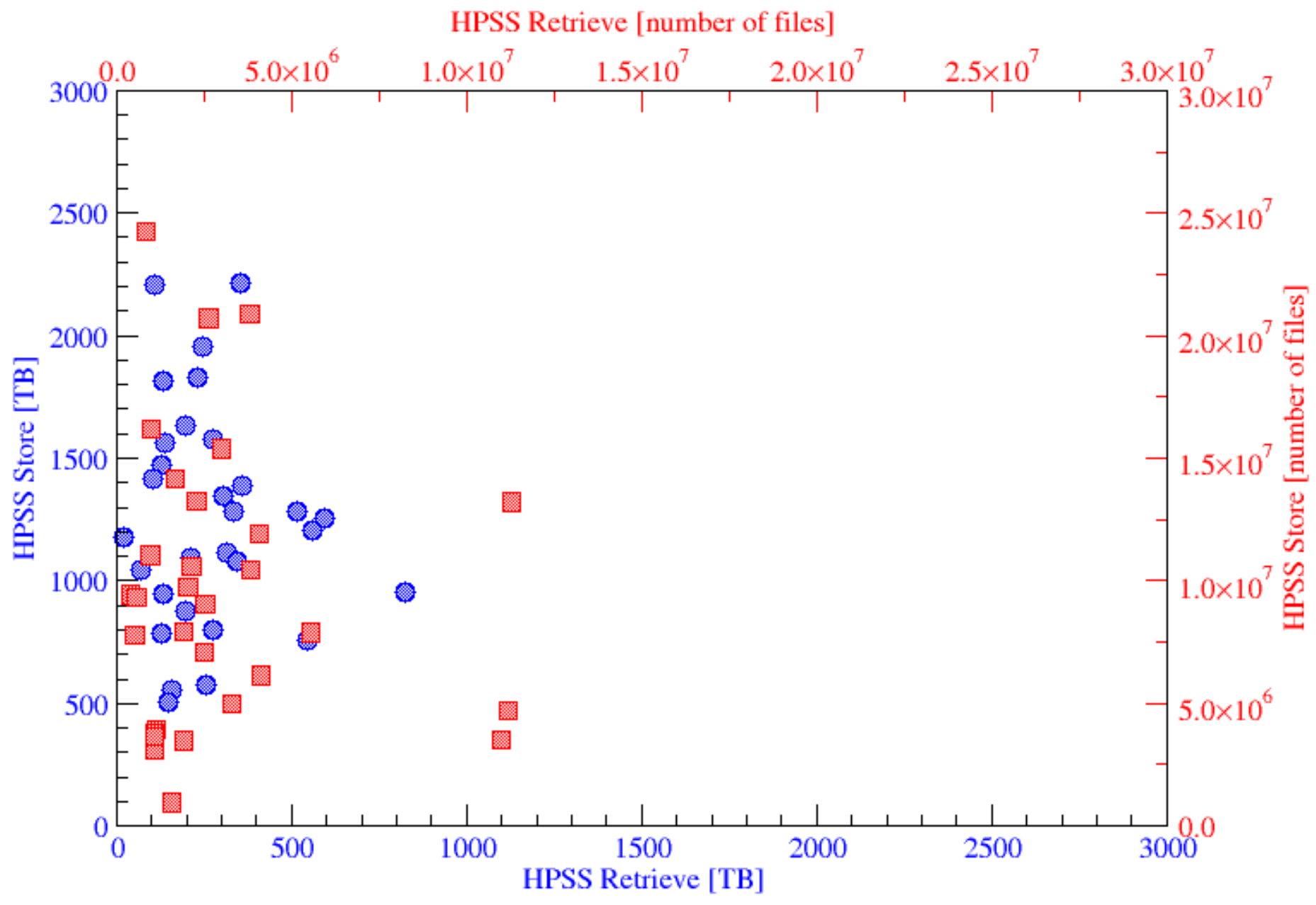

*Figure 7.1-3 Nearline storage activity to (store) and from (retrieve) the (HPSS system), monthly averages, as measured by HPSS, from March, 2014 to August, 2016. Blue circles are volume, red squares are file counts.*

Figure 7.1-1 shows a roughly even balance overall between read and write activity to the three filesystems on Blue Waters (home, projects, and scratch), with a fair amount of fluctuations that show up even in monthly averages. The largest (and least restricted quota) scratch filesystem naturally has the largest volume traffic, with peaks close to 10PB. Figure 7.1-2 shows a fairly strong bias towards monthly data moving into Blue Waters, peaking at nearly 3 PB and 50M files. Figure 7.1-3 shows nearline/backup activity to HPSS, and shows a significant ratio of stores to retrieves.

Note that historical quota information is not available for Blue Waters. Hence we cannot discuss trends in file counts and file sizes for home, projects, and scratch.

## *7.2 I/O Trends*

In this section, an analysis of the Darshan data is presented with the intention to provide insight into the I/O use patterns by Blue Waters users. It is important to note that the Darshan data coverage is limited to a subset of I/O modes as discussed in Section 1.3. Therefore this analysis will miss trends for I/O for jobs that were not instrumented with Darshan.

### 7.2.1 Darshan Data Coverage and Limitations.

The coverage of Blue Waters jobs by Darshan data is shown in Figure 7.2-1. In order for Darshan to be able to track I/O, the target application needs to be compiled with the Darshan library, which is the default on Blue Waters. The user can unload Darshan when building or disable Darshan tracking during the job execution. For these reasons, the Darshan coverage is limited. The records start on March, 2014 and run through October, 2016. Figure 7.2-1 shows the percent of all jobs by job count and by node hours for which Darshan data exists over this time period (blue trace). During most of this time it is less than 20% in both cases.

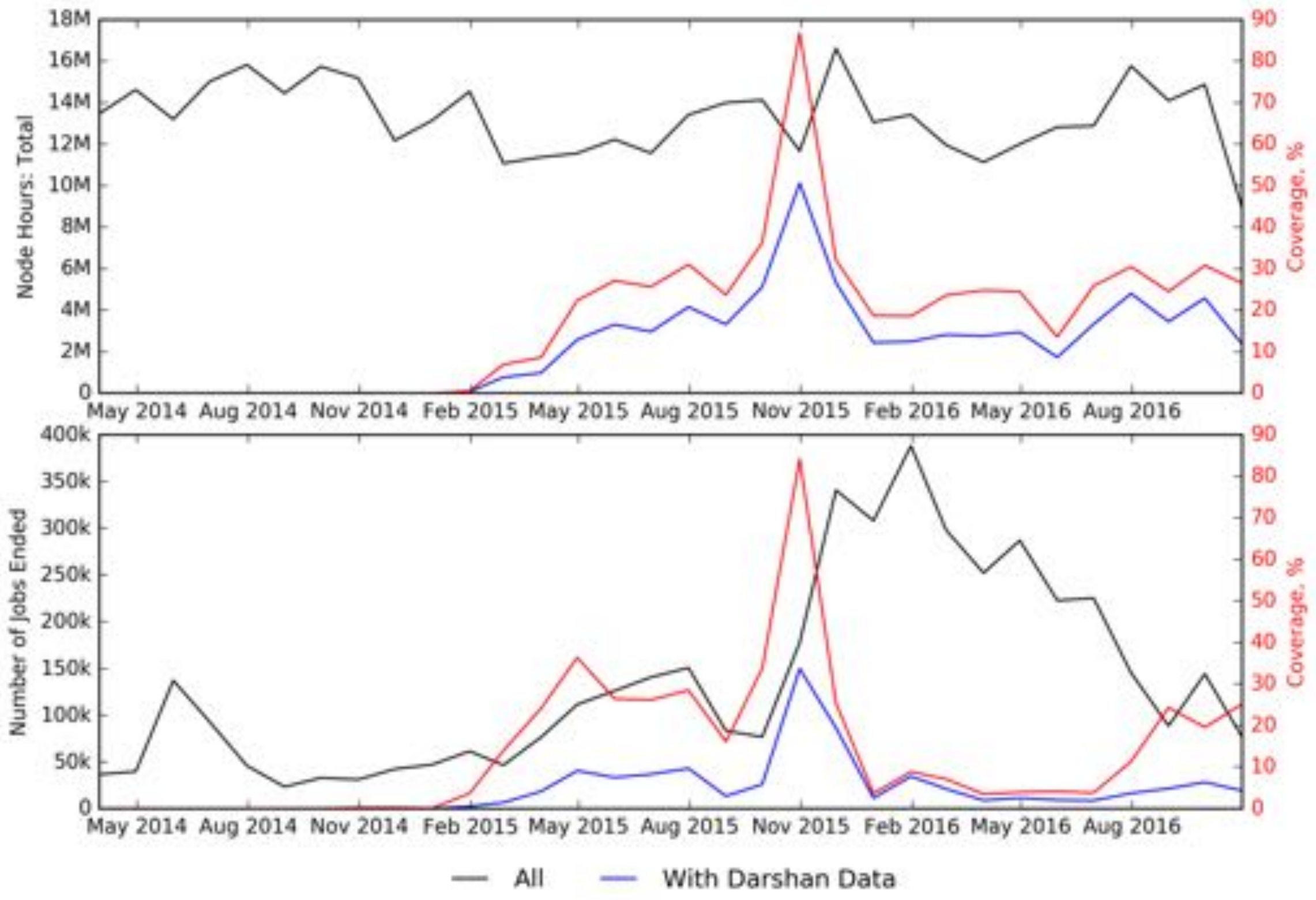

*Figure 7.2-1 Dashan data coverage of Blue Waters jobs.*

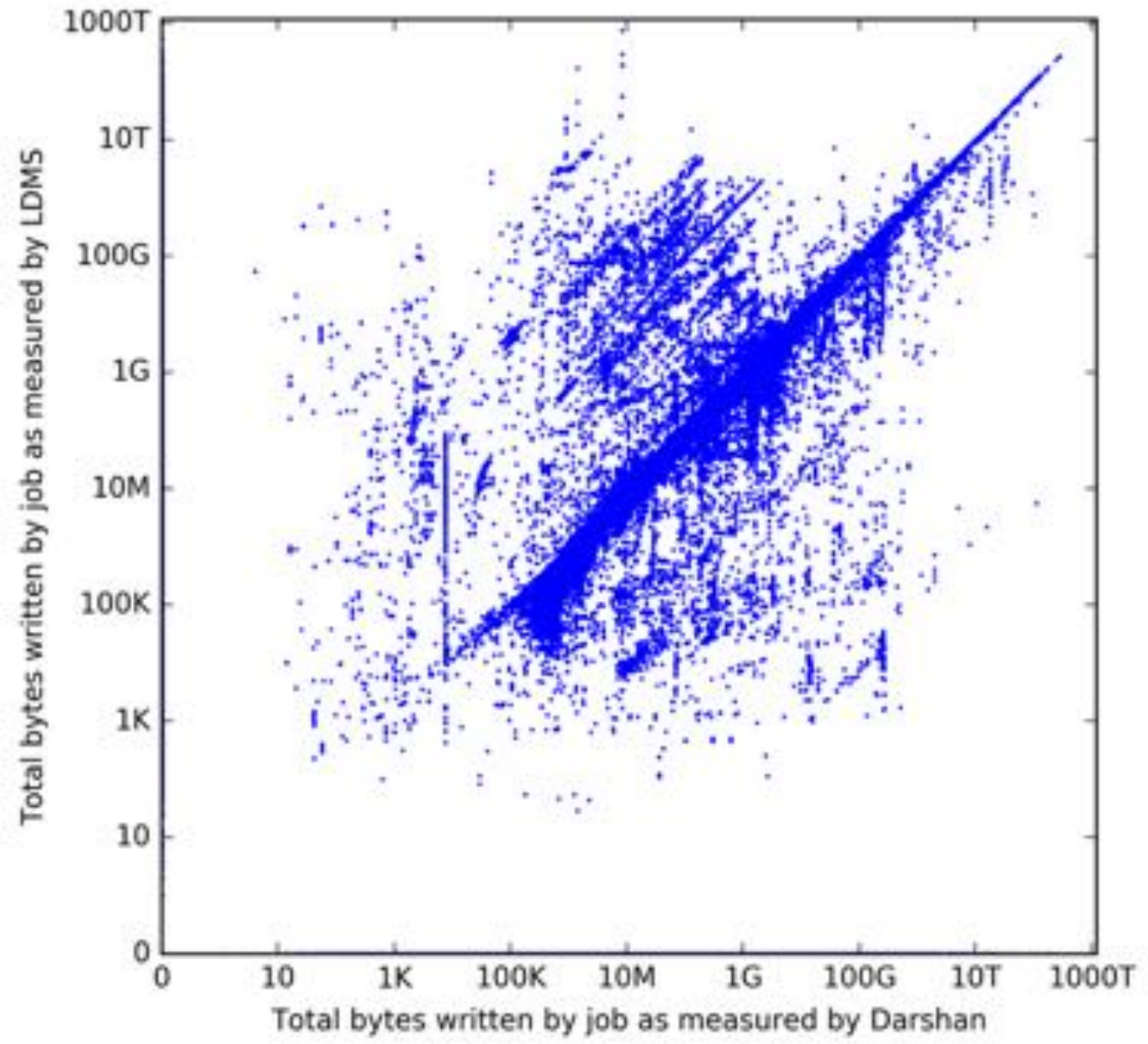

*Figure 7.2-2 Comparison of Darshan and LDMS I/O data. There is significant difference between Darshan and LDMS data: In about 40% of all jobs, one of the tools reports no writes while the other does. In 20% both show no writes and in only 14% of all jobs do both tools report writes within 20% of each other. This discrepancy may be due to the different modes of I/O the tools track.*

### 7.2.2 Darshan I/O Analysis

Figures 7.2-3 and 7.2-4 show the frequency of file openings, bytes written and read as well as I/O rates. This indicates that user jobs span the gamut of the amount of I/O performed from little I/O to very large I/O. It also shows that many jobs use a very large number of files, which is a challenge for many parallel file systems. At the same time, the read and write rates stay very small in comparison with peak performance. Although there are large data transfers to and from file system, the recorded jobs spend a very small fraction of time in file system I/O operation (Figure 7.2-5 and 7.2-6.). Specifically, 90% of the jobs spend less than 0.04% of their runtime on file system read and writes (job comprising 90% of total node hours spend less than 0.025% of run time on file-system I/O operations). An unpublished analysis of Blue Waters Darshan data [13] by Gropp et. al. shows that I/O on Blue Waters and other leadership class systems can be dominated by small files and that only a small number of applications are able to sustain a reasonable fraction of peak performance. Also, even though the common I/O rates are relatively small compared to peak rates, there are sufficient reports of I/O slowness via service request submission to indicate that there are research team impacts even with low average percent of peak usage. This is similar to network usage, where the average utilization is not a good indicator of quality of service since it is really the instantaneous usage relative to peak that is the most important factor for quality of service.

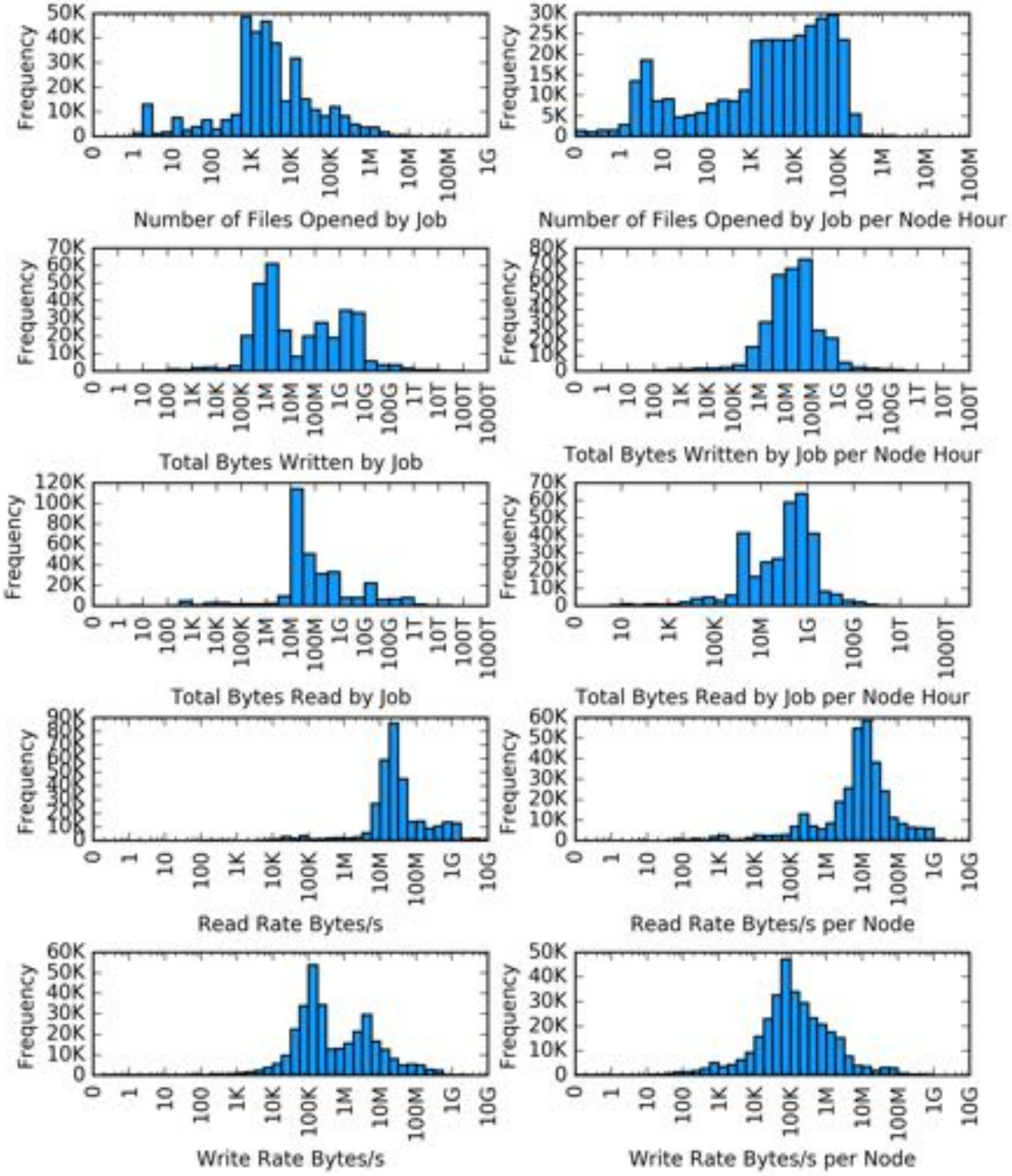

*Figure 7.2-3 Distribution of files opened, total bytes written and read, write and read rates per job (left) and per job per node hour (right). Write and read rates are shown per node (right).*

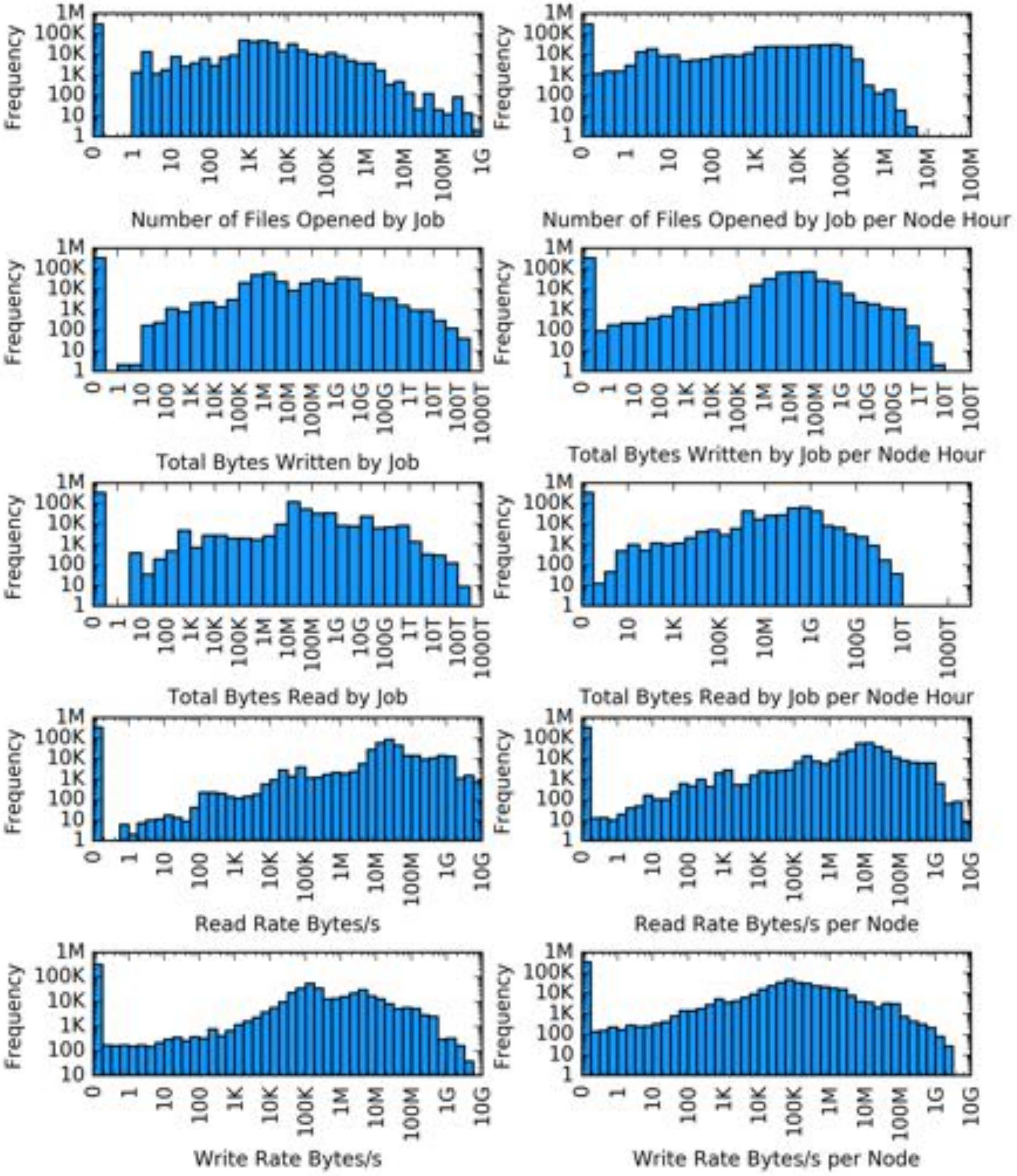

*Figure 7.2-4. Same as Figure 7.2-3 but with log scale on y axis. Distribution of files opened, total bytes written and read, write and read rates per job (left) and per job per node hour (right). Write and read rates are shown per node (right).*

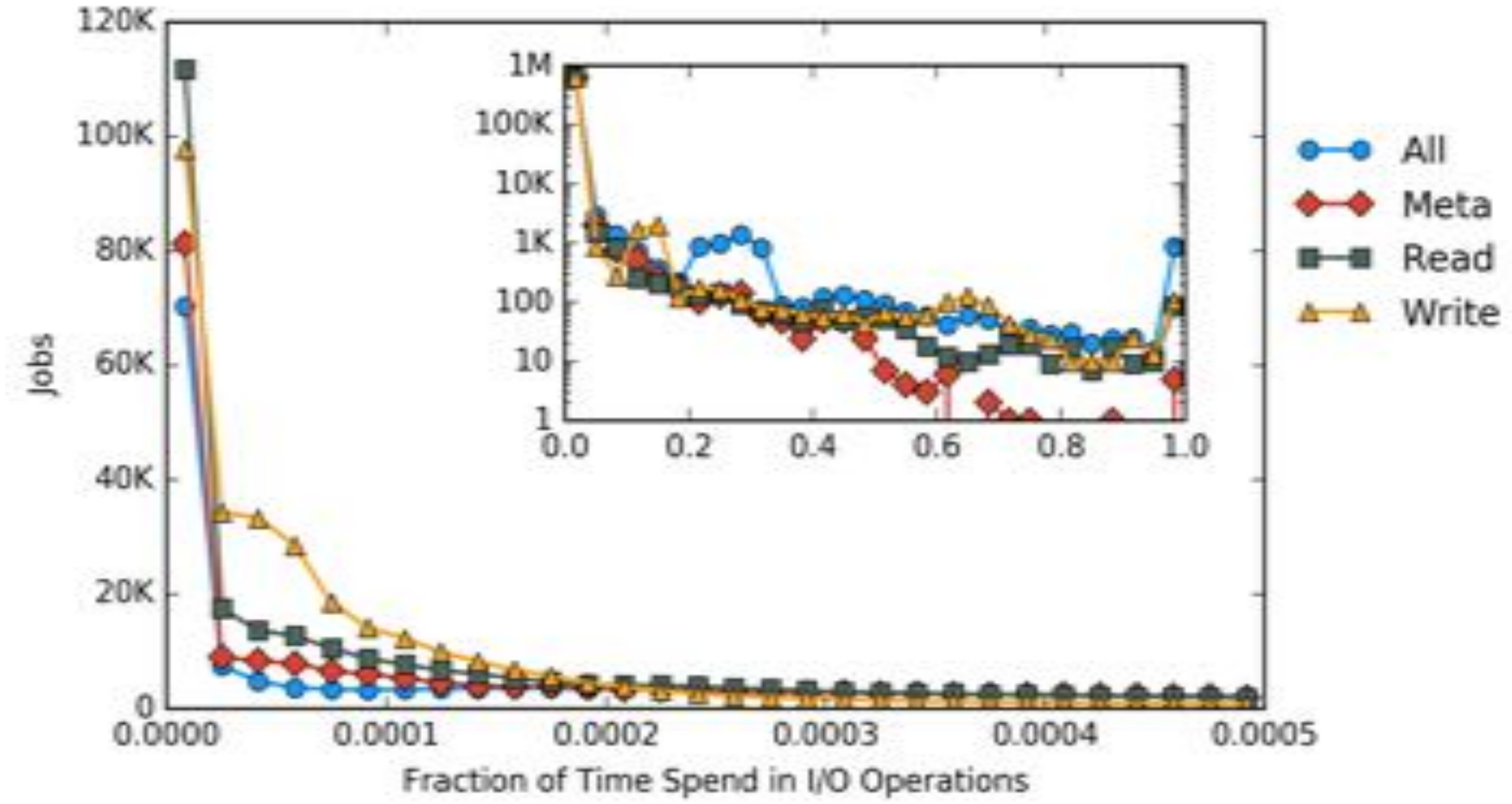

*Figure 7.2-5. Fraction of time spent in file system I/O operations by jobs. Insert shows fraction in range from 0 to 1 and uses a log scale on y-axis.*

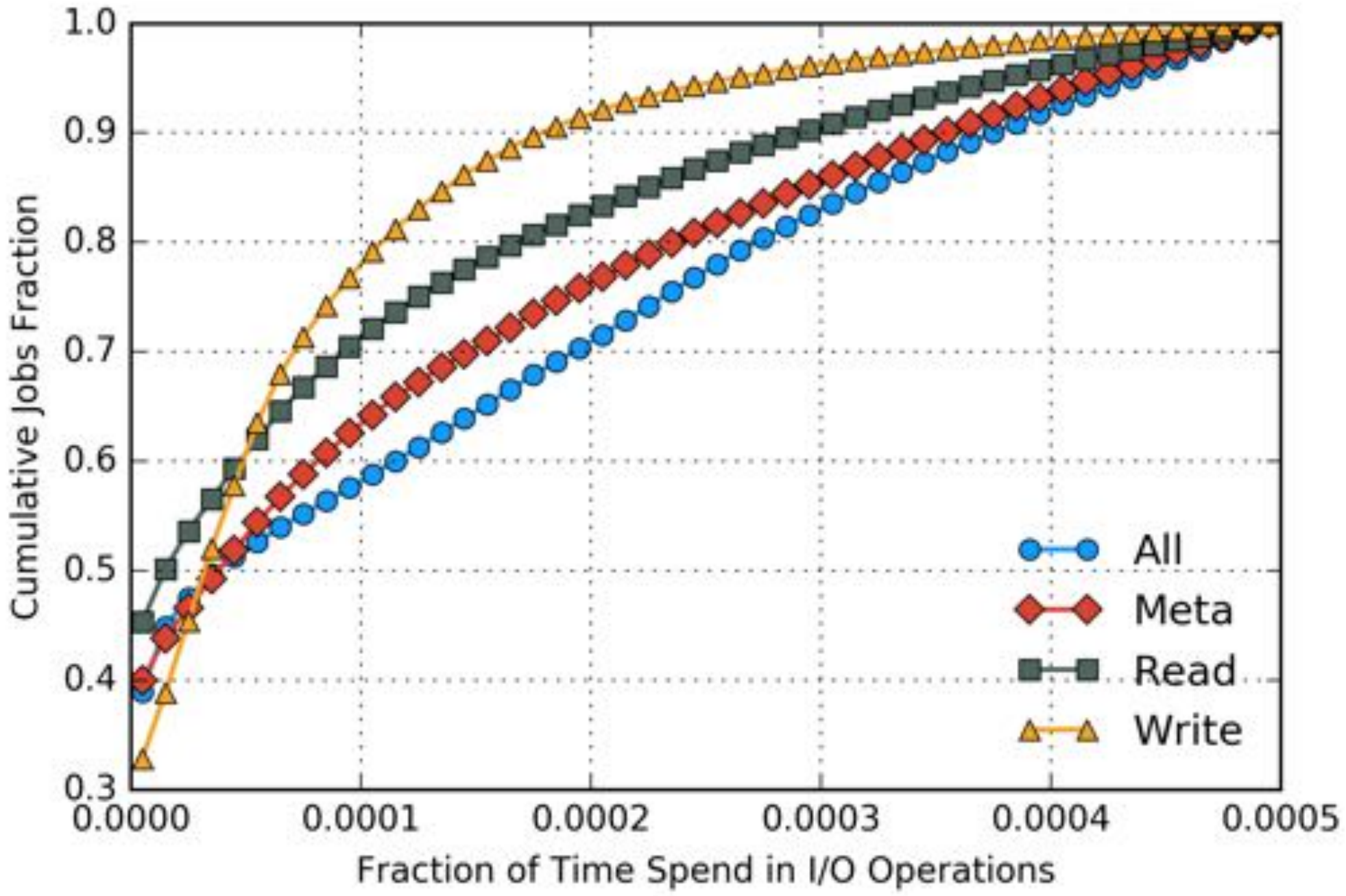

*Figure 7.2-6. Cumulative jobs fraction over time fraction spent in file system I/O operations by jobs. Ninety (90%) of file-system's active jobs as recorded by Darshan spend less than 0.04% of the total execution time in filesystem I/O. Note that some time in meta data operations can be associated with writes because during the file closing, previously buffered data are flushed to filesystem.*

We can gain insight into the distribution of file formats utilized by Blue Waters users by analyzing file open operations. This data is shown in Table 7.2-1. From the table, we find that 20% of the files were written using specialized, parallel-enabled libraries or using MPI-IO calls and HDF5 is widely used on the Blue Waters (15% of all files). This is in agreement with earlier analyses [13].

Different fields of science as well as applications exhibit different performance patterns in file system I/O operations (*Figure 7.2-7 and 7.2-8*) both in terms of means and the distribution width. It is worth noting that the presented information in these Figures should be considered with high degree of skepticism due to sparse coverage of some fields of science and applications. For example, the boxplot for NAMD application corresponds to only 0.02% of all NAMD jobs (node hours based) executed during studied period.

*Table 7.2-1 File opens over time period from March, 2014 to October, 2016.*

| Description | Total Count |
|---|---|
| File stream open operations | 34076822021 |
| non-stream file opens | 14545589590 |
| non-collective MPI opens | 33402468 |
| collective MPI opens | 2774495111 |
| HDF5 opens | 9195810810 |
| Independent Parallel NetCDF opens | 37 |
| Collective Parallel NetCDF opens | 5897582 |

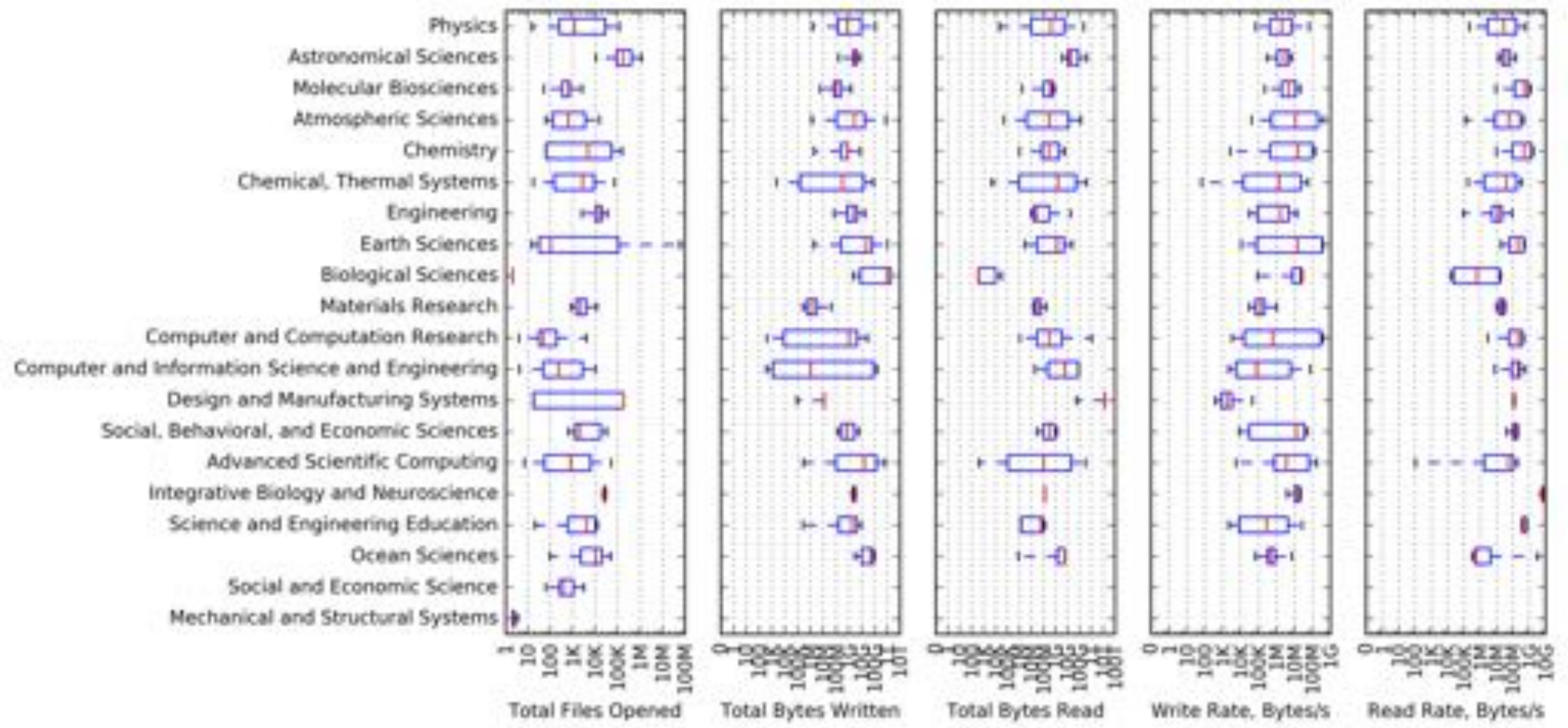

*Figure 7.2-7. Box plot of files opened, total bytes written and read grouped by parent field of science. Left whisker, left box side, middle red line, right box side and right whisker corresponds to values at 5%, 25%,50%,75% and 95% of population.*

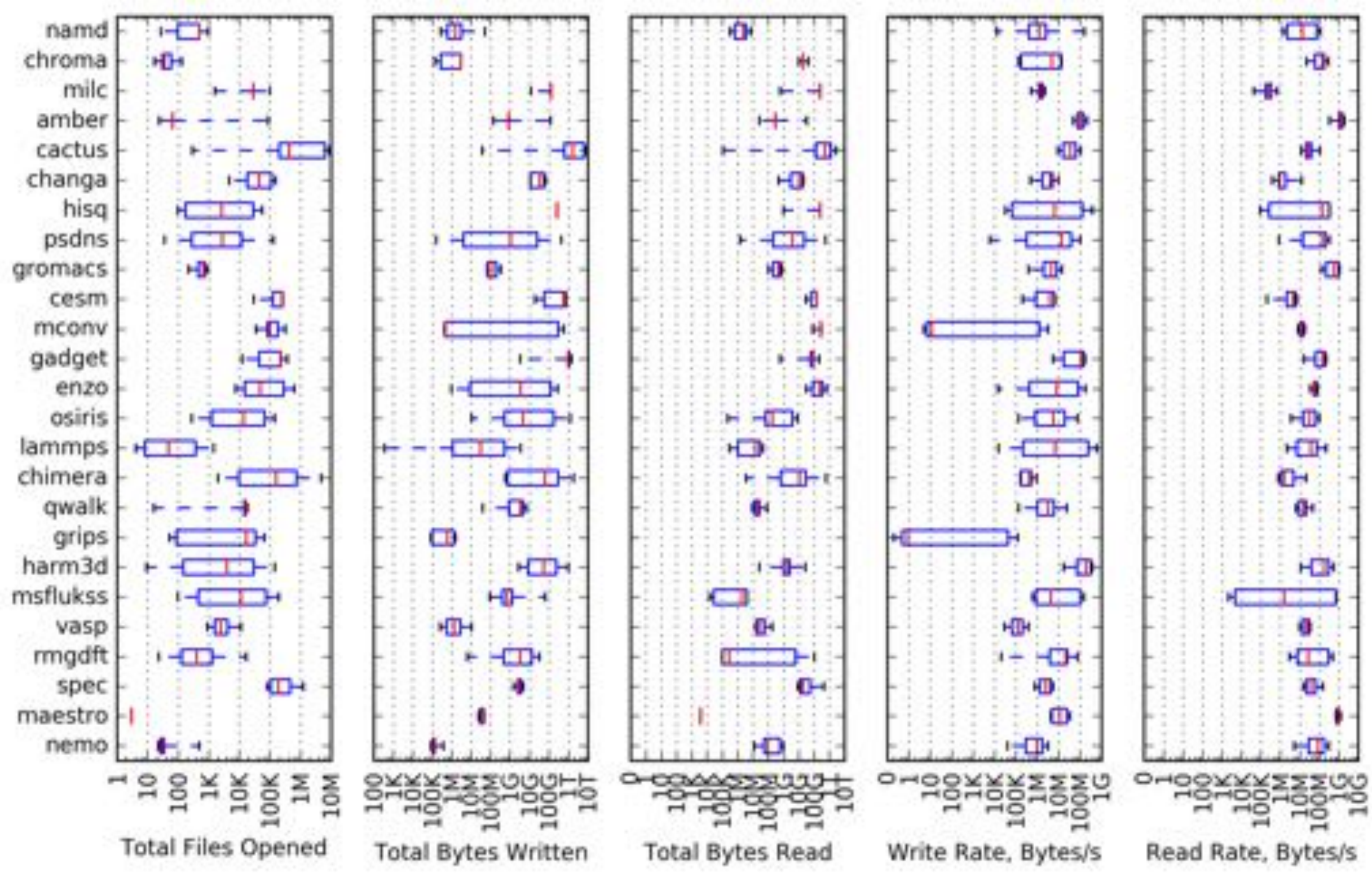

*Figure 7.2-8. Box plot of files opened, total bytes written and read grouped by application. Left whisker, left box side, middle red line, right box side and right whisker corresponds to values at 5%, 25%,50%,75% and 95% of population.*

## 7.3 I/O By Application

Figures 7.3-1 and 7.3-2 show the average rate of data written to the /scratch filesystem over time for a NAMD job and a Cactus job. These data are collected every 60 seconds from the Lustre driver on the compute nodes by LDMS. These two different applications have quite different write behavior. The NAMD job in Figure 7.3-1 shows an approximately uniform write rate throughout the job from a single compute node and no data written by any of the other nodes. The Cactus job shown in Figure 7.3-2 has approximately equal amounts of data written by each node. For this job the majority of the data is written towards the end of the run.

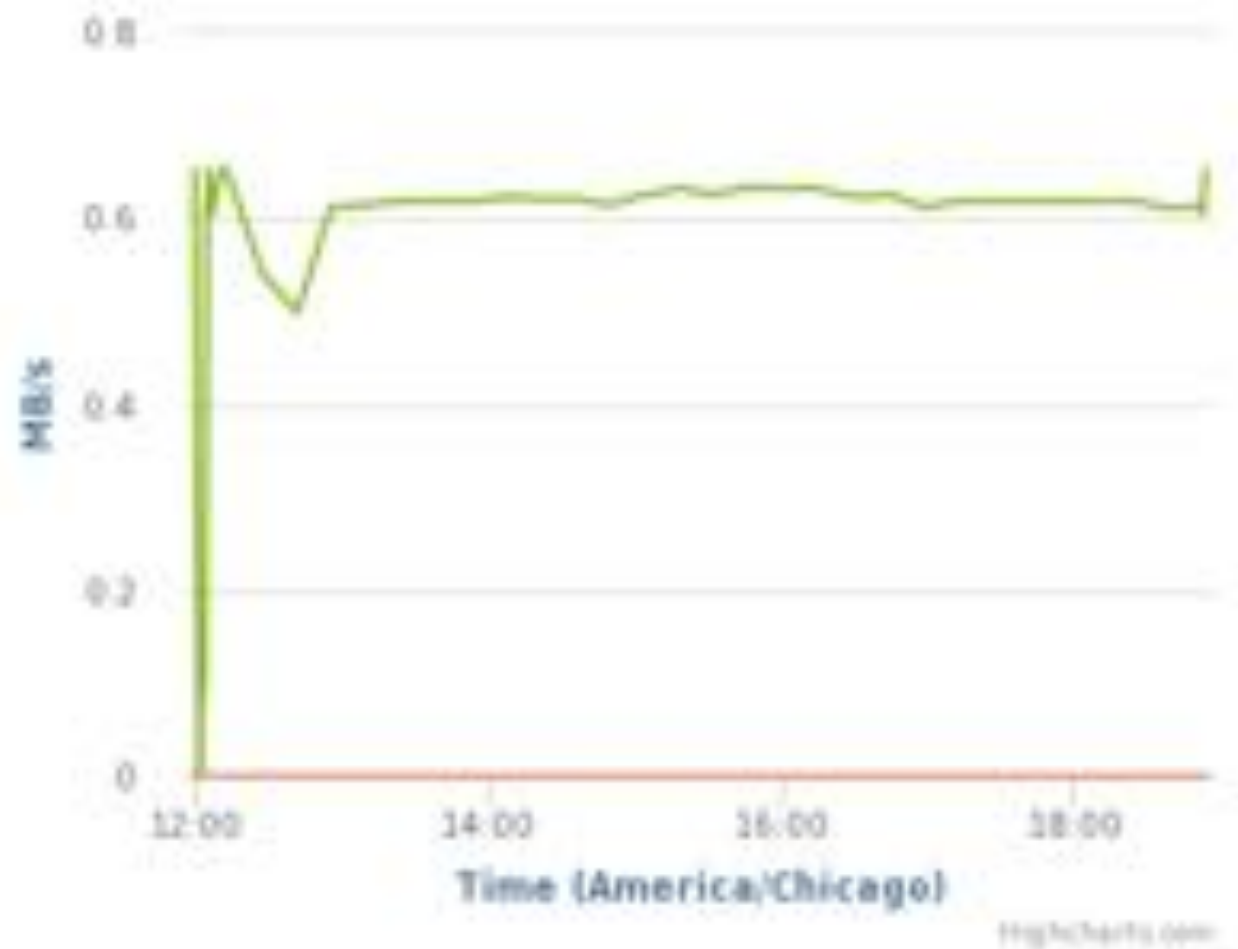

*Figure 7.3-1 Average write rate to the /scratch filesystem by each node for a NAMD job that ran on 25 nodes. The data are written by only a single node. The other 24 nodes have no filesystem writes during the job. Note the hardware counters are only sampled every 60 seconds, so the plot shows the 60 second average data rates.*

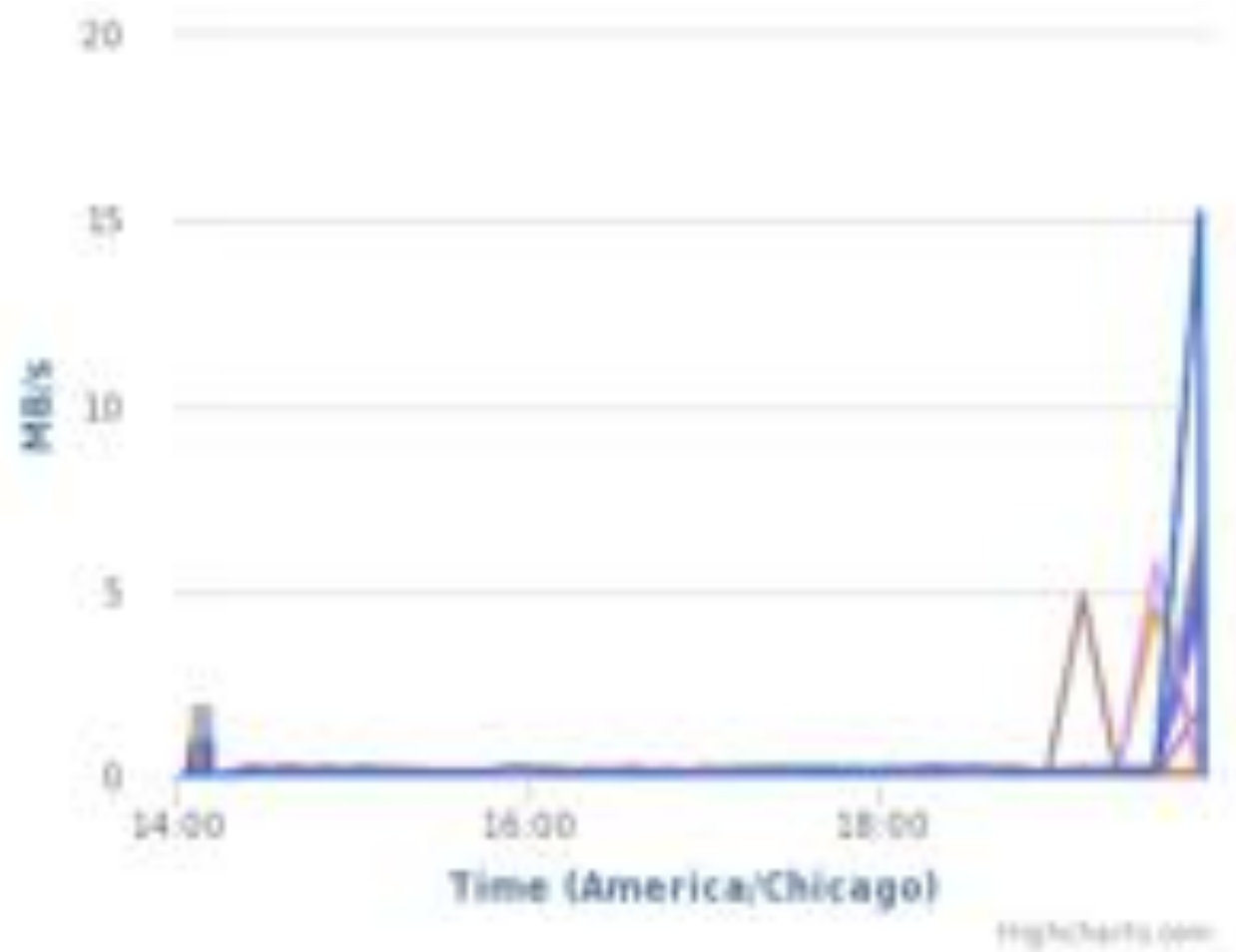

*Figure 7.3-2 Average write rate to the /scratch filesystem by each node for a Cactus job that ran on 16 nodes. Each node in the job wrote approximately the same amount of data. Note, the hardware counters are only sampled every 60 seconds, so the plot shows the 60 second average data rates.*

To further investigate the node-level I/O write distribution of different applications, we computed the coefficient of variation (defined here as the ratio of standard deviation to mean) of the data written to the /scratch filesystem by each node within a job. Figure 7.3-3 shows a cumulative plot of the coefficient of variation for writes to /scratch for different applications. We only plot data for XE node jobs that ran for more than 10 minutes on more than one node and had an overall average data rate of more than 10 KB/s per node. Ten of the top 11 applications with the highest node hours are shown. CHROMA is not shown in this plot as there were no jobs that met the selection criteria. The plot shows there are two main categories of application: ones where the majority of node hours are spent with approximately equal balance of data written by each node during the job and those where there is a large variation between the data written by different nodes in the job. For example, the ChaNGa application has 60% of the node hours with jobs that have a coefficient of variation less than 0.1. This means that compute nodes for these jobs write approximately the same amount of data. Whereas 90% of NAMD jobs have a coefficient of variation greater than 5. This means that there is a large variability between the data written by different nodes in the job.

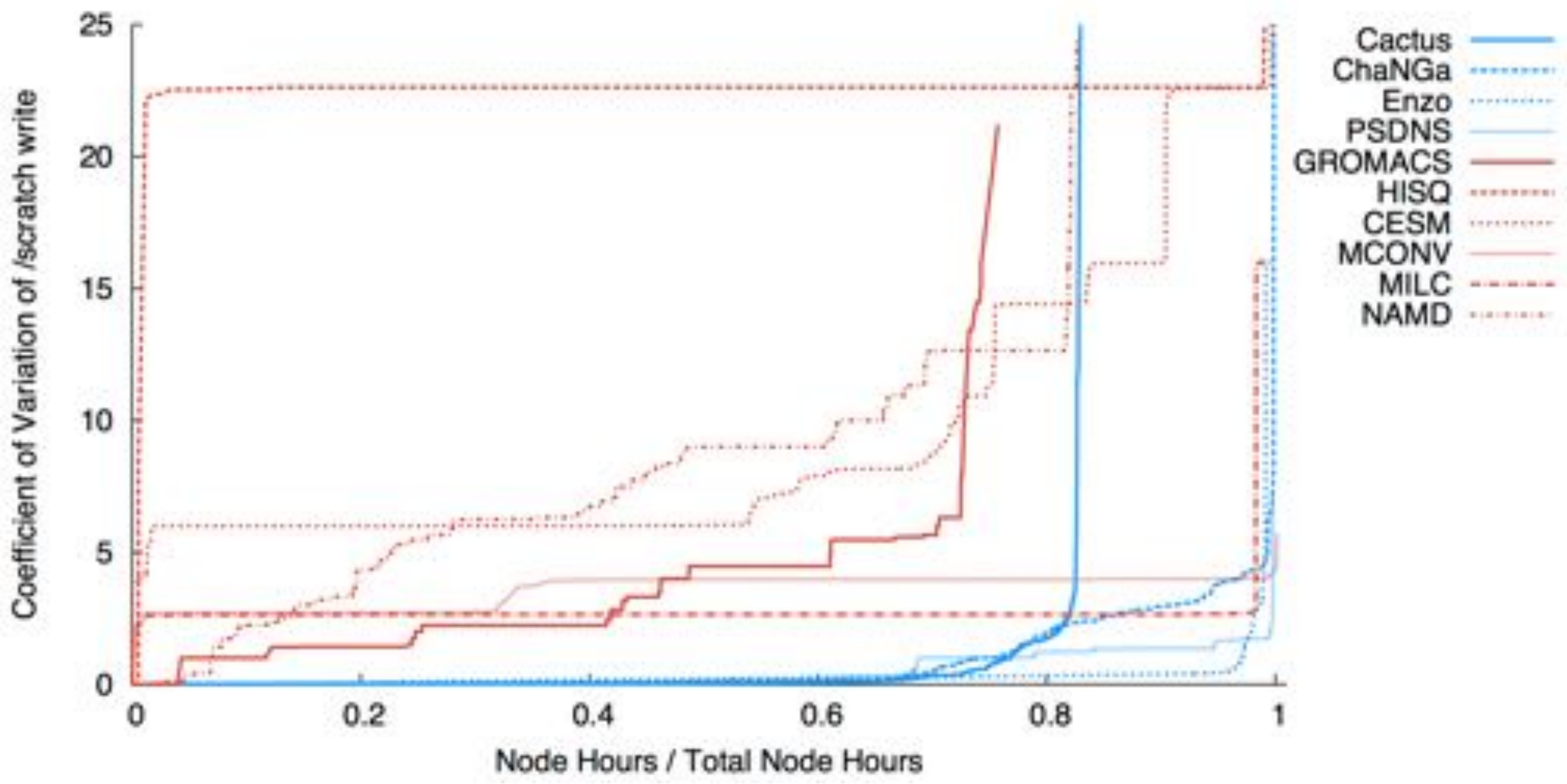

Figure 7.3-3 Cumulative plot of the coefficient of variation of the data written by each node to the /scratch filesystem for different applications. We only plot data for XE node jobs that ran for more than 10 minutes on more than one node and had an overall average data rate of more than 10KB/s per node.

## *7.4 Summary: Storage and I/O*

On average, the Blue Waters' three filesystems have balanced reads/writes ratios, but with large fluctuations. The volume of I/O traffic on the largest (/scratch) filesystem peaks at 10PB per month. Jobs exhibit a wide range of I/O patterns. Overall there is a tendency, independent of science area, for applications to use a very large number of small files rather than using larger files, which is a challenge for many parallel file systems. Possibly due to this, the read and write rates stay significantly below their possible peak performance rates, while the metadata rates may have very high utilization, but this deeper analysis was unable to be completed for this report due to data collection limitations. Nevertheless, even with such relatively small transfer rates and large data transfers to and from the filesystems, the recorded jobs spend a very small fraction of time in filesystem I/O operations (0.04% of runtime for 90% of jobs).

Many jobs utilize specialized libraries for their I/O operations (about 20% use MPI-IO, HDF5, NetCDF, etc.). The I/O variation study across nodes shows that while some application are well balanced across nodes, other are not. However, given a small fraction of time spent in I/O operations it is not a significant limitation for applications in general but may be for specific applications.

# 8.0 APPENDIX I: VALIDATION OF Open XDMoD DATA

In this appendix, we compare usage data taken from the Blue Waters Q3-2016 Report to NSF (not published) with usage data from the Blue Waters Open XDMoD instance running at NCSA in order to validate ingestion of Blue Waters' accounting data into Open XDMoD. The data used to populate the Open XDMoD instance is obtained by pre-processing the Blue Waters' log files and then ingesting the information into Open XDMoD.

Figure 8.0-1 is a plot that compares the charged node hours for the top 44 BW users, taken from the 2016/Q3 Quarterly Report, with the same data taken from Open XDMoD. The plot indicates good agreement in most cases from both data sources.

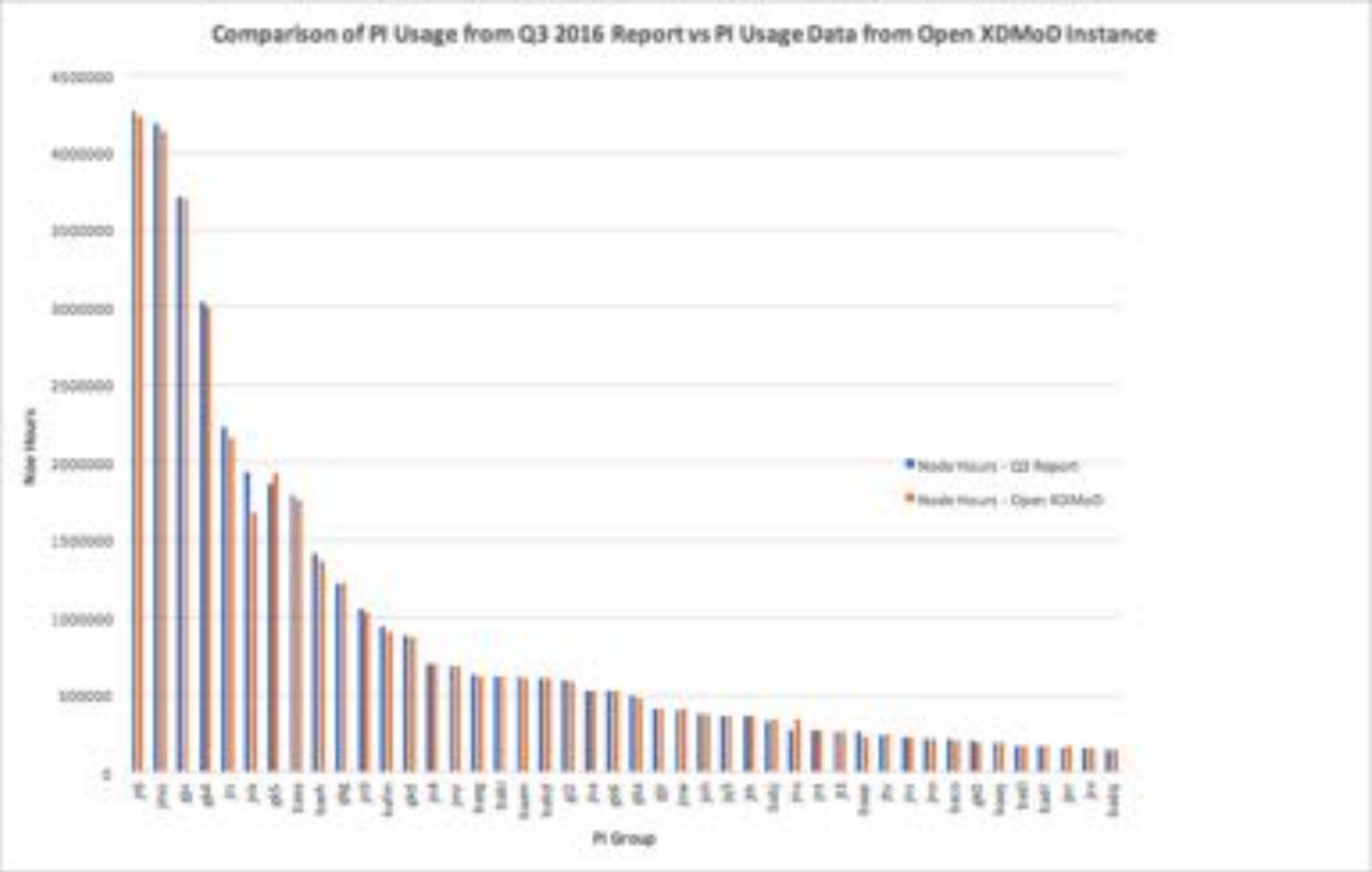

*Figure 8.0-1 Comparison of PI Usage from Q3 2016 Report vs PI Usage Data from Open XDMoD Instance*

Figure 8.0-2 is a plot comparing node hours for the top 25 fields of science in Q3 of 2016 taken from Q3 2016 Quarterly Report to NSF with the same data taken from Open XDMoD. The plots show good agreement between the data sources.

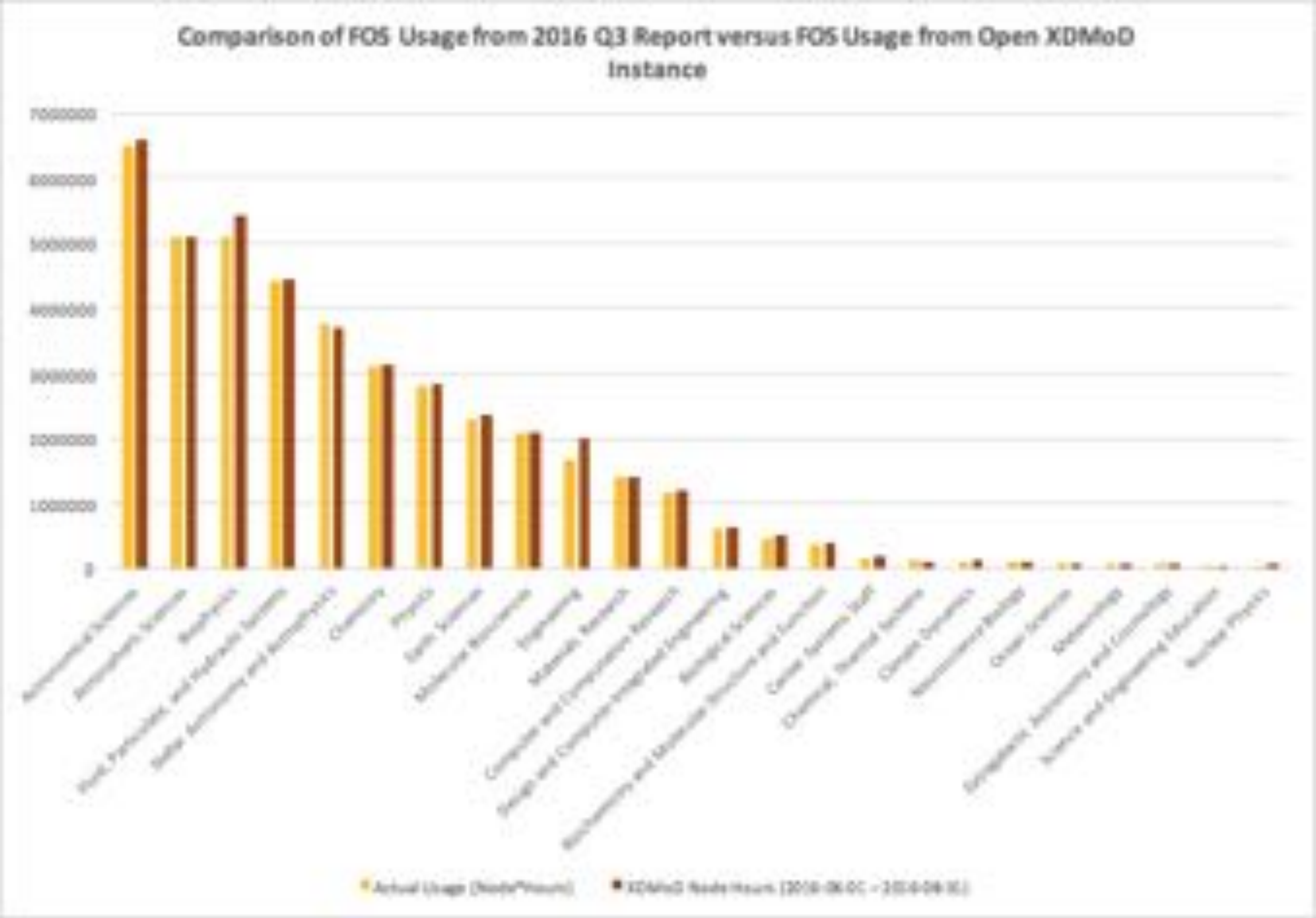

*Figure 8.0-2 Comparison of node hours for the top 25 fields of science taken from the Q3 2016 Report vs Data from the Blue Water Open XDMoD Instance*

## 9.0 Appendix II: DATA COVERAGE ESTIMATES

In this appendix we provide some statistics on the data coverage that should accompany the analysis of the metrics in this report. There are two main categories presented: the availability of source data and the success of the analysis in using this data.

*Table 9.0-1 Source data coverage*

| Metric | Total Jobs | Jobs Covered | % Jobs | Total Hours | Hours Covered | % Hours |
|---|---|---|---|---|---|---|
| APRUN | 4,646,659 | 4,410,785 | 95 | 547,809,848 | 543,197,000 | 99 |
| Darshan | 4,646,659 | 657,422 | 14 | 547,809,848 | 75,169,985 | 14 |
| LDMS | 4,646,659 | 3,304,285 | 71 | 547,809.848 | 382,280,591 | 70 |
| MSR | 4,646,659 | 2,032,068 | 44 | 547,809,848 | 153,723,022 | 28 |

The source data coverage is over the life of the machine and is indicative of the data that is missing at various time periods due to several reasons that are described in the main document. These include early time periods where the full set of data was not being collected as well as bugs in the system that prevented some metrics from being collected.

With a wide variety of data sources, there may be some conflicting measurements within or between them. Below is a list of the node hours for various data conflicts that exist among the data sets. These conflicts have been dealt with in various ways. In some cases, specific measurements can be ignored, some can be reconciled and in other cases the jobs that contain these data must be excluded from the analysis. Specifically, for those jobs where we cannot determine the node type (XE vs. XK), those jobs are excluded from analysis that is specific to the node type.

*Table 9.0-2 Data conflicts*

| Conflict | Hours affected | % Hours |
|---|---|---|
| Duplicate Job IDs | 360038 | 0.1% |
| Unknown Node Type | 4126645 | 0.8% |
| Undetermined Node Memory | 644561 | 0.1% |
| APRUN Times Before Job | 3079047 | 0.6% |
| APRUN Times After Job | 4581964 | 0.8% |
| Negative Lustre Metrics | 748043 | 0.1% |
| Negative Gemini Metrics | 307594 | 0.1% |

From Table 9.0-2 we have several cases to add comments. For a small subset of jobs, the APRUN timing information seems to be corrupted. For a given job ID, the timestamps on the APRUN records were found to be as far as 30 days before the job actually ran or 70 days after the job ran. In some cases we were able to confirm that the APRUN records were correct and simply had bad time stamps. In other cases, there appeared to be records from unrelated jobs with an incorrect job identifier. For these jobs, the counters were ignored.

For the negative performance metrics, we found that for certain intervals, some of the counters reported physically impossible values. When the counters resumed their expected reporting, the large discontinuity resulted in negative values for the overall utilization. For these jobs, the counters were ignored.

As part of the job summarization process, a certain temporal data quantity is required to generate the necessary statistics. We outline below the result of this analysis. This is presented on two forms: overall over the life of the machine, and since 2014-03-28 when LDMS data became available. For jobs that ran before LDMS data was available, no performance analysis can be done. Even when data is available, a sufficient quantity of samples is required to produce meaningful analysis. As such, the largest contributor to the inability to summarize jobs is the existence of jobs that ran for a short amount of time. In reality jobs must run for 3-5 minutes in order to provide enough performance samples to be of use. Jobs that run less than 5 minutes probably indicate some basic issue with the job script.

Descriptions of the labels;

- No Archives
    - No archives exist for any hosts in the job. Mostly pre LDMS jobs
- Not Enough Archives
    - Fewer than 95% of hosts in a job have archives
- Too Short
    - 1 node and <= 3 minutes
- Parallel Too Short
    - > 1 node and < 5 minutes
- pmlogextract Error
    - Failed to find 95% coverage for performance data
- Good
    - Successfully summarized, but may have some metrics missing as described previously

*Table 9.0-3 Summarization coverage for the Blue Waters operational period.*

| Error | Node hours |
|---|---:|
| Not Enough Archives | 10,936,404 |
| No Archives | 137,616,639 |
| Time Too Long | 244,953 |
| Too Short | 6,756 |
| Parallel Too Short | 1,550,960 |
| Pmlogextract error | 8,950,569 |
| Good | 385,299,627 |

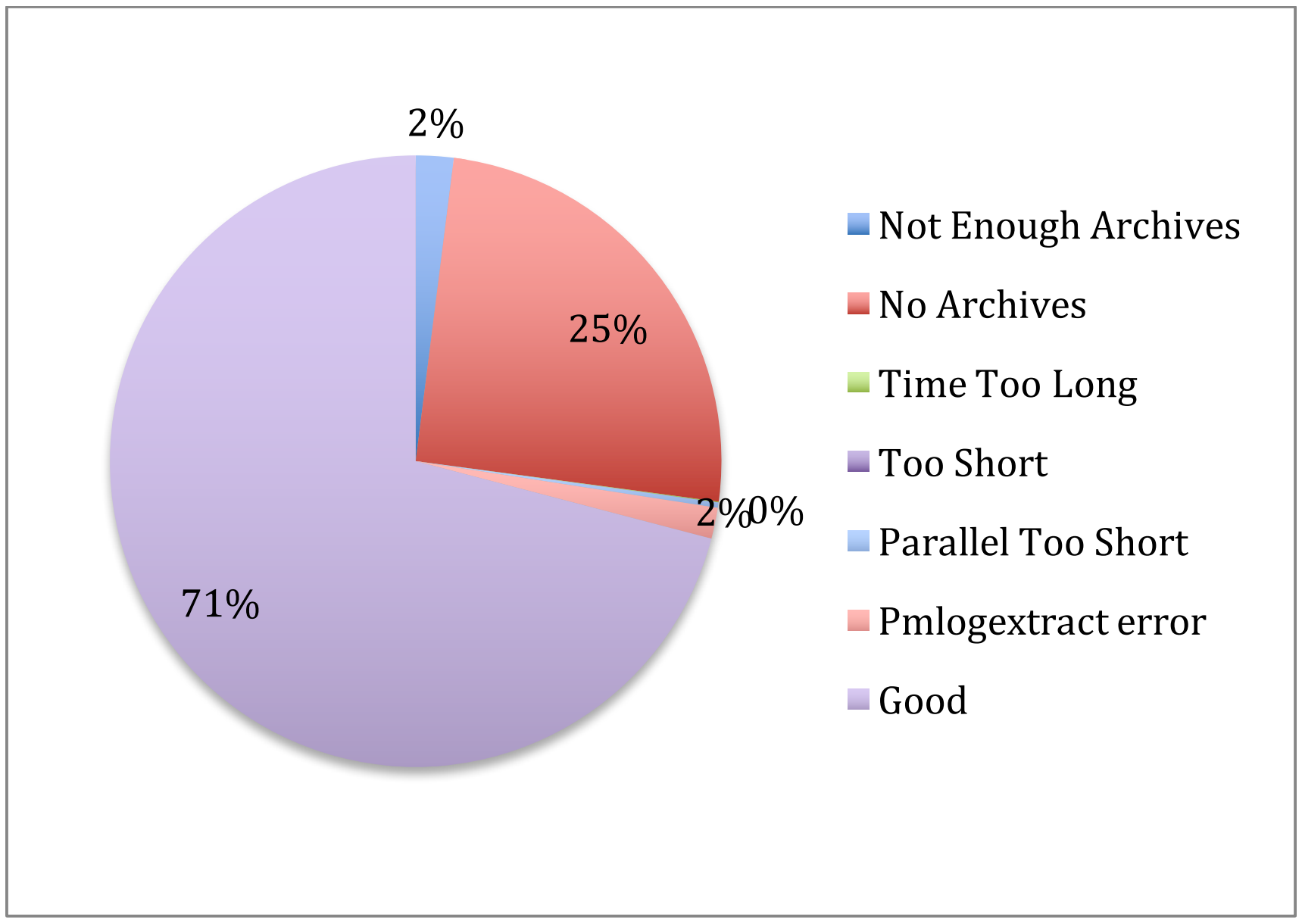

*Figure 9.0-1 Summarization coverage for the operational period of Blue Waters*

Since the many of the job level performance analyses require LDMS data to be useful, the table and chart below show the summarization success during the time period when LDMS data was being collected.

*Table 9.0-4 Summarization coverage for jobs from 2014-03-28 to to 2016-09-30.*

| Error | Node hours |
|---|---|
| Not Enough Archives | 10936404 |
| No Archives | 7039410 |
| Time Too Long | 169371 |
| Too Short | 6284 |
| Parallel Too Short | 862360 |
| Pmlogextract error | 6869102 |
| Good | 385299627 |

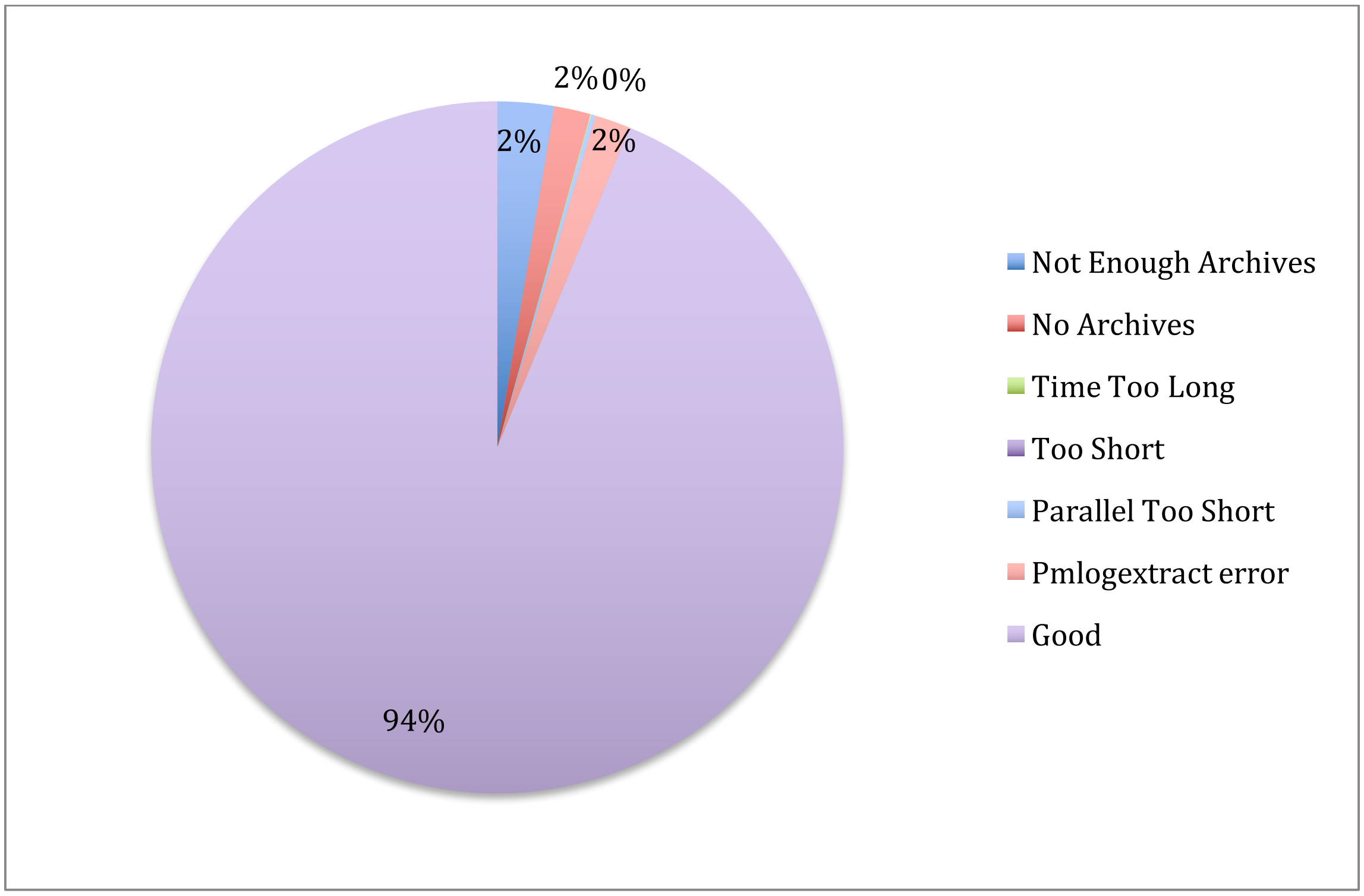

*Figure 9.0-2 Summarization coverage for jobs between 2014-03-28 and 2016-09-30 shows 94% of the jobs have valid data.*

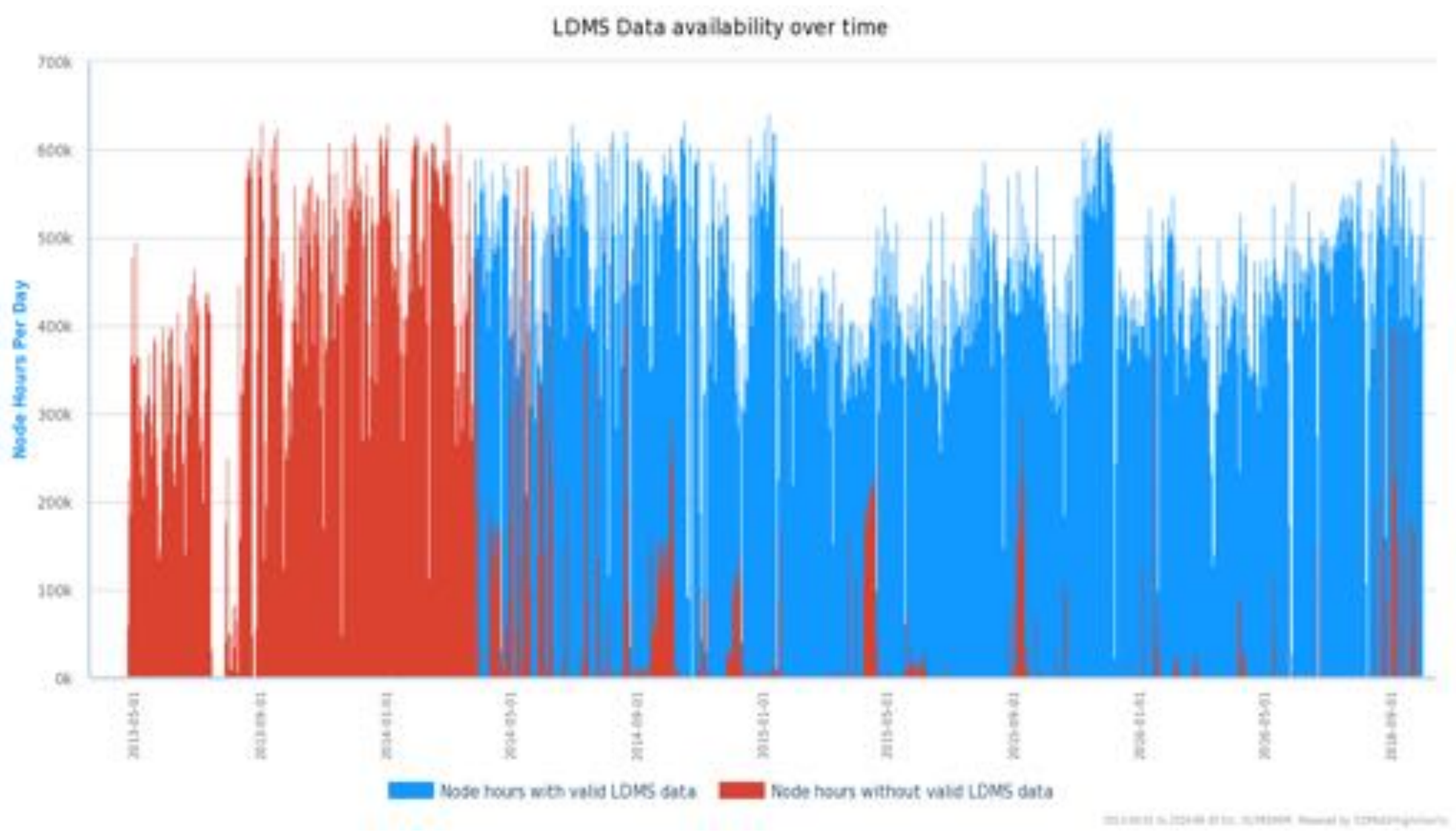

*Figure 9.0-3 Data coverage over time for LDMS data. There is no data before 2014-03-28 and there are short periods of time after this point where the data are not available for a subset of the machine*[5].

[5] The data availability charts are plotted using the information from the job-level summarized data. This means that we only include the contribution from compute nodes that were running jobs. In the case of jobs where not all of the compute nodes have LDMS data, we only generate job level data if more than 95% of the nodes have data available.

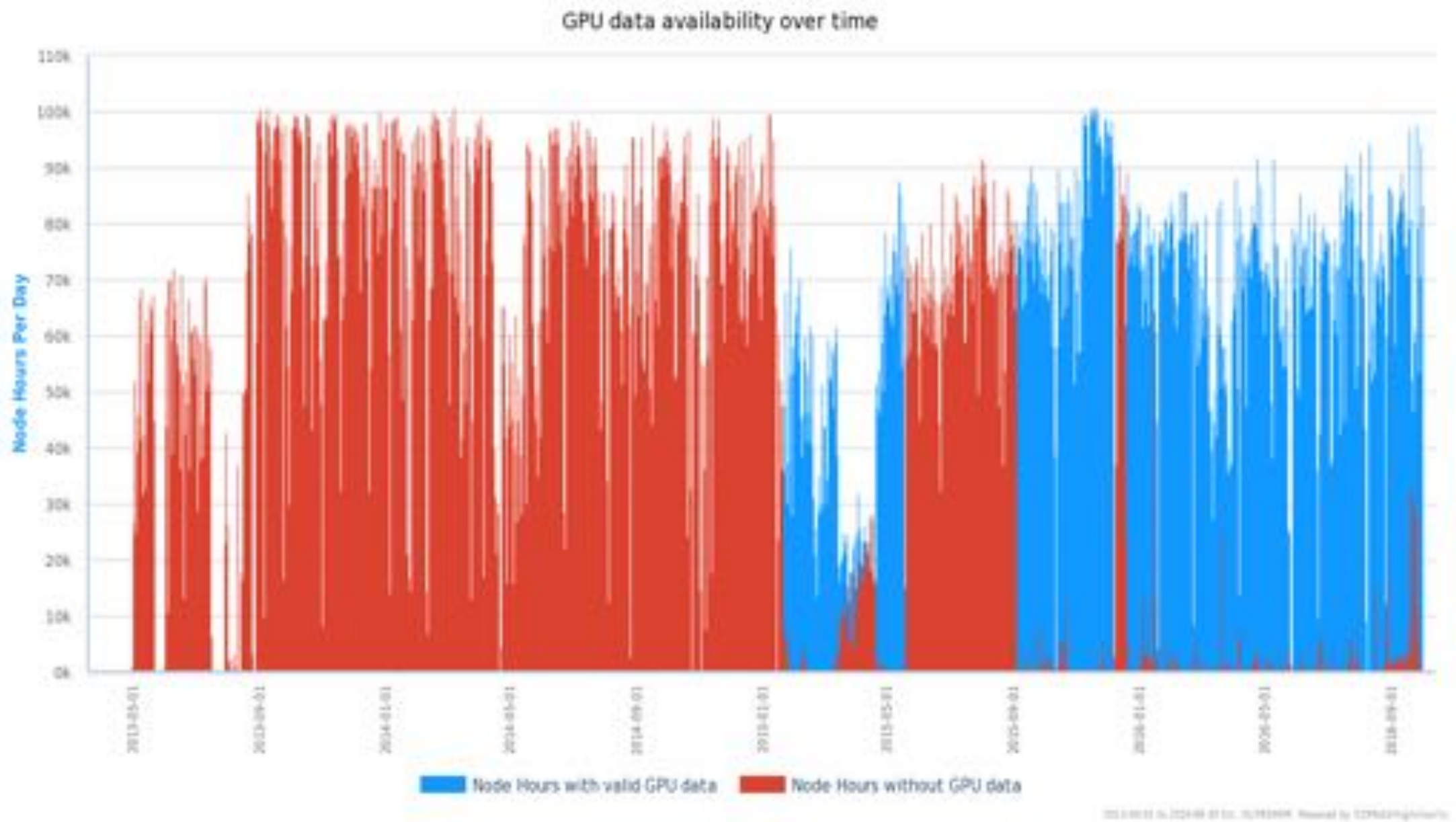

*Figure 9.0-4 Data coverage over time for GPU data. This data only includes the contributions from jobs that ran exclusively on XK nodes*[6]

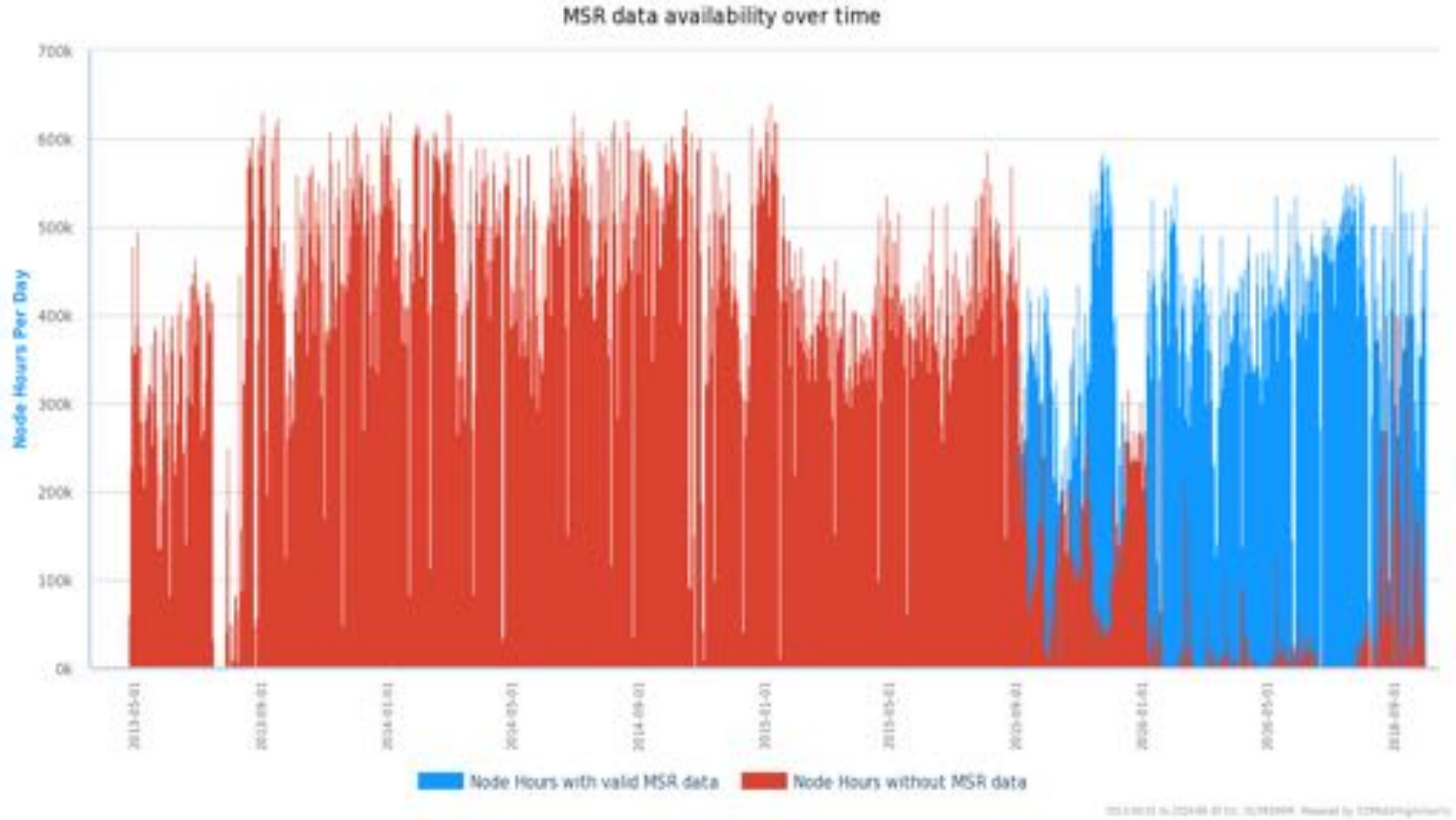

*Figure 9.0-5 Data coverage over time for MSR data.*

[6] Note also that there are a small number of jobs where we do not have the node type information available (from early deployment in 2013).

## 10.0 Appendix III: Application Maximum Memory Per Node

Tables 10.0-1 and 10.0-2 show a complete list of the maximum memory used per XE node averaged over each application and each Parent Science, respectively. Tables 10.0-3 and 10.0-4 show a complete list of the maximum memory used per XK node averaged over each application and Parent Science, respectively. In the “by Application” tables the values are generated by taking the maximum value of the memory usage per node for each job and computing the node hour weighted average for all jobs that ran the same application. The same method is used for the Parent Science tables except the jobs are grouped by parent science rather than by application. Note we show the weighted average values rather than the maximum value of all jobs because the average value is more representative of the overall usage and is less sensitive to outliers.

*Table 10.0-1 Node hour weighted average of the maximum memory usage per node analyzed by Application for XE nodes.*

| Application | Node hour weighted average maximum memory usage (GB) |
|---|---|
| **gadget** | 56.9 |
| **ls-dyna** | 51.3 |
| **cybershake** | 45.5 |
| **dsmc** | 45.4 |
| **rockstar-galaxies** | 42.4 |
| **specfem3d** | 40.7 |
| **linpack** | 40.5 |
| **pegasus** | 40.1 |
| **wsmp** | 39.4 |
| **awp-odc** | 35.3 |
| **nwchem** | 35.1 |
| **nemo** | 34.2 |
| **lsu3shell** | 34.0 |
| **rsqsim** | 32.6 |
| **enzo** | 31.3 |
| **rmgdft** | 31.1 |
| **supremm** | 29.2 |
| **psdns** | 28.2 |
| **neo-rxchf** | 28.2 |
| **hmc** | 27.2 |
| **vpic** | 27.0 |
| **distuf** | 25.8 |
| **milc** | 24.7 |
| **cactus** | 23.2 |
| **python** | 22.6 |
| **a.out** | 22.6 |
| **visit** | 22.3 |
| **3dh** | 22.3 |
| **vmd** | 22.0 |
| **SYSTEM TESTING** | 21.8 |
| **INTERACTIVE** | 21.5 |
| **waveqlab3d** | 21.5 |
| **cmaq_cctm** | 21.4 |
| **star-ccm+** | 20.5 |
| **chroma** | 20.5 |
| **changa** | 20.2 |
| **feap** | 20.1 |
| **chimera** | 20.1 |
| **gamess** | 19.7 |
| **openatom** | 19.7 |
| **uncategorized** | 19.0 |
| **fdl3di** | 18.9 |
| **msflukss** | 18.8 |
| **paraview** | 18.6 |
| **zeus-mp** | 18.5 |
| **harm3d** | 18.4 |
| **ppm** | 17.8 |
| **r** | 17.7 |
| **hpcg** | 17.1 |
| **graph500** | 16.4 |
| **vasp** | 16.4 |
| **cp2k** | 15.9 |
| **swift** | 15.8 |
| **adda** | 15.7 |
| **cesm** | 15.6 |
| **gdal** | 15.1 |
| **wrf** | 14.5 |
| **NA** | 14.1 |
| **spec** | 13.5 |
| **petsc** | 13.0 |
| **rosetta** | 12.9 |
| **hisq** | 12.8 |
| **grips** | 12.5 |
| **q-espresso** | 12.3 |
| **fluent** | 12.1 |
| **qwalk** | 11.8 |
| **castro** | 11.4 |
| **matmult** | 11.4 |
| **setsm** | 11.4 |
| **maestro** | 11.4 |
| **qmcchem** | 11.3 |
| **nek5000** | 11.2 |
| **hpcc** | 11.1 |
| **siesta** | 10.8 |
| **osiris** | 10.7 |
| **arps** | 10.4 |
| **charmm** | 10.4 |
| **citcoms** | 10.4 |
| **sord** | 10.3 |
| **namd** | 10.3 |
| **numactl** | 9.6 |
| **ramses** | 9.1 |
| **episimdemics** | 8.7 |
| **unknown (ran under a debugger)** | 8.6 |
| **cg-md** | 8.5 |
| **imb** | 8.5 |
| **mconv** | 8.3 |
| **INDUSTRY (various applications)** | 8.3 |
| **amber** | 8.0 |
| **berkeleygw** | 7.8 |
| **hoomd** | 7.6 |
| **alya** | 7.4 |
| **openfoam** | 6.6 |
| **ior** | 6.3 |
| **lammps** | 6.1 |
| **gromacs** | 5.7 |
| **orca** | 5.2 |
| **system applications** | 5.1 |
| **matlab** | 4.9 |
| **mdtest** | 4.8 |
| **cgmd** | 4.3 |
| **lattice boltzmann** | 4.2 |
| **moose** | 3.7 |
| **fd3d** | 3.7 |
| **X11 applications** | 3.6 |
| **maker** | 3.4 |
| **charm++** | 3.2 |
| **acesiii** | 3.1 |

*Table 10.0-2 Node hour weighted average of the maximum memory usage per node analyzed by Parent Science for XE nodes.*

| Parent Science | Node hour weighed average of the maximum memory usage (GB) |
|---|---|
| **Humanities/Arts** (uncategorized application) | 62.4 |
| **Humanities** | 53.6 |
| **Mechanical and Structural Systems** | 44.9 |
| **Industrial Partners Research** | 35.9 |
| **Science and Engineering Education** | 30.1 |
| **Chemical, Thermal Systems** | 27.8 |
| **Engineering** | 22.3 |
| **Physics** | 20.8 |
| **Chemistry** | 20.4 |
| **Astronomical Sciences** | 20.1 |
| **Environmental Biology** | 18.8 |
| **Integrative Biology and Neuroscience** | 18.3 |
| **Design and Manufacturing Systems** | 18.2 |
| **Computer and Computation Research** | 17.0 |
| **Biological Sciences** | 16.9 |
| **Materials Research** | 16.6 |
| **Atmospheric Sciences** | 16.3 |
| **Earth Sciences** | 13.4 |
| **Social, Behavioral, and Economic Sciences** | 13.3 |
| **Computer and Information Science and Engineering** | 12.4 |
| **Ocean Sciences** | 9.9 |
| **Advanced Scientific Computing** | 9.4 |
| **Molecular Biosciences** | 9.2 |
| **Social and Economic Science** | 7.4 |
| **Information, Robotics, and Intelligent Systems** | 4.6 |
| **Mathematical Sciences** | 4.0 |

*Table 10.0-3 Node hour weighted average of the maximum memory usage per node analyzed by Application for XK nodes*

| Application | Node hour weighted average of the maximum Memory Usage (GB) |
|---|---|
| feap | 28.1 |
| nwchem | 25.5 |
| milc | 24.7 |
| nemo | 23.0 |
| awp-odc | 21.2 |
| rmgdft | 20.7 |
| wsmp | 19.5 |
| rockstar-galaxies | 19.4 |
| cybershake | 18.8 |
| system applications | 14.6 |
| q-espresso | 14.5 |
| visit | 13.3 |
| cactus | 12.9 |
| qmcchem | 12.8 |
| openatom | 11.7 |
| changa | 10.6 |
| vmd | 10.2 |
| vasp | 9.8 |
| paraview | 9.2 |
| osiris | 8.0 |
| uncategorized | 7.8 |
| hpcg | 7.7 |
| SYSTEM TESTING | 7.5 |
| spec | 7.5 |
| fluent | 7.1 |
| caffe | 7.0 |
| linpack | 6.8 |
| namd | 6.8 |
| NA | 6.0 |
| unknown (ran under a debugger) | 5.9 |
| ramses | 5.7 |
| octopus | 5.5 |
| psdns | 5.4 |
| INTERACTIVE | 5.1 |
| python | 4.3 |
| **cesm** | 4.2 |
| **chroma** | 4.1 |
| **wrf** | 4.1 |
| **siesta** | 4.1 |
| **meld** | 4.0 |
| **hmc** | 3.9 |
| **gromacs** | 3.8 |
| **hoomd** | 3.4 |
| **lammps** | 3.3 |
| **X11 applications** | 3.2 |
| **amber** | 3.0 |
| **matmult** | 2.9 |
| **nek5000** | 2.7 |
| **a.out** | 2.5 |
| **maestro** | 2.2 |
| **petsc** | 2.1 |
| **cgmd** | 1.9 |
| **mdtest** | 1.9 |
| **matlab** | 1.7 |

*Table 10.0-4 Maximum memory usage per node analyzed by Parent Science for XK nodes*

| Parent Science | Node hour weighted average of the Maximum Memory Usage (GB) |
|---|---|
| Atmospheric Sciences | 16.1 |
| Environmental Biology | 11.1 |
| Engineering | 9.6 |
| Physics | 9.6 |
| Advanced Scientific Computing | 7.9 |
| Center Systems Staff | 7.4 |
| Information, Robotics, and Intelligent Systems | 7.0 |
| Materials Research | 6.3 |
| Molecular Biosciences | 6.2 |
| Astronomical Sciences | 5.8 |
| Chemical, Thermal Systems | 5.5 |
| Computer and Computation Research | 4.9 |
| Computer and Information Science and Engineering | 4.3 |
| Biological Sciences | 4.1 |
| Integrative Biology and Neuroscience | 3.7 |
| Earth Sciences | 3.4 |
| Mathematical Sciences | 3.1 |
| Chemistry | 3.0 |
| Science and Engineering Education | 2.8 |
| Social, Behavioral, and Economic Sciences | 2.3 |

# 11.0 Appendix IV: Application Algorithm Classification

Table 11.0-1 lists applications mapped into representative algorithms and was provided by the Blue Waters science team. We have augmented Table 11.0-1 only to include QWalk and qmcchem in the Monte Carlo category.

*Table 11.0-1. Application Algorithm Classification*

| Codes | Science Area | Structured Grids | Unstructured Grids | Dense Matrix | Sparse Matrix | N-Body | Monte Carlo | FFT | I/O |
|---|---|---|---|---|---|---|---|---|---|
| ALYA | Computational Fluie Dynamics/ Computational Mechanic | x | x | x | x | x | | | x |
| AMBER | Molecular Dynamics | X | | X | | X | | X | X |
| AWP-ODC | Earthquakes/Seismology | X | X | | | X | | | X |
| BAM | General Relativity | X | | | X | | | | |
| Cactus | General Relativity | X | | | X | | | | |
| CASTRO | Stellar Atmospheres and Supernovae | X | | | X | | X | | X |
| CESM | Climate and Weather | X | X | | X | | X | | |
| ChaNGa | Cosmology | X | | | X | X | | | |
| Chroma | Quantum Chromo Dynamics | X | | X | X | X | | X | |
| CitcomS | Earthquakes/Seismology | X | X | | | X | | | X |
| CM1 | Climate and Weather | X | X | | X | | X | | |
| CP2K | Material Science | X | | X | | X | | X | X |
| CPMD | Material Science | x | | x | | x | | x | X |
| CPS | Quantum Chromo Dynamics | X | | X | X | X | | X | |
| Dalton | Computational Chemistry | | | x | | x | | | X |
| DISTUF | Turbulence | X | | | | | | X | |
| DL_POLY | Molecular Dynamics | x | | x | | x | | | X |
| ENZO | Cosmology | X | | | X | X | | | |
| EPISIMDEMICS | Social Networks | | | | | | | | |
| Eve | Cellular Evolution | | | | | | | | |
| FDTD (EIAniso) | Atmostpheric Science | X | | X | | | | | X |
| Fluent | Computational Fluid Dynamics | x | x | x | x | | | x | x |
| GAMESS UK | Computational Chemistry | | | x | | x | | | X |
| GAMESS US | Computational Chemistry | | | X | X | X | X | | X |
| Gaussian | Computational Chemistry | | | x | x | x | | | X |
| Gromacs | Molecular Dynamics | X | | X | | X | | X | X |
| HARM3D | General Relativity | X | | | X | | | | |
| HOMME | Climate and Weather | X | X | | X | | X | | |
| LAMMPS | Molecular Dynamics | X | | X | | X | | X | X |
| LazEV | General Relativity | X | | | X | | | | |
| MAESTRO | Stellar Atmospheres and Supernovae | X | | | X | | X | | X |
| MESA | Cellular Evolution | X (1d code) | | | X | | | | |
| MILC | Quantum Chromo Dynamics | X | | X | X | X | | X | |
| MIRANDA | Combustion/Turbulence | X | | | | | | X | |
| Molpro | Computational Chemistry | | | x | x | x | | | X |
| MOOSE | Computational Mechanics | x | x | x | x | | | | x |
| NAMD | Molecular Dynamics | X | | X | | X | | X | X |
| NEK5000 | Computational Fluid Dynamics | x | x | x | x | | | x | x |
| NEMO5 | Material Science | | | X | X | X | X | | |
| NWChem | Computational Chemistry | | | X | X | X | X | | X |
| OMEN | Material Science | | | X | X | X | X | | |
| Osiris | physics | X | | | | X | | X | X |
| PGADGET | Cosmology | X | | | X | X | | | |
| PLSQR | Earthquakes/Seismology | X | X | | | X | | | X |
| PPM | Stellar Atmospheres and Supernovae | X | | | X | | X | | X |
| PSDNS | Turbulence | X | | | | | | X | |
| QMCPACK | Material Science | | | X | X | X | X | | |
| Quantum Espresso | Material Science | X | | X | | | | X | X |
| RMG | Material Science | X | | X | X | | | X | |
| SETSM | Earth Sciences | X | | | | | | | X |
| SIAL/ACES III | Computational Chemistry | | | X | X | X | X | | X |
| Siesta | Material Science | x | | x | | x | | x | X |
| SpEC | General Relativity | X | | X | | | | X | |
| SPECFEM3D | Earthquakes/Seismology | X | X | | | X | | | X |
| Spectre | General Relativity | X | | X | | | | X | |
| Turbomole | Computational Chemistry | | | x | x | x | | | X |
| VASP | Material Science | X | | X | | X | | X | X |
| VPIC | Plasmas/Magnetosphere | X | | | | X | | X | X |
| WRF | Climate and Weather | X | X | | X | | | | |

## 12.0 Appendix V: Fields of Science represented in Blue Water's Portfolio

| Field of Science of Research Carried out on Blue Waters |
|---|
| Advanced Scientific Computing |
| Astronomical Sciences |
| Atmospheric Sciences |
| Biochemistry and Molecular Structure and Function |
| Biological Oceanography |
| Biological Sciences |
| Biophysics |
| Center Systems Staff |
| Chemical, Thermal Systems |
| Chemistry |
| Climate Dynamics |
| Computational Mathematics |
| Computer and Computation Research |
| Computer and Information Science and Engineering |
| Computer Systems Architecture |
| Design and Computer-Integrated Engineering |
| Earth Sciences |
| Elementary Particle Physics |
| Engineering |
| Environmental Biology |
| Extragalactic Astronomy and Cosmology |
| Fluid, Particulate, and Hydraulic Systems |
| Galactic Astronomy |
| Genetics and Nucleic Acids |
| Geophysics |
| Humanities |
| Humanities/Arts |
| Industrial Partners Research |
| Magnetospheric Physics |
| Materials Research |
| Mathematical Sciences |
| Mechanics and Materials |
| Mechanical and Structural Systems |
| Meteorology |
| Methodology, Measurement, and Statistics |
| Molecular Biosciences |
| Neuroscience Biology |
| Nuclear Physics |
| Ocean Sciences |
| Operations Research and Production Systems |
| Performance Evaluation and Benchmarking |
| Physical Chemistry |
| Physics |
| Planetary Astronomy |
| Robotics and Machine Intelligence |
| Science and Engineering Education |
| Social and Economic Science |
| Social, Behavioral, and Economic Sciences |
| Solar Terrestrial Research |
| Stellar Astronomy and Astrophysics |
| Systematic and Population Biology |
| Unknown |
| Vendor |
| Visualization, Graphics, and Image Processing |
| Volcanology and Mantle Geochemistry |